\documentclass[prb,aps,nofootinbib,amssymb,twocolumn,superscriptaddress,10pt]{revtex4-2}

\usepackage{times}
\usepackage{graphicx}
\usepackage{float}
\usepackage{latexsym,amsmath,amssymb,bm,euscript}
\usepackage{color}
\usepackage{subfigure}
\usepackage{epstopdf}
\usepackage[colorlinks=true,linkcolor=blue,citecolor=blue,urlcolor=blue]{hyperref}
\usepackage{hyperref}
\usepackage{comment}
\usepackage{cleveref}
\usepackage{siunitx}
\usepackage{multirow}
\usepackage{svg}
\usepackage{physics}
\svgpath{{./figure/}}
\usepackage{multirow}
\usepackage{bbm}
\usepackage{tabularx}
\usepackage{ltablex}
\usepackage{placeins}
\usepackage{tikz}
\usetikzlibrary{decorations.pathreplacing}
\usepackage{ytableau}
\usepackage{dcolumn}% Align table columns on decimal point
\usepackage{bm}% bold math
\usepackage{array,booktabs,tabularx}
\usepackage[USenglish]{babel}
\usepackage{xcolor}
\usepackage{amsfonts}
\usepackage{chemformula}

\usepackage{placeins}

\usepackage{amstext} % for \text macro
\usepackage{array}   % for \newcolumntype macro
\newcolumntype{C}{>{$}c<{$}} % math-mode version of "l" column type

\makeatletter
\renewcommand\onecolumngrid{% <<<<<<
	\do@columngrid{one}{\@ne}%
	\def\set@footnotewidth{\onecolumngrid}% <<<<<<<<<<<<<<<<
	\def\footnoterule{\kern-6pt\hrule width 1.5in\kern6pt}%
}

\def\ba#1\ea{\begin{align}#1\end{align}}

{\catcode`\|=\active
	\xdef\set{\protect\expandafter\noexpand\csname set \endcsname}
	\expandafter\gdef\csname set \endcsname#1{\mathinner
		{\lbrace\,{\mathcode`\|32768\let|\midvert #1}\,\rbrace}}
	\xdef\Set{\protect\expandafter\noexpand\csname Set \endcsname}
	\expandafter\gdef\csname Set \endcsname#1{\left\{%
		\ifx\SavedDoubleVert\relax \let\SavedDoubleVert\|\fi
		\:{\let\|\SetDoubleVert
			\mathcode`\|32768\let|\SetVert
			#1}\:\right\}}
}
\def\midvert{\egroup\mid\bgroup}
\def\SetVert{\@ifnextchar|{\|\@gobble}% turn || into \|
	{\egroup\;\mid@vertical\;\bgroup}}
\def\SetDoubleVert{\egroup\;\mid@dblvertical\;\bgroup}

\begingroup
\edef\@tempa{\meaning\middle}
\edef\@tempb{\string\middle}
\expandafter \endgroup \ifx\@tempa\@tempb
\def\mid@vertical{\middle|}
\def\mid@dblvertical{\middle\SavedDoubleVert}
\else
\def\mid@vertical{\mskip1mu\vrule\mskip1mu}
\def\mid@dblvertical{\mskip1mu\vrule\mskip2.5mu\vrule\mskip1mu}
\fi

\allowdisplaybreaks

\crefname{appendix}{Appendix}{Appendices}
\crefname{equation}{Eq.}{Eqs.}
\crefname{figure}{Fig.}{Figs.}
\crefname{table}{Table}{Tables}
\crefname{section}{Section}{Sections}
\crefname{enumi}{Case}{Cases}
\creflabelformat{appendix}{[#2#1#3]}
\crefrangelabelformat{figure}{#3#1#4--(#5\crefstripprefix{#1}{#2}#6}
\crefmultiformat{figure}{Figs.~#2#1\xdef\mycreffirstarg{#1}#3}%
 { and~#2#1#3}{, #2#1#3}{ and~#2#1#3}
\makeatletter     
\renewcommand\onecolumngrid{% <<<<<<
\do@columngrid{one}{\@ne}%
\def\set@footnotewidth{\onecolumngrid}% <<<<<<<<<<<<<<<<
\def\footnoterule{\kern-6pt\hrule width 1.5in\kern6pt}%
}

\makeatletter
\AtBeginDocument{\let\LS@rot\@undefined}
\makeatother

\newcommand{\citeBCS}{\cite{zotero-4159,ELC17,VER17}}

\newcommand{\be}[0]{\begin{equation}}
\newcommand{\ee}[0]{\end{equation}}

\def\ba#1\ea{\begin{align}#1\end{align}}  
\newcommand{\up}[0]{\uparrow}
\newcommand{\dn}[0]{\downarrow}
\newcommand{\bmat}[0]{\begin{bmatrix}}
\newcommand{\emat}[0]{\end{bmatrix}}

\def\kk{\mathbf{k}}
\def\qq{\mathbf{q}}
\def\pp{\mathbf{p}}

\def\RR{\mathbf{R}}

\def\rr{\mathbf{r}}

\def\up{\uparrow}

\def\qq{\mathbf{q}}
\def\kk{\mathbf{k}}

\def\pp{\mathbf{p}}
\def\RR{\mathbf{R}}

\def\rr{\mathbf{r}}

\newcommand{\titlePaper}{
A unique topological heavy-fermion system: CeCo$_2$P$_2$ with $P\cdot\mathcal{T}$-protected Kondo effect and nodal-line excitations in the antiferromagnetic phase
}

\newcommand{\PreserveBackslash}[1]{\let\temp=\\#1\let\\=\temp}

\crefname{appendix}{Appendix}{Appendices}
\crefname{equation}{Eq.}{Eqs.}
\crefname{figure}{Fig.}{Figs.}
\crefname{table}{Table}{Tables}
\crefname{section}{Section}{Sections}
\creflabelformat{appendix}{[#2#1#3]}

\makeatletter
\AtBeginDocument{\let\LS@rot\@undefined}
\makeatother

\makeatletter
\renewcommand\onecolumngrid{% <<<<<<
\do@columngrid{one}{\@ne}%
\def\set@footnotewidth{\onecolumngrid}% <<<<<<<<<<<<<<<<
\def\footnoterule{\kern-6pt\hrule width 1.5in\kern6pt}%
}

\newcommand{\citeSI}[1]{(see \cref{#1})}

\allowdisplaybreaks
\begin{document}
\title{CeCo$_2$P$_2$: a unique Co-antiferromagnetic topological heavy-fermion system with $P\cdot\mathcal{T}$-protected Kondo effect and nodal-line excitations}

\author{Haoyu Hu}
\thanks{These authors contributed equally to this work.}
\affiliation{Donostia International Physics Center (DIPC), Paseo Manuel de Lardizábal. 20018, San Sebastián, Spain}
	
	\author{Yi Jiang}
\thanks{These authors contributed equally to this work.}
 \affiliation{Donostia International Physics Center (DIPC), Paseo Manuel de Lardizábal. 20018, San Sebastián, Spain}
 
	\author{Defa Liu}
 \affiliation{School of Physics and Astronomy, Beijing Normal University, Beijing 100875, China}
 \affiliation{Key Laboratory of Multiscale Spin Physics (Minsitry of Education), Beijing Normal University, Beijing 100875, China}

\author{Yulin Chen}
\affiliation{Department of Physics, University of Oxford, Oxford, OX1 3PU, United Kingdom}
\affiliation{School of Physical Science and Technology, ShanghaiTech University, Shanghai 201210, China}

\author{ Alexei M. Tsvelik}
\affiliation{Division of Condensed Matter Physics and Materials Science, Brookhaven National Laboratory, Upton, NY 11973-5000, USA}

\author{Yuanfeng Xu}
\affiliation{Center for Correlated Matter and School of Physics, Zhejiang University, Hangzhou 310058, China}

\author{Kristjan Haule}
\affiliation{Center for Materials Theory, Department of Physics and Astronomy, Rutgers University, Piscataway, NJ 08854, USA}

	\author{B.~Andrei Bernevig}
	\email{bernevig@princeton.edu}
	\affiliation{Department of Physics, Princeton University, Princeton, NJ 08544, USA}
 \affiliation{Donostia International Physics Center (DIPC), Paseo Manuel de Lardizábal. 20018, San Sebastián, Spain}
	\affiliation{IKERBASQUE, Basque Foundation for Science, 48013 Bilbao, Spain}

\begin{abstract}
Based on high-throughput screening \cite{xu2020high} and experimental data \cite{ccp_exp_paper}, we find that CeCo$_2$P$_2$ is unique in heavy-fermion materials: it has a Kondo effect at a high temperature which is nonetheless below a Co- antiferromagnetic ordering temperature. This begs the question: how is the Kondo singlet formed? \emph{All} other magnetic Kondo materials do not first form magnetism on the atoms whose electrons are supposed to screen the local moments.  We theoretically explain these observations and show the multifaceted uniqueness of CeCo$_2$P$_2$: a playground for Kondo, magnetism, flat band, and topological physics. At high temperatures, the itinerant Co $c$ electrons of the system form non-atomic bands with a narrow bandwidth, leading to a high antiferromagnetic transition temperature. We show that the quantum geometry of the bands promotes in-plane ferromagnetism, while the weak dispersion along the $z$ direction facilitates out-of-plane antiferromagnetism. 
At low temperatures, we uncover a novel phase that manifests the coexistence of Co-antiferromagnetism and the Kondo effect,
 linked to the $P\cdot \mathcal{T}$-protected Kramers' doublets and the filling-enforced metallic nature of $c$ electrons in the antiferromagnetic phase. 
Subsequently, the emergence of the Kondo effect, in cooperation with glide-mirror-$z$ symmetry, creates nodal-line excitation near the Fermi energy. 
Our results emphasize the importance of lattice symmetry and quantum geometry, Kondo physics, and magnetism in the understanding of the correlation physics of this unique compound. We also test our theory on the structurally similar compound  LaCo$_2$P$_2$ and show how we are able to understand its vastly different phase diagram. 
\end{abstract}
\maketitle 
\emph{Introduction.} 
The impact of symmetry and topology~\cite{bradlyn2017topological,cano2018building,elcoro2021magnetic} in weakly correlated systems has been thoroughly explored, leading to significant success in both theory and experiment. Numerous topological materials, such as topological insulators~\cite{liu_model_2010,RevModPhys.82.3045,qi2011topological,annurev:/content/journals/10.1146/annurev-conmatphys-031214-014501,doi:10.1126/science.1173034,zhang_topological_2009,benalcazar_quantized_2017,schindler_higher-order_2018, doi:10.7566/JPSJ.82.102001,Tokura2019,three_dim_top_ins,PhysRevLett.98.106803,PhysRevB.76.045302}, topological superconductors~\cite{nadj-perge_observation_2014,li_observation_2021,zhang_observation_2018,benalcazar_electric_2017,bernevig_quantum_2006,vergniory_complete_2019,schindler_higher-order_2018-1,yu_quantized_2010}, and Dirac/Weyl semimetals~\cite{soluyanov_type-ii_2015,wan_topological_2011,lv_experimental_2015,weng_weyl_2015,liu_magnetic_2019,lv_experimental_2021,armitage_weyl_2018,wang_dirac_2012,liu_discovery_2014,bradlyn_beyond_2016,liu_stable_2014,vergniory_complete_2019,huang_observation_2015,wang_three-dimensional_2013,PhysRevLett.108.140405,PhysRevLett.115.126803,PhysRevLett.113.027603,Xiong2015,PhysRevLett.116.186402,Neupane2014,Yan_top_mat,PhysRevLett.107.127205,Yang2015} have been extensively studied and their properties are now comprehensively understood. 
However, understanding the role of symmetry and topology in strongly correlated systems remains a challenging question. 
Recent advances in heavy-fermion materials have demonstrated that the interplay between lattice symmetry and correlation physics could potentially give rise to novel quantum phases, such as topological Kondo insulators~\cite{dzero_topological_2016,dzero_theory_2012,alexandrov_cubic_2013,knolle_excitons_2017,roy_surface_2014,syers_tuning_2015,xu_surface_2013,tran_phase_2012,pirie_imaging_2020,xu_direct_2014,neupane_surface_2013,nakajima_one-dimensional_2016,alexandrov_kondo_2015,kofuji_effects_2021} and Weyl Kondo semi-metals~\cite{dzsaber_giant_2021,grefe_weyl-kondo_2020,lai_weylkondo_2018,chen_topological_2022}.

% Traditionally, correlation physics is explored using simplified models. For instance, heavy-fermion systems are frequently modeled with single-orbital electrons that interact with local moments at identical lattice sites. In contrast, real materials generally possess multiple atoms and orbitals per unit cell. 
% This complexity, when combined with lattice symmetries, can give rise to non-trivial quantum geometry and topology even at the non-interacting level. 
% Upon introducing interactions, these non-trivial properties of the electronic bands could lead to novel correlation effects.

\begin{figure}
    \centering
    \includegraphics[width=0.5\textwidth]{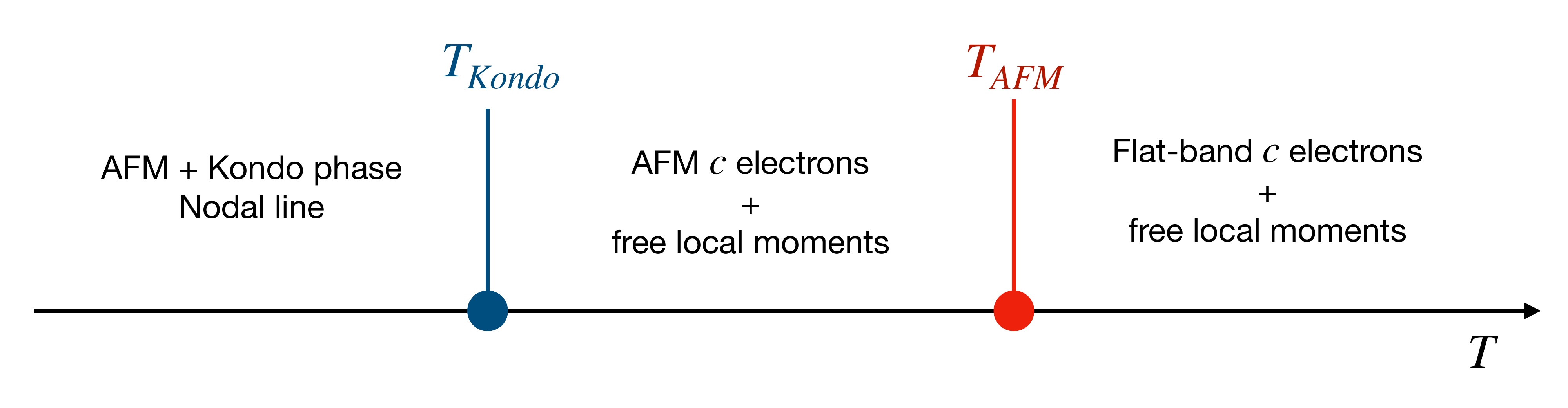}
    \caption{Theoretical and experimental phase diagram of the system. 
    % Above $T_{\text{AFM}}$, the system features paramagnetic $c$ electrons characterized by relatively flat bands and quasi-free $f$ local moments. Below $T_{\text{AFM}}$, yet above $T_{\text{Kondo}}$, $c$ electrons exhibit antiferromagnetic ordering, while $f$ local moments are still paramagnetic. Below $T_{\text{Kondo}}$, the Kondo effect emerges, where the $P\cdot\mathcal{T}$ protected Kramer doublets formed by antiferromagnetic $c$ electrons screen the $f$ local moments.
    }
    \label{fig:main:illus}
\end{figure}

Recent experiments have discovered a unique topological heavy-fermion material, CeCo$_2$P$_2$, where the interplay among quantum geometry, band topology, Kondo effect, and magnetism gives rise to novel quantum phases~\cite{ccp_exp_paper}. 
This system contains two types of electrons: the itinerant $c$ electrons, which include the $d$ orbitals of Co and Ce as well as the $p$ orbitals of P, and the correlated $f$ electrons
% , of the $f$ orbitals 
of Ce. 
The experimental and theoretical phase diagram~\cite{ccp_exp_paper} is shown in Fig.~\ref{fig:main:illus}, which highlights several unique features of the systems. 
Firstly, unlike in conventional heavy-fermion compounds where only the correlated $f$-electrons develop ordering, here $c$-electrons from Co atoms develop antiferromagnetic (AFM) order and the transition temperature ($T_{\text{AFM}} \sim 450$K) is also relatively high.
Secondly, a Kondo effect develops at $T_{\text{Kondo}} \sim 100$K inside the AFM phase, which contrasts with the conventional expectation that magnetism suppresses the Kondo effect. 
{The coexistence of the Kondo effect and magnetism has also been observed in other heavy-fermion materials, including USb$_2$\cite{PhysRevLett.123.106402}, UAs$_2$\cite{PhysRevB.106.125120}, and CeCoGe$_3$\cite{PhysRevB.107.L201104}. In all these compounds, both magnetism and the Kondo effect originate from the $f$-electron states, and the magnetic transition temperature is comparable to or lower than the Kondo temperature. The coexistence is attributed to the distinct behaviors of $f$ electrons associated with different orbitals or different lattice sites\cite{PhysRevLett.123.106402,PhysRevB.106.125120,PhysRevB.107.L201104}. In contrast, in CeCo$_2$P$_2$, the AFM order is formed by the $c$ electrons ($d$ orbitals of Co atoms), rather than the $f$ electrons, and the AFM ordering temperature is much higher than the Kondo temperature. These distinctions suggest a fundamentally different mechanism for the coexistence of Kondo screening and magnetism in CeCo$_2$P$_2$.}
Finally, a topological nodal-line excitation can appear inside the AFM Kondo phase. 
In this work, we provide a theoretical understanding of these experimental observations.
We demonstrate that the AFM order of Co $c$ electrons is driven by their narrow bands, which feature non-trivial quantum geometry. The development of the Kondo effect in the AFM phase arises from the $P \cdot \mathcal{T}$ symmetry, where $P$ represents inversion and $\mathcal{T}$ denotes time-reversal transformation, and the filling-enforced metallic nature of the $c$ electrons. 
The $P \cdot \mathcal{T}$ symmetry facilitates the formation of Kondo singlets involving $c$ electrons from different layers. Concurrently, the metallic nature of the $c$ electrons ensures a non-zero density of states, which is essential for the Kondo screening. 
Additionally, at low temperatures, the emergent Kondo excitation could form glide-mirror-$z$-symmetry-protected nodal-line excitation. 
{We also note that earlier theoretical study\cite{PhysRevB.107.195440} on this compound has primarily focused on surface effects in the AFM phase and Kondo effects in the paramagnetic phase. In contrast, our work provides new insights into the origin of antiferromagnetism, the coexistence of Kondo screening and magnetic order, and the topological properties of the system.}

% Our results provide a comprehensive understanding of how quantum geometry, topology, magnetism, and the Kondo effect drive novel quantum phases in this unique heavy-fermion compound. 
Finally, we also test our theory on a similar compound, LaCo$_2$P$_2$, and explain the large differences between the two materials. 

% Mnetic ordering at a relatively high temperature. 
% At high temperatures, $c$ electrons exhibit relatively flat bands. The non-trivial quantum geometry of these bands, in cooperation with effective Heisenberg coupling, stabilizes a type-A antiferromagnetic (AFM) order below $T_{\text{AFM}} \sim 450$K. 
% Within the antiferromagnetic phase, the Kondo effect emerges at $T_{\text{Kondo}} \sim 75$K. This phenomenon is attributed to the $P\cdot\mathcal{T}$ symmetry of the AFM phase, where $P$ represents inversion and $\mathcal{T}$ denotes time-reversal transformation. 
% The presence of $P\cdot\mathcal{T}$ symmetry indicates the formation of Kramers' doublets by two $c$ electrons from different layers. 
% These Kramers' doublets screen the local moments created by $f$ electrons, thereby inducing the Kondo effect even within the antiferromagnetic phase.
% Following the onset of the Kondo effect, additional $f$-type excitation arises, leading to the reconstruction of the band structures. This additional $f$ excitation in the Kondo phase contributes to the formation of a glide-mirror-$z$ protected nodal line. 
% In summary, the interplay among flat-band magnetism, the $P\cdot\mathcal{T}$-protected Kondo effect, and glide-mirror-$z$ symmetry stabilizes a novel quantum phase at low temperatures where magnetism, the Kondo effect, and non-trivial band topology coexist. 
% Our work not only elucidates the experimental observations in detail but also underscores the crucial role of lattice symmetry, quantum geometry, and topology in correlation physics

\emph{Model.}
The paramagnetic (PM) phase of the CeCo$_2$P$_2$ is described by space group 139 (I4/mmm). 
The Hamiltonian of the system \citeSI{app:sec:model} can be written as
$
H = H_{c} +H_{c,U} +H_f+ H_V$.
$H_c$ and $H_{c,U}$ are the kinetic term and interacting term of $c$ electrons respectively. $H_f$ describes the on-site coupling of $f$ electrons including the Hubbard repulsion, spin-orbit couplings, and crystal field splitting. $H_V$ is the hybridization between $f$ and $c$ electrons. 
We use $f_{\RR,a,i,\sigma}$ and $c_{\RR,a,i,\sigma}$ to describe the $f$ electrons and $c$ electrons respectively, where $\RR$ is the position of the unit cell, $a$ is the sublattice index, $i$ is the orbital index and $\sigma$ is the spin index. 

\begin{figure*}
    \centering    \includegraphics[width=0.98\textwidth]{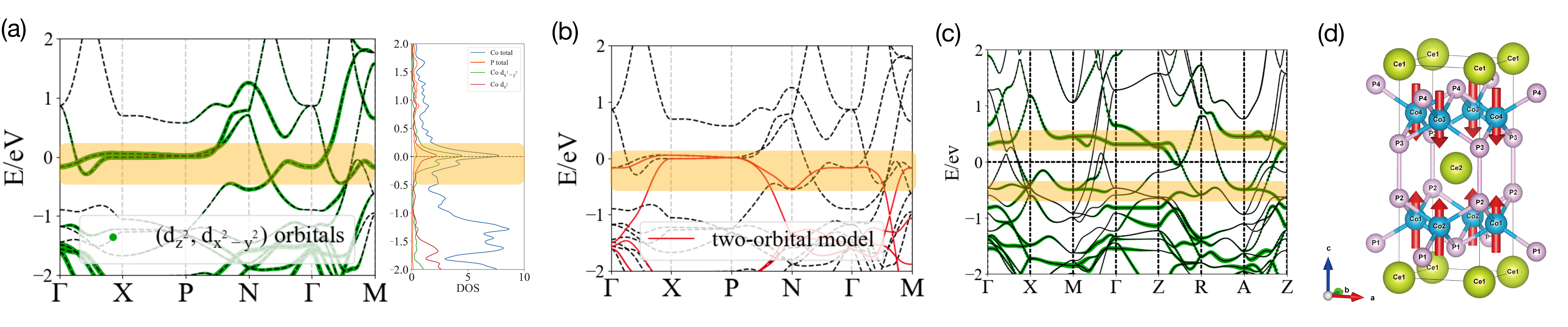}
    \caption{ (a) DFT band structures in the primitive cell and density of states of PM phase. Green dot marks the weight of Co $(d_{z^2},d_{x^2-y^2})$ orbitals. (b) Comparison between band structures of two-orbital model (red lines) and DFT model (dashed black lines). The two-orbital model successfully reproduces the narrow bands near the Fermi energy. (c) DFT band structures of AFM phase. Green dot marks the weight of Co $(d_{z^2},d_{x^2-y^2})$ orbitals. The yellow-shaded region marks the narrow bands. (d) Magnetic structure of the system. }
    \label{fig:main:flat_band_mag}
\end{figure*}

\emph{Narrow bands in the paramagnetic phase.}
In the high-temperature paramagnetic phase, due to the absence of the Kondo effect and strong Coulomb repulsion of $f$ electrons, the low-energy single-particle excitation comes from $c$ electrons. We perform \textit{ab-initio} DFT calculations in the PM phase by treating $f$ electrons as core states. In Fig.~\ref{fig:main:flat_band_mag} (a), we show the band structures and density of states (DOS) obtained from DFT calculations. 
We observe a relatively flat (narrow) band near the Fermi energy formed by $d_{x^2-y^2}$ and $d_{z^2}$ orbitals of Co atoms. 
The relatively flat bands produce an enhanced peak in the DOS. Such flat bands can be modeled by a two-orbital model in the primitive cell that only contains $d_{z^2}$ and $d_{x^2-y^2}$ orbitals of Co atoms (Fig.~\ref{fig:main:flat_band_mag} (b)). The Hamiltonian is $H_c = \sum_{\kk,\sigma}\psi_{\kk,\sigma}^\dag h_{\kk}\psi_{\kk,\sigma}$ \citeSI{app:sec:flat_band} with 
\begin{align}
% H_{c} = &\sum_{\kk,\sigma}\psi_{\kk,\sigma}^\dag h_{\kk} \psi_{\kk,\sigma} \nonumber\\ 
\psi_{\kk} =&[
    c_{\kk,Co_1,d_{z^2}} , c_{\kk,Co_2,d_{z^2}} , c_{\kk,Co_1,d_{x^2-y^2}},
     c_{\kk,Co_2,d_{x^2-y^2}}]\nonumber\\ 
h_{\kk} = & \begin{bmatrix}
    \epsilon_1 &0 \\
    0&\epsilon_2
\end{bmatrix}\tau_0 +\begin{bmatrix}
    0 & n_{\kk} \\
    n_{\kk} & g_{\kk}
\end{bmatrix} \tau_x\nonumber\\ 
g_\kk = &4t_1\cos(\frac{k_1+k_3}{2})\cos(\frac{k_2+k_3}{2})\nonumber\\ 
 n_{\kk} = &-4t_2 \sin(\frac{k_1+k_3}{2})\sin(\frac{k_2+k_3}{2}) 
 \label{eq:2orb_non_int_term}
\end{align}
where $\tau_{x,0}$ are Pauli matrices in the sublattice space.
% $\epsilon_1=-1.33$eV and $\epsilon_2=-0.8$eV are the on-site potentials, $t_1=-0.25$eV is the in-plane intra-orbital hopping of $d_{x^2-y^2}$ orbitals, and $t_2=-0.23$eV is the in-plane inter-orbital hopping between $d_{z^2}$ and $d_{x^2-y^2}$ orbitals. 
Without hopping between two orbitals ($t_2=0$), 
$d_{x^2-y^2}$ orbitals develop dispersive bands near the Fermi energy. After turning on the $t_2$, additional effective hoppings between $d_{x^2-y^2}$ orbitals will be induced by $t_2$ via second-order perturbation effect. This additional contribution reduces the bandwidth of $d_{x^2-y^2}$ band and generates a relatively band \citeSI{app:sec:flat_band}.

\begin{figure*}[htbp]
    \centering
    \includegraphics[width=1.0\textwidth]{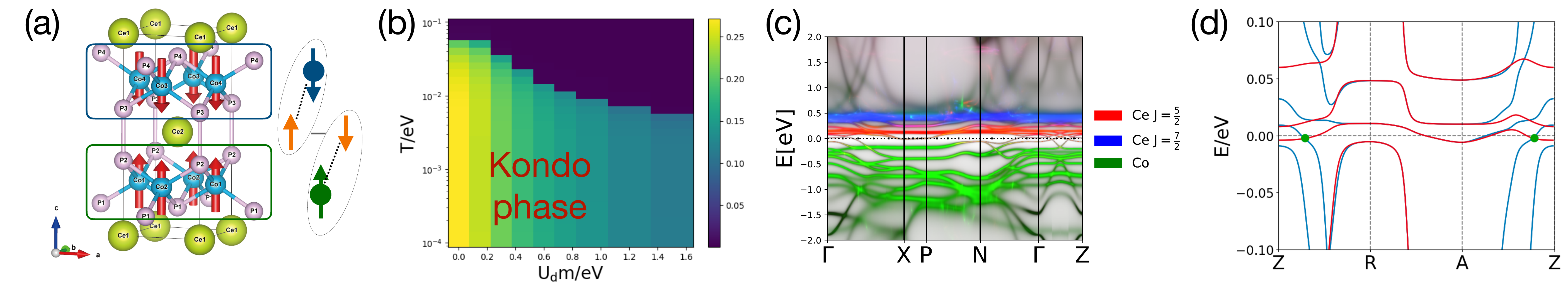}
    \caption{
    (a) Illustration of Kondo-singlet formation in the AFM phase. The spin $\up$ $c$ electrons of Co-P layer $I$ (green) and spin $\dn$ $c$ electrons of Co-P layer $II$ (blue) are degenerate and form the Kondo singlet with the $f$ local moments of Ce (orange).  
    (b) Evolution of the $fc$ hybridization strength $\chi\sim \langle f^\dag c\rangle$ at different strength of Co magnetic orders ($U_dm$) and temperatures $T$.  
   We observe the stability of the Kondo phase, even in the presence of strong magnetic ordering. We can observe the Kondo temperature tends to saturate as we increase $U_d m$.  
    % (c) Band structures of the Kondo phase from mean-field calculations in the conventional cell of the PM phase which is the primitive cell of the AFM phase.
    % % \citeSI{app:sec:kondo_band},
    % The color characterizes the orbital weights of $f$ (red) and $c$ (blue) electrons. We can observe two narrow $f$-electron bands corresponding to the $J=5/2$ and $J=7/2$ $f$-states respectively. 
    {(c) Spectrum of the Kondo phase obtained from embedded DMFT (DFT+DMFT) calculations. Different colors indicate the spectral weight contributions from various electronic components.}
    % % \citeSI{app:sec:kondo_band},}
    (d) Band structures of the Kondo phase in a smaller energy window (obtained from large-$N$ calculations). Red and blue mark the bands with opposite glide-mirror-$z$ eigenvalues. We can observe the formation of the nodal line as marked by green dots. 
    }
    \label{fig:main:kondo_band}
\end{figure*}

\emph{Flat-band magnetism.} 
The narrow band and the corresponding enhanced DOS peak indicate strong magnetic instability. 
To understand the corresponding magnetic order, we first discuss the properties of narrow bands. The narrow bands are non-atomic along $x,y$ directions which can be observed by the large hopping of $d$ orbitals along $x,y$ directions ($\sim 0.3$eV), but is atomic along $z$ directions which can be seen by the small hopping of $d$ orbitals along $z$ directions ($\sim 0.04$eV). 
We now demonstrate this type of flat band will favor a type-A AFM order. 

We first ignore the weak $z$-direction hopping of the system. Then, we have an effective 2D system with non-atomic flat bands near Fermi energy. 
For non-atomic flats band with projected Hubbard interactions, ferromagnetic state has the lowest energy~\cite{mielke_ferromagnetism_1991,tasaki_ferromagnetism_1992,lian_twisted_2021}~\citeSI{sec:app:flat_band_magnetism}.
This mechanism accounts for the in-plane ferromagnetism observed in the system. 
The weak $z$-direction hopping, via second-order perturbation theory~\cite{auerbach2012interacting}, generates an effective AFM coupling~\citeSI{sec:app:flat_band_magnetism}.
% $
% J\sum_{\langle R_z,R_z'\rangle }S_{R_z}^\mu S_{R_z'}^{\mu}
% $. 
% Here, $J \sim t_z^2/U_d$ is positive,
% where $t_z$ represents the hopping strength along the $z$-direction and $U_d$ denotes the Coulomb repulsion of Co $d$-electrons.
% $S_{R_z}$ is the sum of spin operators for the electrons in layer $R_z$. 
This effective AFM coupling combined with the in-plane ferromagnetism leads to a type-A antiferromagnetism with magnetic structure shown in Fig.~\ref{fig:main:illus} (d). The corresponding magnetic group is $P_I4/nnc$ (No. 126.386). 
We confirm our stamtne with both 
% To further demonstrate the stability of the type-A AFM phase, we perform both
mean-field and DFT calculations\citeSI{sec:app:afm_c}. 
% By comparing the energies of various magnetic configurations, we confirm that the type-A AFM phase indeed has the lowest energy. 

We therefore conclude the narrow bands near Fermi energy stabilize a type-A AFM phase. 
% which is consistent with experimental observations~\cite{ccp_exp_paper}. 
We present the band structures of the AFM phase in Fig.~\ref{fig:main:flat_band_mag} (c), where we can observe the splitting of the narrow bands. Finally, we comment that the narrow bandwidth also indicates strong instabilities due to its enhanced density of states. This can be seen from the large energy difference ($\sim 0.24$eV) between the type-A AFM phase and PM phase.
% where the energy of the type-A AFM phase is 0.24eV lower than the PM phase in the DFT calculations. 
This also naturally explains the relatively high AFM transition temperatures. 

After the development of the type-A AFM phase, we need to use the primitive cell of the AFM phase 
which is twice the size of the primitive cell of the PM phase~\citeSI{app:sec:lattice}.

\emph{Comparison with LaCo$_2$P$_2$.} 
We also investigate the magnetism of LaCo$_2$P$_2$, a material closely related to CeCo$_2$P$_2$. Experimentally, LaCo$_2$P$_2$ develops a ferromagnetic order instead of a type-A AFM order~\cite{TIAN201775}. 
The distinction in magnetic structures can be attributed to the differences in the distances between P atoms along the $z$ direction.
The P-P distance in LaCo$_2$P$_2$ is greater than that in CeCo$_2$P$_2$ \cite{TIAN201775}, leading to weaker $z$-direction hopping between the $p_z$ orbitals of P atoms. Consequently, this results in the mirror-even states formed by $p_z@$P in LaCo$_2$P$_2$ being lower in energy and appearing near the Fermi energy~\citeSI{sec:app:LCP}. 
These mirror-even states, with a Wannier center located between two Co layers, interact with Co $d$ electrons from adjacent layers. This interaction induces effective ferromagnetic Ruderman-Kittel-Kasuya-Yosida (RKKY) interactions between the Co $d$ electrons of the neighboring layers, thereby stabilizing the ferromagnetic state~\citeSI{sec:app:LCP}.

\emph{Development of Kondo effect in the antiferromagnetic phase.}
The specific magnetic structure, stabilized by the relatively narrow bands, induces a Kondo effect at low temperatures. Experimentally, a Kondo phase has been observed within the AFM phase, below $T_{\text{Kondo}}\sim 100$K~\cite{ccp_exp_paper}. The coexistence of Co magnetism and the Kondo effect contrasts with conventional expectations and can be attributed to the $P\cdot\mathcal{T}$ symmetry present in the magnetic phase and the filling-enforced metallic nature of the $c$ electrons. 

Conventionally, the magnetic order is expected to suppress the spin-flipping scattering between $f$ and $c$ electrons by polarizing the electrons. 
This spin-flipping scattering is important for the Kondo effect and is directly generated by the Kondo coupling $\sim J_K (S_f^+s_c^- + S_f^-s_c^+)$. Here, $S_f^{\pm}, s_c^{\pm}$ denote the spin ladder operator of $f$ and $c$ electrons respectively. 
Consequently, the magnetic order is expected to suppress the Kondo effect. 

However, it is important to note the specific layered structure and the $P\cdot\mathcal{T}$ symmetry of the system. 
Each Ce layer is sandwiched between one Co-P layer above (layer $I$) and one below (layer $II$), as illustrated in Fig.~\ref{fig:main:kondo_band} (a). 
Within each Co-P layer, electrons exhibit in-plane ferromagnetic ordering, while the spin moments of two adjacent layers are antiparallel. 
If only one of the two Co-P layers is considered, the Kondo effect would be suppressed due to the polarization of electrons within a single layer which also leads to a polarization of the $f$ electrons. However, due to the $P\cdot\mathcal{T}$ symmetry, the spin $\up$ electrons of layer $I$ ($c_{I,\up}$) and spin $\dn$ electrons of layer $II$ ($c_{II,\dn}$) are energetically degenerate and form Kramers' doublets. Moreover, $P\cdot\mathcal{T}$ symmetry also ensures that the magnetic order of $c$ electrons will not induce a magnetic order of $f$ electrons from a Hartree-Fock contribution. 
Then, when the Kramers' doublet states of $c$ electrons appear near the Fermi energy, a Kondo effect can then develop from the formation of the following type of spin-singlet state 
% of the total spin operators $S^\mu = \frac{1}{2}\sum_{\psi \in\{ f,c_I,c_{II}\},\sigma_1,\sigma_2} \psi_{\sigma_1}^\dag \sigma^\mu_{\sigma_1,\sigma_2}\psi_{\sigma_2}$
\begin{equation}
    \frac{1}{\sqrt{2}}(c_{I,\up}^\dag f_{\dn}^\dag - c_{II,\dn}^\dag f_{\up}^\dag )|0\rangle 
\end{equation} 
where $f_{\sigma}$ denotes the corresponding $f$ electrons. An illustration of such Kondo singlet formation has also been shown in Fig.~\ref{fig:main:kondo_band} (a).  

We now show $c$ electrons must be gapless, and the Kramers' doublets must appear at the Fermi energy. In the AFM phase with the magnetic group (126.386), the system must be metallic, due to the connectivity of the bands, if the total filling $n \bmod 4 \ne 0$~\citeSI{sec:app:filling_enforce_metal}. Since there are a total of 78 $c$ electrons and $78 \bmod 4 = 2 \ne 0$, the $c$ electrons are indeed gapless. This implies that the magnetic order cannot create a gap at the Fermi energy. This gapless nature further supports the development of the Kondo effect, as the formation of Kondo singlets requires the $c$ electrons to appear near the Fermi energy~\cite{coleman2015introduction}.

We now demonstrate the stability of the Kondo effect from the mean-field calculations~\cite{PhysRevB.35.5072,coleman2015introduction}. We consider the following Kondo lattice model
\begin{align}
H_{KL} &= H_{c}^{\text{AFM}} +H_f +H_{\text{Kondo}}.
\end{align} 
$H_c^{\text{AFM}}$ represents the kinetic term of $c$ electrons in the AFM phase. $H_f$ is the interacting Hamiltonian of $f$ electrons. $H_{\text{Kondo}}$ is the Kondo coupling terms obtained from Schrieffer-Wolff transformation~\cite{PhysRev.149.491}~\citeSI{sec:app:klm}. 

We begin with a simplified model that includes only the $d_{x^2-y^2},d_{z^2}$ orbitals of Co and the lowest Kramer-doublet states of Ce. We describe $H_{c}^{\text{AFM}}$ by adding the a mean-field term to the non-interacting Hamiltonian $H_c$ (Eq.~\ref{eq:2orb_non_int_term})
\begin{align}
&H_{c}^{\text{AFM}} = H_c + H_{c,\text{MF}},\quad \nonumber\\ 
&H_{c,\text{MF}} = mU_d\sum_{\RR,a,i,\sigma} s(\RR+\rr_a)\bigg(\sigma c_{\RR,a,i,\sigma }^\dag c_{\RR,a,i,\sigma}
\bigg) 
\end{align}
where $U_d$ is the Hubbard interactions of $d$ electrons, $m$ denotes the magnetic moments formed by these electrons, and $s(\RR+\rr_a) = \pm 1$ characterize the direction of the magnetic moment for the type-A AFM phase (\cref{fig:main:flat_band_mag} (d)). 
To explore the interplay between magnetic ordering and the Kondo effect, we treat the product of $m$ and $U_d$ as our tuning parameters and solve the Kondo Hamiltonian at the mean-field level. 
The hybridization fields $\chi \sim \langle f^\dag c\rangle $ \citeSI{sec:app:klm} are used to characterize the Kondo effect, where $|\chi| \ne 0 $ indicates the development of the Kondo effect. 
We present the phase diagram in Fig.~\ref{fig:main:kondo_band} (b). As we increase the strength of magnetic ordering ($mU_d$, with $m$ the magnetic moments and $U_d$ the Hubbard interactions of Co), the Kondo temperature initially decreases but eventually converges to a finite value. 
Thus, irrespective of the strength of the magnetic ordering, there exists a robust mechanism, the $P\cdot\mathcal{T}$-protected Kramers' doublet states, that supports the development of the Kondo effect. 
{In addition, we find that breaking $P\cdot \mathcal{T}$ symmetry suppresses the Kondo effect and can eventually destroy it when the symmetry breaking becomes sufficiently strong \citeSI{sec:app:inv_break}. Therefore, the disorder effect that breaks $P\cdot\mathcal{T}$ symmetry could also suppress the Kondo effect. 
However, given that the hybridization function in the current system is relatively large, we expect that weak disorder will not qualitatively affect the Kondo effect.
} 

To further substantiate the stability of the Kondo effect, we perform both mean-field (or large-$N$) calculations for using the complete electronic model derived from DFT calculations, which includes all orbitals ($d$ orbitals of Ce and Co and $p$ orbitals of P)~\citeSI{sec:app:klm}, {and an embedded dynamical mean-field theory (eDMFT) calculation\cite{RevModPhys.78.865,RevModPhys.68.13,PhysRevLett.115.256402,PhysRevLett.115.196403} \citeSI{sec:app:kondo_afm}.
In both calculations, we have observed a Kondo effect within the AFM phase. The interacting single-particle spectrum obtained from eDMFT is shown in Fig.~\ref{fig:main:kondo_band} (c), where we could observe $f$ electron bands (marked by red) appeared near the Fermi energy. Within eDMFT calculations, we have also observed a relatively high Kondo temperature ($>100$K), which could be attributed to the relatively large hybridization functions \citeSI{sec:app:kondo_afm}. 
}

% We indeed identify a Kondo phase characterized by a non-zero hybridization field $\chi \ne 0$ \citeSI{sec:app:kondo_afm}. 
% The single-particle spectrum of the Kondo phase has been shown in Fig.\ref{fig:main:kondo_band} (c), where we also mark the delocalized $f$ excitations by red\cite{coleman_how_2001,auerbach_kondo_1986}. 
% Furthermore, we observe two sets of $f$ excitations, corresponding to states with total angular momentum $J=\frac{5}{2}$ (near the Fermi energy) and $J=\frac{7}{2}$ (at approximately $0.3$eV). 

\emph{Nodal line in the Kondo phase.}
In our simulations, we have also observed that the correlated $f$-electron excitations form a topologically non-trivial band. 
This non-trivial band topology is illustrated in Fig.\ref{fig:main:kondo_band} (d). 
At the $k_z=1/2$ plane, the glide-mirror-$z$ symmetry protects a nodal line~\cite{fang_topological_2015,fang_topological_2016,yu_topological_2017}.

At the $k_z=1/2$ plane, each band is two-fold degenerate due to the $P\cdot\mathcal{T}$ symmetry. Furthermore, two degenerate bands share the same eigenvalue under glide-mirror-$z$ symmetry, which is either
\begin{equation}
ie^{i\pi(k_x+k_y)} \text{ or } -ie^{i\pi(k_x+k_y)}
\end{equation}
In Fig.~\ref{fig:main:kondo_band} (d), bands with different glide-mirror-$z$ eigenvalues are indicated by different colors. Here, the formation of a nodal line near the Fermi energy is evident between two bands with opposite glide-mirror-$z$ eigenvalues. 

Finally, we explore the surface states of our model by considering the system with an open boundary along the $z$-direction. At any fixed $k_x, k_y$, the system can be treated as an effective one-dimensional (1D) system, with $P\cdot\mathcal{T}$ and $\{M_Z|1/2,1/2,1/2\}$ symmetries. 
We then first analyze the surface states of the three-dimensional (3D) system by investigating those of the effective 1D system.
In such an effective 1D framework, surface states could emerge if the ground state is an obstructed atomic insulator (OAI)~\cite{xu_filling-enforced_2021,xu2021three,song2020twisted,benalcazar2017quantized,song2017d}. However, as we will demonstrate, this system only has trivial surface states.

In the effective 1D system, Co atoms occupy the only maximal Wyckoff positions ($2a$), with $R_z$ coordinates at $1/4$ and $3/4$. Meanwhile, Ce and P atoms are located at non-maximal Wyckoff positions $4b$ with $R_z = 0+z, 0-z, 1/2+z, 1/2-z$. 
The effective 1D system has two types of elementary band representations (EBRs)~\cite{bradlyn2017topological,cano2018building,elcoro2021magnetic,ELC17,VER17}, denoted by $EBR^+$ and $EBR^-$. $EBR^+$ ($EBR^-$) corresponds to placing an $s$-orbital with spin $\up$ ($\dn$) at $R_z=1/4$ and an $s$-orbital with spin $\dn$ ($\up$) at $R_z=3/4$.
By shifting the values of $k_x, k_y$ from inside the nodal line to outside, one of the filled, two-fold degenerate bands with $EBR^-$ becomes empty, and one of the empty, two-fold degenerate bands with $EBR^+$ becomes filled. 
Since Co atoms already occupy the only maximal Wyckoff positions, the effective 1D system does not form an OAI; thus, we do not expect the existence of non-trivial surface states. 
However, given that P and Ce atoms are located at non-maximal Wyckoff positions, they could potentially form molecular orbitals with Wannier centers that are distinct from the atomic positions. If the boundary of the system cuts through these molecular orbitals, trivial surface states could still emerge. 
As demonstrated in our simple model, which features only one $d$ orbital at the Co site and one $f$ orbital at the Ce site \citeSI{app:sec:surface_state}, this mechanism indeed produces surface states. 
Since these surface states are not symmetry-protected in-gap states, we can introduce an additional boundary term to merge them into the bulk spectrum\citeSI{app:sec:surface_state}.

\emph{Summary.}
In this work, we theoretically analyze the phase diagrams of a unique heavy-fermion material CeCo$_2$P$_2$. 
% We explain the relatively high AFM transition temperature, the coexistence of magnetism and Kondo effect, and topological nodal-line excitations.  
% Our findings demonstrate that lattice symmetry, band topology, and quantum geometry not only influence the single-particle properties of the system but also play a pivotal role in understanding correlation physics.
Key results include: (1) the quantum geometry and relatively narrow bands of Co $d$ orbitals stabilize a type-A AFM phase;
(2) the $P\cdot\mathcal{T}$ symmetry and the filling-enforced metallic $c$ electrons offer a robust mechanism that supports the development of the Kondo phase, even within the magnetically ordered phase;
(3) the composite $f$-electron excitations in the Kondo phase exhibit non-trivial band topology characterized by a glide-mirror-$z$-protected nodal line. 
% Our work provides a comprehensive theoretical understanding of the experimental observations in CeCo$_2$P$_2$. It also elucidates the critical roles played by the wave functions of the electronic bands and the lattice symmetry in understanding correlation physics. 

\begin{acknowledgments}
We thank Silke Paschen for the initial collaboration.
H. H. and Y. J. were supported by the European Research Council (ERC) under the European Union’s Horizon 2020 research and innovation program (Grant Agreement No. 101020833). D.L. and Y.X. were supported by the National Natural Science Foundation of China (General Program no. 12374454, 12374163) and the Fundamental Research Funds for the Central Universities (grant no. 226-2024-00200). 
B.A.B was supported by the Gordon and Betty Moore Foundation through Grant No.GBMF8685 towards the Princeton theory program, the Gordon and Betty Moore Foundation’s EPiQS Initia- tive (Grant No. GBMF11070), Office of Naval Research (ONR Grant No. N00014-20-1-2303), Global Collaborative Network Grant at Princeton University, BSF Israel US foundation No. 2018226, NSF-MERSEC (Grant No. MERSEC DMR 2011750), Simons Collaboration on New Frontiers in Superconductivity and the Schmidt Foundation at the Princeton University.
\end{acknowledgments}

\let\oldaddcontentsline\addcontentsline
\renewcommand{\addcontentsline}[3]{}
\bibliographystyle{apsrev4-2}
\bibliography{ref}

\begin{thebibliography}{127}%
\makeatletter
\providecommand \@ifxundefined [1]{%
 \@ifx{#1\undefined}
}%
\providecommand \@ifnum [1]{%
 \ifnum #1\expandafter \@firstoftwo
 \else \expandafter \@secondoftwo
 \fi
}%
\providecommand \@ifx [1]{%
 \ifx #1\expandafter \@firstoftwo
 \else \expandafter \@secondoftwo
 \fi
}%
\providecommand \natexlab [1]{#1}%
\providecommand \enquote  [1]{``#1''}%
\providecommand \bibnamefont  [1]{#1}%
\providecommand \bibfnamefont [1]{#1}%
\providecommand \citenamefont [1]{#1}%
\providecommand \href@noop [0]{\@secondoftwo}%
\providecommand \href [0]{\begingroup \@sanitize@url \@href}%
\providecommand \@href[1]{\@@startlink{#1}\@@href}%
\providecommand \@@href[1]{\endgroup#1\@@endlink}%
\providecommand \@sanitize@url [0]{\catcode `\\12\catcode `\$12\catcode
  `\&12\catcode `\#12\catcode `\^12\catcode `\_12\catcode `\%12\relax}%
\providecommand \@@startlink[1]{}%
\providecommand \@@endlink[0]{}%
\providecommand \url  [0]{\begingroup\@sanitize@url \@url }%
\providecommand \@url [1]{\endgroup\@href {#1}{\urlprefix }}%
\providecommand \urlprefix  [0]{URL }%
\providecommand \Eprint [0]{\href }%
\providecommand \doibase [0]{https://doi.org/}%
\providecommand \selectlanguage [0]{\@gobble}%
\providecommand \bibinfo  [0]{\@secondoftwo}%
\providecommand \bibfield  [0]{\@secondoftwo}%
\providecommand \translation [1]{[#1]}%
\providecommand \BibitemOpen [0]{}%
\providecommand \bibitemStop [0]{}%
\providecommand \bibitemNoStop [0]{.\EOS\space}%
\providecommand \EOS [0]{\spacefactor3000\relax}%
\providecommand \BibitemShut  [1]{\csname bibitem#1\endcsname}%
\let\auto@bib@innerbib\@empty
%</preamble>
\bibitem [{\citenamefont {Xu}\ \emph {et~al.}(2020)\citenamefont {Xu},
  \citenamefont {Elcoro}, \citenamefont {Song}, \citenamefont {Wieder},
  \citenamefont {Vergniory}, \citenamefont {Regnault}, \citenamefont {Chen},
  \citenamefont {Felser},\ and\ \citenamefont {Bernevig}}]{xu2020high}%
  \BibitemOpen
  \bibfield  {author} {\bibinfo {author} {\bibfnamefont {Y.}~\bibnamefont
  {Xu}}, \bibinfo {author} {\bibfnamefont {L.}~\bibnamefont {Elcoro}}, \bibinfo
  {author} {\bibfnamefont {Z.-D.}\ \bibnamefont {Song}}, \bibinfo {author}
  {\bibfnamefont {B.~J.}\ \bibnamefont {Wieder}}, \bibinfo {author}
  {\bibfnamefont {M.}~\bibnamefont {Vergniory}}, \bibinfo {author}
  {\bibfnamefont {N.}~\bibnamefont {Regnault}}, \bibinfo {author}
  {\bibfnamefont {Y.}~\bibnamefont {Chen}}, \bibinfo {author} {\bibfnamefont
  {C.}~\bibnamefont {Felser}},\ and\ \bibinfo {author} {\bibfnamefont {B.~A.}\
  \bibnamefont {Bernevig}},\ }\href@noop {} {\bibfield  {journal} {\bibinfo
  {journal} {Nature}\ }\textbf {\bibinfo {volume} {586}},\ \bibinfo {pages}
  {702} (\bibinfo {year} {2020})}\BibitemShut {NoStop}%
\bibitem [{\citenamefont {Liu}\ \emph {et~al.}(2024)\citenamefont {Liu},
  \citenamefont {Xu}, \citenamefont {Hu}, \citenamefont {Liu}, \citenamefont
  {Ying}, \citenamefont {Lv}, \citenamefont {Jiang}, \citenamefont {Chen},
  \citenamefont {Yang}, \citenamefont {Pei}, \citenamefont {Prabhakaran},
  \citenamefont {Gao}, \citenamefont {Wang}, \citenamefont {Zhang},
  \citenamefont {Meng}, \citenamefont {Thiagarajan}, \citenamefont {Polley},
  \citenamefont {Hashimoto}, \citenamefont {Lu}, \citenamefont {Schröter},
  \citenamefont {Strocov}, \citenamefont {Louat}, \citenamefont {Cacho},
  \citenamefont {Biswas}, \citenamefont {Lee}, \citenamefont {Steadman},
  \citenamefont {Bencok}, \citenamefont {Chen}, \citenamefont {Gu},
  \citenamefont {Hesjeda}, \citenamefont {van~der Laan}, \citenamefont
  {Hosono}, \citenamefont {Yang}, \citenamefont {Liu}, \citenamefont {Yuan},
  \citenamefont {Bernevig},\ and\ \citenamefont {Chen}}]{ccp_exp_paper}%
  \BibitemOpen
  \bibfield  {author} {\bibinfo {author} {\bibfnamefont {D.~F.}\ \bibnamefont
  {Liu}}, \bibinfo {author} {\bibfnamefont {Y.~F.}\ \bibnamefont {Xu}},
  \bibinfo {author} {\bibfnamefont {H.~Y.}\ \bibnamefont {Hu}}, \bibinfo
  {author} {\bibfnamefont {J.~Y.}\ \bibnamefont {Liu}}, \bibinfo {author}
  {\bibfnamefont {T.~P.}\ \bibnamefont {Ying}}, \bibinfo {author}
  {\bibfnamefont {Y.~Y.}\ \bibnamefont {Lv}}, \bibinfo {author} {\bibfnamefont
  {Y.}~\bibnamefont {Jiang}}, \bibinfo {author} {\bibfnamefont
  {C.}~\bibnamefont {Chen}}, \bibinfo {author} {\bibfnamefont {Y.~H.}\
  \bibnamefont {Yang}}, \bibinfo {author} {\bibfnamefont {D.}~\bibnamefont
  {Pei}}, \bibinfo {author} {\bibfnamefont {D.}~\bibnamefont {Prabhakaran}},
  \bibinfo {author} {\bibfnamefont {M.~H.}\ \bibnamefont {Gao}}, \bibinfo
  {author} {\bibfnamefont {J.~J.}\ \bibnamefont {Wang}}, \bibinfo {author}
  {\bibfnamefont {Q.~H.}\ \bibnamefont {Zhang}}, \bibinfo {author}
  {\bibfnamefont {F.~Q.}\ \bibnamefont {Meng}}, \bibinfo {author}
  {\bibfnamefont {B.}~\bibnamefont {Thiagarajan}}, \bibinfo {author}
  {\bibfnamefont {C.}~\bibnamefont {Polley}}, \bibinfo {author} {\bibfnamefont
  {M.}~\bibnamefont {Hashimoto}}, \bibinfo {author} {\bibfnamefont {D.~H.}\
  \bibnamefont {Lu}}, \bibinfo {author} {\bibfnamefont {N.~B.~M.}\ \bibnamefont
  {Schröter}}, \bibinfo {author} {\bibfnamefont {V.~N.}\ \bibnamefont
  {Strocov}}, \bibinfo {author} {\bibfnamefont {A.}~\bibnamefont {Louat}},
  \bibinfo {author} {\bibfnamefont {C.}~\bibnamefont {Cacho}}, \bibinfo
  {author} {\bibfnamefont {D.}~\bibnamefont {Biswas}}, \bibinfo {author}
  {\bibfnamefont {T.~L.}\ \bibnamefont {Lee}}, \bibinfo {author} {\bibfnamefont
  {P.}~\bibnamefont {Steadman}}, \bibinfo {author} {\bibfnamefont
  {P.}~\bibnamefont {Bencok}}, \bibinfo {author} {\bibfnamefont {Y.~B.}\
  \bibnamefont {Chen}}, \bibinfo {author} {\bibfnamefont {L.}~\bibnamefont
  {Gu}}, \bibinfo {author} {\bibfnamefont {T.}~\bibnamefont {Hesjeda}},
  \bibinfo {author} {\bibfnamefont {G.}~\bibnamefont {van~der Laan}}, \bibinfo
  {author} {\bibfnamefont {H.}~\bibnamefont {Hosono}}, \bibinfo {author}
  {\bibfnamefont {L.~X.}\ \bibnamefont {Yang}}, \bibinfo {author}
  {\bibfnamefont {Z.~K.}\ \bibnamefont {Liu}}, \bibinfo {author} {\bibfnamefont
  {H.~Q.}\ \bibnamefont {Yuan}}, \bibinfo {author} {\bibfnamefont {B.~A.}\
  \bibnamefont {Bernevig}},\ and\ \bibinfo {author} {\bibfnamefont {Y.~L.}\
  \bibnamefont {Chen}},\ }\href {https://arxiv.org/abs/2411.13898} {\bibinfo
  {title} {Discovery of an antiferromagnetic topological nodal-line kondo
  semimetal}} (\bibinfo {year} {2024}),\ \Eprint
  {https://arxiv.org/abs/2411.13898} {arXiv:2411.13898 [cond-mat.str-el]}
  \BibitemShut {NoStop}%
\bibitem [{\citenamefont {Bradlyn}\ \emph {et~al.}(2017)\citenamefont
  {Bradlyn}, \citenamefont {Elcoro}, \citenamefont {Cano}, \citenamefont
  {Vergniory}, \citenamefont {Wang}, \citenamefont {Felser}, \citenamefont
  {Aroyo},\ and\ \citenamefont {Bernevig}}]{bradlyn2017topological}%
  \BibitemOpen
  \bibfield  {author} {\bibinfo {author} {\bibfnamefont {B.}~\bibnamefont
  {Bradlyn}}, \bibinfo {author} {\bibfnamefont {L.}~\bibnamefont {Elcoro}},
  \bibinfo {author} {\bibfnamefont {J.}~\bibnamefont {Cano}}, \bibinfo {author}
  {\bibfnamefont {M.~G.}\ \bibnamefont {Vergniory}}, \bibinfo {author}
  {\bibfnamefont {Z.}~\bibnamefont {Wang}}, \bibinfo {author} {\bibfnamefont
  {C.}~\bibnamefont {Felser}}, \bibinfo {author} {\bibfnamefont {M.~I.}\
  \bibnamefont {Aroyo}},\ and\ \bibinfo {author} {\bibfnamefont {B.~A.}\
  \bibnamefont {Bernevig}},\ }\href@noop {} {\bibfield  {journal} {\bibinfo
  {journal} {Nature}\ }\textbf {\bibinfo {volume} {547}},\ \bibinfo {pages}
  {298} (\bibinfo {year} {2017})}\BibitemShut {NoStop}%
\bibitem [{\citenamefont {Cano}\ \emph {et~al.}(2018)\citenamefont {Cano},
  \citenamefont {Bradlyn}, \citenamefont {Wang}, \citenamefont {Elcoro},
  \citenamefont {Vergniory}, \citenamefont {Felser}, \citenamefont {Aroyo},\
  and\ \citenamefont {Bernevig}}]{cano2018building}%
  \BibitemOpen
  \bibfield  {author} {\bibinfo {author} {\bibfnamefont {J.}~\bibnamefont
  {Cano}}, \bibinfo {author} {\bibfnamefont {B.}~\bibnamefont {Bradlyn}},
  \bibinfo {author} {\bibfnamefont {Z.}~\bibnamefont {Wang}}, \bibinfo {author}
  {\bibfnamefont {L.}~\bibnamefont {Elcoro}}, \bibinfo {author} {\bibfnamefont
  {M.~G.}\ \bibnamefont {Vergniory}}, \bibinfo {author} {\bibfnamefont
  {C.}~\bibnamefont {Felser}}, \bibinfo {author} {\bibfnamefont {M.~I.}\
  \bibnamefont {Aroyo}},\ and\ \bibinfo {author} {\bibfnamefont {B.~A.}\
  \bibnamefont {Bernevig}},\ }\href@noop {} {\bibfield  {journal} {\bibinfo
  {journal} {Physical Review B}\ }\textbf {\bibinfo {volume} {97}},\ \bibinfo
  {pages} {035139} (\bibinfo {year} {2018})}\BibitemShut {NoStop}%
\bibitem [{\citenamefont {Elcoro}\ \emph {et~al.}(2021)\citenamefont {Elcoro},
  \citenamefont {Wieder}, \citenamefont {Song}, \citenamefont {Xu},
  \citenamefont {Bradlyn},\ and\ \citenamefont
  {Bernevig}}]{elcoro2021magnetic}%
  \BibitemOpen
  \bibfield  {author} {\bibinfo {author} {\bibfnamefont {L.}~\bibnamefont
  {Elcoro}}, \bibinfo {author} {\bibfnamefont {B.~J.}\ \bibnamefont {Wieder}},
  \bibinfo {author} {\bibfnamefont {Z.}~\bibnamefont {Song}}, \bibinfo {author}
  {\bibfnamefont {Y.}~\bibnamefont {Xu}}, \bibinfo {author} {\bibfnamefont
  {B.}~\bibnamefont {Bradlyn}},\ and\ \bibinfo {author} {\bibfnamefont {B.~A.}\
  \bibnamefont {Bernevig}},\ }\href@noop {} {\bibfield  {journal} {\bibinfo
  {journal} {Nature communications}\ }\textbf {\bibinfo {volume} {12}},\
  \bibinfo {pages} {5965} (\bibinfo {year} {2021})}\BibitemShut {NoStop}%
\bibitem [{\citenamefont {Liu}\ \emph {et~al.}(2010)\citenamefont {Liu},
  \citenamefont {Qi}, \citenamefont {Zhang}, \citenamefont {Dai}, \citenamefont
  {Fang},\ and\ \citenamefont {Zhang}}]{liu_model_2010}%
  \BibitemOpen
  \bibfield  {author} {\bibinfo {author} {\bibfnamefont {C.-X.}\ \bibnamefont
  {Liu}}, \bibinfo {author} {\bibfnamefont {X.-L.}\ \bibnamefont {Qi}},
  \bibinfo {author} {\bibfnamefont {H.}~\bibnamefont {Zhang}}, \bibinfo
  {author} {\bibfnamefont {X.}~\bibnamefont {Dai}}, \bibinfo {author}
  {\bibfnamefont {Z.}~\bibnamefont {Fang}},\ and\ \bibinfo {author}
  {\bibfnamefont {S.-C.}\ \bibnamefont {Zhang}},\ }\href
  {https://doi.org/10.1103/PhysRevB.82.045122} {\bibfield  {journal} {\bibinfo
  {journal} {Physical Review B}\ }\textbf {\bibinfo {volume} {82}},\ \bibinfo
  {pages} {045122} (\bibinfo {year} {2010})},\ \bibinfo {note} {publisher:
  American Physical Society}\BibitemShut {NoStop}%
\bibitem [{\citenamefont {Hasan}\ and\ \citenamefont
  {Kane}(2010)}]{RevModPhys.82.3045}%
  \BibitemOpen
  \bibfield  {author} {\bibinfo {author} {\bibfnamefont {M.~Z.}\ \bibnamefont
  {Hasan}}\ and\ \bibinfo {author} {\bibfnamefont {C.~L.}\ \bibnamefont
  {Kane}},\ }\href {https://doi.org/10.1103/RevModPhys.82.3045} {\bibfield
  {journal} {\bibinfo  {journal} {Rev. Mod. Phys.}\ }\textbf {\bibinfo {volume}
  {82}},\ \bibinfo {pages} {3045} (\bibinfo {year} {2010})}\BibitemShut
  {NoStop}%
\bibitem [{\citenamefont {Qi}\ and\ \citenamefont
  {Zhang}(2011)}]{qi2011topological}%
  \BibitemOpen
  \bibfield  {author} {\bibinfo {author} {\bibfnamefont {X.-L.}\ \bibnamefont
  {Qi}}\ and\ \bibinfo {author} {\bibfnamefont {S.-C.}\ \bibnamefont {Zhang}},\
  }\href@noop {} {\bibfield  {journal} {\bibinfo  {journal} {Reviews of Modern
  Physics}\ }\textbf {\bibinfo {volume} {83}},\ \bibinfo {pages} {1057}
  (\bibinfo {year} {2011})}\BibitemShut {NoStop}%
\bibitem [{\citenamefont {Ando}\ and\ \citenamefont
  {Fu}(2015)}]{annurev:/content/journals/10.1146/annurev-conmatphys-031214-014501}%
  \BibitemOpen
  \bibfield  {author} {\bibinfo {author} {\bibfnamefont {Y.}~\bibnamefont
  {Ando}}\ and\ \bibinfo {author} {\bibfnamefont {L.}~\bibnamefont {Fu}},\
  }\href
  {https://doi.org/https://doi.org/10.1146/annurev-conmatphys-031214-014501}
  {\bibfield  {journal} {\bibinfo  {journal} {Annual Review of Condensed Matter
  Physics}\ }\textbf {\bibinfo {volume} {6}},\ \bibinfo {pages} {361} (\bibinfo
  {year} {2015})}\BibitemShut {NoStop}%
\bibitem [{\citenamefont {Chen}\ \emph {et~al.}(2009)\citenamefont {Chen},
  \citenamefont {Analytis}, \citenamefont {Chu}, \citenamefont {Liu},
  \citenamefont {Mo}, \citenamefont {Qi}, \citenamefont {Zhang}, \citenamefont
  {Lu}, \citenamefont {Dai}, \citenamefont {Fang}, \citenamefont {Zhang},
  \citenamefont {Fisher}, \citenamefont {Hussain},\ and\ \citenamefont
  {Shen}}]{doi:10.1126/science.1173034}%
  \BibitemOpen
  \bibfield  {author} {\bibinfo {author} {\bibfnamefont {Y.~L.}\ \bibnamefont
  {Chen}}, \bibinfo {author} {\bibfnamefont {J.~G.}\ \bibnamefont {Analytis}},
  \bibinfo {author} {\bibfnamefont {J.-H.}\ \bibnamefont {Chu}}, \bibinfo
  {author} {\bibfnamefont {Z.~K.}\ \bibnamefont {Liu}}, \bibinfo {author}
  {\bibfnamefont {S.-K.}\ \bibnamefont {Mo}}, \bibinfo {author} {\bibfnamefont
  {X.~L.}\ \bibnamefont {Qi}}, \bibinfo {author} {\bibfnamefont {H.~J.}\
  \bibnamefont {Zhang}}, \bibinfo {author} {\bibfnamefont {D.~H.}\ \bibnamefont
  {Lu}}, \bibinfo {author} {\bibfnamefont {X.}~\bibnamefont {Dai}}, \bibinfo
  {author} {\bibfnamefont {Z.}~\bibnamefont {Fang}}, \bibinfo {author}
  {\bibfnamefont {S.~C.}\ \bibnamefont {Zhang}}, \bibinfo {author}
  {\bibfnamefont {I.~R.}\ \bibnamefont {Fisher}}, \bibinfo {author}
  {\bibfnamefont {Z.}~\bibnamefont {Hussain}},\ and\ \bibinfo {author}
  {\bibfnamefont {Z.-X.}\ \bibnamefont {Shen}},\ }\href
  {https://doi.org/10.1126/science.1173034} {\bibfield  {journal} {\bibinfo
  {journal} {Science}\ }\textbf {\bibinfo {volume} {325}},\ \bibinfo {pages}
  {178} (\bibinfo {year} {2009})},\ \Eprint
  {https://arxiv.org/abs/https://www.science.org/doi/pdf/10.1126/science.1173034}
  {https://www.science.org/doi/pdf/10.1126/science.1173034} \BibitemShut
  {NoStop}%
\bibitem [{\citenamefont {Zhang}\ \emph {et~al.}(2009)\citenamefont {Zhang},
  \citenamefont {Liu}, \citenamefont {Qi}, \citenamefont {Dai}, \citenamefont
  {Fang},\ and\ \citenamefont {Zhang}}]{zhang_topological_2009}%
  \BibitemOpen
  \bibfield  {author} {\bibinfo {author} {\bibfnamefont {H.}~\bibnamefont
  {Zhang}}, \bibinfo {author} {\bibfnamefont {C.-X.}\ \bibnamefont {Liu}},
  \bibinfo {author} {\bibfnamefont {X.-L.}\ \bibnamefont {Qi}}, \bibinfo
  {author} {\bibfnamefont {X.}~\bibnamefont {Dai}}, \bibinfo {author}
  {\bibfnamefont {Z.}~\bibnamefont {Fang}},\ and\ \bibinfo {author}
  {\bibfnamefont {S.-C.}\ \bibnamefont {Zhang}},\ }\href
  {https://doi.org/10.1038/nphys1270} {\bibfield  {journal} {\bibinfo
  {journal} {Nature Physics}\ }\textbf {\bibinfo {volume} {5}},\ \bibinfo
  {pages} {438} (\bibinfo {year} {2009})},\ \bibinfo {note} {publisher: Nature
  Publishing Group}\BibitemShut {NoStop}%
\bibitem [{\citenamefont {Benalcazar}\ \emph
  {et~al.}(2017{\natexlab{a}})\citenamefont {Benalcazar}, \citenamefont
  {Bernevig},\ and\ \citenamefont {Hughes}}]{benalcazar_quantized_2017}%
  \BibitemOpen
  \bibfield  {author} {\bibinfo {author} {\bibfnamefont {W.~A.}\ \bibnamefont
  {Benalcazar}}, \bibinfo {author} {\bibfnamefont {B.~A.}\ \bibnamefont
  {Bernevig}},\ and\ \bibinfo {author} {\bibfnamefont {T.~L.}\ \bibnamefont
  {Hughes}},\ }\href {https://doi.org/10.1126/science.aah6442} {\bibfield
  {journal} {\bibinfo  {journal} {Science}\ }\textbf {\bibinfo {volume}
  {357}},\ \bibinfo {pages} {61} (\bibinfo {year} {2017}{\natexlab{a}})},\
  \bibinfo {note} {publisher: American Association for the Advancement of
  Science}\BibitemShut {NoStop}%
\bibitem [{\citenamefont {Schindler}\ \emph
  {et~al.}(2018{\natexlab{a}})\citenamefont {Schindler}, \citenamefont {Cook},
  \citenamefont {Vergniory}, \citenamefont {Wang}, \citenamefont {Parkin},
  \citenamefont {Bernevig},\ and\ \citenamefont
  {Neupert}}]{schindler_higher-order_2018}%
  \BibitemOpen
  \bibfield  {author} {\bibinfo {author} {\bibfnamefont {F.}~\bibnamefont
  {Schindler}}, \bibinfo {author} {\bibfnamefont {A.~M.}\ \bibnamefont {Cook}},
  \bibinfo {author} {\bibfnamefont {M.~G.}\ \bibnamefont {Vergniory}}, \bibinfo
  {author} {\bibfnamefont {Z.}~\bibnamefont {Wang}}, \bibinfo {author}
  {\bibfnamefont {S.~S.~P.}\ \bibnamefont {Parkin}}, \bibinfo {author}
  {\bibfnamefont {B.~A.}\ \bibnamefont {Bernevig}},\ and\ \bibinfo {author}
  {\bibfnamefont {T.}~\bibnamefont {Neupert}},\ }\href
  {https://doi.org/10.1126/sciadv.aat0346} {\bibfield  {journal} {\bibinfo
  {journal} {Science Advances}\ }\textbf {\bibinfo {volume} {4}},\ \bibinfo
  {pages} {eaat0346} (\bibinfo {year} {2018}{\natexlab{a}})},\ \bibinfo {note}
  {publisher: American Association for the Advancement of Science}\BibitemShut
  {NoStop}%
\bibitem [{\citenamefont {Ando}(2013)}]{doi:10.7566/JPSJ.82.102001}%
  \BibitemOpen
  \bibfield  {author} {\bibinfo {author} {\bibfnamefont {Y.}~\bibnamefont
  {Ando}},\ }\href {https://doi.org/10.7566/JPSJ.82.102001} {\bibfield
  {journal} {\bibinfo  {journal} {Journal of the Physical Society of Japan}\
  }\textbf {\bibinfo {volume} {82}},\ \bibinfo {pages} {102001} (\bibinfo
  {year} {2013})},\ \Eprint
  {https://arxiv.org/abs/https://doi.org/10.7566/JPSJ.82.102001}
  {https://doi.org/10.7566/JPSJ.82.102001} \BibitemShut {NoStop}%
\bibitem [{\citenamefont {Tokura}\ \emph {et~al.}(2019)\citenamefont {Tokura},
  \citenamefont {Yasuda},\ and\ \citenamefont {Tsukazaki}}]{Tokura2019}%
  \BibitemOpen
  \bibfield  {author} {\bibinfo {author} {\bibfnamefont {Y.}~\bibnamefont
  {Tokura}}, \bibinfo {author} {\bibfnamefont {K.}~\bibnamefont {Yasuda}},\
  and\ \bibinfo {author} {\bibfnamefont {A.}~\bibnamefont {Tsukazaki}},\ }\href
  {https://doi.org/10.1038/s42254-018-0011-5} {\bibfield  {journal} {\bibinfo
  {journal} {Nature Reviews Physics}\ }\textbf {\bibinfo {volume} {1}},\
  \bibinfo {pages} {126} (\bibinfo {year} {2019})}\BibitemShut {NoStop}%
\bibitem [{\citenamefont {Hasan}\ and\ \citenamefont
  {Moore}(2011)}]{three_dim_top_ins}%
  \BibitemOpen
  \bibfield  {author} {\bibinfo {author} {\bibfnamefont {M.~Z.}\ \bibnamefont
  {Hasan}}\ and\ \bibinfo {author} {\bibfnamefont {J.~E.}\ \bibnamefont
  {Moore}},\ }\href
  {https://doi.org/https://doi.org/10.1146/annurev-conmatphys-062910-140432}
  {\bibfield  {journal} {\bibinfo  {journal} {Annual Review of Condensed Matter
  Physics}\ }\textbf {\bibinfo {volume} {2}},\ \bibinfo {pages} {55} (\bibinfo
  {year} {2011})}\BibitemShut {NoStop}%
\bibitem [{\citenamefont {Fu}\ \emph {et~al.}(2007)\citenamefont {Fu},
  \citenamefont {Kane},\ and\ \citenamefont {Mele}}]{PhysRevLett.98.106803}%
  \BibitemOpen
  \bibfield  {author} {\bibinfo {author} {\bibfnamefont {L.}~\bibnamefont
  {Fu}}, \bibinfo {author} {\bibfnamefont {C.~L.}\ \bibnamefont {Kane}},\ and\
  \bibinfo {author} {\bibfnamefont {E.~J.}\ \bibnamefont {Mele}},\ }\href
  {https://doi.org/10.1103/PhysRevLett.98.106803} {\bibfield  {journal}
  {\bibinfo  {journal} {Phys. Rev. Lett.}\ }\textbf {\bibinfo {volume} {98}},\
  \bibinfo {pages} {106803} (\bibinfo {year} {2007})}\BibitemShut {NoStop}%
\bibitem [{\citenamefont {Fu}\ and\ \citenamefont
  {Kane}(2007)}]{PhysRevB.76.045302}%
  \BibitemOpen
  \bibfield  {author} {\bibinfo {author} {\bibfnamefont {L.}~\bibnamefont
  {Fu}}\ and\ \bibinfo {author} {\bibfnamefont {C.~L.}\ \bibnamefont {Kane}},\
  }\href {https://doi.org/10.1103/PhysRevB.76.045302} {\bibfield  {journal}
  {\bibinfo  {journal} {Phys. Rev. B}\ }\textbf {\bibinfo {volume} {76}},\
  \bibinfo {pages} {045302} (\bibinfo {year} {2007})}\BibitemShut {NoStop}%
\bibitem [{\citenamefont {Nadj-Perge}\ \emph {et~al.}(2014)\citenamefont
  {Nadj-Perge}, \citenamefont {Drozdov}, \citenamefont {Li}, \citenamefont
  {Chen}, \citenamefont {Jeon}, \citenamefont {Seo}, \citenamefont {MacDonald},
  \citenamefont {Bernevig},\ and\ \citenamefont
  {Yazdani}}]{nadj-perge_observation_2014}%
  \BibitemOpen
  \bibfield  {author} {\bibinfo {author} {\bibfnamefont {S.}~\bibnamefont
  {Nadj-Perge}}, \bibinfo {author} {\bibfnamefont {I.~K.}\ \bibnamefont
  {Drozdov}}, \bibinfo {author} {\bibfnamefont {J.}~\bibnamefont {Li}},
  \bibinfo {author} {\bibfnamefont {H.}~\bibnamefont {Chen}}, \bibinfo {author}
  {\bibfnamefont {S.}~\bibnamefont {Jeon}}, \bibinfo {author} {\bibfnamefont
  {J.}~\bibnamefont {Seo}}, \bibinfo {author} {\bibfnamefont {A.~H.}\
  \bibnamefont {MacDonald}}, \bibinfo {author} {\bibfnamefont {B.~A.}\
  \bibnamefont {Bernevig}},\ and\ \bibinfo {author} {\bibfnamefont
  {A.}~\bibnamefont {Yazdani}},\ }\href
  {https://doi.org/10.1126/science.1259327} {\bibfield  {journal} {\bibinfo
  {journal} {Science}\ }\textbf {\bibinfo {volume} {346}},\ \bibinfo {pages}
  {602} (\bibinfo {year} {2014})},\ \bibinfo {note} {publisher: American
  Association for the Advancement of Science}\BibitemShut {NoStop}%
\bibitem [{\citenamefont {Li}\ \emph {et~al.}(2021)\citenamefont {Li},
  \citenamefont {Zheng}, \citenamefont {Fang}, \citenamefont {Zhang},
  \citenamefont {Chen}, \citenamefont {Chen}, \citenamefont {Liang},
  \citenamefont {Shi}, \citenamefont {Pei}, \citenamefont {Xu}, \citenamefont
  {Liu}, \citenamefont {Pan}, \citenamefont {Lu}, \citenamefont {Hashimoto},
  \citenamefont {Barinov}, \citenamefont {Jung}, \citenamefont {Cacho},
  \citenamefont {Wang}, \citenamefont {He}, \citenamefont {Fu}, \citenamefont
  {Zhang}, \citenamefont {Huang}, \citenamefont {Yang}, \citenamefont {Liu},\
  and\ \citenamefont {Chen}}]{li_observation_2021}%
  \BibitemOpen
  \bibfield  {author} {\bibinfo {author} {\bibfnamefont {Y.~W.}\ \bibnamefont
  {Li}}, \bibinfo {author} {\bibfnamefont {H.~J.}\ \bibnamefont {Zheng}},
  \bibinfo {author} {\bibfnamefont {Y.~Q.}\ \bibnamefont {Fang}}, \bibinfo
  {author} {\bibfnamefont {D.~Q.}\ \bibnamefont {Zhang}}, \bibinfo {author}
  {\bibfnamefont {Y.~J.}\ \bibnamefont {Chen}}, \bibinfo {author}
  {\bibfnamefont {C.}~\bibnamefont {Chen}}, \bibinfo {author} {\bibfnamefont
  {A.~J.}\ \bibnamefont {Liang}}, \bibinfo {author} {\bibfnamefont {W.~J.}\
  \bibnamefont {Shi}}, \bibinfo {author} {\bibfnamefont {D.}~\bibnamefont
  {Pei}}, \bibinfo {author} {\bibfnamefont {L.~X.}\ \bibnamefont {Xu}},
  \bibinfo {author} {\bibfnamefont {S.}~\bibnamefont {Liu}}, \bibinfo {author}
  {\bibfnamefont {J.}~\bibnamefont {Pan}}, \bibinfo {author} {\bibfnamefont
  {D.~H.}\ \bibnamefont {Lu}}, \bibinfo {author} {\bibfnamefont
  {M.}~\bibnamefont {Hashimoto}}, \bibinfo {author} {\bibfnamefont
  {A.}~\bibnamefont {Barinov}}, \bibinfo {author} {\bibfnamefont {S.~W.}\
  \bibnamefont {Jung}}, \bibinfo {author} {\bibfnamefont {C.}~\bibnamefont
  {Cacho}}, \bibinfo {author} {\bibfnamefont {M.~X.}\ \bibnamefont {Wang}},
  \bibinfo {author} {\bibfnamefont {Y.}~\bibnamefont {He}}, \bibinfo {author}
  {\bibfnamefont {L.}~\bibnamefont {Fu}}, \bibinfo {author} {\bibfnamefont
  {H.~J.}\ \bibnamefont {Zhang}}, \bibinfo {author} {\bibfnamefont {F.~Q.}\
  \bibnamefont {Huang}}, \bibinfo {author} {\bibfnamefont {L.~X.}\ \bibnamefont
  {Yang}}, \bibinfo {author} {\bibfnamefont {Z.~K.}\ \bibnamefont {Liu}},\ and\
  \bibinfo {author} {\bibfnamefont {Y.~L.}\ \bibnamefont {Chen}},\ }\href
  {https://doi.org/10.1038/s41467-021-23076-1} {\bibfield  {journal} {\bibinfo
  {journal} {Nature Communications}\ }\textbf {\bibinfo {volume} {12}},\
  \bibinfo {pages} {2874} (\bibinfo {year} {2021})},\ \bibinfo {note}
  {publisher: Nature Publishing Group}\BibitemShut {NoStop}%
\bibitem [{\citenamefont {Zhang}\ \emph {et~al.}(2018)\citenamefont {Zhang},
  \citenamefont {Yaji}, \citenamefont {Hashimoto}, \citenamefont {Ota},
  \citenamefont {Kondo}, \citenamefont {Okazaki}, \citenamefont {Wang},
  \citenamefont {Wen}, \citenamefont {Gu}, \citenamefont {Ding},\ and\
  \citenamefont {Shin}}]{zhang_observation_2018}%
  \BibitemOpen
  \bibfield  {author} {\bibinfo {author} {\bibfnamefont {P.}~\bibnamefont
  {Zhang}}, \bibinfo {author} {\bibfnamefont {K.}~\bibnamefont {Yaji}},
  \bibinfo {author} {\bibfnamefont {T.}~\bibnamefont {Hashimoto}}, \bibinfo
  {author} {\bibfnamefont {Y.}~\bibnamefont {Ota}}, \bibinfo {author}
  {\bibfnamefont {T.}~\bibnamefont {Kondo}}, \bibinfo {author} {\bibfnamefont
  {K.}~\bibnamefont {Okazaki}}, \bibinfo {author} {\bibfnamefont
  {Z.}~\bibnamefont {Wang}}, \bibinfo {author} {\bibfnamefont {J.}~\bibnamefont
  {Wen}}, \bibinfo {author} {\bibfnamefont {G.~D.}\ \bibnamefont {Gu}},
  \bibinfo {author} {\bibfnamefont {H.}~\bibnamefont {Ding}},\ and\ \bibinfo
  {author} {\bibfnamefont {S.}~\bibnamefont {Shin}},\ }\href
  {https://doi.org/10.1126/science.aan4596} {\bibfield  {journal} {\bibinfo
  {journal} {Science}\ }\textbf {\bibinfo {volume} {360}},\ \bibinfo {pages}
  {182} (\bibinfo {year} {2018})},\ \bibinfo {note} {publisher: American
  Association for the Advancement of Science}\BibitemShut {NoStop}%
\bibitem [{\citenamefont {Benalcazar}\ \emph
  {et~al.}(2017{\natexlab{b}})\citenamefont {Benalcazar}, \citenamefont
  {Bernevig},\ and\ \citenamefont {Hughes}}]{benalcazar_electric_2017}%
  \BibitemOpen
  \bibfield  {author} {\bibinfo {author} {\bibfnamefont {W.~A.}\ \bibnamefont
  {Benalcazar}}, \bibinfo {author} {\bibfnamefont {B.~A.}\ \bibnamefont
  {Bernevig}},\ and\ \bibinfo {author} {\bibfnamefont {T.~L.}\ \bibnamefont
  {Hughes}},\ }\href {https://doi.org/10.1103/PhysRevB.96.245115} {\bibfield
  {journal} {\bibinfo  {journal} {Physical Review B}\ }\textbf {\bibinfo
  {volume} {96}},\ \bibinfo {pages} {245115} (\bibinfo {year}
  {2017}{\natexlab{b}})},\ \bibinfo {note} {publisher: American Physical
  Society}\BibitemShut {NoStop}%
\bibitem [{\citenamefont {Bernevig}\ and\ \citenamefont
  {Zhang}(2006)}]{bernevig_quantum_2006}%
  \BibitemOpen
  \bibfield  {author} {\bibinfo {author} {\bibfnamefont {B.~A.}\ \bibnamefont
  {Bernevig}}\ and\ \bibinfo {author} {\bibfnamefont {S.-C.}\ \bibnamefont
  {Zhang}},\ }\href {https://doi.org/10.1103/PhysRevLett.96.106802} {\bibfield
  {journal} {\bibinfo  {journal} {Physical Review Letters}\ }\textbf {\bibinfo
  {volume} {96}},\ \bibinfo {pages} {106802} (\bibinfo {year} {2006})},\
  \bibinfo {note} {publisher: American Physical Society}\BibitemShut {NoStop}%
\bibitem [{\citenamefont {Vergniory}\ \emph {et~al.}(2019)\citenamefont
  {Vergniory}, \citenamefont {Elcoro}, \citenamefont {Felser}, \citenamefont
  {Regnault}, \citenamefont {Bernevig},\ and\ \citenamefont
  {Wang}}]{vergniory_complete_2019}%
  \BibitemOpen
  \bibfield  {author} {\bibinfo {author} {\bibfnamefont {M.~G.}\ \bibnamefont
  {Vergniory}}, \bibinfo {author} {\bibfnamefont {L.}~\bibnamefont {Elcoro}},
  \bibinfo {author} {\bibfnamefont {C.}~\bibnamefont {Felser}}, \bibinfo
  {author} {\bibfnamefont {N.}~\bibnamefont {Regnault}}, \bibinfo {author}
  {\bibfnamefont {B.~A.}\ \bibnamefont {Bernevig}},\ and\ \bibinfo {author}
  {\bibfnamefont {Z.}~\bibnamefont {Wang}},\ }\href
  {https://doi.org/10.1038/s41586-019-0954-4} {\bibfield  {journal} {\bibinfo
  {journal} {Nature}\ }\textbf {\bibinfo {volume} {566}},\ \bibinfo {pages}
  {480} (\bibinfo {year} {2019})},\ \bibinfo {note} {publisher: Nature
  Publishing Group}\BibitemShut {NoStop}%
\bibitem [{\citenamefont {Schindler}\ \emph
  {et~al.}(2018{\natexlab{b}})\citenamefont {Schindler}, \citenamefont {Wang},
  \citenamefont {Vergniory}, \citenamefont {Cook}, \citenamefont {Murani},
  \citenamefont {Sengupta}, \citenamefont {Kasumov}, \citenamefont {Deblock},
  \citenamefont {Jeon}, \citenamefont {Drozdov}, \citenamefont {Bouchiat},
  \citenamefont {Guéron}, \citenamefont {Yazdani}, \citenamefont {Bernevig},\
  and\ \citenamefont {Neupert}}]{schindler_higher-order_2018-1}%
  \BibitemOpen
  \bibfield  {author} {\bibinfo {author} {\bibfnamefont {F.}~\bibnamefont
  {Schindler}}, \bibinfo {author} {\bibfnamefont {Z.}~\bibnamefont {Wang}},
  \bibinfo {author} {\bibfnamefont {M.~G.}\ \bibnamefont {Vergniory}}, \bibinfo
  {author} {\bibfnamefont {A.~M.}\ \bibnamefont {Cook}}, \bibinfo {author}
  {\bibfnamefont {A.}~\bibnamefont {Murani}}, \bibinfo {author} {\bibfnamefont
  {S.}~\bibnamefont {Sengupta}}, \bibinfo {author} {\bibfnamefont {A.~Y.}\
  \bibnamefont {Kasumov}}, \bibinfo {author} {\bibfnamefont {R.}~\bibnamefont
  {Deblock}}, \bibinfo {author} {\bibfnamefont {S.}~\bibnamefont {Jeon}},
  \bibinfo {author} {\bibfnamefont {I.}~\bibnamefont {Drozdov}}, \bibinfo
  {author} {\bibfnamefont {H.}~\bibnamefont {Bouchiat}}, \bibinfo {author}
  {\bibfnamefont {S.}~\bibnamefont {Guéron}}, \bibinfo {author} {\bibfnamefont
  {A.}~\bibnamefont {Yazdani}}, \bibinfo {author} {\bibfnamefont {B.~A.}\
  \bibnamefont {Bernevig}},\ and\ \bibinfo {author} {\bibfnamefont
  {T.}~\bibnamefont {Neupert}},\ }\href
  {https://doi.org/10.1038/s41567-018-0224-7} {\bibfield  {journal} {\bibinfo
  {journal} {Nature Physics}\ }\textbf {\bibinfo {volume} {14}},\ \bibinfo
  {pages} {918} (\bibinfo {year} {2018}{\natexlab{b}})},\ \bibinfo {note}
  {publisher: Nature Publishing Group}\BibitemShut {NoStop}%
\bibitem [{\citenamefont {Yu}\ \emph {et~al.}(2010)\citenamefont {Yu},
  \citenamefont {Zhang}, \citenamefont {Zhang}, \citenamefont {Zhang},
  \citenamefont {Dai},\ and\ \citenamefont {Fang}}]{yu_quantized_2010}%
  \BibitemOpen
  \bibfield  {author} {\bibinfo {author} {\bibfnamefont {R.}~\bibnamefont
  {Yu}}, \bibinfo {author} {\bibfnamefont {W.}~\bibnamefont {Zhang}}, \bibinfo
  {author} {\bibfnamefont {H.-J.}\ \bibnamefont {Zhang}}, \bibinfo {author}
  {\bibfnamefont {S.-C.}\ \bibnamefont {Zhang}}, \bibinfo {author}
  {\bibfnamefont {X.}~\bibnamefont {Dai}},\ and\ \bibinfo {author}
  {\bibfnamefont {Z.}~\bibnamefont {Fang}},\ }\href
  {https://doi.org/10.1126/science.1187485} {\bibfield  {journal} {\bibinfo
  {journal} {Science}\ }\textbf {\bibinfo {volume} {329}},\ \bibinfo {pages}
  {61} (\bibinfo {year} {2010})},\ \bibinfo {note} {publisher: American
  Association for the Advancement of Science}\BibitemShut {NoStop}%
\bibitem [{\citenamefont {Soluyanov}\ \emph {et~al.}(2015)\citenamefont
  {Soluyanov}, \citenamefont {Gresch}, \citenamefont {Wang}, \citenamefont
  {Wu}, \citenamefont {Troyer}, \citenamefont {Dai},\ and\ \citenamefont
  {Bernevig}}]{soluyanov_type-ii_2015}%
  \BibitemOpen
  \bibfield  {author} {\bibinfo {author} {\bibfnamefont {A.~A.}\ \bibnamefont
  {Soluyanov}}, \bibinfo {author} {\bibfnamefont {D.}~\bibnamefont {Gresch}},
  \bibinfo {author} {\bibfnamefont {Z.}~\bibnamefont {Wang}}, \bibinfo {author}
  {\bibfnamefont {Q.}~\bibnamefont {Wu}}, \bibinfo {author} {\bibfnamefont
  {M.}~\bibnamefont {Troyer}}, \bibinfo {author} {\bibfnamefont
  {X.}~\bibnamefont {Dai}},\ and\ \bibinfo {author} {\bibfnamefont {B.~A.}\
  \bibnamefont {Bernevig}},\ }\href {https://doi.org/10.1038/nature15768}
  {\bibfield  {journal} {\bibinfo  {journal} {Nature}\ }\textbf {\bibinfo
  {volume} {527}},\ \bibinfo {pages} {495} (\bibinfo {year} {2015})},\ \bibinfo
  {note} {publisher: Nature Publishing Group}\BibitemShut {NoStop}%
\bibitem [{\citenamefont {Wan}\ \emph {et~al.}(2011)\citenamefont {Wan},
  \citenamefont {Turner}, \citenamefont {Vishwanath},\ and\ \citenamefont
  {Savrasov}}]{wan_topological_2011}%
  \BibitemOpen
  \bibfield  {author} {\bibinfo {author} {\bibfnamefont {X.}~\bibnamefont
  {Wan}}, \bibinfo {author} {\bibfnamefont {A.~M.}\ \bibnamefont {Turner}},
  \bibinfo {author} {\bibfnamefont {A.}~\bibnamefont {Vishwanath}},\ and\
  \bibinfo {author} {\bibfnamefont {S.~Y.}\ \bibnamefont {Savrasov}},\ }\href
  {https://doi.org/10.1103/PhysRevB.83.205101} {\bibfield  {journal} {\bibinfo
  {journal} {Physical Review B}\ }\textbf {\bibinfo {volume} {83}},\ \bibinfo
  {pages} {205101} (\bibinfo {year} {2011})},\ \bibinfo {note} {publisher:
  American Physical Society}\BibitemShut {NoStop}%
\bibitem [{\citenamefont {Lv}\ \emph {et~al.}(2015)\citenamefont {Lv},
  \citenamefont {Weng}, \citenamefont {Fu}, \citenamefont {Wang}, \citenamefont
  {Miao}, \citenamefont {Ma}, \citenamefont {Richard}, \citenamefont {Huang},
  \citenamefont {Zhao}, \citenamefont {Chen}, \citenamefont {Fang},
  \citenamefont {Dai}, \citenamefont {Qian},\ and\ \citenamefont
  {Ding}}]{lv_experimental_2015}%
  \BibitemOpen
  \bibfield  {author} {\bibinfo {author} {\bibfnamefont {B.}~\bibnamefont
  {Lv}}, \bibinfo {author} {\bibfnamefont {H.}~\bibnamefont {Weng}}, \bibinfo
  {author} {\bibfnamefont {B.}~\bibnamefont {Fu}}, \bibinfo {author}
  {\bibfnamefont {X.}~\bibnamefont {Wang}}, \bibinfo {author} {\bibfnamefont
  {H.}~\bibnamefont {Miao}}, \bibinfo {author} {\bibfnamefont {J.}~\bibnamefont
  {Ma}}, \bibinfo {author} {\bibfnamefont {P.}~\bibnamefont {Richard}},
  \bibinfo {author} {\bibfnamefont {X.}~\bibnamefont {Huang}}, \bibinfo
  {author} {\bibfnamefont {L.}~\bibnamefont {Zhao}}, \bibinfo {author}
  {\bibfnamefont {G.}~\bibnamefont {Chen}}, \bibinfo {author} {\bibfnamefont
  {Z.}~\bibnamefont {Fang}}, \bibinfo {author} {\bibfnamefont {X.}~\bibnamefont
  {Dai}}, \bibinfo {author} {\bibfnamefont {T.}~\bibnamefont {Qian}},\ and\
  \bibinfo {author} {\bibfnamefont {H.}~\bibnamefont {Ding}},\ }\href
  {https://doi.org/10.1103/PhysRevX.5.031013} {\bibfield  {journal} {\bibinfo
  {journal} {Physical Review X}\ }\textbf {\bibinfo {volume} {5}},\ \bibinfo
  {pages} {031013} (\bibinfo {year} {2015})},\ \bibinfo {note} {publisher:
  American Physical Society}\BibitemShut {NoStop}%
\bibitem [{\citenamefont {Weng}\ \emph {et~al.}(2015)\citenamefont {Weng},
  \citenamefont {Fang}, \citenamefont {Fang}, \citenamefont {Bernevig},\ and\
  \citenamefont {Dai}}]{weng_weyl_2015}%
  \BibitemOpen
  \bibfield  {author} {\bibinfo {author} {\bibfnamefont {H.}~\bibnamefont
  {Weng}}, \bibinfo {author} {\bibfnamefont {C.}~\bibnamefont {Fang}}, \bibinfo
  {author} {\bibfnamefont {Z.}~\bibnamefont {Fang}}, \bibinfo {author}
  {\bibfnamefont {B.~A.}\ \bibnamefont {Bernevig}},\ and\ \bibinfo {author}
  {\bibfnamefont {X.}~\bibnamefont {Dai}},\ }\href
  {https://doi.org/10.1103/PhysRevX.5.011029} {\bibfield  {journal} {\bibinfo
  {journal} {Physical Review X}\ }\textbf {\bibinfo {volume} {5}},\ \bibinfo
  {pages} {011029} (\bibinfo {year} {2015})},\ \bibinfo {note} {publisher:
  American Physical Society}\BibitemShut {NoStop}%
\bibitem [{\citenamefont {Liu}\ \emph {et~al.}(2019)\citenamefont {Liu},
  \citenamefont {Liang}, \citenamefont {Liu}, \citenamefont {Xu}, \citenamefont
  {Li}, \citenamefont {Chen}, \citenamefont {Pei}, \citenamefont {Shi},
  \citenamefont {Mo}, \citenamefont {Dudin}, \citenamefont {Kim}, \citenamefont
  {Cacho}, \citenamefont {Li}, \citenamefont {Sun}, \citenamefont {Yang},
  \citenamefont {Liu}, \citenamefont {Parkin}, \citenamefont {Felser},\ and\
  \citenamefont {Chen}}]{liu_magnetic_2019}%
  \BibitemOpen
  \bibfield  {author} {\bibinfo {author} {\bibfnamefont {D.~F.}\ \bibnamefont
  {Liu}}, \bibinfo {author} {\bibfnamefont {A.~J.}\ \bibnamefont {Liang}},
  \bibinfo {author} {\bibfnamefont {E.~K.}\ \bibnamefont {Liu}}, \bibinfo
  {author} {\bibfnamefont {Q.~N.}\ \bibnamefont {Xu}}, \bibinfo {author}
  {\bibfnamefont {Y.~W.}\ \bibnamefont {Li}}, \bibinfo {author} {\bibfnamefont
  {C.}~\bibnamefont {Chen}}, \bibinfo {author} {\bibfnamefont {D.}~\bibnamefont
  {Pei}}, \bibinfo {author} {\bibfnamefont {W.~J.}\ \bibnamefont {Shi}},
  \bibinfo {author} {\bibfnamefont {S.~K.}\ \bibnamefont {Mo}}, \bibinfo
  {author} {\bibfnamefont {P.}~\bibnamefont {Dudin}}, \bibinfo {author}
  {\bibfnamefont {T.}~\bibnamefont {Kim}}, \bibinfo {author} {\bibfnamefont
  {C.}~\bibnamefont {Cacho}}, \bibinfo {author} {\bibfnamefont
  {G.}~\bibnamefont {Li}}, \bibinfo {author} {\bibfnamefont {Y.}~\bibnamefont
  {Sun}}, \bibinfo {author} {\bibfnamefont {L.~X.}\ \bibnamefont {Yang}},
  \bibinfo {author} {\bibfnamefont {Z.~K.}\ \bibnamefont {Liu}}, \bibinfo
  {author} {\bibfnamefont {S.~S.~P.}\ \bibnamefont {Parkin}}, \bibinfo {author}
  {\bibfnamefont {C.}~\bibnamefont {Felser}},\ and\ \bibinfo {author}
  {\bibfnamefont {Y.~L.}\ \bibnamefont {Chen}},\ }\href
  {https://doi.org/10.1126/science.aav2873} {\bibfield  {journal} {\bibinfo
  {journal} {Science}\ }\textbf {\bibinfo {volume} {365}},\ \bibinfo {pages}
  {1282} (\bibinfo {year} {2019})},\ \bibinfo {note} {publisher: American
  Association for the Advancement of Science}\BibitemShut {NoStop}%
\bibitem [{\citenamefont {Lv}\ \emph {et~al.}(2021)\citenamefont {Lv},
  \citenamefont {Qian},\ and\ \citenamefont {Ding}}]{lv_experimental_2021}%
  \BibitemOpen
  \bibfield  {author} {\bibinfo {author} {\bibfnamefont {B.}~\bibnamefont
  {Lv}}, \bibinfo {author} {\bibfnamefont {T.}~\bibnamefont {Qian}},\ and\
  \bibinfo {author} {\bibfnamefont {H.}~\bibnamefont {Ding}},\ }\href
  {https://doi.org/10.1103/RevModPhys.93.025002} {\bibfield  {journal}
  {\bibinfo  {journal} {Reviews of Modern Physics}\ }\textbf {\bibinfo {volume}
  {93}},\ \bibinfo {pages} {025002} (\bibinfo {year} {2021})},\ \bibinfo {note}
  {publisher: American Physical Society}\BibitemShut {NoStop}%
\bibitem [{\citenamefont {Armitage}\ \emph {et~al.}(2018)\citenamefont
  {Armitage}, \citenamefont {Mele},\ and\ \citenamefont
  {Vishwanath}}]{armitage_weyl_2018}%
  \BibitemOpen
  \bibfield  {author} {\bibinfo {author} {\bibfnamefont {N.}~\bibnamefont
  {Armitage}}, \bibinfo {author} {\bibfnamefont {E.}~\bibnamefont {Mele}},\
  and\ \bibinfo {author} {\bibfnamefont {A.}~\bibnamefont {Vishwanath}},\
  }\href {https://doi.org/10.1103/RevModPhys.90.015001} {\bibfield  {journal}
  {\bibinfo  {journal} {Reviews of Modern Physics}\ }\textbf {\bibinfo {volume}
  {90}},\ \bibinfo {pages} {015001} (\bibinfo {year} {2018})}\BibitemShut
  {NoStop}%
\bibitem [{\citenamefont {Wang}\ \emph {et~al.}(2012)\citenamefont {Wang},
  \citenamefont {Sun}, \citenamefont {Chen}, \citenamefont {Franchini},
  \citenamefont {Xu}, \citenamefont {Weng}, \citenamefont {Dai},\ and\
  \citenamefont {Fang}}]{wang_dirac_2012}%
  \BibitemOpen
  \bibfield  {author} {\bibinfo {author} {\bibfnamefont {Z.}~\bibnamefont
  {Wang}}, \bibinfo {author} {\bibfnamefont {Y.}~\bibnamefont {Sun}}, \bibinfo
  {author} {\bibfnamefont {X.-Q.}\ \bibnamefont {Chen}}, \bibinfo {author}
  {\bibfnamefont {C.}~\bibnamefont {Franchini}}, \bibinfo {author}
  {\bibfnamefont {G.}~\bibnamefont {Xu}}, \bibinfo {author} {\bibfnamefont
  {H.}~\bibnamefont {Weng}}, \bibinfo {author} {\bibfnamefont {X.}~\bibnamefont
  {Dai}},\ and\ \bibinfo {author} {\bibfnamefont {Z.}~\bibnamefont {Fang}},\
  }\href {https://doi.org/10.1103/PhysRevB.85.195320} {\bibfield  {journal}
  {\bibinfo  {journal} {Physical Review B}\ }\textbf {\bibinfo {volume} {85}},\
  \bibinfo {pages} {195320} (\bibinfo {year} {2012})},\ \bibinfo {note}
  {publisher: American Physical Society}\BibitemShut {NoStop}%
\bibitem [{\citenamefont {Liu}\ \emph {et~al.}(2014{\natexlab{a}})\citenamefont
  {Liu}, \citenamefont {Zhou}, \citenamefont {Zhang}, \citenamefont {Wang},
  \citenamefont {Weng}, \citenamefont {Prabhakaran}, \citenamefont {Mo},
  \citenamefont {Shen}, \citenamefont {Fang}, \citenamefont {Dai},
  \citenamefont {Hussain},\ and\ \citenamefont {Chen}}]{liu_discovery_2014}%
  \BibitemOpen
  \bibfield  {author} {\bibinfo {author} {\bibfnamefont {Z.~K.}\ \bibnamefont
  {Liu}}, \bibinfo {author} {\bibfnamefont {B.}~\bibnamefont {Zhou}}, \bibinfo
  {author} {\bibfnamefont {Y.}~\bibnamefont {Zhang}}, \bibinfo {author}
  {\bibfnamefont {Z.~J.}\ \bibnamefont {Wang}}, \bibinfo {author}
  {\bibfnamefont {H.~M.}\ \bibnamefont {Weng}}, \bibinfo {author}
  {\bibfnamefont {D.}~\bibnamefont {Prabhakaran}}, \bibinfo {author}
  {\bibfnamefont {S.-K.}\ \bibnamefont {Mo}}, \bibinfo {author} {\bibfnamefont
  {Z.~X.}\ \bibnamefont {Shen}}, \bibinfo {author} {\bibfnamefont
  {Z.}~\bibnamefont {Fang}}, \bibinfo {author} {\bibfnamefont {X.}~\bibnamefont
  {Dai}}, \bibinfo {author} {\bibfnamefont {Z.}~\bibnamefont {Hussain}},\ and\
  \bibinfo {author} {\bibfnamefont {Y.~L.}\ \bibnamefont {Chen}},\ }\href
  {https://doi.org/10.1126/science.1245085} {\bibfield  {journal} {\bibinfo
  {journal} {Science}\ }\textbf {\bibinfo {volume} {343}},\ \bibinfo {pages}
  {864} (\bibinfo {year} {2014}{\natexlab{a}})},\ \bibinfo {note} {publisher:
  American Association for the Advancement of Science}\BibitemShut {NoStop}%
\bibitem [{\citenamefont {Bradlyn}\ \emph {et~al.}(2016)\citenamefont
  {Bradlyn}, \citenamefont {Cano}, \citenamefont {Wang}, \citenamefont
  {Vergniory}, \citenamefont {Felser}, \citenamefont {Cava},\ and\
  \citenamefont {Bernevig}}]{bradlyn_beyond_2016}%
  \BibitemOpen
  \bibfield  {author} {\bibinfo {author} {\bibfnamefont {B.}~\bibnamefont
  {Bradlyn}}, \bibinfo {author} {\bibfnamefont {J.}~\bibnamefont {Cano}},
  \bibinfo {author} {\bibfnamefont {Z.}~\bibnamefont {Wang}}, \bibinfo {author}
  {\bibfnamefont {M.~G.}\ \bibnamefont {Vergniory}}, \bibinfo {author}
  {\bibfnamefont {C.}~\bibnamefont {Felser}}, \bibinfo {author} {\bibfnamefont
  {R.~J.}\ \bibnamefont {Cava}},\ and\ \bibinfo {author} {\bibfnamefont
  {B.~A.}\ \bibnamefont {Bernevig}},\ }\href
  {https://doi.org/10.1126/science.aaf5037} {\bibfield  {journal} {\bibinfo
  {journal} {Science}\ }\textbf {\bibinfo {volume} {353}},\ \bibinfo {pages}
  {aaf5037} (\bibinfo {year} {2016})},\ \bibinfo {note} {publisher: American
  Association for the Advancement of Science}\BibitemShut {NoStop}%
\bibitem [{\citenamefont {Liu}\ \emph {et~al.}(2014{\natexlab{b}})\citenamefont
  {Liu}, \citenamefont {Jiang}, \citenamefont {Zhou}, \citenamefont {Wang},
  \citenamefont {Zhang}, \citenamefont {Weng}, \citenamefont {Prabhakaran},
  \citenamefont {Mo}, \citenamefont {Peng}, \citenamefont {Dudin},
  \citenamefont {Kim}, \citenamefont {Hoesch}, \citenamefont {Fang},
  \citenamefont {Dai}, \citenamefont {Shen}, \citenamefont {Feng},
  \citenamefont {Hussain},\ and\ \citenamefont {Chen}}]{liu_stable_2014}%
  \BibitemOpen
  \bibfield  {author} {\bibinfo {author} {\bibfnamefont {Z.~K.}\ \bibnamefont
  {Liu}}, \bibinfo {author} {\bibfnamefont {J.}~\bibnamefont {Jiang}}, \bibinfo
  {author} {\bibfnamefont {B.}~\bibnamefont {Zhou}}, \bibinfo {author}
  {\bibfnamefont {Z.~J.}\ \bibnamefont {Wang}}, \bibinfo {author}
  {\bibfnamefont {Y.}~\bibnamefont {Zhang}}, \bibinfo {author} {\bibfnamefont
  {H.~M.}\ \bibnamefont {Weng}}, \bibinfo {author} {\bibfnamefont
  {D.}~\bibnamefont {Prabhakaran}}, \bibinfo {author} {\bibfnamefont {S.-K.}\
  \bibnamefont {Mo}}, \bibinfo {author} {\bibfnamefont {H.}~\bibnamefont
  {Peng}}, \bibinfo {author} {\bibfnamefont {P.}~\bibnamefont {Dudin}},
  \bibinfo {author} {\bibfnamefont {T.}~\bibnamefont {Kim}}, \bibinfo {author}
  {\bibfnamefont {M.}~\bibnamefont {Hoesch}}, \bibinfo {author} {\bibfnamefont
  {Z.}~\bibnamefont {Fang}}, \bibinfo {author} {\bibfnamefont {X.}~\bibnamefont
  {Dai}}, \bibinfo {author} {\bibfnamefont {Z.~X.}\ \bibnamefont {Shen}},
  \bibinfo {author} {\bibfnamefont {D.~L.}\ \bibnamefont {Feng}}, \bibinfo
  {author} {\bibfnamefont {Z.}~\bibnamefont {Hussain}},\ and\ \bibinfo {author}
  {\bibfnamefont {Y.~L.}\ \bibnamefont {Chen}},\ }\href
  {https://doi.org/10.1038/nmat3990} {\bibfield  {journal} {\bibinfo  {journal}
  {Nature Materials}\ }\textbf {\bibinfo {volume} {13}},\ \bibinfo {pages}
  {677} (\bibinfo {year} {2014}{\natexlab{b}})},\ \bibinfo {note} {publisher:
  Nature Publishing Group}\BibitemShut {NoStop}%
\bibitem [{\citenamefont {Huang}\ \emph {et~al.}(2015)\citenamefont {Huang},
  \citenamefont {Zhao}, \citenamefont {Long}, \citenamefont {Wang},
  \citenamefont {Chen}, \citenamefont {Yang}, \citenamefont {Liang},
  \citenamefont {Xue}, \citenamefont {Weng}, \citenamefont {Fang},
  \citenamefont {Dai},\ and\ \citenamefont {Chen}}]{huang_observation_2015}%
  \BibitemOpen
  \bibfield  {author} {\bibinfo {author} {\bibfnamefont {X.}~\bibnamefont
  {Huang}}, \bibinfo {author} {\bibfnamefont {L.}~\bibnamefont {Zhao}},
  \bibinfo {author} {\bibfnamefont {Y.}~\bibnamefont {Long}}, \bibinfo {author}
  {\bibfnamefont {P.}~\bibnamefont {Wang}}, \bibinfo {author} {\bibfnamefont
  {D.}~\bibnamefont {Chen}}, \bibinfo {author} {\bibfnamefont {Z.}~\bibnamefont
  {Yang}}, \bibinfo {author} {\bibfnamefont {H.}~\bibnamefont {Liang}},
  \bibinfo {author} {\bibfnamefont {M.}~\bibnamefont {Xue}}, \bibinfo {author}
  {\bibfnamefont {H.}~\bibnamefont {Weng}}, \bibinfo {author} {\bibfnamefont
  {Z.}~\bibnamefont {Fang}}, \bibinfo {author} {\bibfnamefont {X.}~\bibnamefont
  {Dai}},\ and\ \bibinfo {author} {\bibfnamefont {G.}~\bibnamefont {Chen}},\
  }\href {https://doi.org/10.1103/PhysRevX.5.031023} {\bibfield  {journal}
  {\bibinfo  {journal} {Physical Review X}\ }\textbf {\bibinfo {volume} {5}},\
  \bibinfo {pages} {031023} (\bibinfo {year} {2015})},\ \bibinfo {note}
  {publisher: American Physical Society}\BibitemShut {NoStop}%
\bibitem [{\citenamefont {Wang}\ \emph {et~al.}(2013)\citenamefont {Wang},
  \citenamefont {Weng}, \citenamefont {Wu}, \citenamefont {Dai},\ and\
  \citenamefont {Fang}}]{wang_three-dimensional_2013}%
  \BibitemOpen
  \bibfield  {author} {\bibinfo {author} {\bibfnamefont {Z.}~\bibnamefont
  {Wang}}, \bibinfo {author} {\bibfnamefont {H.}~\bibnamefont {Weng}}, \bibinfo
  {author} {\bibfnamefont {Q.}~\bibnamefont {Wu}}, \bibinfo {author}
  {\bibfnamefont {X.}~\bibnamefont {Dai}},\ and\ \bibinfo {author}
  {\bibfnamefont {Z.}~\bibnamefont {Fang}},\ }\href
  {https://doi.org/10.1103/PhysRevB.88.125427} {\bibfield  {journal} {\bibinfo
  {journal} {Physical Review B}\ }\textbf {\bibinfo {volume} {88}},\ \bibinfo
  {pages} {125427} (\bibinfo {year} {2013})},\ \bibinfo {note} {publisher:
  American Physical Society}\BibitemShut {NoStop}%
\bibitem [{\citenamefont {Young}\ \emph {et~al.}(2012)\citenamefont {Young},
  \citenamefont {Zaheer}, \citenamefont {Teo}, \citenamefont {Kane},
  \citenamefont {Mele},\ and\ \citenamefont {Rappe}}]{PhysRevLett.108.140405}%
  \BibitemOpen
  \bibfield  {author} {\bibinfo {author} {\bibfnamefont {S.~M.}\ \bibnamefont
  {Young}}, \bibinfo {author} {\bibfnamefont {S.}~\bibnamefont {Zaheer}},
  \bibinfo {author} {\bibfnamefont {J.~C.~Y.}\ \bibnamefont {Teo}}, \bibinfo
  {author} {\bibfnamefont {C.~L.}\ \bibnamefont {Kane}}, \bibinfo {author}
  {\bibfnamefont {E.~J.}\ \bibnamefont {Mele}},\ and\ \bibinfo {author}
  {\bibfnamefont {A.~M.}\ \bibnamefont {Rappe}},\ }\href
  {https://doi.org/10.1103/PhysRevLett.108.140405} {\bibfield  {journal}
  {\bibinfo  {journal} {Phys. Rev. Lett.}\ }\textbf {\bibinfo {volume} {108}},\
  \bibinfo {pages} {140405} (\bibinfo {year} {2012})}\BibitemShut {NoStop}%
\bibitem [{\citenamefont {Young}\ and\ \citenamefont
  {Kane}(2015)}]{PhysRevLett.115.126803}%
  \BibitemOpen
  \bibfield  {author} {\bibinfo {author} {\bibfnamefont {S.~M.}\ \bibnamefont
  {Young}}\ and\ \bibinfo {author} {\bibfnamefont {C.~L.}\ \bibnamefont
  {Kane}},\ }\href {https://doi.org/10.1103/PhysRevLett.115.126803} {\bibfield
  {journal} {\bibinfo  {journal} {Phys. Rev. Lett.}\ }\textbf {\bibinfo
  {volume} {115}},\ \bibinfo {pages} {126803} (\bibinfo {year}
  {2015})}\BibitemShut {NoStop}%
\bibitem [{\citenamefont {Borisenko}\ \emph {et~al.}(2014)\citenamefont
  {Borisenko}, \citenamefont {Gibson}, \citenamefont {Evtushinsky},
  \citenamefont {Zabolotnyy}, \citenamefont {B\"uchner},\ and\ \citenamefont
  {Cava}}]{PhysRevLett.113.027603}%
  \BibitemOpen
  \bibfield  {author} {\bibinfo {author} {\bibfnamefont {S.}~\bibnamefont
  {Borisenko}}, \bibinfo {author} {\bibfnamefont {Q.}~\bibnamefont {Gibson}},
  \bibinfo {author} {\bibfnamefont {D.}~\bibnamefont {Evtushinsky}}, \bibinfo
  {author} {\bibfnamefont {V.}~\bibnamefont {Zabolotnyy}}, \bibinfo {author}
  {\bibfnamefont {B.}~\bibnamefont {B\"uchner}},\ and\ \bibinfo {author}
  {\bibfnamefont {R.~J.}\ \bibnamefont {Cava}},\ }\href
  {https://doi.org/10.1103/PhysRevLett.113.027603} {\bibfield  {journal}
  {\bibinfo  {journal} {Phys. Rev. Lett.}\ }\textbf {\bibinfo {volume} {113}},\
  \bibinfo {pages} {027603} (\bibinfo {year} {2014})}\BibitemShut {NoStop}%
\bibitem [{\citenamefont {Xiong}\ \emph {et~al.}(2015)\citenamefont {Xiong},
  \citenamefont {Kushwaha}, \citenamefont {Liang}, \citenamefont {Krizan},
  \citenamefont {Hirschberger}, \citenamefont {Wang}, \citenamefont {Cava},\
  and\ \citenamefont {Ong}}]{Xiong2015}%
  \BibitemOpen
  \bibfield  {author} {\bibinfo {author} {\bibfnamefont {J.}~\bibnamefont
  {Xiong}}, \bibinfo {author} {\bibfnamefont {S.~K.}\ \bibnamefont {Kushwaha}},
  \bibinfo {author} {\bibfnamefont {T.}~\bibnamefont {Liang}}, \bibinfo
  {author} {\bibfnamefont {J.~W.}\ \bibnamefont {Krizan}}, \bibinfo {author}
  {\bibfnamefont {M.}~\bibnamefont {Hirschberger}}, \bibinfo {author}
  {\bibfnamefont {W.}~\bibnamefont {Wang}}, \bibinfo {author} {\bibfnamefont
  {R.~J.}\ \bibnamefont {Cava}},\ and\ \bibinfo {author} {\bibfnamefont
  {N.~P.}\ \bibnamefont {Ong}},\ }\href
  {https://doi.org/10.1126/science.aac6089} {\bibfield  {journal} {\bibinfo
  {journal} {Science}\ }\textbf {\bibinfo {volume} {350}},\ \bibinfo {pages}
  {413} (\bibinfo {year} {2015})}\BibitemShut {NoStop}%
\bibitem [{\citenamefont {Wieder}\ \emph {et~al.}(2016)\citenamefont {Wieder},
  \citenamefont {Kim}, \citenamefont {Rappe},\ and\ \citenamefont
  {Kane}}]{PhysRevLett.116.186402}%
  \BibitemOpen
  \bibfield  {author} {\bibinfo {author} {\bibfnamefont {B.~J.}\ \bibnamefont
  {Wieder}}, \bibinfo {author} {\bibfnamefont {Y.}~\bibnamefont {Kim}},
  \bibinfo {author} {\bibfnamefont {A.~M.}\ \bibnamefont {Rappe}},\ and\
  \bibinfo {author} {\bibfnamefont {C.~L.}\ \bibnamefont {Kane}},\ }\href
  {https://doi.org/10.1103/PhysRevLett.116.186402} {\bibfield  {journal}
  {\bibinfo  {journal} {Phys. Rev. Lett.}\ }\textbf {\bibinfo {volume} {116}},\
  \bibinfo {pages} {186402} (\bibinfo {year} {2016})}\BibitemShut {NoStop}%
\bibitem [{\citenamefont {Neupane}\ \emph {et~al.}(2014)\citenamefont
  {Neupane}, \citenamefont {Xu}, \citenamefont {Sankar}, \citenamefont
  {Alidoust}, \citenamefont {Bian}, \citenamefont {Liu}, \citenamefont
  {Belopolski}, \citenamefont {Chang}, \citenamefont {Jeng}, \citenamefont
  {Lin}, \citenamefont {Bansil}, \citenamefont {Chou},\ and\ \citenamefont
  {Hasan}}]{Neupane2014}%
  \BibitemOpen
  \bibfield  {author} {\bibinfo {author} {\bibfnamefont {M.}~\bibnamefont
  {Neupane}}, \bibinfo {author} {\bibfnamefont {S.-Y.}\ \bibnamefont {Xu}},
  \bibinfo {author} {\bibfnamefont {R.}~\bibnamefont {Sankar}}, \bibinfo
  {author} {\bibfnamefont {N.}~\bibnamefont {Alidoust}}, \bibinfo {author}
  {\bibfnamefont {G.}~\bibnamefont {Bian}}, \bibinfo {author} {\bibfnamefont
  {C.}~\bibnamefont {Liu}}, \bibinfo {author} {\bibfnamefont {I.}~\bibnamefont
  {Belopolski}}, \bibinfo {author} {\bibfnamefont {T.-R.}\ \bibnamefont
  {Chang}}, \bibinfo {author} {\bibfnamefont {H.-T.}\ \bibnamefont {Jeng}},
  \bibinfo {author} {\bibfnamefont {H.}~\bibnamefont {Lin}}, \bibinfo {author}
  {\bibfnamefont {A.}~\bibnamefont {Bansil}}, \bibinfo {author} {\bibfnamefont
  {F.}~\bibnamefont {Chou}},\ and\ \bibinfo {author} {\bibfnamefont {M.~Z.}\
  \bibnamefont {Hasan}},\ }\href {https://doi.org/10.1038/ncomms4786}
  {\bibfield  {journal} {\bibinfo  {journal} {Nature Communications}\ }\textbf
  {\bibinfo {volume} {5}},\ \bibinfo {pages} {3786} (\bibinfo {year}
  {2014})}\BibitemShut {NoStop}%
\bibitem [{\citenamefont {Yan}\ and\ \citenamefont
  {Felser}(2017)}]{Yan_top_mat}%
  \BibitemOpen
  \bibfield  {author} {\bibinfo {author} {\bibfnamefont {B.}~\bibnamefont
  {Yan}}\ and\ \bibinfo {author} {\bibfnamefont {C.}~\bibnamefont {Felser}},\
  }\href
  {https://doi.org/https://doi.org/10.1146/annurev-conmatphys-031016-025458}
  {\bibfield  {journal} {\bibinfo  {journal} {Annual Review of Condensed Matter
  Physics}\ }\textbf {\bibinfo {volume} {8}},\ \bibinfo {pages} {337} (\bibinfo
  {year} {2017})}\BibitemShut {NoStop}%
\bibitem [{\citenamefont {Burkov}\ and\ \citenamefont
  {Balents}(2011)}]{PhysRevLett.107.127205}%
  \BibitemOpen
  \bibfield  {author} {\bibinfo {author} {\bibfnamefont {A.~A.}\ \bibnamefont
  {Burkov}}\ and\ \bibinfo {author} {\bibfnamefont {L.}~\bibnamefont
  {Balents}},\ }\href {https://doi.org/10.1103/PhysRevLett.107.127205}
  {\bibfield  {journal} {\bibinfo  {journal} {Phys. Rev. Lett.}\ }\textbf
  {\bibinfo {volume} {107}},\ \bibinfo {pages} {127205} (\bibinfo {year}
  {2011})}\BibitemShut {NoStop}%
\bibitem [{\citenamefont {Yang}\ \emph {et~al.}(2015)\citenamefont {Yang},
  \citenamefont {Liu}, \citenamefont {Sun}, \citenamefont {Peng}, \citenamefont
  {Yang}, \citenamefont {Zhang}, \citenamefont {Zhou}, \citenamefont {Zhang},
  \citenamefont {Guo}, \citenamefont {Rahn}, \citenamefont {Prabhakaran},
  \citenamefont {Hussain}, \citenamefont {Mo}, \citenamefont {Felser},
  \citenamefont {Yan},\ and\ \citenamefont {Chen}}]{Yang2015}%
  \BibitemOpen
  \bibfield  {author} {\bibinfo {author} {\bibfnamefont {L.~X.}\ \bibnamefont
  {Yang}}, \bibinfo {author} {\bibfnamefont {Z.~K.}\ \bibnamefont {Liu}},
  \bibinfo {author} {\bibfnamefont {Y.}~\bibnamefont {Sun}}, \bibinfo {author}
  {\bibfnamefont {H.}~\bibnamefont {Peng}}, \bibinfo {author} {\bibfnamefont
  {H.~F.}\ \bibnamefont {Yang}}, \bibinfo {author} {\bibfnamefont
  {T.}~\bibnamefont {Zhang}}, \bibinfo {author} {\bibfnamefont
  {B.}~\bibnamefont {Zhou}}, \bibinfo {author} {\bibfnamefont {Y.}~\bibnamefont
  {Zhang}}, \bibinfo {author} {\bibfnamefont {Y.~F.}\ \bibnamefont {Guo}},
  \bibinfo {author} {\bibfnamefont {M.}~\bibnamefont {Rahn}}, \bibinfo {author}
  {\bibfnamefont {D.}~\bibnamefont {Prabhakaran}}, \bibinfo {author}
  {\bibfnamefont {Z.}~\bibnamefont {Hussain}}, \bibinfo {author} {\bibfnamefont
  {S.-K.}\ \bibnamefont {Mo}}, \bibinfo {author} {\bibfnamefont
  {C.}~\bibnamefont {Felser}}, \bibinfo {author} {\bibfnamefont
  {B.}~\bibnamefont {Yan}},\ and\ \bibinfo {author} {\bibfnamefont {Y.~L.}\
  \bibnamefont {Chen}},\ }\href {https://doi.org/10.1038/nphys3425} {\bibfield
  {journal} {\bibinfo  {journal} {Nature Physics}\ }\textbf {\bibinfo {volume}
  {11}},\ \bibinfo {pages} {728} (\bibinfo {year} {2015})}\BibitemShut
  {NoStop}%
\bibitem [{\citenamefont {Dzero}\ \emph {et~al.}(2016)\citenamefont {Dzero},
  \citenamefont {Xia}, \citenamefont {Galitski},\ and\ \citenamefont
  {Coleman}}]{dzero_topological_2016}%
  \BibitemOpen
  \bibfield  {author} {\bibinfo {author} {\bibfnamefont {M.}~\bibnamefont
  {Dzero}}, \bibinfo {author} {\bibfnamefont {J.}~\bibnamefont {Xia}}, \bibinfo
  {author} {\bibfnamefont {V.}~\bibnamefont {Galitski}},\ and\ \bibinfo
  {author} {\bibfnamefont {P.}~\bibnamefont {Coleman}},\ }\href
  {https://doi.org/10.1146/annurev-conmatphys-031214-014749} {\bibfield
  {journal} {\bibinfo  {journal} {Annual Review of Condensed Matter Physics}\
  }\textbf {\bibinfo {volume} {7}},\ \bibinfo {pages} {249} (\bibinfo {year}
  {2016})},\ \bibinfo {note} {publisher: Annual Reviews}\BibitemShut {NoStop}%
\bibitem [{\citenamefont {Dzero}\ \emph {et~al.}(2012)\citenamefont {Dzero},
  \citenamefont {Sun}, \citenamefont {Coleman},\ and\ \citenamefont
  {Galitski}}]{dzero_theory_2012}%
  \BibitemOpen
  \bibfield  {author} {\bibinfo {author} {\bibfnamefont {M.}~\bibnamefont
  {Dzero}}, \bibinfo {author} {\bibfnamefont {K.}~\bibnamefont {Sun}}, \bibinfo
  {author} {\bibfnamefont {P.}~\bibnamefont {Coleman}},\ and\ \bibinfo {author}
  {\bibfnamefont {V.}~\bibnamefont {Galitski}},\ }\href
  {https://doi.org/10.1103/PhysRevB.85.045130} {\bibfield  {journal} {\bibinfo
  {journal} {Physical Review B}\ }\textbf {\bibinfo {volume} {85}},\ \bibinfo
  {pages} {045130} (\bibinfo {year} {2012})},\ \bibinfo {note} {publisher:
  American Physical Society}\BibitemShut {NoStop}%
\bibitem [{\citenamefont {Alexandrov}\ \emph {et~al.}(2013)\citenamefont
  {Alexandrov}, \citenamefont {Dzero},\ and\ \citenamefont
  {Coleman}}]{alexandrov_cubic_2013}%
  \BibitemOpen
  \bibfield  {author} {\bibinfo {author} {\bibfnamefont {V.}~\bibnamefont
  {Alexandrov}}, \bibinfo {author} {\bibfnamefont {M.}~\bibnamefont {Dzero}},\
  and\ \bibinfo {author} {\bibfnamefont {P.}~\bibnamefont {Coleman}},\ }\href
  {https://doi.org/10.1103/PhysRevLett.111.226403} {\bibfield  {journal}
  {\bibinfo  {journal} {Physical Review Letters}\ }\textbf {\bibinfo {volume}
  {111}},\ \bibinfo {pages} {226403} (\bibinfo {year} {2013})},\ \bibinfo
  {note} {publisher: American Physical Society}\BibitemShut {NoStop}%
\bibitem [{\citenamefont {Knolle}\ and\ \citenamefont
  {Cooper}(2017)}]{knolle_excitons_2017}%
  \BibitemOpen
  \bibfield  {author} {\bibinfo {author} {\bibfnamefont {J.}~\bibnamefont
  {Knolle}}\ and\ \bibinfo {author} {\bibfnamefont {N.~R.}\ \bibnamefont
  {Cooper}},\ }\href {https://doi.org/10.1103/PhysRevLett.118.096604}
  {\bibfield  {journal} {\bibinfo  {journal} {Physical Review Letters}\
  }\textbf {\bibinfo {volume} {118}},\ \bibinfo {pages} {096604} (\bibinfo
  {year} {2017})},\ \bibinfo {note} {publisher: American Physical
  Society}\BibitemShut {NoStop}%
\bibitem [{\citenamefont {Roy}\ \emph {et~al.}(2014)\citenamefont {Roy},
  \citenamefont {Sau}, \citenamefont {Dzero},\ and\ \citenamefont
  {Galitski}}]{roy_surface_2014}%
  \BibitemOpen
  \bibfield  {author} {\bibinfo {author} {\bibfnamefont {B.}~\bibnamefont
  {Roy}}, \bibinfo {author} {\bibfnamefont {J.~D.}\ \bibnamefont {Sau}},
  \bibinfo {author} {\bibfnamefont {M.}~\bibnamefont {Dzero}},\ and\ \bibinfo
  {author} {\bibfnamefont {V.}~\bibnamefont {Galitski}},\ }\href
  {https://doi.org/10.1103/PhysRevB.90.155314} {\bibfield  {journal} {\bibinfo
  {journal} {Physical Review B}\ }\textbf {\bibinfo {volume} {90}},\ \bibinfo
  {pages} {155314} (\bibinfo {year} {2014})},\ \bibinfo {note} {publisher:
  American Physical Society}\BibitemShut {NoStop}%
\bibitem [{\citenamefont {Syers}\ \emph {et~al.}(2015)\citenamefont {Syers},
  \citenamefont {Kim}, \citenamefont {Fuhrer},\ and\ \citenamefont
  {Paglione}}]{syers_tuning_2015}%
  \BibitemOpen
  \bibfield  {author} {\bibinfo {author} {\bibfnamefont {P.}~\bibnamefont
  {Syers}}, \bibinfo {author} {\bibfnamefont {D.}~\bibnamefont {Kim}}, \bibinfo
  {author} {\bibfnamefont {M.~S.}\ \bibnamefont {Fuhrer}},\ and\ \bibinfo
  {author} {\bibfnamefont {J.}~\bibnamefont {Paglione}},\ }\href
  {https://doi.org/10.1103/PhysRevLett.114.096601} {\bibfield  {journal}
  {\bibinfo  {journal} {Physical Review Letters}\ }\textbf {\bibinfo {volume}
  {114}},\ \bibinfo {pages} {096601} (\bibinfo {year} {2015})},\ \bibinfo
  {note} {publisher: American Physical Society}\BibitemShut {NoStop}%
\bibitem [{\citenamefont {Xu}\ \emph {et~al.}(2013)\citenamefont {Xu},
  \citenamefont {Shi}, \citenamefont {Biswas}, \citenamefont {Matt},
  \citenamefont {Dhaka}, \citenamefont {Huang}, \citenamefont {Plumb},
  \citenamefont {Radović}, \citenamefont {Dil}, \citenamefont {Pomjakushina},
  \citenamefont {Conder}, \citenamefont {Amato}, \citenamefont {Salman},
  \citenamefont {Paul}, \citenamefont {Mesot}, \citenamefont {Ding},\ and\
  \citenamefont {Shi}}]{xu_surface_2013}%
  \BibitemOpen
  \bibfield  {author} {\bibinfo {author} {\bibfnamefont {N.}~\bibnamefont
  {Xu}}, \bibinfo {author} {\bibfnamefont {X.}~\bibnamefont {Shi}}, \bibinfo
  {author} {\bibfnamefont {P.~K.}\ \bibnamefont {Biswas}}, \bibinfo {author}
  {\bibfnamefont {C.~E.}\ \bibnamefont {Matt}}, \bibinfo {author}
  {\bibfnamefont {R.~S.}\ \bibnamefont {Dhaka}}, \bibinfo {author}
  {\bibfnamefont {Y.}~\bibnamefont {Huang}}, \bibinfo {author} {\bibfnamefont
  {N.~C.}\ \bibnamefont {Plumb}}, \bibinfo {author} {\bibfnamefont
  {M.}~\bibnamefont {Radović}}, \bibinfo {author} {\bibfnamefont {J.~H.}\
  \bibnamefont {Dil}}, \bibinfo {author} {\bibfnamefont {E.}~\bibnamefont
  {Pomjakushina}}, \bibinfo {author} {\bibfnamefont {K.}~\bibnamefont
  {Conder}}, \bibinfo {author} {\bibfnamefont {A.}~\bibnamefont {Amato}},
  \bibinfo {author} {\bibfnamefont {Z.}~\bibnamefont {Salman}}, \bibinfo
  {author} {\bibfnamefont {D.~M.}\ \bibnamefont {Paul}}, \bibinfo {author}
  {\bibfnamefont {J.}~\bibnamefont {Mesot}}, \bibinfo {author} {\bibfnamefont
  {H.}~\bibnamefont {Ding}},\ and\ \bibinfo {author} {\bibfnamefont
  {M.}~\bibnamefont {Shi}},\ }\href
  {https://doi.org/10.1103/PhysRevB.88.121102} {\bibfield  {journal} {\bibinfo
  {journal} {Physical Review B}\ }\textbf {\bibinfo {volume} {88}},\ \bibinfo
  {pages} {121102} (\bibinfo {year} {2013})},\ \bibinfo {note} {publisher:
  American Physical Society}\BibitemShut {NoStop}%
\bibitem [{\citenamefont {Tran}\ \emph {et~al.}(2012)\citenamefont {Tran},
  \citenamefont {Takimoto},\ and\ \citenamefont {Kim}}]{tran_phase_2012}%
  \BibitemOpen
  \bibfield  {author} {\bibinfo {author} {\bibfnamefont {M.-T.}\ \bibnamefont
  {Tran}}, \bibinfo {author} {\bibfnamefont {T.}~\bibnamefont {Takimoto}},\
  and\ \bibinfo {author} {\bibfnamefont {K.-S.}\ \bibnamefont {Kim}},\ }\href
  {https://doi.org/10.1103/PhysRevB.85.125128} {\bibfield  {journal} {\bibinfo
  {journal} {Physical Review B}\ }\textbf {\bibinfo {volume} {85}},\ \bibinfo
  {pages} {125128} (\bibinfo {year} {2012})},\ \bibinfo {note} {publisher:
  American Physical Society}\BibitemShut {NoStop}%
\bibitem [{\citenamefont {Pirie}\ \emph {et~al.}(2020)\citenamefont {Pirie},
  \citenamefont {Liu}, \citenamefont {Soumyanarayanan}, \citenamefont {Chen},
  \citenamefont {He}, \citenamefont {Yee}, \citenamefont {Rosa}, \citenamefont
  {Thompson}, \citenamefont {Kim}, \citenamefont {Fisk}, \citenamefont {Wang},
  \citenamefont {Paglione}, \citenamefont {Morr}, \citenamefont {Hamidian},\
  and\ \citenamefont {Hoffman}}]{pirie_imaging_2020}%
  \BibitemOpen
  \bibfield  {author} {\bibinfo {author} {\bibfnamefont {H.}~\bibnamefont
  {Pirie}}, \bibinfo {author} {\bibfnamefont {Y.}~\bibnamefont {Liu}}, \bibinfo
  {author} {\bibfnamefont {A.}~\bibnamefont {Soumyanarayanan}}, \bibinfo
  {author} {\bibfnamefont {P.}~\bibnamefont {Chen}}, \bibinfo {author}
  {\bibfnamefont {Y.}~\bibnamefont {He}}, \bibinfo {author} {\bibfnamefont
  {M.~M.}\ \bibnamefont {Yee}}, \bibinfo {author} {\bibfnamefont {P.~F.~S.}\
  \bibnamefont {Rosa}}, \bibinfo {author} {\bibfnamefont {J.~D.}\ \bibnamefont
  {Thompson}}, \bibinfo {author} {\bibfnamefont {D.-J.}\ \bibnamefont {Kim}},
  \bibinfo {author} {\bibfnamefont {Z.}~\bibnamefont {Fisk}}, \bibinfo {author}
  {\bibfnamefont {X.}~\bibnamefont {Wang}}, \bibinfo {author} {\bibfnamefont
  {J.}~\bibnamefont {Paglione}}, \bibinfo {author} {\bibfnamefont {D.~K.}\
  \bibnamefont {Morr}}, \bibinfo {author} {\bibfnamefont {M.~H.}\ \bibnamefont
  {Hamidian}},\ and\ \bibinfo {author} {\bibfnamefont {J.~E.}\ \bibnamefont
  {Hoffman}},\ }\href {https://doi.org/10.1038/s41567-019-0700-8} {\bibfield
  {journal} {\bibinfo  {journal} {Nature Physics}\ }\textbf {\bibinfo {volume}
  {16}},\ \bibinfo {pages} {52} (\bibinfo {year} {2020})},\ \bibinfo {note}
  {publisher: Nature Publishing Group}\BibitemShut {NoStop}%
\bibitem [{\citenamefont {Xu}\ \emph {et~al.}(2014)\citenamefont {Xu},
  \citenamefont {Biswas}, \citenamefont {Dil}, \citenamefont {Dhaka},
  \citenamefont {Landolt}, \citenamefont {Muff}, \citenamefont {Matt},
  \citenamefont {Shi}, \citenamefont {Plumb}, \citenamefont {Radović},
  \citenamefont {Pomjakushina}, \citenamefont {Conder}, \citenamefont {Amato},
  \citenamefont {Borisenko}, \citenamefont {Yu}, \citenamefont {Weng},
  \citenamefont {Fang}, \citenamefont {Dai}, \citenamefont {Mesot},
  \citenamefont {Ding},\ and\ \citenamefont {Shi}}]{xu_direct_2014}%
  \BibitemOpen
  \bibfield  {author} {\bibinfo {author} {\bibfnamefont {N.}~\bibnamefont
  {Xu}}, \bibinfo {author} {\bibfnamefont {P.~K.}\ \bibnamefont {Biswas}},
  \bibinfo {author} {\bibfnamefont {J.~H.}\ \bibnamefont {Dil}}, \bibinfo
  {author} {\bibfnamefont {R.~S.}\ \bibnamefont {Dhaka}}, \bibinfo {author}
  {\bibfnamefont {G.}~\bibnamefont {Landolt}}, \bibinfo {author} {\bibfnamefont
  {S.}~\bibnamefont {Muff}}, \bibinfo {author} {\bibfnamefont {C.~E.}\
  \bibnamefont {Matt}}, \bibinfo {author} {\bibfnamefont {X.}~\bibnamefont
  {Shi}}, \bibinfo {author} {\bibfnamefont {N.~C.}\ \bibnamefont {Plumb}},
  \bibinfo {author} {\bibfnamefont {M.}~\bibnamefont {Radović}}, \bibinfo
  {author} {\bibfnamefont {E.}~\bibnamefont {Pomjakushina}}, \bibinfo {author}
  {\bibfnamefont {K.}~\bibnamefont {Conder}}, \bibinfo {author} {\bibfnamefont
  {A.}~\bibnamefont {Amato}}, \bibinfo {author} {\bibfnamefont {S.~V.}\
  \bibnamefont {Borisenko}}, \bibinfo {author} {\bibfnamefont {R.}~\bibnamefont
  {Yu}}, \bibinfo {author} {\bibfnamefont {H.-M.}\ \bibnamefont {Weng}},
  \bibinfo {author} {\bibfnamefont {Z.}~\bibnamefont {Fang}}, \bibinfo {author}
  {\bibfnamefont {X.}~\bibnamefont {Dai}}, \bibinfo {author} {\bibfnamefont
  {J.}~\bibnamefont {Mesot}}, \bibinfo {author} {\bibfnamefont
  {H.}~\bibnamefont {Ding}},\ and\ \bibinfo {author} {\bibfnamefont
  {M.}~\bibnamefont {Shi}},\ }\href {https://doi.org/10.1038/ncomms5566}
  {\bibfield  {journal} {\bibinfo  {journal} {Nature Communications}\ }\textbf
  {\bibinfo {volume} {5}},\ \bibinfo {pages} {4566} (\bibinfo {year} {2014})},\
  \bibinfo {note} {publisher: Nature Publishing Group}\BibitemShut {NoStop}%
\bibitem [{\citenamefont {Neupane}\ \emph {et~al.}(2013)\citenamefont
  {Neupane}, \citenamefont {Alidoust}, \citenamefont {Xu}, \citenamefont
  {Kondo}, \citenamefont {Ishida}, \citenamefont {Kim}, \citenamefont {Liu},
  \citenamefont {Belopolski}, \citenamefont {Jo}, \citenamefont {Chang},
  \citenamefont {Jeng}, \citenamefont {Durakiewicz}, \citenamefont {Balicas},
  \citenamefont {Lin}, \citenamefont {Bansil}, \citenamefont {Shin},
  \citenamefont {Fisk},\ and\ \citenamefont {Hasan}}]{neupane_surface_2013}%
  \BibitemOpen
  \bibfield  {author} {\bibinfo {author} {\bibfnamefont {M.}~\bibnamefont
  {Neupane}}, \bibinfo {author} {\bibfnamefont {N.}~\bibnamefont {Alidoust}},
  \bibinfo {author} {\bibfnamefont {S.-Y.}\ \bibnamefont {Xu}}, \bibinfo
  {author} {\bibfnamefont {T.}~\bibnamefont {Kondo}}, \bibinfo {author}
  {\bibfnamefont {Y.}~\bibnamefont {Ishida}}, \bibinfo {author} {\bibfnamefont
  {D.~J.}\ \bibnamefont {Kim}}, \bibinfo {author} {\bibfnamefont
  {C.}~\bibnamefont {Liu}}, \bibinfo {author} {\bibfnamefont {I.}~\bibnamefont
  {Belopolski}}, \bibinfo {author} {\bibfnamefont {Y.~J.}\ \bibnamefont {Jo}},
  \bibinfo {author} {\bibfnamefont {T.-R.}\ \bibnamefont {Chang}}, \bibinfo
  {author} {\bibfnamefont {H.-T.}\ \bibnamefont {Jeng}}, \bibinfo {author}
  {\bibfnamefont {T.}~\bibnamefont {Durakiewicz}}, \bibinfo {author}
  {\bibfnamefont {L.}~\bibnamefont {Balicas}}, \bibinfo {author} {\bibfnamefont
  {H.}~\bibnamefont {Lin}}, \bibinfo {author} {\bibfnamefont {A.}~\bibnamefont
  {Bansil}}, \bibinfo {author} {\bibfnamefont {S.}~\bibnamefont {Shin}},
  \bibinfo {author} {\bibfnamefont {Z.}~\bibnamefont {Fisk}},\ and\ \bibinfo
  {author} {\bibfnamefont {M.~Z.}\ \bibnamefont {Hasan}},\ }\href
  {https://doi.org/10.1038/ncomms3991} {\bibfield  {journal} {\bibinfo
  {journal} {Nature Communications}\ }\textbf {\bibinfo {volume} {4}},\
  \bibinfo {pages} {2991} (\bibinfo {year} {2013})},\ \bibinfo {note}
  {publisher: Nature Publishing Group}\BibitemShut {NoStop}%
\bibitem [{\citenamefont {Nakajima}\ \emph {et~al.}(2016)\citenamefont
  {Nakajima}, \citenamefont {Syers}, \citenamefont {Wang}, \citenamefont
  {Wang},\ and\ \citenamefont {Paglione}}]{nakajima_one-dimensional_2016}%
  \BibitemOpen
  \bibfield  {author} {\bibinfo {author} {\bibfnamefont {Y.}~\bibnamefont
  {Nakajima}}, \bibinfo {author} {\bibfnamefont {P.}~\bibnamefont {Syers}},
  \bibinfo {author} {\bibfnamefont {X.}~\bibnamefont {Wang}}, \bibinfo {author}
  {\bibfnamefont {R.}~\bibnamefont {Wang}},\ and\ \bibinfo {author}
  {\bibfnamefont {J.}~\bibnamefont {Paglione}},\ }\href
  {https://doi.org/10.1038/nphys3555} {\bibfield  {journal} {\bibinfo
  {journal} {Nature Physics}\ }\textbf {\bibinfo {volume} {12}},\ \bibinfo
  {pages} {213} (\bibinfo {year} {2016})},\ \bibinfo {note} {publisher: Nature
  Publishing Group}\BibitemShut {NoStop}%
\bibitem [{\citenamefont {Alexandrov}\ \emph {et~al.}(2015)\citenamefont
  {Alexandrov}, \citenamefont {Coleman},\ and\ \citenamefont
  {Erten}}]{alexandrov_kondo_2015}%
  \BibitemOpen
  \bibfield  {author} {\bibinfo {author} {\bibfnamefont {V.}~\bibnamefont
  {Alexandrov}}, \bibinfo {author} {\bibfnamefont {P.}~\bibnamefont
  {Coleman}},\ and\ \bibinfo {author} {\bibfnamefont {O.}~\bibnamefont
  {Erten}},\ }\href {https://doi.org/10.1103/PhysRevLett.114.177202} {\bibfield
   {journal} {\bibinfo  {journal} {Physical Review Letters}\ }\textbf {\bibinfo
  {volume} {114}},\ \bibinfo {pages} {177202} (\bibinfo {year} {2015})},\
  \bibinfo {note} {publisher: American Physical Society}\BibitemShut {NoStop}%
\bibitem [{\citenamefont {Kofuji}\ \emph {et~al.}(2021)\citenamefont {Kofuji},
  \citenamefont {Michishita},\ and\ \citenamefont
  {Peters}}]{kofuji_effects_2021}%
  \BibitemOpen
  \bibfield  {author} {\bibinfo {author} {\bibfnamefont {A.}~\bibnamefont
  {Kofuji}}, \bibinfo {author} {\bibfnamefont {Y.}~\bibnamefont {Michishita}},\
  and\ \bibinfo {author} {\bibfnamefont {R.}~\bibnamefont {Peters}},\ }\href
  {https://doi.org/10.1103/PhysRevB.104.085151} {\bibfield  {journal} {\bibinfo
   {journal} {Physical Review B}\ }\textbf {\bibinfo {volume} {104}},\ \bibinfo
  {pages} {085151} (\bibinfo {year} {2021})},\ \bibinfo {note} {publisher:
  American Physical Society}\BibitemShut {NoStop}%
\bibitem [{\citenamefont {Dzsaber}\ \emph {et~al.}(2021)\citenamefont
  {Dzsaber}, \citenamefont {Yan}, \citenamefont {Taupin}, \citenamefont
  {Eguchi}, \citenamefont {Prokofiev}, \citenamefont {Shiroka}, \citenamefont
  {Blaha}, \citenamefont {Rubel}, \citenamefont {Grefe}, \citenamefont {Lai},
  \citenamefont {Si},\ and\ \citenamefont {Paschen}}]{dzsaber_giant_2021}%
  \BibitemOpen
  \bibfield  {author} {\bibinfo {author} {\bibfnamefont {S.}~\bibnamefont
  {Dzsaber}}, \bibinfo {author} {\bibfnamefont {X.}~\bibnamefont {Yan}},
  \bibinfo {author} {\bibfnamefont {M.}~\bibnamefont {Taupin}}, \bibinfo
  {author} {\bibfnamefont {G.}~\bibnamefont {Eguchi}}, \bibinfo {author}
  {\bibfnamefont {A.}~\bibnamefont {Prokofiev}}, \bibinfo {author}
  {\bibfnamefont {T.}~\bibnamefont {Shiroka}}, \bibinfo {author} {\bibfnamefont
  {P.}~\bibnamefont {Blaha}}, \bibinfo {author} {\bibfnamefont
  {O.}~\bibnamefont {Rubel}}, \bibinfo {author} {\bibfnamefont {S.~E.}\
  \bibnamefont {Grefe}}, \bibinfo {author} {\bibfnamefont {H.-H.}\ \bibnamefont
  {Lai}}, \bibinfo {author} {\bibfnamefont {Q.}~\bibnamefont {Si}},\ and\
  \bibinfo {author} {\bibfnamefont {S.}~\bibnamefont {Paschen}},\ }\href
  {https://doi.org/10.1073/pnas.2013386118} {\bibfield  {journal} {\bibinfo
  {journal} {Proceedings of the National Academy of Sciences}\ }\textbf
  {\bibinfo {volume} {118}},\ \bibinfo {pages} {e2013386118} (\bibinfo {year}
  {2021})},\ \bibinfo {note} {publisher: Proceedings of the National Academy of
  Sciences}\BibitemShut {NoStop}%
\bibitem [{\citenamefont {Grefe}\ \emph {et~al.}(2020)\citenamefont {Grefe},
  \citenamefont {Lai}, \citenamefont {Paschen},\ and\ \citenamefont
  {Si}}]{grefe_weyl-kondo_2020}%
  \BibitemOpen
  \bibfield  {author} {\bibinfo {author} {\bibfnamefont {S.~E.}\ \bibnamefont
  {Grefe}}, \bibinfo {author} {\bibfnamefont {H.-H.}\ \bibnamefont {Lai}},
  \bibinfo {author} {\bibfnamefont {S.}~\bibnamefont {Paschen}},\ and\ \bibinfo
  {author} {\bibfnamefont {Q.}~\bibnamefont {Si}},\ }\href
  {https://doi.org/10.1103/PhysRevB.101.075138} {\bibfield  {journal} {\bibinfo
   {journal} {Physical Review B}\ }\textbf {\bibinfo {volume} {101}},\ \bibinfo
  {pages} {075138} (\bibinfo {year} {2020})},\ \bibinfo {note} {publisher:
  American Physical Society}\BibitemShut {NoStop}%
\bibitem [{\citenamefont {Lai}\ \emph {et~al.}(2018)\citenamefont {Lai},
  \citenamefont {Grefe}, \citenamefont {Paschen},\ and\ \citenamefont
  {Si}}]{lai_weylkondo_2018}%
  \BibitemOpen
  \bibfield  {author} {\bibinfo {author} {\bibfnamefont {H.-H.}\ \bibnamefont
  {Lai}}, \bibinfo {author} {\bibfnamefont {S.~E.}\ \bibnamefont {Grefe}},
  \bibinfo {author} {\bibfnamefont {S.}~\bibnamefont {Paschen}},\ and\ \bibinfo
  {author} {\bibfnamefont {Q.}~\bibnamefont {Si}},\ }\href
  {https://doi.org/10.1073/pnas.1715851115} {\bibfield  {journal} {\bibinfo
  {journal} {Proceedings of the National Academy of Sciences}\ }\textbf
  {\bibinfo {volume} {115}},\ \bibinfo {pages} {93} (\bibinfo {year} {2018})},\
  \bibinfo {note} {publisher: Proceedings of the National Academy of
  Sciences}\BibitemShut {NoStop}%
\bibitem [{\citenamefont {Chen}\ \emph {et~al.}(2022)\citenamefont {Chen},
  \citenamefont {Setty}, \citenamefont {Hu}, \citenamefont {Vergniory},
  \citenamefont {Grefe}, \citenamefont {Fischer}, \citenamefont {Yan},
  \citenamefont {Eguchi}, \citenamefont {Prokofiev}, \citenamefont {Paschen},
  \citenamefont {Cano},\ and\ \citenamefont {Si}}]{chen_topological_2022}%
  \BibitemOpen
  \bibfield  {author} {\bibinfo {author} {\bibfnamefont {L.}~\bibnamefont
  {Chen}}, \bibinfo {author} {\bibfnamefont {C.}~\bibnamefont {Setty}},
  \bibinfo {author} {\bibfnamefont {H.}~\bibnamefont {Hu}}, \bibinfo {author}
  {\bibfnamefont {M.~G.}\ \bibnamefont {Vergniory}}, \bibinfo {author}
  {\bibfnamefont {S.~E.}\ \bibnamefont {Grefe}}, \bibinfo {author}
  {\bibfnamefont {L.}~\bibnamefont {Fischer}}, \bibinfo {author} {\bibfnamefont
  {X.}~\bibnamefont {Yan}}, \bibinfo {author} {\bibfnamefont {G.}~\bibnamefont
  {Eguchi}}, \bibinfo {author} {\bibfnamefont {A.}~\bibnamefont {Prokofiev}},
  \bibinfo {author} {\bibfnamefont {S.}~\bibnamefont {Paschen}}, \bibinfo
  {author} {\bibfnamefont {J.}~\bibnamefont {Cano}},\ and\ \bibinfo {author}
  {\bibfnamefont {Q.}~\bibnamefont {Si}},\ }\href
  {https://doi.org/10.1038/s41567-022-01743-4} {\bibfield  {journal} {\bibinfo
  {journal} {Nature Physics}\ }\textbf {\bibinfo {volume} {18}},\ \bibinfo
  {pages} {1341} (\bibinfo {year} {2022})},\ \bibinfo {note} {publisher: Nature
  Publishing Group}\BibitemShut {NoStop}%
\bibitem [{\citenamefont {Chen}\ \emph {et~al.}(2019)\citenamefont {Chen},
  \citenamefont {Luo}, \citenamefont {Xie}, \citenamefont {Li}, \citenamefont
  {Ji}, \citenamefont {Zhou}, \citenamefont {Huang}, \citenamefont {Zhang},
  \citenamefont {Feng}, \citenamefont {Zhang}, \citenamefont {Huang},
  \citenamefont {Hao}, \citenamefont {Liu}, \citenamefont {Zhu}, \citenamefont
  {Liu}, \citenamefont {Zhang}, \citenamefont {Lai}, \citenamefont {Si},\ and\
  \citenamefont {Tan}}]{PhysRevLett.123.106402}%
  \BibitemOpen
  \bibfield  {author} {\bibinfo {author} {\bibfnamefont {Q.~Y.}\ \bibnamefont
  {Chen}}, \bibinfo {author} {\bibfnamefont {X.~B.}\ \bibnamefont {Luo}},
  \bibinfo {author} {\bibfnamefont {D.~H.}\ \bibnamefont {Xie}}, \bibinfo
  {author} {\bibfnamefont {M.~L.}\ \bibnamefont {Li}}, \bibinfo {author}
  {\bibfnamefont {X.~Y.}\ \bibnamefont {Ji}}, \bibinfo {author} {\bibfnamefont
  {R.}~\bibnamefont {Zhou}}, \bibinfo {author} {\bibfnamefont {Y.~B.}\
  \bibnamefont {Huang}}, \bibinfo {author} {\bibfnamefont {W.}~\bibnamefont
  {Zhang}}, \bibinfo {author} {\bibfnamefont {W.}~\bibnamefont {Feng}},
  \bibinfo {author} {\bibfnamefont {Y.}~\bibnamefont {Zhang}}, \bibinfo
  {author} {\bibfnamefont {L.}~\bibnamefont {Huang}}, \bibinfo {author}
  {\bibfnamefont {Q.~Q.}\ \bibnamefont {Hao}}, \bibinfo {author} {\bibfnamefont
  {Q.}~\bibnamefont {Liu}}, \bibinfo {author} {\bibfnamefont {X.~G.}\
  \bibnamefont {Zhu}}, \bibinfo {author} {\bibfnamefont {Y.}~\bibnamefont
  {Liu}}, \bibinfo {author} {\bibfnamefont {P.}~\bibnamefont {Zhang}}, \bibinfo
  {author} {\bibfnamefont {X.~C.}\ \bibnamefont {Lai}}, \bibinfo {author}
  {\bibfnamefont {Q.}~\bibnamefont {Si}},\ and\ \bibinfo {author}
  {\bibfnamefont {S.~Y.}\ \bibnamefont {Tan}},\ }\href
  {https://doi.org/10.1103/PhysRevLett.123.106402} {\bibfield  {journal}
  {\bibinfo  {journal} {Phys. Rev. Lett.}\ }\textbf {\bibinfo {volume} {123}},\
  \bibinfo {pages} {106402} (\bibinfo {year} {2019})}\BibitemShut {NoStop}%
\bibitem [{\citenamefont {Ji}\ \emph {et~al.}(2022)\citenamefont {Ji},
  \citenamefont {Luo}, \citenamefont {Chen}, \citenamefont {Feng},
  \citenamefont {Hao}, \citenamefont {Liu}, \citenamefont {Zhang},
  \citenamefont {Liu}, \citenamefont {Wang}, \citenamefont {Tan},\ and\
  \citenamefont {Lai}}]{PhysRevB.106.125120}%
  \BibitemOpen
  \bibfield  {author} {\bibinfo {author} {\bibfnamefont {X.}~\bibnamefont
  {Ji}}, \bibinfo {author} {\bibfnamefont {X.}~\bibnamefont {Luo}}, \bibinfo
  {author} {\bibfnamefont {Q.}~\bibnamefont {Chen}}, \bibinfo {author}
  {\bibfnamefont {W.}~\bibnamefont {Feng}}, \bibinfo {author} {\bibfnamefont
  {Q.}~\bibnamefont {Hao}}, \bibinfo {author} {\bibfnamefont {Q.}~\bibnamefont
  {Liu}}, \bibinfo {author} {\bibfnamefont {Y.}~\bibnamefont {Zhang}}, \bibinfo
  {author} {\bibfnamefont {Y.}~\bibnamefont {Liu}}, \bibinfo {author}
  {\bibfnamefont {X.}~\bibnamefont {Wang}}, \bibinfo {author} {\bibfnamefont
  {S.}~\bibnamefont {Tan}},\ and\ \bibinfo {author} {\bibfnamefont
  {X.}~\bibnamefont {Lai}},\ }\href
  {https://doi.org/10.1103/PhysRevB.106.125120} {\bibfield  {journal} {\bibinfo
   {journal} {Phys. Rev. B}\ }\textbf {\bibinfo {volume} {106}},\ \bibinfo
  {pages} {125120} (\bibinfo {year} {2022})}\BibitemShut {NoStop}%
\bibitem [{\citenamefont {Li}\ \emph {et~al.}(2023)\citenamefont {Li},
  \citenamefont {Ye}, \citenamefont {Hu}, \citenamefont {Fang}, \citenamefont
  {Xiao}, \citenamefont {Wu}, \citenamefont {Shan}, \citenamefont {Singh},
  \citenamefont {Balakrishnan}, \citenamefont {Shen}, \citenamefont {Yang},
  \citenamefont {Cao}, \citenamefont {Plumb}, \citenamefont {Smidman},
  \citenamefont {Shi}, \citenamefont {Kroha}, \citenamefont {Yuan},
  \citenamefont {Steglich},\ and\ \citenamefont {Liu}}]{PhysRevB.107.L201104}%
  \BibitemOpen
  \bibfield  {author} {\bibinfo {author} {\bibfnamefont {P.}~\bibnamefont
  {Li}}, \bibinfo {author} {\bibfnamefont {H.}~\bibnamefont {Ye}}, \bibinfo
  {author} {\bibfnamefont {Y.}~\bibnamefont {Hu}}, \bibinfo {author}
  {\bibfnamefont {Y.}~\bibnamefont {Fang}}, \bibinfo {author} {\bibfnamefont
  {Z.}~\bibnamefont {Xiao}}, \bibinfo {author} {\bibfnamefont {Z.}~\bibnamefont
  {Wu}}, \bibinfo {author} {\bibfnamefont {Z.}~\bibnamefont {Shan}}, \bibinfo
  {author} {\bibfnamefont {R.~P.}\ \bibnamefont {Singh}}, \bibinfo {author}
  {\bibfnamefont {G.}~\bibnamefont {Balakrishnan}}, \bibinfo {author}
  {\bibfnamefont {D.}~\bibnamefont {Shen}}, \bibinfo {author} {\bibfnamefont
  {Y.-f.}\ \bibnamefont {Yang}}, \bibinfo {author} {\bibfnamefont
  {C.}~\bibnamefont {Cao}}, \bibinfo {author} {\bibfnamefont {N.~C.}\
  \bibnamefont {Plumb}}, \bibinfo {author} {\bibfnamefont {M.}~\bibnamefont
  {Smidman}}, \bibinfo {author} {\bibfnamefont {M.}~\bibnamefont {Shi}},
  \bibinfo {author} {\bibfnamefont {J.}~\bibnamefont {Kroha}}, \bibinfo
  {author} {\bibfnamefont {H.}~\bibnamefont {Yuan}}, \bibinfo {author}
  {\bibfnamefont {F.}~\bibnamefont {Steglich}},\ and\ \bibinfo {author}
  {\bibfnamefont {Y.}~\bibnamefont {Liu}},\ }\href
  {https://doi.org/10.1103/PhysRevB.107.L201104} {\bibfield  {journal}
  {\bibinfo  {journal} {Phys. Rev. B}\ }\textbf {\bibinfo {volume} {107}},\
  \bibinfo {pages} {L201104} (\bibinfo {year} {2023})}\BibitemShut {NoStop}%
\bibitem [{\citenamefont {Zhang}\ \emph {et~al.}(2023)\citenamefont {Zhang},
  \citenamefont {Zeng}, \citenamefont {Xu}, \citenamefont {Chen},\ and\
  \citenamefont {Ji}}]{PhysRevB.107.195440}%
  \BibitemOpen
  \bibfield  {author} {\bibinfo {author} {\bibfnamefont {S.-K.}\ \bibnamefont
  {Zhang}}, \bibinfo {author} {\bibfnamefont {Z.-Y.}\ \bibnamefont {Zeng}},
  \bibinfo {author} {\bibfnamefont {Y.-J.}\ \bibnamefont {Xu}}, \bibinfo
  {author} {\bibfnamefont {X.-R.}\ \bibnamefont {Chen}},\ and\ \bibinfo
  {author} {\bibfnamefont {G.-F.}\ \bibnamefont {Ji}},\ }\href
  {https://doi.org/10.1103/PhysRevB.107.195440} {\bibfield  {journal} {\bibinfo
   {journal} {Phys. Rev. B}\ }\textbf {\bibinfo {volume} {107}},\ \bibinfo
  {pages} {195440} (\bibinfo {year} {2023})}\BibitemShut {NoStop}%
\bibitem [{\citenamefont {Mielke}(1991)}]{mielke_ferromagnetism_1991}%
  \BibitemOpen
  \bibfield  {author} {\bibinfo {author} {\bibfnamefont {A.}~\bibnamefont
  {Mielke}},\ }\href {https://doi.org/10.1088/0305-4470/24/14/018} {\bibfield
  {journal} {\bibinfo  {journal} {Journal of Physics A: Mathematical and
  General}\ }\textbf {\bibinfo {volume} {24}},\ \bibinfo {pages} {3311}
  (\bibinfo {year} {1991})}\BibitemShut {NoStop}%
\bibitem [{\citenamefont {Tasaki}(1992)}]{tasaki_ferromagnetism_1992}%
  \BibitemOpen
  \bibfield  {author} {\bibinfo {author} {\bibfnamefont {H.}~\bibnamefont
  {Tasaki}},\ }\href {https://doi.org/10.1103/PhysRevLett.69.1608} {\bibfield
  {journal} {\bibinfo  {journal} {Physical Review Letters}\ }\textbf {\bibinfo
  {volume} {69}},\ \bibinfo {pages} {1608} (\bibinfo {year} {1992})},\ \bibinfo
  {note} {publisher: American Physical Society}\BibitemShut {NoStop}%
\bibitem [{\citenamefont {Lian}\ \emph {et~al.}(2021)\citenamefont {Lian},
  \citenamefont {Song}, \citenamefont {Regnault}, \citenamefont {Efetov},
  \citenamefont {Yazdani},\ and\ \citenamefont {Bernevig}}]{lian_twisted_2021}%
  \BibitemOpen
  \bibfield  {author} {\bibinfo {author} {\bibfnamefont {B.}~\bibnamefont
  {Lian}}, \bibinfo {author} {\bibfnamefont {Z.-D.}\ \bibnamefont {Song}},
  \bibinfo {author} {\bibfnamefont {N.}~\bibnamefont {Regnault}}, \bibinfo
  {author} {\bibfnamefont {D.~K.}\ \bibnamefont {Efetov}}, \bibinfo {author}
  {\bibfnamefont {A.}~\bibnamefont {Yazdani}},\ and\ \bibinfo {author}
  {\bibfnamefont {B.~A.}\ \bibnamefont {Bernevig}},\ }\href
  {https://doi.org/10.1103/PhysRevB.103.205414} {\bibfield  {journal} {\bibinfo
   {journal} {Physical Review B}\ }\textbf {\bibinfo {volume} {103}},\ \bibinfo
  {pages} {205414} (\bibinfo {year} {2021})}\BibitemShut {NoStop}%
\bibitem [{\citenamefont {Auerbach}(2012)}]{auerbach2012interacting}%
  \BibitemOpen
  \bibfield  {author} {\bibinfo {author} {\bibfnamefont {A.}~\bibnamefont
  {Auerbach}},\ }\href@noop {} {\emph {\bibinfo {title} {Interacting electrons
  and quantum magnetism}}}\ (\bibinfo  {publisher} {Springer Science \&
  Business Media},\ \bibinfo {year} {2012})\BibitemShut {NoStop}%
\bibitem [{\citenamefont {Tian}\ \emph {et~al.}(2017)\citenamefont {Tian},
  \citenamefont {Kong}, \citenamefont {Liu}, \citenamefont {Zhang},
  \citenamefont {He},\ and\ \citenamefont {Zhang}}]{TIAN201775}%
  \BibitemOpen
  \bibfield  {author} {\bibinfo {author} {\bibfnamefont {Y.}~\bibnamefont
  {Tian}}, \bibinfo {author} {\bibfnamefont {Y.}~\bibnamefont {Kong}}, \bibinfo
  {author} {\bibfnamefont {K.}~\bibnamefont {Liu}}, \bibinfo {author}
  {\bibfnamefont {A.}~\bibnamefont {Zhang}}, \bibinfo {author} {\bibfnamefont
  {R.}~\bibnamefont {He}},\ and\ \bibinfo {author} {\bibfnamefont
  {Q.}~\bibnamefont {Zhang}},\ }\href
  {https://doi.org/https://doi.org/10.1016/j.physb.2017.02.016} {\bibfield
  {journal} {\bibinfo  {journal} {Physica B: Condensed Matter}\ }\textbf
  {\bibinfo {volume} {512}},\ \bibinfo {pages} {75} (\bibinfo {year}
  {2017})}\BibitemShut {NoStop}%
\bibitem [{\citenamefont {Coleman}(2015)}]{coleman2015introduction}%
  \BibitemOpen
  \bibfield  {author} {\bibinfo {author} {\bibfnamefont {P.}~\bibnamefont
  {Coleman}},\ }\href@noop {} {\emph {\bibinfo {title} {Introduction to
  many-body physics}}}\ (\bibinfo  {publisher} {Cambridge University Press},\
  \bibinfo {year} {2015})\BibitemShut {NoStop}%
\bibitem [{\citenamefont {Coleman}(1987)}]{PhysRevB.35.5072}%
  \BibitemOpen
  \bibfield  {author} {\bibinfo {author} {\bibfnamefont {P.}~\bibnamefont
  {Coleman}},\ }\href {https://doi.org/10.1103/PhysRevB.35.5072} {\bibfield
  {journal} {\bibinfo  {journal} {Phys. Rev. B}\ }\textbf {\bibinfo {volume}
  {35}},\ \bibinfo {pages} {5072} (\bibinfo {year} {1987})}\BibitemShut
  {NoStop}%
\bibitem [{\citenamefont {Schrieffer}\ and\ \citenamefont
  {Wolff}(1966)}]{PhysRev.149.491}%
  \BibitemOpen
  \bibfield  {author} {\bibinfo {author} {\bibfnamefont {J.~R.}\ \bibnamefont
  {Schrieffer}}\ and\ \bibinfo {author} {\bibfnamefont {P.~A.}\ \bibnamefont
  {Wolff}},\ }\href {https://doi.org/10.1103/PhysRev.149.491} {\bibfield
  {journal} {\bibinfo  {journal} {Phys. Rev.}\ }\textbf {\bibinfo {volume}
  {149}},\ \bibinfo {pages} {491} (\bibinfo {year} {1966})}\BibitemShut
  {NoStop}%
\bibitem [{\citenamefont {Kotliar}\ \emph {et~al.}(2006)\citenamefont
  {Kotliar}, \citenamefont {Savrasov}, \citenamefont {Haule}, \citenamefont
  {Oudovenko}, \citenamefont {Parcollet},\ and\ \citenamefont
  {Marianetti}}]{RevModPhys.78.865}%
  \BibitemOpen
  \bibfield  {author} {\bibinfo {author} {\bibfnamefont {G.}~\bibnamefont
  {Kotliar}}, \bibinfo {author} {\bibfnamefont {S.~Y.}\ \bibnamefont
  {Savrasov}}, \bibinfo {author} {\bibfnamefont {K.}~\bibnamefont {Haule}},
  \bibinfo {author} {\bibfnamefont {V.~S.}\ \bibnamefont {Oudovenko}}, \bibinfo
  {author} {\bibfnamefont {O.}~\bibnamefont {Parcollet}},\ and\ \bibinfo
  {author} {\bibfnamefont {C.~A.}\ \bibnamefont {Marianetti}},\ }\href
  {https://doi.org/10.1103/RevModPhys.78.865} {\bibfield  {journal} {\bibinfo
  {journal} {Rev. Mod. Phys.}\ }\textbf {\bibinfo {volume} {78}},\ \bibinfo
  {pages} {865} (\bibinfo {year} {2006})}\BibitemShut {NoStop}%
\bibitem [{\citenamefont {Georges}\ \emph {et~al.}(1996)\citenamefont
  {Georges}, \citenamefont {Kotliar}, \citenamefont {Krauth},\ and\
  \citenamefont {Rozenberg}}]{RevModPhys.68.13}%
  \BibitemOpen
  \bibfield  {author} {\bibinfo {author} {\bibfnamefont {A.}~\bibnamefont
  {Georges}}, \bibinfo {author} {\bibfnamefont {G.}~\bibnamefont {Kotliar}},
  \bibinfo {author} {\bibfnamefont {W.}~\bibnamefont {Krauth}},\ and\ \bibinfo
  {author} {\bibfnamefont {M.~J.}\ \bibnamefont {Rozenberg}},\ }\href
  {https://doi.org/10.1103/RevModPhys.68.13} {\bibfield  {journal} {\bibinfo
  {journal} {Rev. Mod. Phys.}\ }\textbf {\bibinfo {volume} {68}},\ \bibinfo
  {pages} {13} (\bibinfo {year} {1996})}\BibitemShut {NoStop}%
\bibitem [{\citenamefont {Haule}\ and\ \citenamefont
  {Birol}(2015)}]{PhysRevLett.115.256402}%
  \BibitemOpen
  \bibfield  {author} {\bibinfo {author} {\bibfnamefont {K.}~\bibnamefont
  {Haule}}\ and\ \bibinfo {author} {\bibfnamefont {T.}~\bibnamefont {Birol}},\
  }\href {https://doi.org/10.1103/PhysRevLett.115.256402} {\bibfield  {journal}
  {\bibinfo  {journal} {Phys. Rev. Lett.}\ }\textbf {\bibinfo {volume} {115}},\
  \bibinfo {pages} {256402} (\bibinfo {year} {2015})}\BibitemShut {NoStop}%
\bibitem [{\citenamefont {Haule}(2015)}]{PhysRevLett.115.196403}%
  \BibitemOpen
  \bibfield  {author} {\bibinfo {author} {\bibfnamefont {K.}~\bibnamefont
  {Haule}},\ }\href {https://doi.org/10.1103/PhysRevLett.115.196403} {\bibfield
   {journal} {\bibinfo  {journal} {Phys. Rev. Lett.}\ }\textbf {\bibinfo
  {volume} {115}},\ \bibinfo {pages} {196403} (\bibinfo {year}
  {2015})}\BibitemShut {NoStop}%
\bibitem [{\citenamefont {Fang}\ \emph {et~al.}(2015)\citenamefont {Fang},
  \citenamefont {Chen}, \citenamefont {Kee},\ and\ \citenamefont
  {Fu}}]{fang_topological_2015}%
  \BibitemOpen
  \bibfield  {author} {\bibinfo {author} {\bibfnamefont {C.}~\bibnamefont
  {Fang}}, \bibinfo {author} {\bibfnamefont {Y.}~\bibnamefont {Chen}}, \bibinfo
  {author} {\bibfnamefont {H.-Y.}\ \bibnamefont {Kee}},\ and\ \bibinfo {author}
  {\bibfnamefont {L.}~\bibnamefont {Fu}},\ }\href
  {https://doi.org/10.1103/PhysRevB.92.081201} {\bibfield  {journal} {\bibinfo
  {journal} {Physical Review B}\ }\textbf {\bibinfo {volume} {92}},\ \bibinfo
  {pages} {081201} (\bibinfo {year} {2015})},\ \bibinfo {note} {publisher:
  American Physical Society}\BibitemShut {NoStop}%
\bibitem [{\citenamefont {Fang}\ \emph {et~al.}(2016)\citenamefont {Fang},
  \citenamefont {Weng}, \citenamefont {Dai},\ and\ \citenamefont
  {Fang}}]{fang_topological_2016}%
  \BibitemOpen
  \bibfield  {author} {\bibinfo {author} {\bibfnamefont {C.}~\bibnamefont
  {Fang}}, \bibinfo {author} {\bibfnamefont {H.}~\bibnamefont {Weng}}, \bibinfo
  {author} {\bibfnamefont {X.}~\bibnamefont {Dai}},\ and\ \bibinfo {author}
  {\bibfnamefont {Z.}~\bibnamefont {Fang}},\ }\href
  {https://doi.org/10.1088/1674-1056/25/11/117106} {\bibfield  {journal}
  {\bibinfo  {journal} {Chinese Physics B}\ }\textbf {\bibinfo {volume} {25}},\
  \bibinfo {pages} {117106} (\bibinfo {year} {2016})},\ \bibinfo {note}
  {publisher: IOP Publishing}\BibitemShut {NoStop}%
\bibitem [{\citenamefont {Yu}\ \emph {et~al.}(2017)\citenamefont {Yu},
  \citenamefont {Fang}, \citenamefont {Dai},\ and\ \citenamefont
  {Weng}}]{yu_topological_2017}%
  \BibitemOpen
  \bibfield  {author} {\bibinfo {author} {\bibfnamefont {R.}~\bibnamefont
  {Yu}}, \bibinfo {author} {\bibfnamefont {Z.}~\bibnamefont {Fang}}, \bibinfo
  {author} {\bibfnamefont {X.}~\bibnamefont {Dai}},\ and\ \bibinfo {author}
  {\bibfnamefont {H.}~\bibnamefont {Weng}},\ }\href
  {https://doi.org/10.1007/s11467-016-0630-1} {\bibfield  {journal} {\bibinfo
  {journal} {Frontiers of Physics}\ }\textbf {\bibinfo {volume} {12}},\
  \bibinfo {pages} {127202} (\bibinfo {year} {2017})}\BibitemShut {NoStop}%
\bibitem [{\citenamefont {Xu}\ \emph {et~al.}(2024)\citenamefont {Xu},
  \citenamefont {Elcoro}, \citenamefont {Song}, \citenamefont {Vergniory},
  \citenamefont {Felser}, \citenamefont {Parkin}, \citenamefont {Regnault},
  \citenamefont {Ma{\~n}es},\ and\ \citenamefont
  {Bernevig}}]{xu_filling-enforced_2021}%
  \BibitemOpen
  \bibfield  {author} {\bibinfo {author} {\bibfnamefont {Y.}~\bibnamefont
  {Xu}}, \bibinfo {author} {\bibfnamefont {L.}~\bibnamefont {Elcoro}}, \bibinfo
  {author} {\bibfnamefont {Z.-D.}\ \bibnamefont {Song}}, \bibinfo {author}
  {\bibfnamefont {M.}~\bibnamefont {Vergniory}}, \bibinfo {author}
  {\bibfnamefont {C.}~\bibnamefont {Felser}}, \bibinfo {author} {\bibfnamefont
  {S.~S.}\ \bibnamefont {Parkin}}, \bibinfo {author} {\bibfnamefont
  {N.}~\bibnamefont {Regnault}}, \bibinfo {author} {\bibfnamefont {J.~L.}\
  \bibnamefont {Ma{\~n}es}},\ and\ \bibinfo {author} {\bibfnamefont {B.~A.}\
  \bibnamefont {Bernevig}},\ }\href@noop {} {\bibfield  {journal} {\bibinfo
  {journal} {Physical Review B}\ }\textbf {\bibinfo {volume} {109}},\ \bibinfo
  {pages} {165139} (\bibinfo {year} {2024})}\BibitemShut {NoStop}%
\bibitem [{\citenamefont {Xu}\ \emph {et~al.}(2021{\natexlab{a}})\citenamefont
  {Xu}, \citenamefont {Elcoro}, \citenamefont {Li}, \citenamefont {Song},
  \citenamefont {Regnault}, \citenamefont {Yang}, \citenamefont {Sun},
  \citenamefont {Parkin}, \citenamefont {Felser},\ and\ \citenamefont
  {Bernevig}}]{xu2021three}%
  \BibitemOpen
  \bibfield  {author} {\bibinfo {author} {\bibfnamefont {Y.}~\bibnamefont
  {Xu}}, \bibinfo {author} {\bibfnamefont {L.}~\bibnamefont {Elcoro}}, \bibinfo
  {author} {\bibfnamefont {G.}~\bibnamefont {Li}}, \bibinfo {author}
  {\bibfnamefont {Z.-D.}\ \bibnamefont {Song}}, \bibinfo {author}
  {\bibfnamefont {N.}~\bibnamefont {Regnault}}, \bibinfo {author}
  {\bibfnamefont {Q.}~\bibnamefont {Yang}}, \bibinfo {author} {\bibfnamefont
  {Y.}~\bibnamefont {Sun}}, \bibinfo {author} {\bibfnamefont {S.}~\bibnamefont
  {Parkin}}, \bibinfo {author} {\bibfnamefont {C.}~\bibnamefont {Felser}},\
  and\ \bibinfo {author} {\bibfnamefont {B.~A.}\ \bibnamefont {Bernevig}},\
  }\href@noop {} {\bibfield  {journal} {\bibinfo  {journal} {arXiv preprint
  arXiv:2111.02433}\ } (\bibinfo {year} {2021}{\natexlab{a}})}\BibitemShut
  {NoStop}%
\bibitem [{\citenamefont {Song}\ \emph {et~al.}(2020)\citenamefont {Song},
  \citenamefont {Elcoro},\ and\ \citenamefont {Bernevig}}]{song2020twisted}%
  \BibitemOpen
  \bibfield  {author} {\bibinfo {author} {\bibfnamefont {Z.-D.}\ \bibnamefont
  {Song}}, \bibinfo {author} {\bibfnamefont {L.}~\bibnamefont {Elcoro}},\ and\
  \bibinfo {author} {\bibfnamefont {B.~A.}\ \bibnamefont {Bernevig}},\
  }\href@noop {} {\bibfield  {journal} {\bibinfo  {journal} {Science}\ }\textbf
  {\bibinfo {volume} {367}},\ \bibinfo {pages} {794} (\bibinfo {year}
  {2020})}\BibitemShut {NoStop}%
\bibitem [{\citenamefont {Benalcazar}\ \emph
  {et~al.}(2017{\natexlab{c}})\citenamefont {Benalcazar}, \citenamefont
  {Bernevig},\ and\ \citenamefont {Hughes}}]{benalcazar2017quantized}%
  \BibitemOpen
  \bibfield  {author} {\bibinfo {author} {\bibfnamefont {W.~A.}\ \bibnamefont
  {Benalcazar}}, \bibinfo {author} {\bibfnamefont {B.~A.}\ \bibnamefont
  {Bernevig}},\ and\ \bibinfo {author} {\bibfnamefont {T.~L.}\ \bibnamefont
  {Hughes}},\ }\href@noop {} {\bibfield  {journal} {\bibinfo  {journal}
  {Science}\ }\textbf {\bibinfo {volume} {357}},\ \bibinfo {pages} {61}
  (\bibinfo {year} {2017}{\natexlab{c}})}\BibitemShut {NoStop}%
\bibitem [{\citenamefont {Song}\ \emph {et~al.}(2017)\citenamefont {Song},
  \citenamefont {Fang},\ and\ \citenamefont {Fang}}]{song2017d}%
  \BibitemOpen
  \bibfield  {author} {\bibinfo {author} {\bibfnamefont {Z.}~\bibnamefont
  {Song}}, \bibinfo {author} {\bibfnamefont {Z.}~\bibnamefont {Fang}},\ and\
  \bibinfo {author} {\bibfnamefont {C.}~\bibnamefont {Fang}},\ }\href@noop {}
  {\bibfield  {journal} {\bibinfo  {journal} {Physical review letters}\
  }\textbf {\bibinfo {volume} {119}},\ \bibinfo {pages} {246402} (\bibinfo
  {year} {2017})}\BibitemShut {NoStop}%
\bibitem [{\citenamefont {Elcoro}\ \emph {et~al.}(2017)\citenamefont {Elcoro},
  \citenamefont {Bradlyn}, \citenamefont {Wang}, \citenamefont {Vergniory},
  \citenamefont {Cano}, \citenamefont {Felser}, \citenamefont {Bernevig},
  \citenamefont {Orobengoa}, \citenamefont {de~la Flor},\ and\ \citenamefont
  {Aroyo}}]{ELC17}%
  \BibitemOpen
  \bibfield  {author} {\bibinfo {author} {\bibfnamefont {L.}~\bibnamefont
  {Elcoro}}, \bibinfo {author} {\bibfnamefont {B.}~\bibnamefont {Bradlyn}},
  \bibinfo {author} {\bibfnamefont {Z.}~\bibnamefont {Wang}}, \bibinfo {author}
  {\bibfnamefont {M.~G.}\ \bibnamefont {Vergniory}}, \bibinfo {author}
  {\bibfnamefont {J.}~\bibnamefont {Cano}}, \bibinfo {author} {\bibfnamefont
  {C.}~\bibnamefont {Felser}}, \bibinfo {author} {\bibfnamefont {B.~A.}\
  \bibnamefont {Bernevig}}, \bibinfo {author} {\bibfnamefont {D.}~\bibnamefont
  {Orobengoa}}, \bibinfo {author} {\bibfnamefont {G.}~\bibnamefont {de~la
  Flor}},\ and\ \bibinfo {author} {\bibfnamefont {M.~I.}\ \bibnamefont
  {Aroyo}},\ }\href {https://doi.org/10.1107/S1600576717011712} {\bibfield
  {journal} {\bibinfo  {journal} {J. Appl. Cryst.}\ }\textbf {\bibinfo {volume}
  {50}},\ \bibinfo {pages} {1457} (\bibinfo {year} {2017})}\BibitemShut
  {NoStop}%
\bibitem [{\citenamefont {Vergniory}\ \emph {et~al.}(2017)\citenamefont
  {Vergniory}, \citenamefont {Elcoro}, \citenamefont {Wang}, \citenamefont
  {Cano}, \citenamefont {Felser}, \citenamefont {Aroyo}, \citenamefont
  {Bernevig},\ and\ \citenamefont {Bradlyn}}]{VER17}%
  \BibitemOpen
  \bibfield  {author} {\bibinfo {author} {\bibfnamefont {M.~G.}\ \bibnamefont
  {Vergniory}}, \bibinfo {author} {\bibfnamefont {L.}~\bibnamefont {Elcoro}},
  \bibinfo {author} {\bibfnamefont {Z.}~\bibnamefont {Wang}}, \bibinfo {author}
  {\bibfnamefont {J.}~\bibnamefont {Cano}}, \bibinfo {author} {\bibfnamefont
  {C.}~\bibnamefont {Felser}}, \bibinfo {author} {\bibfnamefont {M.~I.}\
  \bibnamefont {Aroyo}}, \bibinfo {author} {\bibfnamefont {B.~A.}\ \bibnamefont
  {Bernevig}},\ and\ \bibinfo {author} {\bibfnamefont {B.}~\bibnamefont
  {Bradlyn}},\ }\href {https://doi.org/10.1103/PhysRevE.96.023310} {\bibfield
  {journal} {\bibinfo  {journal} {Phys. Rev. E}\ }\textbf {\bibinfo {volume}
  {96}},\ \bibinfo {pages} {023310} (\bibinfo {year} {2017})}\BibitemShut
  {NoStop}%
\bibitem [{\citenamefont {Kresse}\ and\ \citenamefont
  {Furthm{\"u}ller}(1996{\natexlab{a}})}]{kresse1996efficiency}%
  \BibitemOpen
  \bibfield  {author} {\bibinfo {author} {\bibfnamefont {G.}~\bibnamefont
  {Kresse}}\ and\ \bibinfo {author} {\bibfnamefont {J.}~\bibnamefont
  {Furthm{\"u}ller}},\ }\href@noop {} {\bibfield  {journal} {\bibinfo
  {journal} {Computational materials science}\ }\textbf {\bibinfo {volume}
  {6}},\ \bibinfo {pages} {15} (\bibinfo {year}
  {1996}{\natexlab{a}})}\BibitemShut {NoStop}%
\bibitem [{\citenamefont {Kresse}\ and\ \citenamefont
  {Hafner}(1993{\natexlab{a}})}]{kresse1993ab1}%
  \BibitemOpen
  \bibfield  {author} {\bibinfo {author} {\bibfnamefont {G.}~\bibnamefont
  {Kresse}}\ and\ \bibinfo {author} {\bibfnamefont {J.}~\bibnamefont
  {Hafner}},\ }\href@noop {} {\bibfield  {journal} {\bibinfo  {journal}
  {Physical Review B}\ }\textbf {\bibinfo {volume} {48}},\ \bibinfo {pages}
  {13115} (\bibinfo {year} {1993}{\natexlab{a}})}\BibitemShut {NoStop}%
\bibitem [{\citenamefont {Kresse}\ and\ \citenamefont
  {Hafner}(1993{\natexlab{b}})}]{kresse1993ab2}%
  \BibitemOpen
  \bibfield  {author} {\bibinfo {author} {\bibfnamefont {G.}~\bibnamefont
  {Kresse}}\ and\ \bibinfo {author} {\bibfnamefont {J.}~\bibnamefont
  {Hafner}},\ }\href@noop {} {\bibfield  {journal} {\bibinfo  {journal}
  {Physical review B}\ }\textbf {\bibinfo {volume} {47}},\ \bibinfo {pages}
  {558} (\bibinfo {year} {1993}{\natexlab{b}})}\BibitemShut {NoStop}%
\bibitem [{\citenamefont {Kresse}\ and\ \citenamefont
  {Hafner}(1994)}]{kresse1994ab}%
  \BibitemOpen
  \bibfield  {author} {\bibinfo {author} {\bibfnamefont {G.}~\bibnamefont
  {Kresse}}\ and\ \bibinfo {author} {\bibfnamefont {J.}~\bibnamefont
  {Hafner}},\ }\href@noop {} {\bibfield  {journal} {\bibinfo  {journal}
  {Physical Review B}\ }\textbf {\bibinfo {volume} {49}},\ \bibinfo {pages}
  {14251} (\bibinfo {year} {1994})}\BibitemShut {NoStop}%
\bibitem [{\citenamefont {Kresse}\ and\ \citenamefont
  {Furthm{\"u}ller}(1996{\natexlab{b}})}]{kresse1996efficient}%
  \BibitemOpen
  \bibfield  {author} {\bibinfo {author} {\bibfnamefont {G.}~\bibnamefont
  {Kresse}}\ and\ \bibinfo {author} {\bibfnamefont {J.}~\bibnamefont
  {Furthm{\"u}ller}},\ }\href@noop {} {\bibfield  {journal} {\bibinfo
  {journal} {Physical review B}\ }\textbf {\bibinfo {volume} {54}},\ \bibinfo
  {pages} {11169} (\bibinfo {year} {1996}{\natexlab{b}})}\BibitemShut {NoStop}%
\bibitem [{\citenamefont {Perdew}\ \emph {et~al.}(1996)\citenamefont {Perdew},
  \citenamefont {Burke},\ and\ \citenamefont
  {Ernzerhof}}]{perdew1996generalized}%
  \BibitemOpen
  \bibfield  {author} {\bibinfo {author} {\bibfnamefont {J.~P.}\ \bibnamefont
  {Perdew}}, \bibinfo {author} {\bibfnamefont {K.}~\bibnamefont {Burke}},\ and\
  \bibinfo {author} {\bibfnamefont {M.}~\bibnamefont {Ernzerhof}},\ }\href@noop
  {} {\bibfield  {journal} {\bibinfo  {journal} {Physical review letters}\
  }\textbf {\bibinfo {volume} {77}},\ \bibinfo {pages} {3865} (\bibinfo {year}
  {1996})}\BibitemShut {NoStop}%
\bibitem [{\citenamefont {Marzari}\ and\ \citenamefont
  {Vanderbilt}(1997)}]{marzari1997maximally}%
  \BibitemOpen
  \bibfield  {author} {\bibinfo {author} {\bibfnamefont {N.}~\bibnamefont
  {Marzari}}\ and\ \bibinfo {author} {\bibfnamefont {D.}~\bibnamefont
  {Vanderbilt}},\ }\href@noop {} {\bibfield  {journal} {\bibinfo  {journal}
  {Physical review B}\ }\textbf {\bibinfo {volume} {56}},\ \bibinfo {pages}
  {12847} (\bibinfo {year} {1997})}\BibitemShut {NoStop}%
\bibitem [{\citenamefont {Souza}\ \emph {et~al.}(2001)\citenamefont {Souza},
  \citenamefont {Marzari},\ and\ \citenamefont
  {Vanderbilt}}]{souza2001maximally}%
  \BibitemOpen
  \bibfield  {author} {\bibinfo {author} {\bibfnamefont {I.}~\bibnamefont
  {Souza}}, \bibinfo {author} {\bibfnamefont {N.}~\bibnamefont {Marzari}},\
  and\ \bibinfo {author} {\bibfnamefont {D.}~\bibnamefont {Vanderbilt}},\
  }\href@noop {} {\bibfield  {journal} {\bibinfo  {journal} {Physical Review
  B}\ }\textbf {\bibinfo {volume} {65}},\ \bibinfo {pages} {035109} (\bibinfo
  {year} {2001})}\BibitemShut {NoStop}%
\bibitem [{\citenamefont {Marzari}\ \emph {et~al.}(2012)\citenamefont
  {Marzari}, \citenamefont {Mostofi}, \citenamefont {Yates}, \citenamefont
  {Souza},\ and\ \citenamefont {Vanderbilt}}]{marzari2012maximally}%
  \BibitemOpen
  \bibfield  {author} {\bibinfo {author} {\bibfnamefont {N.}~\bibnamefont
  {Marzari}}, \bibinfo {author} {\bibfnamefont {A.~A.}\ \bibnamefont
  {Mostofi}}, \bibinfo {author} {\bibfnamefont {J.~R.}\ \bibnamefont {Yates}},
  \bibinfo {author} {\bibfnamefont {I.}~\bibnamefont {Souza}},\ and\ \bibinfo
  {author} {\bibfnamefont {D.}~\bibnamefont {Vanderbilt}},\ }\href@noop {}
  {\bibfield  {journal} {\bibinfo  {journal} {Reviews of Modern Physics}\
  }\textbf {\bibinfo {volume} {84}},\ \bibinfo {pages} {1419} (\bibinfo {year}
  {2012})}\BibitemShut {NoStop}%
\bibitem [{\citenamefont {Pizzi}\ \emph {et~al.}(2020)\citenamefont {Pizzi},
  \citenamefont {Vitale}, \citenamefont {Arita}, \citenamefont {Bl{\"u}gel},
  \citenamefont {Freimuth}, \citenamefont {G{\'e}ranton}, \citenamefont
  {Gibertini}, \citenamefont {Gresch}, \citenamefont {Johnson}, \citenamefont
  {Koretsune} \emph {et~al.}}]{pizzi2020wannier90}%
  \BibitemOpen
  \bibfield  {author} {\bibinfo {author} {\bibfnamefont {G.}~\bibnamefont
  {Pizzi}}, \bibinfo {author} {\bibfnamefont {V.}~\bibnamefont {Vitale}},
  \bibinfo {author} {\bibfnamefont {R.}~\bibnamefont {Arita}}, \bibinfo
  {author} {\bibfnamefont {S.}~\bibnamefont {Bl{\"u}gel}}, \bibinfo {author}
  {\bibfnamefont {F.}~\bibnamefont {Freimuth}}, \bibinfo {author}
  {\bibfnamefont {G.}~\bibnamefont {G{\'e}ranton}}, \bibinfo {author}
  {\bibfnamefont {M.}~\bibnamefont {Gibertini}}, \bibinfo {author}
  {\bibfnamefont {D.}~\bibnamefont {Gresch}}, \bibinfo {author} {\bibfnamefont
  {C.}~\bibnamefont {Johnson}}, \bibinfo {author} {\bibfnamefont
  {T.}~\bibnamefont {Koretsune}}, \emph {et~al.},\ }\href@noop {} {\bibfield
  {journal} {\bibinfo  {journal} {Journal of Physics: Condensed Matter}\
  }\textbf {\bibinfo {volume} {32}},\ \bibinfo {pages} {165902} (\bibinfo
  {year} {2020})}\BibitemShut {NoStop}%
\bibitem [{\citenamefont {Wu}\ \emph {et~al.}(2018)\citenamefont {Wu},
  \citenamefont {Zhang}, \citenamefont {Song}, \citenamefont {Troyer},\ and\
  \citenamefont {Soluyanov}}]{wu2018wanniertools}%
  \BibitemOpen
  \bibfield  {author} {\bibinfo {author} {\bibfnamefont {Q.}~\bibnamefont
  {Wu}}, \bibinfo {author} {\bibfnamefont {S.}~\bibnamefont {Zhang}}, \bibinfo
  {author} {\bibfnamefont {H.-F.}\ \bibnamefont {Song}}, \bibinfo {author}
  {\bibfnamefont {M.}~\bibnamefont {Troyer}},\ and\ \bibinfo {author}
  {\bibfnamefont {A.~A.}\ \bibnamefont {Soluyanov}},\ }\href@noop {} {\bibfield
   {journal} {\bibinfo  {journal} {Computer Physics Communications}\ }\textbf
  {\bibinfo {volume} {224}},\ \bibinfo {pages} {405} (\bibinfo {year}
  {2018})}\BibitemShut {NoStop}%
\bibitem [{\citenamefont {Gao}\ \emph {et~al.}(2021)\citenamefont {Gao},
  \citenamefont {Wu}, \citenamefont {Persson},\ and\ \citenamefont
  {Wang}}]{gao2021irvsp}%
  \BibitemOpen
  \bibfield  {author} {\bibinfo {author} {\bibfnamefont {J.}~\bibnamefont
  {Gao}}, \bibinfo {author} {\bibfnamefont {Q.}~\bibnamefont {Wu}}, \bibinfo
  {author} {\bibfnamefont {C.}~\bibnamefont {Persson}},\ and\ \bibinfo {author}
  {\bibfnamefont {Z.}~\bibnamefont {Wang}},\ }\href@noop {} {\bibfield
  {journal} {\bibinfo  {journal} {Computer Physics Communications}\ }\textbf
  {\bibinfo {volume} {261}},\ \bibinfo {pages} {107760} (\bibinfo {year}
  {2021})}\BibitemShut {NoStop}%
\bibitem [{\citenamefont {Zhang}\ \emph {et~al.}(2022)\citenamefont {Zhang},
  \citenamefont {Yu}, \citenamefont {Liu},\ and\ \citenamefont
  {Yao}}]{zhang2022magnetictb}%
  \BibitemOpen
  \bibfield  {author} {\bibinfo {author} {\bibfnamefont {Z.}~\bibnamefont
  {Zhang}}, \bibinfo {author} {\bibfnamefont {Z.-M.}\ \bibnamefont {Yu}},
  \bibinfo {author} {\bibfnamefont {G.-B.}\ \bibnamefont {Liu}},\ and\ \bibinfo
  {author} {\bibfnamefont {Y.}~\bibnamefont {Yao}},\ }\href@noop {} {\bibfield
  {journal} {\bibinfo  {journal} {Computer Physics Communications}\ }\textbf
  {\bibinfo {volume} {270}},\ \bibinfo {pages} {108153} (\bibinfo {year}
  {2022})}\BibitemShut {NoStop}%
\bibitem [{\citenamefont {Jiang}\ \emph {et~al.}(2023)\citenamefont {Jiang},
  \citenamefont {Hu}, \citenamefont {Călugăru}, \citenamefont {Felser},
  \citenamefont {Blanco-Canosa}, \citenamefont {Weng}, \citenamefont {Xu},\
  and\ \citenamefont {Bernevig}}]{jiang2023kagome}%
  \BibitemOpen
  \bibfield  {author} {\bibinfo {author} {\bibfnamefont {Y.}~\bibnamefont
  {Jiang}}, \bibinfo {author} {\bibfnamefont {H.}~\bibnamefont {Hu}}, \bibinfo
  {author} {\bibfnamefont {D.}~\bibnamefont {Călugăru}}, \bibinfo {author}
  {\bibfnamefont {C.}~\bibnamefont {Felser}}, \bibinfo {author} {\bibfnamefont
  {S.}~\bibnamefont {Blanco-Canosa}}, \bibinfo {author} {\bibfnamefont
  {H.}~\bibnamefont {Weng}}, \bibinfo {author} {\bibfnamefont {Y.}~\bibnamefont
  {Xu}},\ and\ \bibinfo {author} {\bibfnamefont {B.~A.}\ \bibnamefont
  {Bernevig}},\ }\href@noop {} {\bibinfo {title} {Kagome materials ii: Sg 191:
  Fege as a lego building block for the entire 1:6:6 series: hidden d-orbital
  decoupling of flat band sectors, effective models and interaction
  hamiltonians}} (\bibinfo {year} {2023}),\ \Eprint
  {https://arxiv.org/abs/2311.09290} {arXiv:2311.09290 [cond-mat.str-el]}
  \BibitemShut {NoStop}%
\bibitem [{\citenamefont {Herzog-Arbeitman}\ \emph {et~al.}(2022)\citenamefont
  {Herzog-Arbeitman}, \citenamefont {Chew}, \citenamefont {Huhtinen},
  \citenamefont {Törmä},\ and\ \citenamefont
  {Bernevig}}]{herzogarbeitman2022manybodysuperconductivitytopologicalflat}%
  \BibitemOpen
  \bibfield  {author} {\bibinfo {author} {\bibfnamefont {J.}~\bibnamefont
  {Herzog-Arbeitman}}, \bibinfo {author} {\bibfnamefont {A.}~\bibnamefont
  {Chew}}, \bibinfo {author} {\bibfnamefont {K.-E.}\ \bibnamefont {Huhtinen}},
  \bibinfo {author} {\bibfnamefont {P.}~\bibnamefont {Törmä}},\ and\ \bibinfo
  {author} {\bibfnamefont {B.~A.}\ \bibnamefont {Bernevig}},\ }\href
  {https://arxiv.org/abs/2209.00007} {\bibinfo {title} {Many-body
  superconductivity in topological flat bands}} (\bibinfo {year} {2022}),\
  \Eprint {https://arxiv.org/abs/2209.00007} {arXiv:2209.00007
  [cond-mat.str-el]} \BibitemShut {NoStop}%
\bibitem [{\citenamefont {Hu}\ and\ \citenamefont
  {Bernevig}()}]{flat_band_mag}%
  \BibitemOpen
  \bibfield  {author} {\bibinfo {author} {\bibfnamefont {H.}~\bibnamefont
  {Hu}}\ and\ \bibinfo {author} {\bibfnamefont {B.~A.}\ \bibnamefont
  {Bernevig}},\ }\href@noop {} {\bibinfo  {journal} {to appear}\ }\BibitemShut
  {NoStop}%
\bibitem [{\citenamefont {Aryasetiawan}\ \emph {et~al.}(2004)\citenamefont
  {Aryasetiawan}, \citenamefont {Imada}, \citenamefont {Georges}, \citenamefont
  {Kotliar}, \citenamefont {Biermann},\ and\ \citenamefont
  {Lichtenstein}}]{aryasetiawan2004frequency}%
  \BibitemOpen
\bibfield  {journal} {  }\bibfield  {author} {\bibinfo {author} {\bibfnamefont
  {F.}~\bibnamefont {Aryasetiawan}}, \bibinfo {author} {\bibfnamefont
  {M.}~\bibnamefont {Imada}}, \bibinfo {author} {\bibfnamefont
  {A.}~\bibnamefont {Georges}}, \bibinfo {author} {\bibfnamefont
  {G.}~\bibnamefont {Kotliar}}, \bibinfo {author} {\bibfnamefont
  {S.}~\bibnamefont {Biermann}},\ and\ \bibinfo {author} {\bibfnamefont
  {A.~I.}\ \bibnamefont {Lichtenstein}},\ }\href
  {https://doi.org/10.1103/PhysRevB.70.195104} {\bibfield  {journal} {\bibinfo
  {journal} {Phys. Rev. B}\ }\textbf {\bibinfo {volume} {70}},\ \bibinfo
  {pages} {195104} (\bibinfo {year} {2004})}\BibitemShut {NoStop}%
\bibitem [{\citenamefont {Aryasetiawan}\ \emph {et~al.}(2006)\citenamefont
  {Aryasetiawan}, \citenamefont {Karlsson}, \citenamefont {Jepsen},\ and\
  \citenamefont {Sch{\"o}nberger}}]{aryasetiawan2006calculations}%
  \BibitemOpen
  \bibfield  {author} {\bibinfo {author} {\bibfnamefont {F.}~\bibnamefont
  {Aryasetiawan}}, \bibinfo {author} {\bibfnamefont {K.}~\bibnamefont
  {Karlsson}}, \bibinfo {author} {\bibfnamefont {O.}~\bibnamefont {Jepsen}},\
  and\ \bibinfo {author} {\bibfnamefont {U.}~\bibnamefont {Sch{\"o}nberger}},\
  }\href@noop {} {\bibfield  {journal} {\bibinfo  {journal} {Physical Review
  B}\ }\textbf {\bibinfo {volume} {74}},\ \bibinfo {pages} {125106} (\bibinfo
  {year} {2006})}\BibitemShut {NoStop}%
\bibitem [{\citenamefont {Miyake}\ \emph {et~al.}(2009)\citenamefont {Miyake},
  \citenamefont {Aryasetiawan},\ and\ \citenamefont {Imada}}]{miyake2009ab}%
  \BibitemOpen
  \bibfield  {author} {\bibinfo {author} {\bibfnamefont {T.}~\bibnamefont
  {Miyake}}, \bibinfo {author} {\bibfnamefont {F.}~\bibnamefont
  {Aryasetiawan}},\ and\ \bibinfo {author} {\bibfnamefont {M.}~\bibnamefont
  {Imada}},\ }\href@noop {} {\bibfield  {journal} {\bibinfo  {journal}
  {Physical Review B}\ }\textbf {\bibinfo {volume} {80}},\ \bibinfo {pages}
  {155134} (\bibinfo {year} {2009})}\BibitemShut {NoStop}%
\bibitem [{\citenamefont {Hu}\ \emph {et~al.}(2023{\natexlab{a}})\citenamefont
  {Hu}, \citenamefont {Bernevig},\ and\ \citenamefont
  {Tsvelik}}]{PhysRevLett.131.026502}%
  \BibitemOpen
  \bibfield  {author} {\bibinfo {author} {\bibfnamefont {H.}~\bibnamefont
  {Hu}}, \bibinfo {author} {\bibfnamefont {B.~A.}\ \bibnamefont {Bernevig}},\
  and\ \bibinfo {author} {\bibfnamefont {A.~M.}\ \bibnamefont {Tsvelik}},\
  }\href {https://doi.org/10.1103/PhysRevLett.131.026502} {\bibfield  {journal}
  {\bibinfo  {journal} {Phys. Rev. Lett.}\ }\textbf {\bibinfo {volume} {131}},\
  \bibinfo {pages} {026502} (\bibinfo {year} {2023}{\natexlab{a}})}\BibitemShut
  {NoStop}%
\bibitem [{\citenamefont {Ruderman}\ and\ \citenamefont
  {Kittel}(1954)}]{RKKY_1}%
  \BibitemOpen
  \bibfield  {author} {\bibinfo {author} {\bibfnamefont {M.~A.}\ \bibnamefont
  {Ruderman}}\ and\ \bibinfo {author} {\bibfnamefont {C.}~\bibnamefont
  {Kittel}},\ }\href {https://doi.org/10.1103/PhysRev.96.99} {\bibfield
  {journal} {\bibinfo  {journal} {Phys. Rev.}\ }\textbf {\bibinfo {volume}
  {96}},\ \bibinfo {pages} {99} (\bibinfo {year} {1954})}\BibitemShut {NoStop}%
\bibitem [{\citenamefont {Kasuya}(1956)}]{RKKY_2}%
  \BibitemOpen
  \bibfield  {author} {\bibinfo {author} {\bibfnamefont {T.}~\bibnamefont
  {Kasuya}},\ }\href {https://doi.org/10.1143/PTP.16.45} {\bibfield  {journal}
  {\bibinfo  {journal} {Progress of Theoretical Physics}\ }\textbf {\bibinfo
  {volume} {16}},\ \bibinfo {pages} {45} (\bibinfo {year} {1956})},\ \Eprint
  {https://arxiv.org/abs/https://academic.oup.com/ptp/article-pdf/16/1/45/5266722/16-1-45.pdf}
  {https://academic.oup.com/ptp/article-pdf/16/1/45/5266722/16-1-45.pdf}
  \BibitemShut {NoStop}%
\bibitem [{\citenamefont {Yosida}(1957)}]{RKKY_3}%
  \BibitemOpen
  \bibfield  {author} {\bibinfo {author} {\bibfnamefont {K.}~\bibnamefont
  {Yosida}},\ }\href {https://doi.org/10.1103/PhysRev.106.893} {\bibfield
  {journal} {\bibinfo  {journal} {Phys. Rev.}\ }\textbf {\bibinfo {volume}
  {106}},\ \bibinfo {pages} {893} (\bibinfo {year} {1957})}\BibitemShut
  {NoStop}%
\bibitem [{\citenamefont {Hu}\ \emph {et~al.}(2023{\natexlab{b}})\citenamefont
  {Hu}, \citenamefont {Rai}, \citenamefont {Crippa}, \citenamefont
  {Herzog-Arbeitman}, \citenamefont {C\ifmmode \u{a}\else
  \u{a}\fi{}lug\ifmmode~\u{a}\else \u{a}\fi{}ru}, \citenamefont {Wehling},
  \citenamefont {Sangiovanni}, \citenamefont {Valent\'{\i}}, \citenamefont
  {Tsvelik},\ and\ \citenamefont {Bernevig}}]{PhysRevLett.131.166501}%
  \BibitemOpen
  \bibfield  {author} {\bibinfo {author} {\bibfnamefont {H.}~\bibnamefont
  {Hu}}, \bibinfo {author} {\bibfnamefont {G.}~\bibnamefont {Rai}}, \bibinfo
  {author} {\bibfnamefont {L.}~\bibnamefont {Crippa}}, \bibinfo {author}
  {\bibfnamefont {J.}~\bibnamefont {Herzog-Arbeitman}}, \bibinfo {author}
  {\bibfnamefont {D.}~\bibnamefont {C\ifmmode \u{a}\else
  \u{a}\fi{}lug\ifmmode~\u{a}\else \u{a}\fi{}ru}}, \bibinfo {author}
  {\bibfnamefont {T.}~\bibnamefont {Wehling}}, \bibinfo {author} {\bibfnamefont
  {G.}~\bibnamefont {Sangiovanni}}, \bibinfo {author} {\bibfnamefont
  {R.}~\bibnamefont {Valent\'{\i}}}, \bibinfo {author} {\bibfnamefont {A.~M.}\
  \bibnamefont {Tsvelik}},\ and\ \bibinfo {author} {\bibfnamefont {B.~A.}\
  \bibnamefont {Bernevig}},\ }\href
  {https://doi.org/10.1103/PhysRevLett.131.166501} {\bibfield  {journal}
  {\bibinfo  {journal} {Phys. Rev. Lett.}\ }\textbf {\bibinfo {volume} {131}},\
  \bibinfo {pages} {166501} (\bibinfo {year} {2023}{\natexlab{b}})}\BibitemShut
  {NoStop}%
\bibitem [{\citenamefont {Petrovic}\ \emph {et~al.}(2001)\citenamefont
  {Petrovic}, \citenamefont {Pagliuso}, \citenamefont {Hundley}, \citenamefont
  {Movshovich}, \citenamefont {Sarrao}, \citenamefont {Thompson}, \citenamefont
  {Fisk},\ and\ \citenamefont {Monthoux}}]{petrovic2001heavy}%
  \BibitemOpen
  \bibfield  {author} {\bibinfo {author} {\bibfnamefont {C.}~\bibnamefont
  {Petrovic}}, \bibinfo {author} {\bibfnamefont {P.}~\bibnamefont {Pagliuso}},
  \bibinfo {author} {\bibfnamefont {M.~F.}\ \bibnamefont {Hundley}}, \bibinfo
  {author} {\bibfnamefont {R.}~\bibnamefont {Movshovich}}, \bibinfo {author}
  {\bibfnamefont {J.~L.}\ \bibnamefont {Sarrao}}, \bibinfo {author}
  {\bibfnamefont {J.~D.}\ \bibnamefont {Thompson}}, \bibinfo {author}
  {\bibfnamefont {Z.}~\bibnamefont {Fisk}},\ and\ \bibinfo {author}
  {\bibfnamefont {P.}~\bibnamefont {Monthoux}},\ }\href@noop {} {\bibfield
  {journal} {\bibinfo  {journal} {Journal of Physics: Condensed Matter}\
  }\textbf {\bibinfo {volume} {13}},\ \bibinfo {pages} {L337} (\bibinfo {year}
  {2001})}\BibitemShut {NoStop}%
\bibitem [{\citenamefont {Kirchner}\ \emph {et~al.}(2020)\citenamefont
  {Kirchner}, \citenamefont {Paschen}, \citenamefont {Chen}, \citenamefont
  {Wirth}, \citenamefont {Feng}, \citenamefont {Thompson},\ and\ \citenamefont
  {Si}}]{RevModPhys.92.011002}%
  \BibitemOpen
  \bibfield  {author} {\bibinfo {author} {\bibfnamefont {S.}~\bibnamefont
  {Kirchner}}, \bibinfo {author} {\bibfnamefont {S.}~\bibnamefont {Paschen}},
  \bibinfo {author} {\bibfnamefont {Q.}~\bibnamefont {Chen}}, \bibinfo {author}
  {\bibfnamefont {S.}~\bibnamefont {Wirth}}, \bibinfo {author} {\bibfnamefont
  {D.}~\bibnamefont {Feng}}, \bibinfo {author} {\bibfnamefont {J.~D.}\
  \bibnamefont {Thompson}},\ and\ \bibinfo {author} {\bibfnamefont
  {Q.}~\bibnamefont {Si}},\ }\href
  {https://doi.org/10.1103/RevModPhys.92.011002} {\bibfield  {journal}
  {\bibinfo  {journal} {Rev. Mod. Phys.}\ }\textbf {\bibinfo {volume} {92}},\
  \bibinfo {pages} {011002} (\bibinfo {year} {2020})}\BibitemShut {NoStop}%
\bibitem [{\citenamefont {Reehuis}\ \emph {et~al.}(1998)\citenamefont
  {Reehuis}, \citenamefont {Jeitschko}, \citenamefont {Kotzyba}, \citenamefont
  {Zimmer},\ and\ \citenamefont {Hu}}]{reehuis1998antiferromagnetic}%
  \BibitemOpen
  \bibfield  {author} {\bibinfo {author} {\bibfnamefont {M.}~\bibnamefont
  {Reehuis}}, \bibinfo {author} {\bibfnamefont {W.}~\bibnamefont {Jeitschko}},
  \bibinfo {author} {\bibfnamefont {G.}~\bibnamefont {Kotzyba}}, \bibinfo
  {author} {\bibfnamefont {B.}~\bibnamefont {Zimmer}},\ and\ \bibinfo {author}
  {\bibfnamefont {X.}~\bibnamefont {Hu}},\ }\href@noop {} {\bibfield  {journal}
  {\bibinfo  {journal} {Journal of alloys and compounds}\ }\textbf {\bibinfo
  {volume} {266}},\ \bibinfo {pages} {54} (\bibinfo {year} {1998})}\BibitemShut
  {NoStop}%
\bibitem [{\citenamefont {Reehuis}\ and\ \citenamefont
  {Jeitschko}(1990)}]{REEHUIS1990961}%
  \BibitemOpen
  \bibfield  {author} {\bibinfo {author} {\bibfnamefont {M.}~\bibnamefont
  {Reehuis}}\ and\ \bibinfo {author} {\bibfnamefont {W.}~\bibnamefont
  {Jeitschko}},\ }\href
  {https://doi.org/https://doi.org/10.1016/0022-3697(90)90039-I} {\bibfield
  {journal} {\bibinfo  {journal} {Journal of Physics and Chemistry of Solids}\
  }\textbf {\bibinfo {volume} {51}},\ \bibinfo {pages} {961} (\bibinfo {year}
  {1990})}\BibitemShut {NoStop}%
\bibitem [{zot()}]{zotero-4159}%
  \BibitemOpen
  \href@noop {} {\bibinfo {title} {Bilbao {{Crystallographic Server}}}},\
  \bibinfo {howpublished} {https://www.cryst.ehu.es/}\BibitemShut {NoStop}%
\bibitem [{\citenamefont {Peng}\ \emph {et~al.}(2022)\citenamefont {Peng},
  \citenamefont {Jiang}, \citenamefont {Fang}, \citenamefont {Weng},\ and\
  \citenamefont {Fang}}]{peng2022topological}%
  \BibitemOpen
  \bibfield  {author} {\bibinfo {author} {\bibfnamefont {B.}~\bibnamefont
  {Peng}}, \bibinfo {author} {\bibfnamefont {Y.}~\bibnamefont {Jiang}},
  \bibinfo {author} {\bibfnamefont {Z.}~\bibnamefont {Fang}}, \bibinfo {author}
  {\bibfnamefont {H.}~\bibnamefont {Weng}},\ and\ \bibinfo {author}
  {\bibfnamefont {C.}~\bibnamefont {Fang}},\ }\href@noop {} {\bibfield
  {journal} {\bibinfo  {journal} {Physical Review B}\ }\textbf {\bibinfo
  {volume} {105}},\ \bibinfo {pages} {235138} (\bibinfo {year}
  {2022})}\BibitemShut {NoStop}%
\bibitem [{\citenamefont {Qi}\ \emph {et~al.}(2008)\citenamefont {Qi},
  \citenamefont {Hughes},\ and\ \citenamefont {Zhang}}]{qi2008topological}%
  \BibitemOpen
  \bibfield  {author} {\bibinfo {author} {\bibfnamefont {X.-L.}\ \bibnamefont
  {Qi}}, \bibinfo {author} {\bibfnamefont {T.~L.}\ \bibnamefont {Hughes}},\
  and\ \bibinfo {author} {\bibfnamefont {S.-C.}\ \bibnamefont {Zhang}},\
  }\href@noop {} {\bibfield  {journal} {\bibinfo  {journal} {Physical Review
  B}\ }\textbf {\bibinfo {volume} {78}},\ \bibinfo {pages} {195424} (\bibinfo
  {year} {2008})}\BibitemShut {NoStop}%
\bibitem [{\citenamefont {Fang}\ \emph {et~al.}(2012)\citenamefont {Fang},
  \citenamefont {Gilbert},\ and\ \citenamefont {Bernevig}}]{fang2012bulk}%
  \BibitemOpen
  \bibfield  {author} {\bibinfo {author} {\bibfnamefont {C.}~\bibnamefont
  {Fang}}, \bibinfo {author} {\bibfnamefont {M.~J.}\ \bibnamefont {Gilbert}},\
  and\ \bibinfo {author} {\bibfnamefont {B.~A.}\ \bibnamefont {Bernevig}},\
  }\href@noop {} {\bibfield  {journal} {\bibinfo  {journal} {Physical Review
  B}\ }\textbf {\bibinfo {volume} {86}},\ \bibinfo {pages} {115112} (\bibinfo
  {year} {2012})}\BibitemShut {NoStop}%
\bibitem [{\citenamefont {Xu}\ \emph {et~al.}(2021{\natexlab{b}})\citenamefont
  {Xu}, \citenamefont {Elcoro}, \citenamefont {Song}, \citenamefont
  {Vergniory}, \citenamefont {Felser}, \citenamefont {Parkin}, \citenamefont
  {Regnault}, \citenamefont {Ma{\~n}es},\ and\ \citenamefont
  {Bernevig}}]{xu2021filling}%
  \BibitemOpen
  \bibfield  {author} {\bibinfo {author} {\bibfnamefont {Y.}~\bibnamefont
  {Xu}}, \bibinfo {author} {\bibfnamefont {L.}~\bibnamefont {Elcoro}}, \bibinfo
  {author} {\bibfnamefont {Z.-D.}\ \bibnamefont {Song}}, \bibinfo {author}
  {\bibfnamefont {M.}~\bibnamefont {Vergniory}}, \bibinfo {author}
  {\bibfnamefont {C.}~\bibnamefont {Felser}}, \bibinfo {author} {\bibfnamefont
  {S.~S.}\ \bibnamefont {Parkin}}, \bibinfo {author} {\bibfnamefont
  {N.}~\bibnamefont {Regnault}}, \bibinfo {author} {\bibfnamefont {J.~L.}\
  \bibnamefont {Ma{\~n}es}},\ and\ \bibinfo {author} {\bibfnamefont {B.~A.}\
  \bibnamefont {Bernevig}},\ }\href@noop {} {\bibfield  {journal} {\bibinfo
  {journal} {arXiv preprint arXiv:2106.10276}\ } (\bibinfo {year}
  {2021}{\natexlab{b}})}\BibitemShut {NoStop}%
\bibitem [{\citenamefont {Wang}\ \emph {et~al.}(2016)\citenamefont {Wang},
  \citenamefont {Alexandradinata}, \citenamefont {Cava},\ and\ \citenamefont
  {Bernevig}}]{Wang2016}%
  \BibitemOpen
  \bibfield  {author} {\bibinfo {author} {\bibfnamefont {Z.}~\bibnamefont
  {Wang}}, \bibinfo {author} {\bibfnamefont {A.}~\bibnamefont
  {Alexandradinata}}, \bibinfo {author} {\bibfnamefont {R.~J.}\ \bibnamefont
  {Cava}},\ and\ \bibinfo {author} {\bibfnamefont {B.~A.}\ \bibnamefont
  {Bernevig}},\ }\href {https://doi.org/10.1038/nature17410} {\bibfield
  {journal} {\bibinfo  {journal} {Nature}\ }\textbf {\bibinfo {volume} {532}},\
  \bibinfo {pages} {189} (\bibinfo {year} {2016})}\BibitemShut {NoStop}%
\bibitem [{\citenamefont {Bernevig}\ \emph {et~al.}(2021)\citenamefont
  {Bernevig}, \citenamefont {Lian}, \citenamefont {Cowsik}, \citenamefont
  {Xie}, \citenamefont {Regnault},\ and\ \citenamefont
  {Song}}]{PhysRevB.103.205415}%
  \BibitemOpen
  \bibfield  {author} {\bibinfo {author} {\bibfnamefont {B.~A.}\ \bibnamefont
  {Bernevig}}, \bibinfo {author} {\bibfnamefont {B.}~\bibnamefont {Lian}},
  \bibinfo {author} {\bibfnamefont {A.}~\bibnamefont {Cowsik}}, \bibinfo
  {author} {\bibfnamefont {F.}~\bibnamefont {Xie}}, \bibinfo {author}
  {\bibfnamefont {N.}~\bibnamefont {Regnault}},\ and\ \bibinfo {author}
  {\bibfnamefont {Z.-D.}\ \bibnamefont {Song}},\ }\href
  {https://doi.org/10.1103/PhysRevB.103.205415} {\bibfield  {journal} {\bibinfo
   {journal} {Phys. Rev. B}\ }\textbf {\bibinfo {volume} {103}},\ \bibinfo
  {pages} {205415} (\bibinfo {year} {2021})}\BibitemShut {NoStop}%
\end{thebibliography}%
\let\addcontentsline\oldaddcontentsline

\clearpage

\renewcommand{\thetable}{S\arabic{table}}
\renewcommand{\thefigure}{S\arabic{figure}}
\renewcommand{\theequation}{S\arabic{section}.\arabic{equation}}
\onecolumngrid
\pagebreak
\thispagestyle{empty}
\newpage
\begin{center}
	\textbf{\large Supplemental Material: \titlePaper}\\[.2cm]
\end{center}

\appendix
\renewcommand{\thesection}{\Roman{section}}
\tableofcontents
\let\oldaddcontentsline\addcontentsline
\newpage

\section{Experimental observations} 
In this section, we list the experimental observations of the CeCo$_2$P$_2$~\cite{ccp_exp_paper}. 
\begin{itemize}
    \item The system develops a type-A antiferromagnetic (AFM) order at $T_{\text{AFM}}\sim 450$K. The magnetic ordering comes from the electrons of Co. The magnetic structure has been shown in Fig.~\ref{fig:lat_mag_structure}.  

\item 
From XPS and XAS measurement ~\cite{ccp_exp_paper}, Ce atoms have been found to behave as 
\ba 
\text{Ce}^{3+}:\quad 4\text{f}^1 5\text{d}^0 5\text{s}^0
\ea 
In other words, there is one $f$ electron per Ce atom. Therefore, $f$ electron at Ce atom behaves as a local moment. 
\item 
Both resistivity and ARPES measurements suggest the development of the Kondo effect below $T\sim 100$K. 

\item 
At low temperatures (below the Kondo temperature), ARPES measurement identifies a nodal line at $k_z=1/2$ planes. 

\end{itemize}

In this work, we will provide a comprehensive understanding of the above experimental observations.

\section{Lattice structure}
\label{app:sec:lattice}
\subsection{Primitive cell}
The paramagnetic (PM) phase (at $T>T_{\text{AFM}}$) has space group 139. The unit cell (primitive cell) of the paramagnetic phase has the following lattice vectors
\ba 
&\bm{a_1} = (-1.9475\AA\quad 1.9475\AA \quad  4.802\AA) \nonumber \\ 
&\bm{a_2} = (1.9475\AA\quad  -1.9475\AA \quad  4.802\AA)\nonumber \\ 
&\bm{a_3} = (1.9475\AA\quad  1.9475\AA \quad  -4.802\AA) 
\ea 
The reciprocal lattice vectors are
\ba 
&\bm{b_1} = 2\pi (0 \quad 0.2567\AA^{-1} \quad 0.1041\AA^{-1})\nonumber\\
&\bm{b_2} = 2\pi ( 0.2567\AA^{-1} \quad 0\quad 0.1041\AA^{-1})\nonumber\\
&\bm{b_3} = 2\pi ( 0.2567\AA^{-1} \quad 0.2567\AA^{-1} \quad 0)
\label{eq:app:rec_vec_prim_vell}
\ea 

There are two Co atoms two P atoms and one Ce atom per unit cell located at 
\ba 
\text{Co}\quad:\quad &(3/4, 1/4, 1/2),\quad (1/4, 3/4, 1/2)  \nonumber\\
\text{Ce}\quad:\quad &(0,0,0) \nonumber\\
\text{P}\quad:\quad & (0.628, 0.628,0),\quad (0.372,0.372,0)
\ea 

The relevant orbitals we considered in this work are: (1) five $d$ orbitals and seven $f$ orbitals of Ce atoms; (2) five $d$ orbitals of Co atoms; (3) three $p$ orbitals of P atoms. The $d$ orbitals, $p$ orbitals, and $f$ orbitals are
\ba 
&d_{z^2},d_{xz}, d_{yz}, d_{x^2-y^2}, d_{xy}\nonumber\\ 
&p_z, p_x,p_y\nonumber\\ 
&f_{z^3}, f_{xz^2},f_{yz^2}, f_{z(x^2-y^2)}, f_{xyz}, f_{x(x^2-3y^2)}, f_{y(3x^2-y^2)}
\ea 

The high symmetry points of the primitive cell are
\ba 
\Gamma = (0,0,0),\quad M= (\frac{-1}{2},\frac{1}{2},\frac{1}{2}),\quad X= (0,0,\frac{1}{2}),\quad P = (\frac{1}{4},\frac{1}{4},\frac{1}{4}),\quad N= (0,\frac{1}{2},0)
\label{eq:app:high_sym_prim}
\ea

\subsection{Conventional cell}
The system develops an AFM order below $T_{\text{AFM}}\approx 450$K. The AFM phase is described by the magnetic group 126.386 ($P_I4/nnc$). 
The unit cell of the AFM phase corresponds to the conventional cell of the PM phase. For future reference, we denote the unit cell of the AFM phase as the conventional cell, and the unit cell of the PM phase as the primitive cell. 

The lattice vectors of the conventional cell are 
\ba 
\bm{A_1} = \bm{a_2}+\bm{a_3} ,\quad \bm{A_2} =\bm{a_3}+\bm{a_1} ,\quad \bm{A_3} = \bm{a_1}+\bm{a_2}
\ea 

The reciprocal lattice vectors of the conventional cell ($\bm{B_i}$) and primitive cell ($\bm{b_i}$) have the following relations 
\ba 
\bm{B_1} = \frac{1}{2}(-\bm{b_1}+\bm{b_2}+\bm{b_3}),\quad \bm{B_2} =\frac{1}{2}(\bm{b_1}-\bm{b_2}+\bm{b_3}) ,\quad \bm{B_3} = \frac{1}{2}(\bm{b_1}+\bm{b_2}-\bm{b_3}) 
\label{eq:app:recp_vec_conv}
\ea

In the conventional cell, the number of atoms is doubled. The atoms are located at (in the unit of $\bm{A}_{1},\bm{A}_{2},\bm{A}_{3}$)
\ba 
\text{Co}\quad:\quad &(0,  \frac{1}{2}, \frac{1}{4} ),\quad (\frac{1}{2}, 0, \frac{1}{4}),\quad (\frac{1}{2}, 0,\frac{3}{4}),\quad (0, \frac{1}{2}, \frac{3}{4})  \nonumber\\
\text{Ce}\quad:\quad &(0,0,0), \quad (\frac{1}{2},\frac{1}{2},\frac{1}{2}) \nonumber\\
\text{P}\quad:\quad &(\frac{1}{2},\frac{1}{2},z), \quad (0,0,\frac{1}{2}-z),\quad (0,0,\frac{1}{2}+z),\quad  (\frac{1}{2},\frac{1}{2}, 1-z) ,\quad z= 0.128.
\ea

The high symmetry points of the conventional cell are
% (in terms of $B_i$)
\ba 
&\Gamma = (0,0,0),\quad X= (0,\frac{1}{2},0),\quad M= (\frac{1}{2},\frac{1}{2},0)\nonumber\\ 
&Z = (0,0,\frac{1}{2}),\quad R =(0,\frac{1}{2},\frac{1}{2}), 
\quad 
A= (\frac{1}{2},\frac{1}{2},\frac{1}{2})
\label{eq:app:high_sym_conv}
\ea

\begin{figure}
    \centering
    \includegraphics[width=0.5\textwidth]{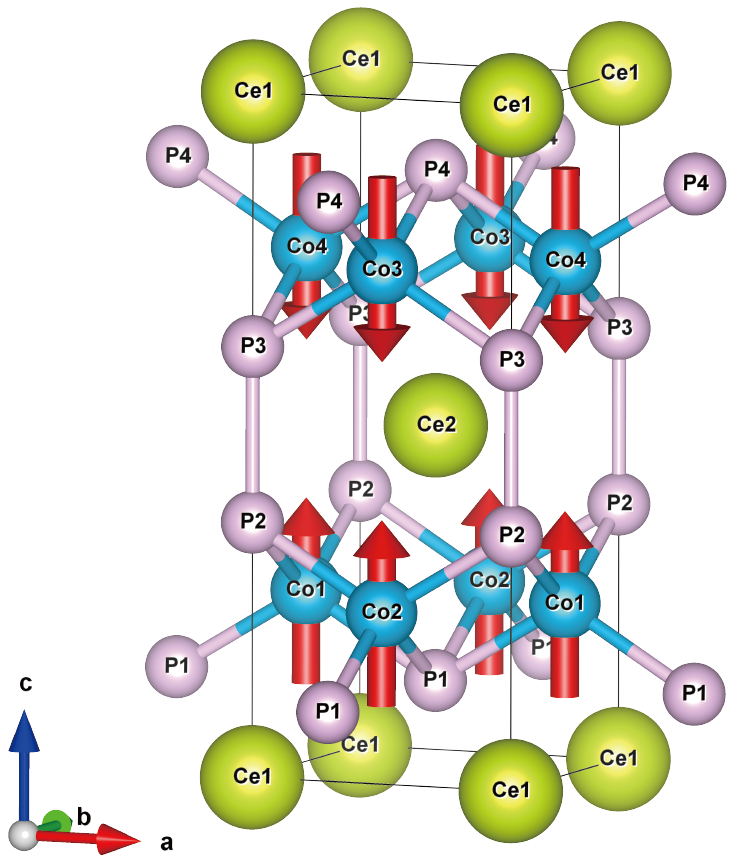}
    \caption{Lattice structures and magnetic structures. }
    \label{fig:lat_mag_structure}
\end{figure}

\section{Model and Hamiltonian}
\label{app:sec:model}
The electrons in the system can be separated into the correlated $f$ electrons and itinerant $c$ electrons. The $f$ electrons correspond to the $f$ orbitals of Ce. The $c$ electrons correspond to the $d$ orbitals of Ce, $d$ orbitals of Co, and $p$ orbitals of P. We use $f_{\RR,a,i,\sigma}$ and $c_{\RR,a,i,\sigma}$ to describe the $f$ electrons and $c$ electrons respectively, where $\RR$ is the position of the unit cell, $a$ is the sublattice index, $i$ is the orbital index and $\sigma$ is the spin index. We will clarify in each section whether a conventional cell or a primitive cell is being used.
The Hamiltonian takes the form of a periodic Anderson model
\ba 
\label{eq:app:pam}
H &= H_c +H_{V}+ H_{c,U} +H_f .
\ea 
 $H_c$ describes the hopping of $c$ electrons 
\ba 
H_c& =  \sum_{\kk, ab,ij,\sigma} \epsilon_{\kk,ab,ij}c_{\kk,a,i,\sigma}^\dag c_{\kk,b,j,\sigma} .
\ea 
$H_V$ describes the hybridization between $f$ and $c$ electrons 
\ba 
H_V &= \sum_{\RR,\Delta\RR, ab,ij,\sigma}V^c_{ab,ij}(\Delta\RR) f_{\RR,a,i,\sigma}^\dag c_{\RR+\Delta\RR,b,j,\sigma} +\text{h.c.}
\label{eq:app:hyb_term_c}
\ea 
$H_{c,U}$ denotes the on-site interactions of $c$ electrons, which takes the form of multi-orbital Hubbard interactions
\ba 
H_{c,U}= &\frac{1}{2}\sum_{\RR,a, (i,\sigma)\ne (j,\sigma')}U_{ij,a} : c_{\RR,a,i,\sigma}^\dag c_{\RR,a,i,\sigma} c_{\RR,a,j,\sigma'}^\dag c_{\RR,a,j,\sigma'}: \nonumber \\ 
&
-\frac{1}{2}\sum_{\RR,a,i,j,\sigma,\sigma'}J_{ij,a}: c_{\RR,a,i,\sigma}^\dag c_{\RR,a,i,\sigma'} c_{\RR,a,j,\sigma'}^\dag c_{\RR,a,j,\sigma}:
+ \frac{1}{2}\sum_{\RR,a,i,j,\sigma,\sigma's}
J_{ij,a}:c_{\RR,a,i,\sigma}^\dag c_{\RR,a,i,\sigma'}^\dag c_{\RR,a,j,\sigma'}c_{\RR,a,j,\sigma}:
\label{eq:app:h_int_c}
\ea 
where $U_{ij,a}$ are the density-density interactions and $J_{ij,a}$ are the Hund's couplings. The values of these interactions will be derived from DFT calculations. For a given operator $O$, we have defined the normal ordering of the operator as $:O:$.
% where $\langle  \rangle_0$ is the expectation value taking with respect to the ground state of the non-interacting Hamiltonian $H_c$. 

$H_f$ describes all the on-site coupling terms of $f$ electrons, including the crystal-field splittings, spin-orbital couplings, and Hubbard repulsion. It takes the form of 
\ba 
H_{f} = \sum_{\RR,a} H_{loc}[f_{\RR,a}^\dag , f_{\RR,a}]
\label{eq:app:ham_f}
\ea 
where $H_{loc}[f^\dag ,f]$ describes the local Hamiltonian of $f$ electrons.  $ H_{loc}[f^\dag, f]$ can be decomposed into three parts, the Hubbard interactions, crystal field splitting, and spin-orbit coupling (SOC)
\ba 
&H_{loc}[f^\dag ,f ] = H_{U}[f^\dag, f] + H_{CEF}[f^\dag, f] +H_{soc}[f^\dag, f]\nonumber\\ 
&H_U[f^\dag, f] = \frac{U}{2} \bigg( \sum_{i,\sigma} f_{i,\sigma}^\dag f_{i,\sigma}  -n_0 \bigg) ^2 \nonumber\\ 
&H_{CEF}[f^\dag, f] = \sum_{ij}\epsilon_{f,ij}f_{i,\sigma}^\dag f_{j,\sigma} \nonumber\\ 
&H_{soc}[f,f^\dag] = \lambda \sum_{\mu}L^\mu S^\mu 
\label{eq:app:h_f_atomic}
\ea 
Here, since there is only one electron in the $f$ shell of Ce, we take $n_0 =1$. In the limit of large $U$, the filling of $f$ electrons will be fixed to be $n_0=1$.  $\epsilon_{f,ij}$ describes crystal field splitting of $f$ electrons. $\lambda$ is the strength of spin-orbit coupling where $L^\mu, S^\mu$ the angular-momentum and spin operators of $f$ electrons respectively.

\section{DFT calculation details}
% The parameters of the model are obtained from DFT calculations. 
We use the Vienna ab-initio Simulation Package (VASP)\cite{kresse1996efficiency, kresse1993ab1, kresse1993ab2, kresse1994ab, kresse1996efficient} to perform the \textit{ab-initio} calculations in this work, where the generalized gradient approximation of Perdew-Burke-Ernzerhof (PBE) exchange-correlation potential\cite{perdew1996generalized} is adopted. An $11 \times11 \times11$ uniform $k$-mesh and an energy cutoff of 400 eV are used for self-consistency computations. 
The maximally localized Wannier functions are obtained using WANNIER90\cite{marzari1997maximally, souza2001maximally, marzari2012maximally, pizzi2020wannier90}. The Wannier TB models are symmetrized using \textit{Wannhr\_symm} in \textit{WannierTools}\cite{wu2018wanniertools}. The irreducible representations are computed using \textit{IR2TB}\cite{gao2021irvsp}. The minimal TB model is built with \textit{MagneticTB}\cite{zhang2022magnetictb}.

\section{Relatively flat bands in the paramagnetic phase}
\label{app:sec:flat_band}
We discuss the band structure of the high-temperature paramagnetic phase. In this phase, due to the absence of the Kondo effect and large Hubbard interactions, $f$ electrons will not contribute to the low-energy single-particle excitation. The band structures are described by only $c$ electrons. We therefore obtain the band structures by treating $f$ electron as core states (fully occupied $f$ shell) and performing DFT calculations. 

\begin{figure}
    \centering
    \includegraphics[width=0.8\textwidth]{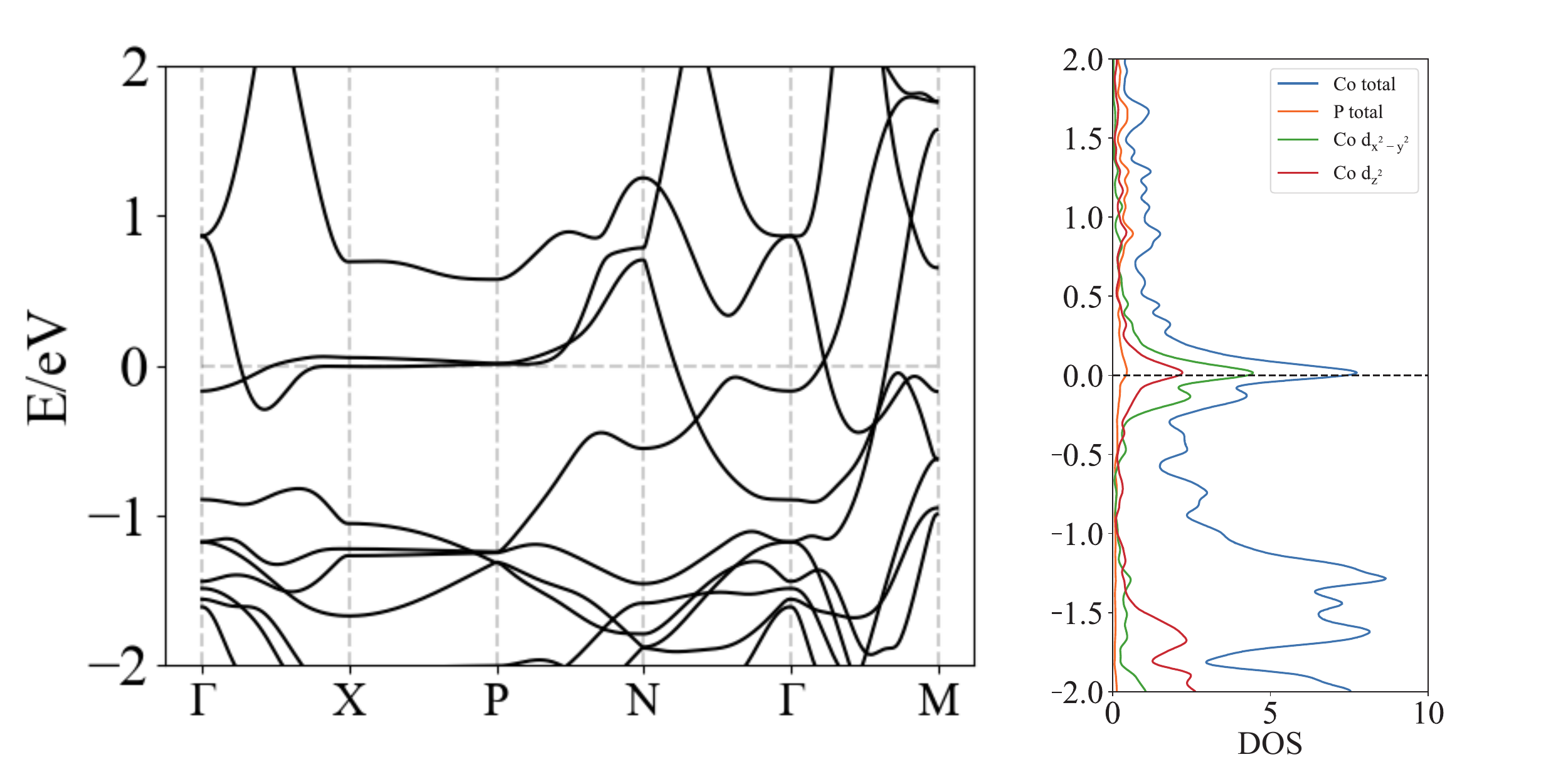}
    \caption{ DFT-calculated band structure and density of states (DOS) of PM phase where $f$ electrons have been treated as core states.}
    \label{fig:app:dft_pm_band}
\end{figure}

\begin{figure}
    \centering
    \includegraphics[width=0.9\textwidth]{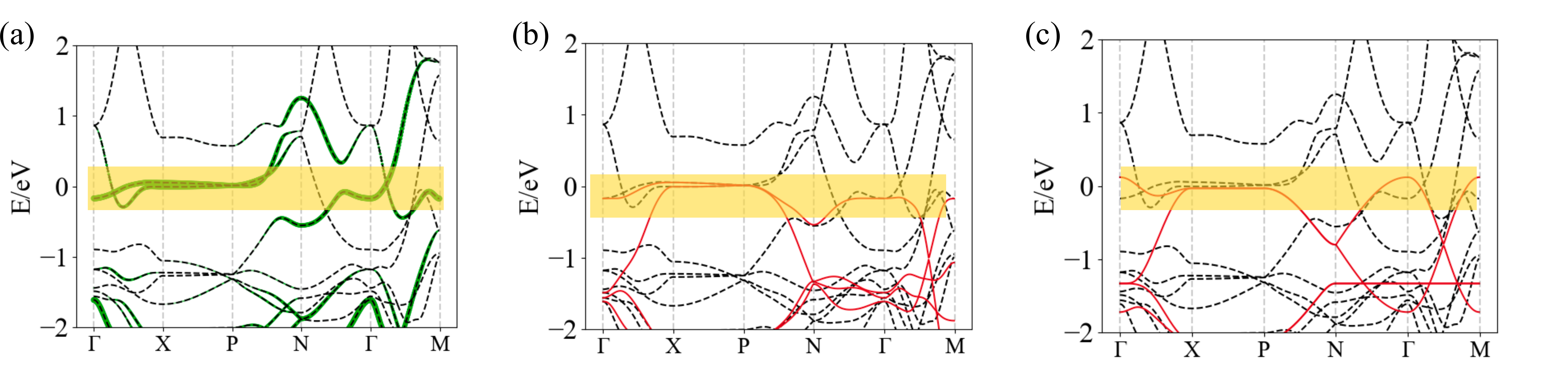}
    \caption{(a) DFT band structure. The green dots mark the orbital weights of the $d_{x^2-y^2},d_{z^2}$ orbitals of Co. Yellow areas mark the relative flat band. (b) Comparison between DFT band structures (black dashed lines) and the band structures obtained from two-orbital models (red solid lines). (c) Comparison between the DFT band structure (black dashed lines) and the band structure obtained from the simplified two-orbital model (red solid lines) given in Eq.~\ref{eq:app:simplified_tbp}. 
    % Blue dots in all three figures mark the Dirac node. 
    }
    \label{fig:app:flat_band}
\end{figure}

\begin{figure}
    \centering
    \includegraphics[width=0.5\textwidth]{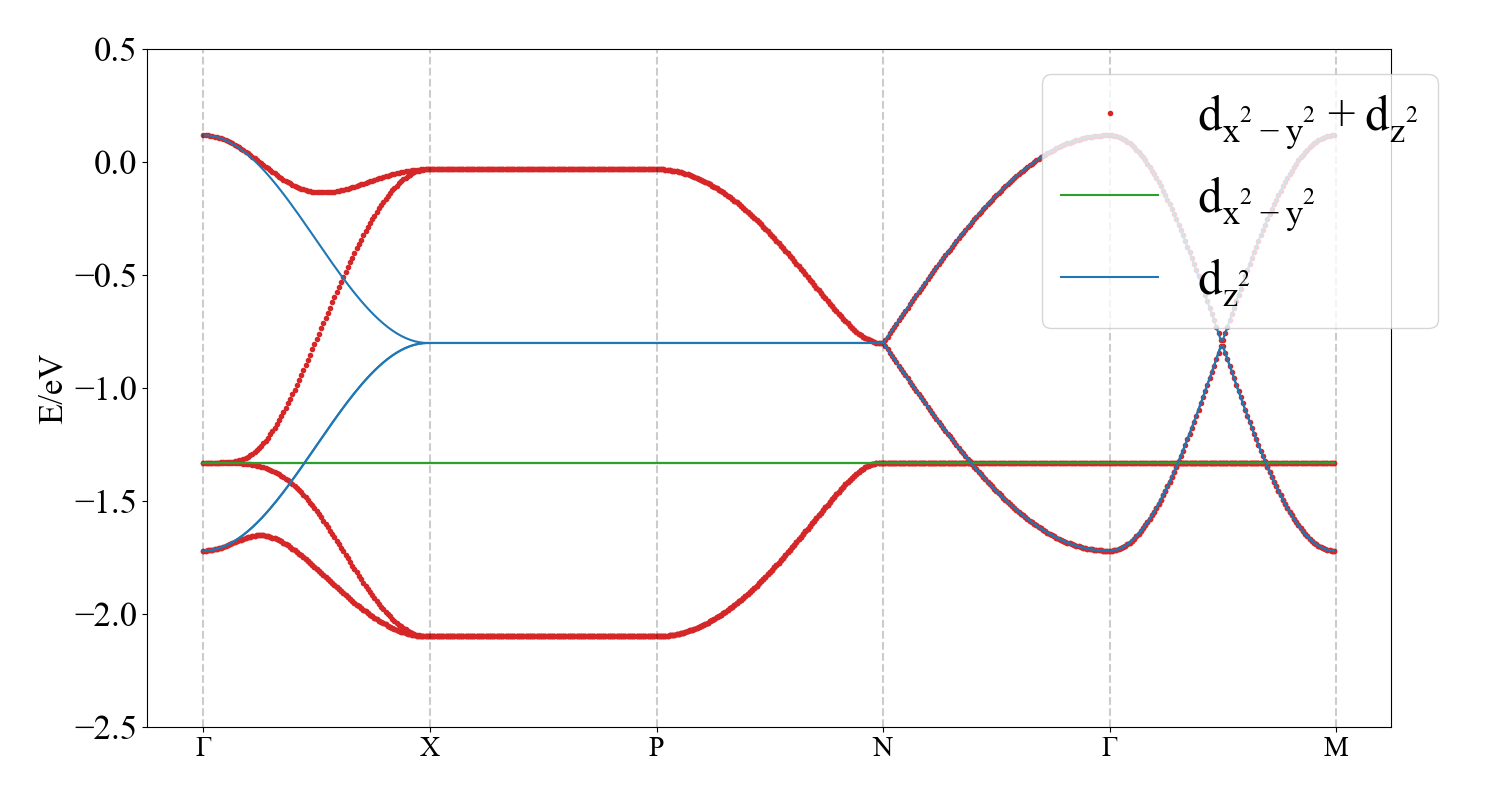}
    \caption{The band structure of the simplified two-orbital model is shown in red. The green and blue lines represent the band structures of the $d_{x^2-y^2}$ and $d_{z^2}$ orbitals, respectively, in the absence of hybridization between these orbitals ($t_2=0$). }
    \label{fig:app:two_orb_model}
\end{figure}

In Fig.~\ref{fig:app:dft_pm_band}, we show the band structures and the density of states (DOS) obtained from DFT calculations. We can observe an enhanced DOS peak appearing near the Fermi energy coming from the relatively flat band near the Fermi energy. This flat band is mostly formed by the $d_{x^2-y^2},d_{z^2}$ orbitals of Co.

We introduce the following three types of tight-binding models to model the relatively flat band in the PM phase (see Fig.~\ref{fig:app:flat_band} for the comparison between different models): 
\begin{itemize}
    \item \textbf{DFT model}: tight-binding model of PM phase obtained from DFT calculations (see Fig.~\ref{fig:app:dft_pm_band} and Fig.~\ref{fig:app:flat_band} (a) for band structure). 
    \item \textbf{Two-orbital model}: tight-binding model obtained by only including the hoppings of $d_{x^2-y^2}@\text{Co},d_{z^2}@\text{Co}$ orbitals of DFT model (see Fig.~\ref{fig:app:flat_band} (b) for band structure).  
    \item \textbf{Simplified two-orbital model}: the nearest-neighbor tight-binding model obtained by only keeping the dominant nearest-neighbor hopping of the two-orbital model (see Fig.~\ref{fig:app:flat_band} (c) for band structure). 
\end{itemize}

We now discuss each model. In the DFT model, the enhanced DOS comes from $d_{x^2-y^2}$ and $d_{z^2}$ orbitals of Co (Fig.~\ref{fig:app:dft_pm_band}, Fig.~\ref{fig:app:flat_band} (a)). 
This indicates the relatively flat band can be approximately described by a model that only contains $d_{x^2-y^2}, d_{z^2}$ orbitals of Co atoms. Thus, we build a two-orbital model by only keeping the hopping of these two orbitals in the DFT model. The resulting band structure gives a good description of the flat band near Fermi energy as shown in Fig.~\ref{fig:app:flat_band} (b).

To understand the origin of the flat band, we find the two-orbital model can be further simplified into the following tight-binding model by only keeping the dominant nearest-neighbor hopping
\ba 
\label{eq:app:simplified_tbp}
&H_c = \sum_{\kk,\sigma} \psi_{\kk,\sigma}^\dag 
h_\kk \psi_{\kk,\sigma} \nonumber\\ 
&\psi_\kk = 
\begin{bmatrix}
    c_{\kk,Co_1,z^2,\sigma} & c_{\kk,Co_2,z^2,\sigma} & c_{\kk,Co_1,x^2-y^2,\sigma} & c_{\kk,Co_2,x^2-y^2,\sigma}
\end{bmatrix} \nonumber\\ 
&h_\kk = 
\begin{bmatrix}
    \epsilon_1 & 0 & 0 & -4t_2\sin(\frac{k_1+k_3}{2})\sin(\frac{k_2+k_3}{2}) \\ 
    0 & \epsilon_1 & -4t_2\sin(\frac{k_1+k_3}{2})\sin(\frac{k_2+k_3}{2})& 0 \\ 
    0& -4t_2\sin(\frac{k_1+k_3}{2})\sin(\frac{k_2+k_3}{2}) & \epsilon_2 
    % + 2r_2(\cos(k_1+k_3)+\cos(k_2+k_3) ) 
    & 
   g_\kk 
    \\ 
   -4t_2\sin(\frac{k_1+k_3}{2})\sin(\frac{k_2+k_3}{2})& 0 &  g_\kk^* & \epsilon_2
   % + 2r_2(\cos(k_1+k_3) + \cos(k_2+k_3) )
\end{bmatrix}\nonumber\\ 
&g_\kk = 4t_1\cos(\frac{k_1+k_3}{2})\cos(\frac{k_2+k_3}{2})  
% +t_3'\bigg[e^{-\frac{i}{2}(k_x+3k_y+2k_z)} +e^{-\frac{i}{2}(k_x-k_y+2k_z)} +e^{-\frac{i}{2}(k_x-k_y-k_z) } + e^{-\frac{i}{2}(-3k_x-k_y-2k_z)}\bigg]
\ea 
where the parameters are
\ba 
&\epsilon_1 = -1.33eV,\quad \epsilon_2 = -0.8eV\nonumber\\ 
&t_2 = -0.25eV,\quad t_1 =-0.23eV\nonumber\\ 
% &t_3' = 0.03eV,\quad r_2 = -0.05eV
\ea 
$t_1$ denotes the nearest-neighbor (in-plane) hopping between $d_{x^2-y^2}$ orbitals. $t_2$ denotes the nearest-neighbor (in-plane) hopping between $d_{x^2-y^2}$ and $d_{z^2}$ orbitals. $\epsilon_1,\epsilon_2$ are the on-site energies of $d_{z^2},d_{x^2-y^2}$ orbitals respectively. The nearest-neighbor hopping of $d_{z^2}$ orbitals is relatively weak and has been dropped. We observe that the amplitudes of the $d_{z^2}$ Wannier orbital are primarily concentrated along the $z$-axis, resulting in weak in-plane hopping of the $d_{z^2}$ orbital. Furthermore, due to the large separation between Co atoms in different layers, the hopping along the $z$-direction is also relatively weak and has been neglected. 
The band structure of the simplified two-orbital model is given in Fig.~\ref{fig:app:flat_band} (c).

We now diagonalize the non-interacting Hamiltonian ~\ref{eq:app:simplified_tbp}.  
We first introduce the following new basis, 
\ba 
&
\begin{bmatrix}
    \tilde{c}_{\kk,1,x^2-y^2,\sigma}\\
    \tilde{c}_{\kk,2,x^2-y^2,\sigma}  \\
    \tilde{c}_{\kk,1,z^2,\sigma}\\
    \tilde{c}_{\kk,2,z^2,\sigma}  
\end{bmatrix} 
=\frac{1}{\sqrt{2}}
\begin{bmatrix}
    1 & g_\kk/|g_\kk| & 0 &0  \\ 
    1 & - g_\kk |g_\kk| & 0 &0  \\ 
    0 & 0 & 1& g_\kk/|g_\kk| \\
    0 & 0 &  1& -g_\kk/|g_\kk| 
\end{bmatrix}
\begin{bmatrix}
  {c}_{\kk,Co_1,x^2-y^2,\sigma}\\
  {c}_{\kk,Co_2,x^2-y^2,\sigma}  \\
  {c}_{\kk,Co_1,z^2,\sigma}\\
  {c}_{\kk,Co_2,z^2,\sigma}  
\end{bmatrix} 
\ea 
where $g_\kk = g_\kk^*$, and $g_\kk/|g_\kk| = \text{sign}(g_\kk)$. 
In the new basis, the Hamiltonians of $d_{x^2-y^2}$ block and $d_{z^2}$ block have been diagonalized. There is still a coupling between the two orbitals. The new Hamiltonian becomes 
\ba 
H_c = &\sum_{\kk,\sigma} \bigg[ \epsilon_2  
+|g_\kk| \bigg] \tilde{c}_{\kk,1,x^2-y^2,\sigma}^\dag  \tilde{c}_{\kk,1,x^2-y^2,\sigma} + \bigg[ \epsilon_2 
-
|g_\kk| \bigg]\tilde{c}_{\kk,2,x^2-y^2,\sigma}^\dag  \tilde{c}_{\kk,2,x^2-y^2,\sigma} 
+\sum_{\kk,i,\sigma} \epsilon_1 \tilde{c}_{\kk,i,z^2,\sigma}^\dag \tilde{c}_{\kk,i,z^2,\sigma}
\nonumber\\ 
& +\sum_{\kk,\sigma} \bigg\{ 4t_2\sin(\frac{k_1+k_3}{2})\sin(\frac{k_2+k_3}{2})
\begin{bmatrix}
    \tilde{c}_{\kk,1,z^2,\sigma}^\dag & \tilde{c}^\dag_{\kk,2,z^2,\sigma}
\end{bmatrix}
\begin{bmatrix}
   \frac{ -g_\kk }{|g_\kk|}& 0  \\
   0  &  \frac{ g_\kk}{|g_\kk|}
\end{bmatrix} \begin{bmatrix}
    \tilde{c}_{\kk,1,x^2-y^2,\sigma}\\
    \tilde{c}_{\kk,2,x^2-y^2,\sigma}  
\end{bmatrix} 
+\text{h.c.}\bigg\}
\label{eq:simplified_hc_new_basis}
\ea 
We first consider the case of $t_2=0$, where the hybridization between $d_{x^2-y^2}$ orbitals and $d_{z^2}$ orbitals vanishes. In this case, $d_{x^2-y^2}$ orbitals have a relatively large dispersion due to the nearest-neighbor hopping $t_1$, and $d_{z^2}$ produces atomic flat bands (but far away from Fermi energy at $\epsilon_1$) as shown in Fig.~\ref{fig:app:two_orb_model}. We now show that, after turning on the hybridization, the hybridization term produces a relatively flat band (Fig.~\ref{fig:app:two_orb_model}).

The hybridization term $t_2$ introduces coupling between $\tilde{c}_{\kk,1,x^2-y^2,\sigma}$ and $\tilde{c}_{\kk,1,z^2,\sigma}$, and coupling between $\tilde{c}_{\kk,2,x^2-y^2,\sigma}$ and $\tilde{c}_{\kk,2,z^2,\sigma}$. Since, $\tilde{c}_{\kk,1,x^2-y^2,\sigma}$ is more close to the Fermi energy at $t_2=0$ limit, we focus on the ($\tilde{c}_{\kk,1,x^2-y^2,\sigma}$, $\tilde{c}_{\kk,1,z^2,\sigma}$) sector. 
The corresponding Hamiltonian can be written as (from Eq.~\ref{eq:simplified_hc_new_basis})
\ba 
&H_c'= \sum_{\kk,\sigma} 
\begin{bmatrix}
    \tilde{c}_{\kk,1,x^2-y^2,\sigma}^\dag & \tilde{c}_{\kk,1,z^2,\sigma}^\dag 
\end{bmatrix}
\begin{bmatrix}
   \Delta + \epsilon_\kk &v_\kk \\ 
   v_\kk  & \epsilon_1 
\end{bmatrix}\begin{bmatrix}
    \tilde{c}_{\kk,1,x^2-y^2,\sigma} \\  \tilde{c}_{\kk,1,z^2,\sigma}
\end{bmatrix}\nonumber\\ 
&v_\kk = -4t_2\sin(\frac{k_1+k_3}{2})\sin(\frac{k_2+k_3}{2}) 
    \frac{ g_\kk}{|g_\kk|}\nonumber\\
& \Delta = \epsilon_2 + \text{mean}_{\kk \in BZ}|g_\kk|  \approx -0.34eV
 \nonumber\\ 
&\epsilon_\kk =
% 2r_2(\cos(k_x+k_z) + \cos(k_y+k_z) ) +
% \bigg[
|g_\kk| - \text{mean}_{\kk \in BZ}|g_\kk| 
% \bigg] 
\label{eq:simplified_hc_new_basis_upper_band}
\ea 
where we have subtracted the mean value of $|g_\kk|$, such that we can separate the dispersion of $ \tilde{c}_{\kk,1,x^2-y^2,\sigma}$ orbitals into a constant part $\Delta$ and a dispersive part $\epsilon_\kk$ (with $\text{mean}_{\kk\in BZ} \epsilon_\kk = 0$). 
The dispersions of Hamiltonian shown in Eq.~\ref{eq:simplified_hc_new_basis_upper_band} are
\ba 
E_{1/2,\kk} = \frac{\epsilon_1 +\Delta+\epsilon_\kk}{2} \pm  \sqrt{ (\frac{\epsilon_1 -\Delta-\epsilon_\kk}{2} )^2 + v_\kk ^2  }
\ea 

We then show that, the upper band with dispersion $E_{1,\kk}$ gives a relatively flat band. 
We notice that $\text{max}\epsilon_\kk - \text{min}\epsilon_\kk = 0.9eV ,\text{max}|v_\kk|-\text{min}|v_\kk|=1.0eV, \epsilon_1-\Delta=1.7$. We therefore perform an expansion by treating $\epsilon_\kk, v_\kk$ as small parameters and find
\ba 
E_{1,\kk} \approx \Delta  +\epsilon_\kk + \frac{v_\kk^2}{|\Delta-\epsilon_1|}
\label{eq:approxmate_disp_flat_band}
\ea 
$\epsilon_\kk$ describes the original dispersions of $d_{x^2-y^2}$ orbitals (see Eq.~\ref{eq:simplified_hc_new_basis_upper_band}), and $\frac{v_\kk^2}{|\Delta-\epsilon_1|}$ denotes the contribution from the $d_{x^2-y^2}$-$d_{z^2}$ hybridization term. We now show that, by combining the contribution of two terms, we could generate a relatively flat band. We check the dispersion at high symmetry points
\ba 
&E_{1,\Gamma=(0,0,0)} \approx \epsilon_2 +  4| t_1| \nonumber\\ 
&E_{1,M=(\frac{1}{2},\frac{1}{2},-\frac{1}{2})}  \approx  \epsilon_2 + 4|t_1|  \nonumber\\ 
&E_{1,X=(0,0,\frac{1}{2})}  \approx  \epsilon_2 + \frac{16t_2^2}{|\Delta-\epsilon_1|} \nonumber\\ 
&E_{1,P=(\frac{1}{4},\frac{1}{4},\frac{1}{4})}  \approx \epsilon_2+  \frac{16t_2^2}{|\Delta-\epsilon_1|} \nonumber\\ 
&E_{1,N=(0,\frac{1}{2},0)}  \approx  \epsilon_2 
\ea 
If we turn off the hybridization term $t_2=0$, the energies of $X,P$ points are much lower than the energy of $\Gamma,M$ points (with energy difference $4|t_1|$). However, these energy differences are reduced by the hybridization. As a consequence, the energies of $\Gamma,M,X,P$ points become similar.  
If we take
\ba 
\frac{|\Delta-\epsilon_1||t_1|}{4t_2^2}  = 1
\label{eq:flat_band_cond}
\ea 
we can realize a relatively flat band with four out of five high symmetry points having similar energy 
\ba 
E_{\Gamma} \approx E_{M} \approx E_X \approx E_P
\ea 
In practice, using the DFT numerical values,  we find the system is close to the condition in Eq.~\ref{eq:flat_band_cond} with
$\frac{|\Delta-\epsilon_1| |t_1|}{4 t_2^2}= 0.91 \approx 1$.
% structures of simplified two-orbital model indeed give a relatively flat band (Fig.~\ref{fig:app:flat_band} (c)). 

Besides the high symmetry points, we also consider the dispersion along the high-symmetry line $\Gamma-X$. From Eq.~\ref{eq:approxmate_disp_flat_band}, we find 
\ba 
&\Gamma\text{-}X \quad:\quad  \{(0,0,x) |x \in [0,\frac{1}{2}]\} \nonumber \\
&
E_{1,(0,0,x)} \approx \epsilon_2 + |4t_1| \cos(\frac{x}{2})^2 + \frac{16t_2^2}{|\Delta-\epsilon_1|} \sin( \frac{x}{2})^4 = \epsilon_2 + |4t_1| 
- |4t_1| \sin(\frac{x}{2})^2 + \frac{16t_2^2}{|\Delta-\epsilon_1|} \sin( \frac{x}{2})^4 
\ea 
We can find the bandwidth along $\Gamma$-$X$ line
\ba 
&\text{max}_{x}E_{1,(0,0,x)} = E_{1,(0,0,0)} = \epsilon_2 + 4|t_1| \nonumber\\ 
&\text{min}_{x}E_{1,(0,0,x)} = E_{1,(0,0,y)}\bigg|_{\sin(y/2) = \frac{4|t_2|^2}{|t_1||\Delta-\epsilon_1|}} = \epsilon_2+4|t_2|  -\frac{|\Delta-\epsilon_1|}{4}\nonumber\\ 
&\text{max}_{x}E_{1,(0,0,x)}-\text{min}_{x}E_{1,(0,0,x)}
 = \frac{|\Delta-\epsilon_1|}{4}\approx 0.24\text{eV}
\ea 
where we observe the bandwidth has been reduced compared to the original bandwidth (4$t_2\sim 1$eV) of the $d_{x^2-y^2}$ orbital along $\Gamma$-$X$ line in the zero-hybridization limit ($t_2=0$). 
In addition, since the simplified two-orbital model has vanishing $z$-direction hopping, the band is flat along $\bm{b}_1+\bm{b}_2-\bm{b}_3$ direction ($k_z$ direction).

In summary, we conclude that, due to the hybridization (or hopping) between $d_{x^2-y^2}$ and $d_{z^2}$ orbitals, a relatively flat band has been generated near the Fermi energy. 

To confirm the validity of the simplified two-orbital model, we also compare the flat-band wavefunctions obtained from the simplified two-orbital model and the DFT model. We find a good match between the two models, where the overlap of the wavefunctions reaches $84.7\%$~\cite{jiang2023kagome}.

% We now compare the band structures of the simplified two-orbital model, two-orbital model, and DFT model (Fig.~\ref{fig:app:flat_band}). The energy of the flat band at high-symmetry point $N$ is also close to the energy of the flat band at $\Gamma,X,P,M$ points in the two-orbital and DFT models. This is because the long-range hopping (which has been dropped in the simplified two-orbital model) raises the energy of $N$ point and further flattens the relatively flat band in both the two-orbital model and DFT model. 

\begin{figure}
    \centering
    \includegraphics[width=0.5\textwidth]{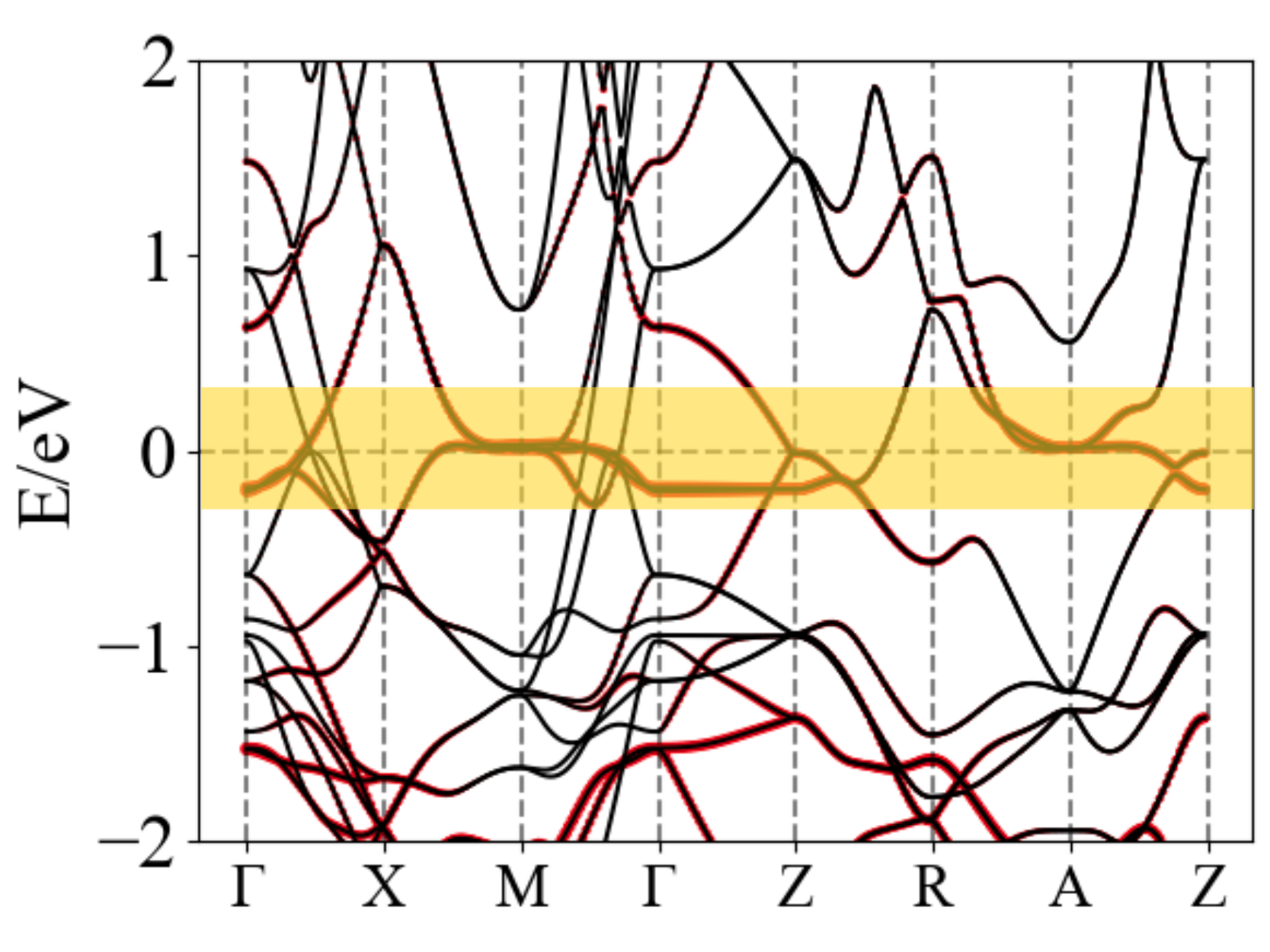}
    \caption{ Dispersion in the conventional cell. Red dots label the orbital weights of $d_{x^2-y^2},d_{z^2}$ orbitals. 
    }
    \label{fig:app:flat_band_conv}
\end{figure}

Finally, we also show the band structure in the conventional cell (Fig.~\ref{fig:app:flat_band_conv}). The Brillouin zone of the conventional cell is folded, and we can still observe the relatively flat bands near Fermi energy formed by $d_{x^2-y^2},d_{z^2}$ orbitals. 
% In Fig.~\ref{fig:app:flat_band_conv} (b), we also mark two bands that have the largest $d_{x^2-y^2},d_{z^2}$ orbital weights within the energy window $[-1eV,1eV]$. We can observe they are relatively flat, and they contribute 70$\%$ of the DOS peak near Fermi energy (Fig.~\ref{fig:app:flat_band_conv} (c) ). 

% In all three models, the irreducible representations of the flat bands at $\Gamma$ and $M$ points are $\Gamma_1^-$ and $M_3^+$ (Fig.~\ref{fig:app:flat_band}). These two irreducible representations are not compatible along the following $\Gamma$-$M$ line~\citeBCS, which indicates the formation of the node.
% \ba 
% \Gamma\text{-}M \text{ line}: 
% (k_1,k_2,k_3) \in \{ (-\frac{x}{2},\frac{x}{2},\frac{x}{2})| x \in [0,1]\}
% \label{eq:line_with_node}
% \ea 
% The node has been marked by blue in Fig.~\ref{fig:app:flat_band}. However, the energies of the nodes in Fig.~\ref{fig:app:flat_band} (a), (b), (c) are not the same. This is because, in our simplified two-orbital model and two-orbital model, we dropped certain $d$ and $p$ orbitals that contribute to the dispersive bands above the Fermi energy, which could affect the position of the node. 

\section{Antiferromagnetism of $c$ electrons}
\label{sec:app:afm_c}
The existence of the relatively flat band leads to magnetic instability as we lower the temperature. 
Since the magnetic transition in the experiment happens at a relatively high temperature $T\sim 450$K. At such a high temperature, the Kondo effect is not present, and we can approximately treat $f$ and $c$ electrons as two decoupled degrees of freedom. We therefore only focus on the $c$ electrons during our study of antiferromagnetism.

Before we discuss the corresponding instability, we first discuss the properties of flat bands. 
% Our current tight-binding model (introduced in Sec.~\ref{app:sec:flat_band}) is defined in the primitive cell. 
We notice that the lattice has a layered structure (Fig.~\ref{fig:lat_mag_structure}). 
In our simplified two-orbital model (Eq.~\ref{eq:app:simplified_tbp}), $t_1$ denotes the in-plane ($xy$ plane) hopping of $d_{x^2-y^2}$ orbitals, and $t_2$ denotes the in-plane ($xy$ plane) hopping between $d_{x^2-y^2}$ and $d_{z^2}$ orbitals. 
Moreover, the system is decoupled along $z$ directions in the simplified two-orbital model. 
This indicates the relatively flat bands we discussed in Sec.~\ref{app:sec:flat_band} are atomic along $z$ direction but non-atomic along $xy$ directions.

% Flat bands that is atomic along $z$ direction and non-atomic along $xy$ direction will lead to a type-A antiferromagnetism. 
The non-atomic nature along $xy$ direction indicates the in-plane ferromagnetism~\cite{herzogarbeitman2022manybodysuperconductivitytopologicalflat}. The atomic nature along $z$ direction combined with the weak dispersion along $z$ direction leads to an out-of-plane antiferromagnetism~\cite{flat_band_mag}. 
We thus expect our current relatively flat band to stabilize a type-A antiferromagnetism (in-plane ferromagnetism and out-of-plane antiferromagnetism) with the magnetic structure shown in Fig.~\ref{fig:lat_mag_structure}.

To understand the antiferromagnetism, we have performed the following calculations
\begin{itemize}
    \item Mean-field calculations. 
    % \item Mean-field calculations using the PM band structures obtained from the DFT model. 
    \item A direct DFT calculation in the AFM phase where $f$ electrons are treated as core states. 
    \item Analytical calculations on the toy model (see Sec.~\ref{sec:app:flat_band_magnetism} for the definition of the model and the details of the calculations).
\end{itemize}
As we will show later in this section, three calculations are qualitatively consistent with each other.

\subsection{Mean-field theory}
We first discuss the procedure of mean-field calculations. We consider the following interacting Hamiltonian of $c$ electrons
\ba 
H_{c,int} = H_c +H_{c,U}
\ea 
The interaction terms are 
\ba 
H_{c,U}= &\frac{1}{2}\sum_{\RR,a, (i,\sigma)\ne (j,\sigma')}U_{ij,a} : c_{\RR,a,i,\sigma}^\dag c_{\RR,a,i,\sigma} c_{\RR,a,j,\sigma'}^\dag c_{\RR,a,j,\sigma'}: \nonumber \\ 
&
-\frac{1}{2}\sum_{\RR,a,i,j,\sigma,\sigma'}J_{ij,a}: c_{\RR,a,i,\sigma}^\dag c_{\RR,a,i,\sigma'} c_{\RR,a,j,\sigma'}^\dag c_{\RR,a,j,\sigma}:
+ \frac{1}{2}\sum_{\RR,a,i,j,\sigma,\sigma's}
J_{ij,a}:c_{\RR,a,i,\sigma}^\dag c_{\RR,a,i,\sigma'}^\dag c_{\RR,a,j,\sigma'}c_{\RR,a,j,\sigma}:
\ea 
The interaction matrices for the five $d$ orbitals of each Co atom are obtained from
the \textit{ab-initio} calculations using constraint random phase approximation~\cite{aryasetiawan2004frequency, aryasetiawan2006calculations, miyake2009ab} and are given below
\ba 
U_a = 
\begin{bmatrix}
    3.79eV & 2.78eV & 2.78 eV & 2.33 eV & 2.35eV \\
    2.78eV & 3.35 eV & 2.33 eV & 2.41 eV & 2.45 eV\\
    2.78 eV & 2.33 eV & 3.35eV & 2.41eV & 2.45eV \\
    2.33eV & 2.41eV & 2.41eV & 3.37eV & 2.96eV \\ 
    2.35eV & 2.45eV & 2.45 eV & 2.96eV & 3.68eV
\end{bmatrix}_{ij},\quad 
J_a = 
\begin{bmatrix}
   0 eV & 0.39eV &  0.39 eV & 0.64 eV & 0.68eV \\
    0.39eV & 0 eV& 0.48 eV & 0.51 eV & 0.53 eV\\
    0.39 eV & 0.48eV & 0 eV & 0.51eV & 0.53eV \\
    0.64eV & 0.51eV & 0.51eV &  0 eV& 0.28eV \\ 
    0.68eV & 0.53eV & 0.53 eV & 0.28eV & 0 eV
\end{bmatrix}_{ij}
\ea 
To perform the mean-field coupling, we first rewrite the interaction term $H_{U,c}$ into the following generic formula
\ba 
H_{U,c} = \frac{1}{2} \sum_{\RR,a} V_{\alpha\beta\gamma\delta} :c_{\RR,a ,\alpha}^\dag c^\dag_{\RR,a ,\beta} c_{\RR,a ,\gamma} c_{\RR,a ,\delta}:
\ea 
where we use $\alpha,\beta,\gamma,\delta$ to denote both the spin and orbital indices, and the on-site interactions are characterized by the tensor $V$.
% The above generic interaction term can be written as (after expanding the normal ordering term)
% \ba 
% H_{U,c} =& \frac{1}{2} \sum_{\RR,\alpha\beta\gamma\delta} V_{\alpha\beta\gamma\delta} :c_{\RR,a ,\alpha}^\dag c^\dag_{\RR,a ,\beta} c_{\RR,a ,\gamma} c_{\RR,a ,\delta}:
% \ea
The Hartree-Fock (mean-field) decoupling of the four-fermion interactions term reads
\ba 
H_{U,c}^{MF} =& \frac{1}{2} \sum_{\RR,\alpha\beta\gamma\delta} V_{\alpha\beta\gamma\delta} \bigg\{-
O_{\RR,a, \alpha\delta}O_{\RR,a,\beta,\gamma} +O_{\RR,a,\alpha\gamma}O_{\RR,a,\beta,\delta } \nonumber\\ 
&+\bigg[ 
O_{\RR,a,\alpha\delta} :c_{\RR,a ,\beta}^\dag c_{\RR,a ,\gamma}:
- O_{\RR,a, \alpha \gamma} :c_{\RR,a ,\beta}^\dag c_{\RR,a ,\delta}:+\text{h.c.}
\bigg] \bigg\}
\label{eq:mf_decoup_c}
\ea 
where we have introduced the mean-fields
\ba 
O_{\RR,a,\alpha\beta} = \langle :c_{\RR,a,\alpha}^\dag c_{\RR,a,\beta}:\rangle 
\label{app:eq:mean_field_afm}
\ea 
The expectation value is taken with respect to the mean-field Hamiltonian
\ba 
H_{c}^{MF} = H_{c} +H_{U,c}^{MF}
\ea 
The mean fields (Eq.~\ref{app:eq:mean_field_afm}) will be self-consistently determined. The final mean-field Hamiltonian then describes the dispersion of $c$ electrons in the AFM phase
% Then, we obtain the following mean-field Hamiltonian that describes the behaviors of $c$ electrons 
\ba 
H^{AFM}_c = \sum_{\kk,ab,ij,\sigma\sigma'}\epsilon^{AFM}_{\kk,ab,ij,\sigma\sigma'}c_{\kk,a,i,\sigma}^\dag c_{\kk,b,j,\sigma'} +E_0
\label{eq:mf_ham_c}
\ea 
where $\epsilon^{AFM}_{\kk,ab,ij}$ includes the contribution from the non-interacting Hamiltonian and also the contribution from mean-fields decoupling (Eq.~\ref{eq:mf_decoup_c}).We will later specify which model was considered in the mean-field calculations. $E_0$ is the constant term from mean-field decoupling (Eq.~\ref{eq:mf_decoup_c}).

\subsection{Mean-field and DFT results of the AFM phase}
To describe the magnetic order, it is more convenient to use the conventional cell. For what follows, we will always take the conventional cell as our unit cell. In addition, the conventional cell of the PM phase is the primitive cell of the type-A AFM phase. As we will show later in this section, the type-A AFM state is also the ground state with the lowest energy.  

\begin{figure}
    \centering
    \includegraphics[width=0.5\textwidth]{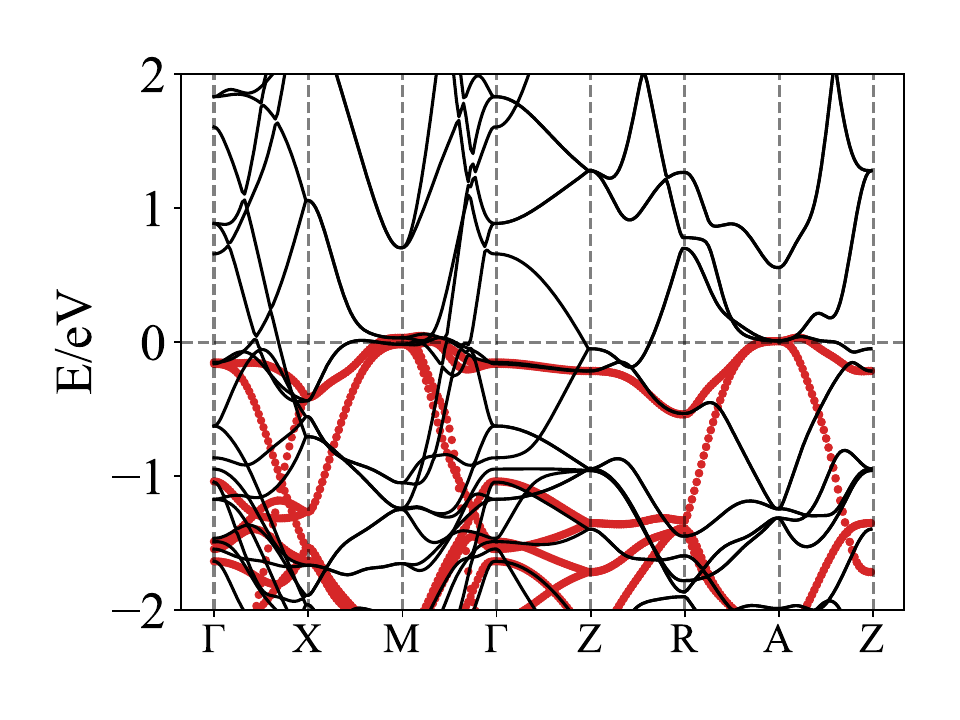}
    \caption{Comparison between DFT bands of $c$ electrons (black) and the bands from the two-orbital model (red) in the conventional cell in the PM phase. In the conventional cell, the two-orbital model produces 8 two-fold-degenerate bands. We observe that the two-orbital model reproduces the relatively flat bands near the Fermi energy. }
    \label{fig:pm_comp_c_band}
\end{figure}

We first show the band structure of the PM phase in the conventional cell. As shown in Fig.~\ref{fig:pm_comp_c_band}, we can, again, observe the two-orbital model reproduce the relatively flat bands near the Fermi energy. 
We then discuss the mean-field results of the two-orbital model.

% As we discussed at the beginning of this section, our relatively flat bands introduce a type-A AFM order. To demonstrate the type-A AFM order, we calculate the mean-field energy of various types of magnetic order. 

To describe the magnetic order, we note that there are four Co per unit cell (conventional cell) located at (see Fig.~\ref{fig:lat_mag_structure})
\ba 
(0,\frac{1}{2},\frac{1}{4}),\quad (\frac{1}{2},0,\frac{1}{4}),
(\frac{1}{2},0,\frac{3}{4}),\quad (0,\frac{1}{2},\frac{3}{4})
\ea 
To realize different magnetic ordering, we take different initializations of the mean-fields $O^{init}$. 
We consider four types of magnetic order as shown below 
\begin{itemize}
    \item Type-A AFM
    \ba 
    O^{init}_{\RR,a,(i\sigma, j\sigma')} =
    \begin{cases}
        n_0\sigma \delta_{i,j}\delta_{\sigma,\sigma'} & \text{, if $a\in \{Co_1,Co_2\}$ } \\
       - n_0\sigma \delta_{i,j}\delta_{\sigma,\sigma'} & \text{, if $a\in \{Co_3,Co_4\}$ } \\
       0 & \text{, otherwise}
    \end{cases}
    \ea 
     \item FM
    \ba 
    O^{init}_{\RR,a,(i\sigma, j\sigma')} =
    \begin{cases}
        n_0\sigma \delta_{i,j}\delta_{\sigma,\sigma'} & \text{, if $a\in \{Co_1,Co_2,Co_3,Co_4\}$ } \\
       0 & \text{, otherwise}
    \end{cases}
    \ea 
    \item In-plane AFM and out-of-plane AFM
    \ba 
    O^{init}_{\RR,a,(i\sigma, j\sigma')} =
    \begin{cases}
        n_0\sigma \delta_{i,j}\delta_{\sigma,\sigma'} & \text{, if $a\in \{Co_1,Co_3\}$ }  \\ 
        -n_0\sigma \delta_{i,j}\delta_{\sigma,\sigma'} & \text{, if $a\in \{Co_2,Co_4\}$ }  \\ 
       0 & \text{, otherwise}
    \end{cases}
    \ea 
    \item In-plane AFM and out-of-plane FM
    \ba 
    O^{init}_{\RR,a,(i\sigma, j\sigma')} =
    \begin{cases}
        n_0\sigma \delta_{i,j}\delta_{\sigma,\sigma'} & \text{, if $a\in \{Co_1,Co_4\}$ }  \\ 
        -n_0\sigma \delta_{i,j}\delta_{\sigma,\sigma'} & \text{, if $a\in \{Co_2,Co_3\}$ }  \\ 
       0 & \text{, otherwise}
    \end{cases}
    \ea 
\end{itemize}
The mean-field energies of the above states are 
\begin{center}
\begin{tabular}{c|c|c|c|c}
   Order & Type-A AFM & FM & In-plane AFM and out-of-plane AFM & In-plane AFM and out-of-plane FM   \\
   \hline 
  $(E-E_0)$/eV&  0& 0.006&0.184 &0.153
\end{tabular}
\end{center}
where we have subtracted the energy of the ground state $E_0$ (the energy of the type-A AFM phase).

We can observe that the type-A AFM phase is favored. Moreover, we can observe states with in-plane ferromagnetism (type-A AFM and FM) have much lower energy than in-plane antiferromagnetic states ($\sim 0.1$eV lower). The energy difference between type-A AFM and FM is small ($\sim 0.01$eV). This is because the $z$-direction correlation comes from the out-of-plane coupling. Since the out-of-plane coupling is weak, the energy difference between type-A AFM and FM is relatively small. 

\begin{figure}
    \centering
    \includegraphics[width=0.7\textwidth]{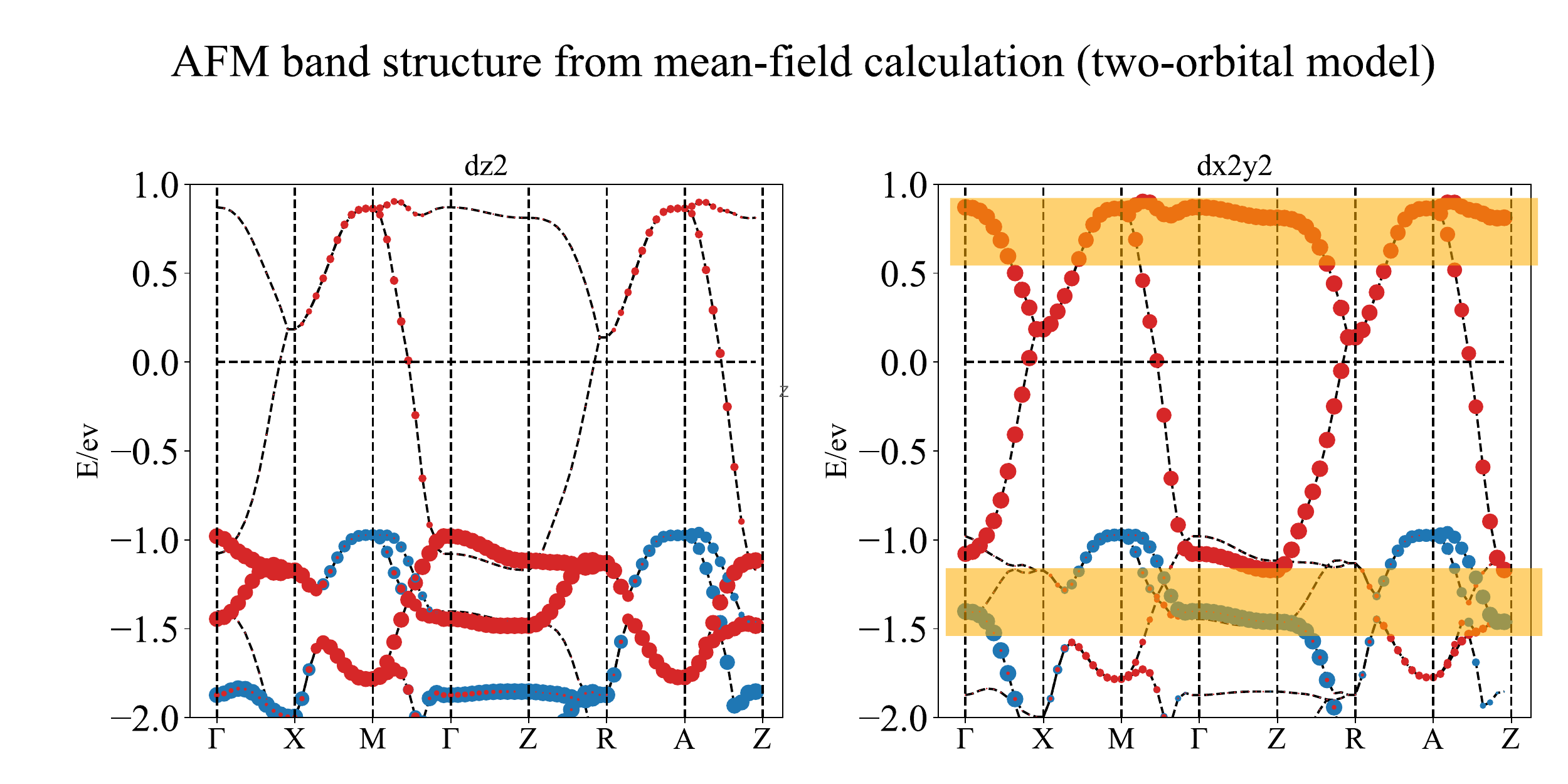}
    \caption{Dispersion of AFM phase obtained by solving tow-orbital model via mean-field approximations at low temperatures $T=0.001$eV. Red color labels bands from spin $\dn$ electrons of Co$_1$, Co$_2$ atoms and spin $\up$ electrons of Co$_3$, Co$_4$ atoms. Blue color labels bands from spin $\up$ electrons of Co$_1$, Co$_2$ atoms and spin $\dn$ electrons of Co$_3$, Co$_4$ atoms. We can observe the splitting of the relatively flat bands due to the magnetism (yellow-shaded region). Sizes of the dots characterize the orbital weights of two relevant $d$ orbitals. We observe that the polarized flat bands mostly come from the $d_{x^2-y^2}$ orbitals, which is also consistent with DFT results (Fig.~\ref{fig:AFM_bs}).}
    \label{fig:disp_afm_simp}
\end{figure}

\begin{figure}
    \centering
    \includegraphics[width=0.5\textwidth]{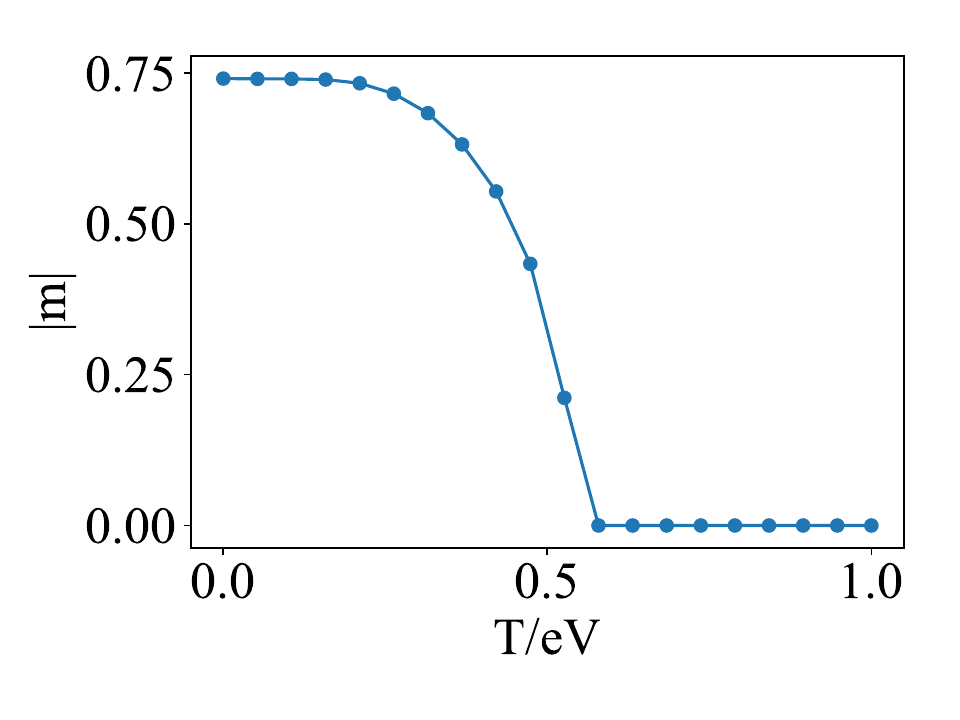}
    \caption{Evolution of order parameter $m$ (Eq.~\ref{eq:mag_op}) as a function of temperature. }
    \label{fig:afm_op_vs_T}
\end{figure}

In Fig.~\ref{fig:disp_afm_simp}, we also show the mean-field band structures of the type-A AFM phase. 
We can also observe the magnetic order splits the relatively flat bands. 
We also perform calculations of the type-A AFM phase at different temperatures, where the evolution of order parameters as a function of temperature is given in Fig.~\ref{fig:afm_op_vs_T}. The order parameter is defined as the average magnetic momentum of each Co atom
\ba 
m = &\frac{1}{4N} \sum_{\RR,\sigma}\sum_{i}\bigg\langle  \bigg[ \sigma c_{\RR, Co1, i, \sigma}^\dag c_{\RR, Co1, i, \sigma}
+ \sigma c_{\RR, Co2, i, \sigma}^\dag c_{\RR, Co2, i, \sigma}\nonumber\\ 
&
- \sigma c_{\RR, Co3, i, \sigma}^\dag c_{\RR, Co3, i, \sigma}
- \sigma c_{\RR, Co4, i, \sigma}^\dag c_{\RR, Co4, i, \sigma}\bigg]\bigg\rangle 
\label{eq:mag_op}
\ea 
% We find a relatively high AFM transition temperature $T\sim 0.6$eV. However, it is worth mentioning that the mean-field calculations ignore the fluctuations of order parameters and usually overestimate the transition temperatures. 

\begin{figure}
    \centering
    \includegraphics[width=0.8\textwidth]{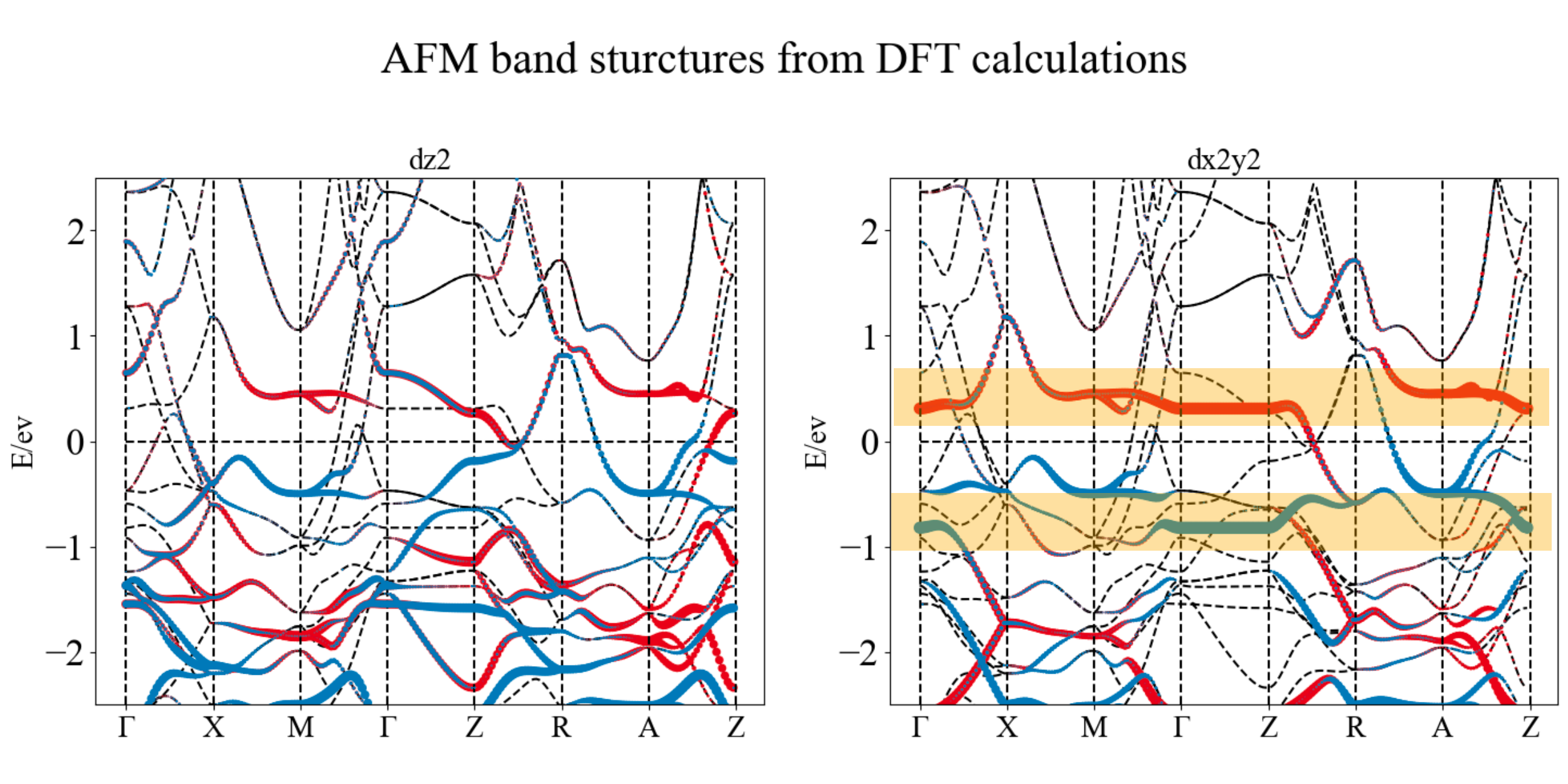}
    \caption{
    % (Top) Band structures of AFM phase obtained from mean-field (MF) calculations using paramagnetic DFT model. (Bottom) 
    Direct DFT calculations in the AFM phase. The yellow regions mark the positions of the relatively flat bands. Red color labels bands from spin $\dn$ electrons of Co$_1$, Co$_2$ atoms and spin $\up$ electrons of Co$_3$, Co$_4$ atoms. Blue color labels bands from spin $\up$ electrons of Co$_1$, Co$_2$ atoms and spin $\dn$ electrons of Co$_3$, Co$_4$ atoms. Sizes of the dots characterize the orbital weights of two relevant $d$ orbitals. }
    \label{fig:AFM_bs}
\end{figure}

Finally, we directly perform the DFT calculations in the magnetic phase which include all the $spd$ orbitals and treat $f$ electrons as core states. 
Again, we observe the type-A AFM phase has the lowest energy. The energies of various magnetic states are given below
\begin{center}
\begin{tabular}{c|c|c|c|c}
   Order & Type-A AFM & FM & In-plane AFM and out-of-plane AFM & In-plane AFM and out-of-plane FM   \\
   \hline 
   $(E-E_0)$/eV&  0& 0.013 &0.236 & 0.238
\end{tabular}
\end{center}
where we have subtracted the energy of the ground state $E_0$ (the energy of the type-A AFM phase).
The DFT band structures of the type-A AFM phase are shown in Fig.~\ref{fig:AFM_bs}.

Finally, in Sec.~\ref{sec:app:flat_band_magnetism}, we also employ perturbation theory to analytically demonstrate the stability of the AFM phase. Thus, all our calculations consistently support each other and indicate the stability of the AFM phase.

\section{Magnetism of LaCo$_2$P$_2$}
\label{sec:app:LCP}
\begin{figure}
    \centering
    \includegraphics[width=0.8\textwidth]{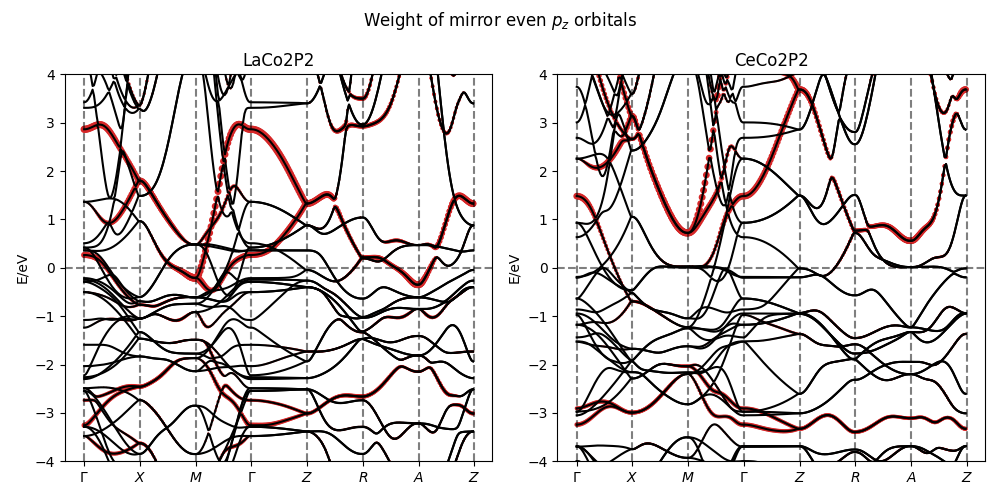}
    \caption{Band structures of LaCo$_2$P$_2$ and CeCo$_2$P$_2$ in the conventional cell. The red dots mark the orbital weights of the mirror even $p_z$ orbitals.}
    \label{fig:pz_orb_ce_la}
\end{figure}

We also study the magnetism in a related compound LaCo$_2$P$_2$. There are two main differences between LaCo$_2$P$_2$ and CeCo$_2$P$_2$. First, La holds zero $f$ electrons, which indicates the absence of Kondo behavior. 
% Then, LaCo$_2$P$_2$ is similar to the CeCo$_2$P$_2$ at high temperature (above Kondo temperature). 
Second, due to the change of the distance between P atoms, $p_z$@P electron bands cross the Fermi energy in LaCo$_2$P$_2$ (see Fig.~\ref{fig:pz_orb_ce_la}). The same bands are above the Fermi energy in CeCo$_2$P$_2$. As a consequence, these additional bands introduce effective RKKY interactions and stabilize ferromagnetic states. Here we comment that, in the CeCo$_2$P$_2$, the $p_z@P$ orbitals are far from the Fermi energy and will not contribute to effective RKKY interactions. 

\begin{figure}
    \centering
    \includegraphics[width=0.4\textwidth]{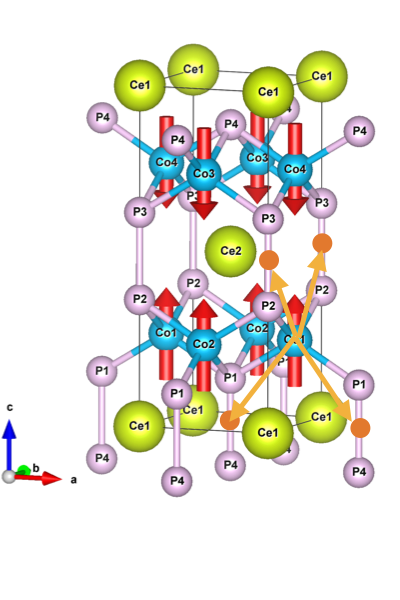}
    \caption{Orange dots mark the Wannier center of the mirror-even (mirror-odd) orbital of $p_z@$P (Eq.~\ref{eq:mirror_eo_orbital_pz}). The orange arrows mark the nearest-neighbor hopping between the mirror-even orbitals and $d_{x^2-y^2}$ orbitals of Co$_1$ atom (Eq.~\ref{eq:hop_pz_dx2y2}).}
    \label{fig:pz_d_coupling}
\end{figure}

We first discuss the band structures of LaCo$_2$P$_2$ and CeCo$_2$P$_2$. We define the mirror-even and mirror-odd basis of $p_z@$P orbitals in the conventional cell (Fig.~\ref{fig:pz_d_coupling})
\ba 
&c_{\RR, e/o, 1, \sigma} = 
\frac{1}{\sqrt{2}}\bigg( c_{\RR,P_1,p_z,\sigma} \pm  c_{\RR-(0,0,1),P_4,p_z,\sigma}\bigg) 
\nonumber\\ 
&c_{\RR, e/o, 2, \sigma} = \frac{1}{\sqrt{2}}(c_{\RR,P_3,p_z,\sigma} \pm  c_{\RR,P_2,p_z,\sigma
}) 
\label{eq:mirror_eo_orbital_pz}
\ea 
We show the orbital weights of the mirror-even states in ~\cref{fig:pz_orb_ce_la}. We can observe that the mirror even states in the LaCo$_2$P$_2$ are close to the Fermi energy. This is because, the nearest-neighbor hopping $t_{p_z,z}$ of $p_z@P$ orbitals along $z$ directions are $2.13$eV for LaCo$_2$P$_2$ and $2.65$eV for CeCo$_2$P$_2$. This hopping, after projecting to the mirror-even orbitals, behaves as a mass term
\ba 
t_{p_z,z}\sum_{\RR,i=1,2,\sigma} c_{\RR,e,i,\sigma}^\dag c_{\RR,e,i,\sigma} 
\ea 
Therefore, by reducing the values of $t_{p_z,z}$ in LaCo$_2$P$_2$, the energy of $c_{\RR,e,i,\sigma}$ orbitals is also lower.

We next show the RKKY-like interactions induced by the coupling to the mirror-even $p_z$ orbitals generate effective ferromagnetic interactions between Co $d$ atoms and then stabilize a ferromagnetic order in LaCo$_2$P$_2$. Since magnetism is mostly contributed by the $d_{x^2-y^2}$ orbitals of Co, we first investigate the hopping between $p_z@$P and $d_{x^2-y^2}@$Co.  The $d_{x^2-y^2}@$Co couples to the four nearest-neighbor $p_z@$P orbitals with strength $t_{p_z,d_{x^2-y^2}}  =-0.487$eV. 
Then the nearest-neighbor coupling between mirror-even states and $d_{x^2-y^2}$ electrons reads (Fig.~\ref{fig:pz_d_coupling})
\ba 
&\sum_{\RR,\sigma}\frac{1}{\sqrt{2}}t_{p_z,d_{x^2-y^2}} 
\bigg[ c_{\RR,Co_1,d_{x^2-y^2}}^\dag 
\bigg( c_{\RR,e,1,\sigma} + c_{\RR+(-1,0,0),e,1,\sigma } +c_{\RR,e,2,\sigma} +c_{\RR+(0,1,0),e,2,\sigma}
\bigg) \nonumber\\ 
&+c_{\RR,Co_2,d_{x^2-y^2}}^\dag
\bigg( 
c_{\RR,e,1,\sigma} + c_{\RR+(1,0,0),e,1,\sigma} +c_{\RR,e,2,\sigma} + c_{\RR+(0,-1,0),e,2,\sigma}
\bigg) \nonumber\\ 
&+c_{\RR,Co_3,d_{x^2-y^2}}^\dag 
\bigg( c_{\RR+(0,0,1),e,1,\sigma} +c_{\RR+(1,0,1),e,1,\sigma} + c_{\RR,e,2,\sigma} +c_{\RR+(0,-1,0), e,2,\sigma}
\bigg)\nonumber\\ 
&
+c_{\RR,Co_4,d_{x^2-y^2}}^\dag 
\bigg( c_{\RR+(0,0,1),e,1,\sigma} +c_{\RR+(0,1,1),e,1,\sigma} + c_{\RR,e,2,\sigma} +c_{\RR+(-1,0,0), e,2,\sigma}
\bigg) +\text{h.c.}\bigg] 
\label{eq:hop_pz_dx2y2}
\ea 
We can then observe that, locally, each $c_{\RR,Co_i,d_{x^2-y^2}}$ orbitals couple to a molecular orbital formed by the mirror-even states of $p_z$ electrons. We can thus define
 \ba 
&w_{\RR,e,1,\sigma} = \frac{1}{2}
\bigg( c_{\RR,e,1,\sigma} + c_{\RR+(-1,0,0),e,1,\sigma } +c_{\RR,e,2,\sigma} +c_{\RR+(0,1,0),e,2,\sigma}
\bigg) 
\nonumber\\ 
&w_{\RR,e,2,\sigma} = \frac{1}{2}
\bigg( 
c_{\RR,e,1,\sigma} + c_{\RR+(1,0,0),e,1,\sigma} +c_{\RR,e,2,\sigma} + c_{\RR+(0,-1,0),e,2,\sigma}
\bigg) \nonumber\\ 
&w_{\RR,e, 3,\sigma} = \frac{1}{2}
\bigg( c_{\RR+(0,0,1),e,1,\sigma} +c_{\RR+(1,0,1),e,1,\sigma} + c_{\RR,e,2,\sigma} +c_{\RR+(0,-1,0), e,2,\sigma}
\bigg)\nonumber\\ 
&
w_{\RR,e,4,\sigma} = \frac{1}{2}
\bigg( c_{\RR+(0,0,1),e,1,\sigma} +c_{\RR+(0,1,1),e,1,\sigma} + c_{\RR,e,2,\sigma} +c_{\RR+(-1,0,0),e,2,\sigma}
\bigg) 
 \ea 
and then Eq.~\ref{eq:hop_pz_dx2y2} can be written as
\ba 
\sum_{\RR,\sigma}\sum_{i=1,2,3,4}\sqrt{2} t_{p_z,d_{x^2-y^2}} 
[ c_{\RR,Co_i,d_{x^2-y^2}}^\dag w_{\RR,e,i,\sigma}+\text{h.c.}] 
\ea 
  
This hopping term then introduces an effective spin-spin or Kondo-like coupling between $w_{\RR,b,i,\sigma}$ and $c_{\RR,Co_i,d_{x^2-y^2}}$ (see also Appendix \ref{sec:app:klm})
\ba 
J_K \sum_{\RR, i,\mu, \sigma_1,\sigma_2,\sigma_3,\sigma_4} w_{\RR,b,i,\sigma_1}^\dag {\sigma^\mu_{\sigma_1,\sigma_2}}w_{\RR,b,i,\sigma_2}
c^\dag_{\RR,Co_i,d_{x^2-y^2},\sigma_3}\frac{\sigma^\mu_{\sigma_3,\sigma_4}}{2}   c_{\RR,Co_i,d_{x^2-y^2},\sigma_4}
\ea 
where $J_K = 4|t_{p_z,d_{x^2-y^2}}|^2/|U_{d}|$ is the coupling strength. 
$U_{d}$ is the strength of Hubbard repulsion of Co $d$ electrons. 
We observe $d$ electron develops magnetic order and has strong interactions $U_d\sim 4$eV (from DFT calculation). This indicates $d$ electrons are correlated. However, in general, $p_z$ orbitals are itinerant and weakly correlated, since $p$ orbitals have a very strong hopping ($\sim 2$eV). Therefore, $d$ orbitals and $p$ orbitals are coupled via an effective Kondo type of coupling. 
% After integrating out the $w_{\RR,e,4,\sigma}$ states, we obtain effective RKKY interactions between the local moments of Co $d$ electrons. We now show that, such an effective RKKY interaction introduces a ferromagnetic correlation between each Co planes, which then stabilize a ferromagnetic state in LaCo$_2$P$_2$. 
The particle-hole fluctuations of the $p$ orbitals then generate effective RKKY interactions between $d$ orbitals via the Kondo coupling (see Appendix S5 of Ref.~\cite{PhysRevLett.131.026502} for the derivations of RKKY interactions and also Refs.~\cite{RKKY_1,RKKY_2,RKKY_3}) are  
\ba 
&\sum_{\RR,\RR'} J^{RKKY}_{\RR-\RR',ij}S_{\RR,Co_i, d_{z^2}}^\mu S_{\RR',Co_i, d_{z^2}}^\nu \nonumber\\ 
& J^{RKKY}_{\RR-\RR',ij}= - \frac{1}{2} J_K^2\chi_{ij,\RR-\RR'}
\ea 
where $S_{\RR,Co_i,d_{z^2}} = \sum_{\sigma_1,\sigma_2}c_{\RR,Co_i,d_{x^2-y^2},\sigma_1}\frac{\sigma^\mu_{\sigma_1,\sigma_2}}{2}  c_{\RR,Co_i,d_{x^2-y^2},\sigma_2}^\dag$ is the spin operator of $d_{x^2-y^2}$ orbitals of Co$_i$ atom and the susceptibility is defined as
\ba 
&\chi_{ij,\RR-\RR'} = -\int_{0}^\beta G_{ij}(\RR-\RR', \tau) G_{ji}(\RR'-\RR,-\tau) d\tau  \nonumber\\ 
&G_{ij}(\RR,\tau)=  -\langle T_\tau w_{\RR,b,i,\sigma}(\tau) w_{\RR',b,j,\sigma}^\dag(0) \rangle 
\ea 
Since we have a strong in-plane ferromagnetic coupling due to the non-atomic nature of the flat band, we now aim to understand what types of out-of-plane coupling the RKKY interaction generates. 
We can estimate the nearest-neighbor RKKY coupling along $z$ directions, which is equivalent to calculate
\ba 
J_z = -\frac{J_K^2}{2} \chi_{14,\RR=0} =  -\frac{J_K^2}{2} \chi_{23,\RR=0}
\ea 
By numerical calculations, we find $\chi_{14,\RR=0}=\chi_{23,\RR=0} = 0.018$eV$^{-1} > 0 $. We observe that, Co$_4$ (Co$_3$) atom is on top of the Co$_1$ (Co$_2$) atom. Then the nearest-neighbor RKKY interactions along $z$ directions are proportional to $\chi_{12,\RR=0}$ or $\chi_{23,\RR=0}$. These interactions are ferromagnetic and favor the formation of ferromagnetism along $z$ directions. 

% We could also understand this ferromagnetic coupling from the lattice structures. 
% The mirror-even states of the $p_z@$P orbitals are located between two Co planes. Each mirror-even state couples anti-ferromagnetically to the $d$ electrons of two nearest-neighbor Co planes, since the effective Kondo-like coupling $J_K$ is antiferromagnetic. Therefore, the spins of $d$ electrons of two Co planes tend to anti-align with the spin of the same mirror-even state. This introduces an effective ferromagnetic coupling between two Co planes and stabilizes a ferromagnetic order.

To further confirm the stability of the ferromagnetic state, we calculate the energy of the ferromagnetic state and type-A antiferromagnetic state via DFT calculations. We indeed find that the energy of the ferromagnetic state is 6.2 meV lower than the type-A antiferromagnetic state.

\section{Atomic limit of $f$ electron}
\label{sec:f_atom}
In this appendix, we discuss the atomic physics of $f$ electrons. We pick out $f$ electrons at site $\RR_0$ and sublattice $a$
\ba 
f_{i,\sigma} = f_{\RR_0, a, i,\sigma}
\ea 
Since all Ce atoms are equivalent, the choice of $\RR_0,a$ does not affect the atomic physics. The single-site Hamiltonian of $f$ electrons is given in Eq.~\ref{eq:app:h_f_atomic} and is also shown below
\ba 
&H_{loc}[f^\dag ,f ] = H_{U}[f^\dag, f] + H_{CEF}[f^\dag, f] +H_{soc}[f^\dag, f]\nonumber\\ 
&H_U[f^\dag, f] = \frac{U}{2} \bigg( \sum_{i,\sigma} f_{i,\sigma}^\dag f_{i,\sigma}  -n_0 \bigg) ^2 \nonumber\\ 
&H_{CEF}[f^\dag, f] = \sum_{ij}\epsilon_{f,ij}f_{i,\sigma}^\dag f_{j,\sigma} \nonumber\\ 
&H_{soc}[f,f^\dag] = \lambda \sum_{\mu}L^\mu S^\mu .
\label{eq:ham_atom_f}
\ea 
$H_U$ is the on-site Hubbard interaction, $H_{CEF}$ is the crystal field splitting and $H_{soc}$ is the spin-orbit coupling (Eq.~\ref{eq:soc_ham}). Seven $f$ orbitals are labeled by
\ba 
f_1 = f_{z^3},\quad f_{2} = f_{xz^2}, \quad f_{3}= f_{yz^2}, \quad f_4 = f_{z(x^2-y^2)},\quad f_{5} = f_{xyz},\quad f_6 = f_{x(x^2-3y^2)},\quad f_{7} = f_{y(3x^2-y^2)}
\ea 
The site symmetry group of Ce is 4/mmm. 
After including the spin index, the orbitals form the following irreducible representations of the site symmetry group (see Tab.~\ref{tab:chacracter_tab_4mmm})
\ba 
&f_{1,\up}^\dag,f_{1,\dn}^\dag:\overline{\Gamma_9}\nonumber\\ 
&f_{4,\up}^\dag,f_{4,\dn}^\dag:\overline{\Gamma_8}\nonumber\\
&f_{5,\up}^\dag,f_{5,\dn}^\dag:\overline{\Gamma_8}\nonumber \\
&\frac{f_{2,\up}^\dag -i f_{3,\up}^\dag}{\sqrt{2}},\frac{f_{2,\dn}^\dag + if_{3,\dn}^\dag}{\sqrt{2}}:\overline{\Gamma_9}\nonumber\\
&\frac{f_{2,\dn}^\dag - if_{3,\dn}^\dag}{\sqrt{2}},\frac{f_{2,\up}^\dag + if_{3,\up}^\dag}{\sqrt{2}}:\overline{\Gamma_8}\nonumber\\
&\frac{f_{6,\up}^\dag -if_{7,\up}^\dag}{\sqrt{2}},\frac{f_{6,\dn}^\dag +i f_{7,\dn}^\dag}{\sqrt{2}}:\overline{\Gamma_8}\nonumber\\
&\frac{f_{6,\dn}^\dag -if_{7,\dn}^\dag}{\sqrt{2}},\frac{f_{6,\up}^\dag + if_{7,\up}^\dag}{\sqrt{2}}:\overline{\Gamma_9}\nonumber\\
\ea 

\begin{table}[]
    \centering
    \begin{tabular}{c|c|c|c|c}
        Irrep & $2_{001}$ & $4^+_{001}$ & $2_{010}$ & $-1$ \\
        \hline 
        $\bar{\Gamma_8}$ &0  & $-\sqrt{3}$ & 0 &-2  \\
        $\bar{\Gamma_9}$ &0 & $\sqrt{2}$ & 0 &-2 \\
    \end{tabular}
    \caption{Character table of 4/mmm.}
    \label{tab:chacracter_tab_4mmm}
\end{table}

For future convenience, we also show the relation between the wavefunction of $f_{1,...,7}$ and spherical harmonic wave functions below:
\begin{center}
\begin{tabular}{c|c|c}
Orbital & Wavefunction & Harmonics \\ 
\hline
   $f_1=f_{z^3}$  & $\frac{\sqrt{7}}{4\sqrt{\pi}} (5\cos^3(\theta)-3\cos(\theta))$  &$Y_3^0$ \\
     $f_2=f_{xz^2}$  & $\frac{\sqrt{21}}{4\sqrt{2\pi}} (5\cos^2(\theta)-1)\sin(\theta)\cos(\phi)$  & $(Y_3^{-1}-Y_3^1)/\sqrt{2}$\\ 
     $f_3=f_{yz^2}$   & $\frac{\sqrt{21}}{4\sqrt{2\pi}} (5\cos^2(\theta)-1)\sin(\theta)\sin(\phi)$  & $i(Y_3^{-1}+Y_3^1)/\sqrt{2}$\\ 
    $f_4=f_{z(x^2-y^2)}$  &  $\frac{\sqrt{105}}{4\sqrt{\pi}} \sin(\theta)^2\cos(\theta)\cos(2\phi)$ & $\frac{1}{\sqrt{2}}(Y_3^{-2}+Y_3^2)$ \\ 
    $f_5=f_{xyz}$  & $\frac{\sqrt{105}}{4\sqrt{\pi}} \sin(\theta)^2\cos(\theta)\sin(2\phi)$ &$\frac{i}{\sqrt{2}}(Y_3^{-2}-Y_3^2)$\\ 
    $f_6=f_{x(x^2-3y^2)}$  & $\frac{\sqrt{35}}{4\sqrt{2\pi}} \sin^3(\theta)(\cos^2(\phi)-3\sin^2(\phi))\cos(\phi)$ & $\frac{1}{\sqrt{2}}(Y_3^{-3} -Y_3^3)$\\ 
    $f_7=f_{y(3x^2-y^2)}$  &$\frac{\sqrt{35}}{4\sqrt{2\pi}} \sin^3(\theta)(3\cos^2(\phi)-\sin^2(\phi))\cos(\phi)$ & $\frac{i}{\sqrt{2}}(Y_3^{-3}+Y_3^3)$
\end{tabular}
\end{center}
$Y_L^m$ denotes the spherical Harmonic function which is also the eigenfunctions of the angular-momentum operator along $z$-direction ($L^z$). Based on the eigenvectors of $L^z$, we can introduce the following new electron operators $g_{l,\sigma}$ 
\ba 
&g_{0,\sigma}^\dag = f_{1,\sigma}^\dag \nonumber\\ 
&g_{1,\sigma}^\dag = \frac{1}{\sqrt{2}}(-f_{2,\sigma}^\dag -if_{3,\sigma}^\dag ),\quad 
g_{-1,\sigma}^\dag = \frac{1}{\sqrt{2}}(f_{2,\sigma}^\dag -i f_{3,\sigma}^\dag )\nonumber\\ 
&g_{2,\sigma}^\dag = \frac{1}{\sqrt{2}}(f_{4,\sigma}^\dag +if_{5,\sigma}^\dag),\quad 
g_{-2,\sigma}^\dag = \frac{1}{\sqrt{2}}(f_{4,\sigma}^\dag -if_{5,\sigma}^\dag)\nonumber\\
&g_{3,\sigma}^\dag = \frac{1}{\sqrt{2}}(-f_{6,\sigma}^\dag -i f_{7,\sigma}^\dag),\quad 
g_{-3,\sigma}^\dag =\frac{1}{\sqrt{2}}(f_{6,\sigma}^\dag -if_{7,\sigma}^\dag)
\label{eq:app:gbasis}
\ea 
where $g_{l,\sigma}^\dag$ creates an $f$ electron with $L^z$ eigenvalue $l$ and spin $\sigma$.

In what follows, we will discuss the eigenstates of the atomic Hamiltonian of $f$ electrons. The Hubbard interaction ($\sim 10$eV) is the largest energy scale which forces the filling of $f$ electrons to be $1$ for each site. The spin-orbit coupling ($\sim 0.1$eV) is the second-largest energy scale which introduces an energy splitting between states with different total angular momentum. The crystal field splitting ($\sim 0.01$eV) is the smallest energy scale which further splits the energies of the states with the same total angular momentum.

\subsection{Hubbard interactions and charge $\pm 1$ excitation}
We first discuss the effect of Hubbard interactions
\ba 
H_{U}[f,f^\dag] = \frac{U}{2}(n^f-n_0)^2
\label{eq:Hubbard_f}
\ea 
where 
\ba 
n^f = \sum_{i,\sigma} f_{i,\sigma}^\dag f_{i,\sigma} 
\ea 
Since the system is in the Kondo limit with one $f$ electron per Ce atoms, the charge fluctuations of $f$ electron are suppressed by the Hubbard $U$. We then take $n_0 = 1$ and treat $U$ as the largest energy scale of the problem. The ground states of the $H_U$ are described by the atomic states of $f$ electron with exactly one $f$ electron. The excitation energy of charge $\pm 1$ excitation is $U/2$. 
Here, we also comment that, to simplify the problem, we do not include the orbital dependency of the Hubbard repulsion.

\subsection{Spin-orbit coupling}
\label{sec:app:spin_obit_coupling}
The on-site Hubbard interaction fixes the filling of $f$ electron to be one per Ce atom. This leads to a $14$-fold degenerate one-electron ground states ($7\times 2$ $f$ electrons). The degeneracy of the ground states will be lifted by the spin-orbit coupling and crystal fields.
From our DFT calculations by treating $f$ electrons as valence electrons, we find the strength of the spin-orbit coupling is
\ba 
\lambda = 0.092\text{eV} . 
\ea 
Since spin-orbit coupling is usually larger than the crystal field splitting, we first consider the effect of $H_{soc}$. In the $g_{l,\sigma}$ basis (Eq.~\ref{eq:app:gbasis}), the spin-orbit coupling can be written as
\ba 
&H_{soc} [f,f^\dag]= \lambda \sum_\mu L^\mu S^\mu \nonumber\\ 
&S^\mu = \sum_{l,\sigma\sigma'} g_{l,\sigma}^\dag \frac{\sigma^\mu_{\sigma\sigma'}}{2}g_{l,\sigma'}\nonumber\\ 
&L^\mu = \sum_{l,l',\sigma\sigma'} g_{l,\sigma}^\dag M^{L,\mu}_{ll'} g_{l',\sigma}
\label{eq:soc_ham}
\ea 
where $\sigma^\mu$ are Pauli matrices and the angular-momentum matrices $M^{L,\mu}_{ll'}$ are defined as
\ba 
&M^{L,z}_{ll'} = \delta_{l,l'}l \nonumber\\ 
&M^{L,x} = \frac{1}{2}(M^{L,+} +M^{L,-}),\quad M^{L,y} =\frac{1}{2i}(M^{L,+}-M^{L,-})\nonumber\\ 
&M^{L,+}_{ll'} = \delta_{l'+1,l}\sqrt{L(L+1)-l'(l'+1)},\quad M^{L,-}_{ll'} = \delta_{l'-1,l}\sqrt{L(L+1)-l'(l'-1)},\quad L=3 
\ea

After diagonalizing $H_{soc}$, we find the following eigenvalues (for the states with particle number $1$)
\ba 
&E_0 = -2\lambda ,\quad \text{6-fold degeneracy with total angular momentum }J=5/2 \nonumber\\
&E_1 = 3/2\lambda ,\quad \text{8-fold degeneracy  with total angular momentum }J=7/2 
\label{eq:app:energy_split_J}
\ea 
where $E_1-E_0 \sim 0.322$eV. 
% Written explicitly, the ground states of $H_{soc}$ with energy $E_0=-2\lambda$ are
% \ba 
% &|1\rangle = \frac{1}{\sqrt{7}}(- g^\dag_{-2,\dn} + \sqrt{6}g^\dag_{-3,\up} )|0\rangle ,\quad 
% |2\rangle = \frac{1}{\sqrt{7}}(- g^\dag_{2,\up} +\sqrt{6}g_{3,\dn}^\dag  ) |0\rangle \nonumber\\ 
% &|3\rangle = \frac{1}{\sqrt{7}}(-\sqrt{2}g_{-1,\dn}^\dag  + \sqrt{5} g_{-2,\up}^\dag)|0\rangle ,\quad 
% |4\rangle = \frac{1}{\sqrt{7}}(-\sqrt{2}g_{1,\up}^\dag  + \sqrt{5} g_{2,\dn}^\dag)|0\rangle \nonumber\\ 
% &  |5\rangle = \frac{1}{\sqrt{7}}(-\sqrt{3} g_{0,\dn}^\dag + 2g_{-1,\up}^\dag )|0\rangle ,\quad 
% |6\rangle = \frac{1}{\sqrt{7}}(-\sqrt{3} g_{0,\up}^\dag + 2g_{1,\dn}^\dag )|0\rangle
% \label{eq:app:atomic_J_5_2_states}
% \ea 
% where the corresponding irreducible representations are
% \ba 
% &|1\rangle, |2\rangle: \overline{\Gamma_8}\nonumber\\ 
% &|3\rangle, |4\rangle: \overline{\Gamma_8}\nonumber\\ 
% &|5\rangle, |6\rangle: \overline{\Gamma_9}
% \ea

\subsection{Crystal field splitting}
We next consider the crystal field splitting. The symmetry-allowed crystal field splitting can be obtained by only allowing hybridization between orbitals that form the same irreducible representation of the site-symmetry group and have the same spin indices. The crystal field splitting then takes the form of 
\ba 
H_{CEF}[f,f^\dag] =& \sum_\sigma \bigg[\epsilon_1 f_{1,\sigma}^\dag f_{1,\sigma} + \epsilon_2 (f_{2,\sigma}^\dag f_{2,\sigma}+ 
f_{3,\sigma}^\dag f_{3,\sigma}) 
+\epsilon_4 f_{4,\sigma}^\dag f_{4,\sigma}
+\epsilon_5 f_{5,\sigma}^\dag f_{5,\sigma} +
+\epsilon_6(f_{6,\sigma}^\dag f_{6,\sigma} + f_{7,\sigma}^\dag f_{7,\sigma}^\dag) 
\nonumber\\ 
&+\epsilon_7 \bigg( -f_{2,\sigma}^\dag f_{6,\sigma} + f_{3,\sigma}^\dag f_{7,\sigma}+\text{h.c.}\bigg)
\bigg] \nonumber\\ 
=&\sum_\sigma \bigg[ \epsilon_1 g_{0,\sigma}^\dag g_{0,\sigma}
+\epsilon_2(g_{1,\sigma}^\dag g_{1,\sigma} + g_{-1,\sigma}^\dag g_{-1,\sigma})
+\epsilon_6(g_{3,\sigma}^\dag g_{3,\sigma}+g_{-3,\sigma}^\dag g_{-3,\sigma}) \nonumber\\ 
&
+\frac{\epsilon_4+\epsilon_5}{2} (g_{2,\sigma}^\dag g_{2,\sigma}+g_{-2,\sigma}^\dag g_{-2,\sigma})
+\frac{\epsilon_4-\epsilon_5}{2} (g_{2,\sigma}^\dag g_{-2,\sigma}+g_{-2,\sigma}^\dag g_{2,\sigma}) \nonumber\\ 
&+\epsilon_7\bigg( 
g_{-1,\sigma}^\dag g_{3,\sigma} +g_{1,\sigma}^\dag g_{-3,\sigma} +\text{h.c.}
\bigg)
\bigg]  
\ea 
where we have (from our DFT calculations by treating $f$ as valence electrons)
\ba 
\epsilon_1 = 0.0346eV,\quad \epsilon_2= -0.0501eV,\quad \epsilon_4 =-0.0400eV,\quad \epsilon_{5} = 0.0885eV,\quad \epsilon_{6} = 0.00876eV
,\quad 
\epsilon_7 = 0.0469eV
. 
\ea 
$H_{CEF}$ will lift the degeneracy of the 6-fold degenerate states with angular momentum $J=\frac{5}{2}$.
It will also lift the degeneracy of the 8-fold degenerate states with angular momentum $J=\frac{7}{2}$. 

For future reference, we now use $\eta_i$ to denote the electrons corresponding to the $i$-th eigenstates of the whole atomic Hamiltonian with energy $E_i$
\ba 
(H_U[f,f^\dag] + H_{soc}[f,f^\dag] +H_{CEF}[f,f^\dag])\eta_i^\dag |0\rangle = E_i \eta_i^\dag |0\rangle ,\quad i=1,...,14
\label{eq:app:eigen_value_atom_ham}
\ea 
where $\eta_i$ is a linear combination of $f$ electron operators and can be obtained by solving the atomic Hamiltonian (Eq.~\ref{eq:ham_atom_f}). 
We note that the local Hamiltonian also has time-reversal symmetry, thus all the eigenstates with particle number $1$ are two-fold degenerate. Without loss of generality, we take $E_{2i-1} = E_{2i}$ to characterize such degeneracy. 

% Since the splitting of the crystal field is at the order of $0.1$eV and is smaller than the gap induced by the spin-orbit coupling which is $(3/2\lambda+2\lambda) \sim 0.35$eV, we can effectively treat crystal field splitting as small perturbation. Then the $f$ states can be in general separated into two groups:
% \ba 
% &\{\eta_{i}\}_{i=1,...,6} \nonumber\\ 
% &\{\eta_{i}\}_{i=7,...,14} 
% \ea 
% The first group corresponds to the $J=5/2$ states with energy $\sim -2\lambda$, and the second group corresponds to the $J=7/2$ states with energy $\sim \frac{3\lambda}{2}$. 
After introducing the crystal field splitting, the total angular momentum $J$ is no longer a good quantum number because the crystal field term introduces hybridization between $J=5/2$ states and $J=7/2$ states. However, since crystal field splitting is relatively small compared to the spin-orbit coupling, the eigenvalues of the local Hamiltonian can still be separated into two groups with energy difference around $\frac{3}{2}\lambda -(-2\lambda) = \frac{7}{2}\lambda$ (Eq.~\ref{eq:app:energy_split_J}). Therefore, we still use $J=5/2$ and $J=7/2$ to denote two groups of $\eta$ states. 

% We now consider the effect of crystal field splitting on the six-lowest energy states with $J=5/2$. Via first-order perturbation theory, we find
% \ba 
% &E_1 =E_2 \approx
% -2\lambda + \frac{1}{7}(3\epsilon_1 + 4\epsilon_{2})
% =-0.1978eV, \quad \overline{\Gamma_9}
% % ,\quad J^z = \pm \frac{5}{2}
% \nonumber \\
% &E_3=E_4 \approx 
%  -2\lambda + 
% \frac{1}{7}(A -B) 
%  = -0.1816eV 
% , \quad \overline{\Gamma_8}
% \nonumber \\ 
% &E_5 =E_6 \approx  -2\lambda + 
% \frac{1}{7}(A +B)  =-0.1725eV,\quad \overline{\Gamma_8}
% \nonumber\\ 
% &A = \epsilon_2 +3\epsilon_6 + 3(\epsilon_4+\epsilon_5)/2 ,\quad  
% B = \sqrt{(\frac{ \sqrt{5}\epsilon_4-\sqrt{5}\epsilon_5 +4\sqrt{3}\epsilon_7}{2})^2 + (\epsilon_2 -3\epsilon_6 +\epsilon_4 +\epsilon_5)^2} 
% \label{eq:local_energy}
% \ea 
% where we have also provided the corresponding irreducible representations. 
% In the first-order perturbation theory, we find the energy difference between the lowest-energy states and the first-excited states is around $16$meV. The energy difference between ground states and second excited states is around $25$meV. 

Numerically, we directly solve Eq.~\ref{eq:app:eigen_value_atom_ham}  and obtain the energy levels of the atomic problem with one $f$ electron. The numerical values of $E_i$ are
\ba 
J= \frac{5}{2}\quad:\quad 
\begin{cases}
    E_{1} = E_2 = -0.2073eV\\
E_3 = E_4 = -0.1983eV\\
E_5 = E_6 = -0.1771eV\\ 
\end{cases}
\nonumber\\ 
J =\frac{7}{2}\quad:\quad 
\begin{cases}
    &E_7 = E_8 = 0.1117eV\\
&E_9 = E_{10} = 0.1196eV\\
&E_{11}= E_{12} = 0.1727eV\\
&E_{13}= E_{14} = 0.1789eV
\end{cases}
\label{eq:app:atomic_state}
\ea 
We notice that with one $f$ electron, the energy contribution from $H_U$ (Eq.~\ref{eq:Hubbard_f}) is zero since $n_0=1$. 
We observe the energy difference $E_3-E_1 \approx 9$meV and $E_5-E_1 \approx 30 $meV. The energy differences between $J=\frac{7}{2}$ states and the ground state is around $\text{mean}_{j=7,...,14}(E_j - E_1)=353.1$meV.

\section{Kondo lattice model} 
\label{sec:app:klm}
Due to the large Hubbard interactions, the charge fluctuations of $f$ electrons are suppressed. There will be exactly one $f$ electron for each Ce atom. In other words, $f$ electrons now behave as local moments. We can perform a Schrieffer-Wolff transformation (as we describe in Sec.~\ref{sec:app:sw_transf}) and map the original periodic Anderson mode to a Kondo lattice model. Since the Kondo effect is developed below the $T_{\text{AFM}}$, for what follows, we will use the paramagnetic conventional cell, which is the magnetic primitive cell, to describe the system. 

\subsection{Kondo coupling}
We start with the following periodic Anderson model 
\ba 
H = H_{c}^{\text{AFM}} + H_V + H_f
\ea 
where 
\ba 
H_{c}^{\text{AFM}} =\sum_{\kk,abij,\sigma}\epsilon_{\kk, ai,bj}^{AFM}c_{\kk,ai\sigma}^\dag c_{\kk,bj\sigma }
\ea 
is the single-particle Hamiltonian of the $c$ electrons in the AFM phase, $H_V$ is the $fc$ hybridization term (Eq.~\ref{eq:app:hyb_term_c}), and $H_f$ describes the atomic Hamiltonian of $f$ electrons (Eq.~\ref{eq:app:ham_f}).

To perform the Schrieffer-Wolff transformation, we first separate the Hilbert space into low-energy subspace and high-energy subspace. 
We let the low-energy states be the states with $1$ $f$ electrons per Ce atom, and let the high-energy states be the states that violate the aforementioned filling constraints. The low-energy and high-energy subspaces are spanned by the low-energy and high-energy states respectively. 
% There could also be the states with $3$ or more $f$ electrons at the Ce atom. Since these states are in general at much higher energy, we directly drop their contributions.  

% The low-energy subspace is spanned by the state with singly occupied $f$ states. 
We let $\eta^\dag_{\RR,a,i}$ to characterize the eigenstates of the interacting Hamiltonian of $f$ electrons $H_{loc}$ (Eq.~\ref{eq:ham_atom_f}). In Eq.~\ref{eq:app:eigen_value_atom_ham}, we have introduced $\eta_i$ for the single-site/atomic problem. Here, similarly, we let $\eta_{\RR,a,i}^\dag$ to denote the one-electron eigenstates of the local Hamiltonian of each site (Eq.~\ref{eq:app:eigen_value_atom_ham})
\ba 
&\eta^\dag_{\RR,a, i} |\RR, a,0\rangle = |\RR,a, i\rangle ,\quad i=1,...,14\nonumber\\ 
&H_{loc}[f_{\RR,a}^\dag, f_{\RR,a}]|\RR,a,i\rangle =E_i |\RR,a,i\rangle
\label{eq:app:sw_low_en_space}
\ea
 $\RR$ is the position of unit cell and $a$ is the sublattice. $|\RR,a,i\rangle$ is the eigenstate of $H_{loc}$ with one $f$ electrons at Ce$_a$ atom in the unit cell $\RR$. $|\RR,a,0\rangle$ denotes the state with zero $f$ electron at Ce$_a$ atom in the unit cell $\RR$. 
 $\eta_{\RR,a,i}$ is the superposition of $f$ electron operators that creates the eigenstate of $H_{loc}$ and can be written as
\ba 
\eta_{\RR,a,i} = \sum_{j,\sigma}U^\eta_{i,j\sigma} f_{\RR,a,j,\sigma}
\ea 
The matrix $U_{i,j\sigma}^\eta$ can be obtained by numerically solving the atomic Hamiltonian $H_{loc}$ in the one-electron sector.

The high-energy states we considered are empty states $(|\RR,a,0\rangle)$ and doubly-occupied states  $(|\RR,a,i,j\rangle)$ which corresponds to the charge $\mp 1 $ excitation of the local Hamiltonian
\ba 
&|\RR,a, 0\rangle,\quad H_{loc}[f_{\RR,a}^\dag,f_{\RR,a}]|\RR,0\rangle \approx E_H|0\rangle \nonumber\\ 
&|\RR, a,i,j\rangle = \eta_{\RR,i}^\dag \eta_{\RR,j}^\dag|\RR,a, 0\rangle ,\quad H_{loc}[f_{\RR,a}^\dag,f_{\RR,a}]|\RR,a,i,j\rangle \approx E_H|\RR,a,i,j\rangle 
% ,\quad i>j, 
\nonumber\\ 
&E_H=U/2 
% & |\RR, i,j\rangle = h_{\RR,i}^\dag \eta^\dag_{\RR,i}|\RR,0\rangle sss
\label{eq:app:sw_high_en_space}
\ea 
where the fermion anti-commutation relation also indicates
\ba 
|\RR,a,i,j\rangle =-|\RR,a,j,i\rangle 
\ea 
Since $U$ is the largest scale, we ignore the effect of spin-orbit coupling and crystal field splitting on the charge $\pm 1$ excitation and let the excitation energy be $E_H=\frac{U}{2}$.

In the $\eta$ basis, we can also rewrite the $fc$ hybridization term (Eq.~\ref{eq:app:hyb_term_c}) as
\ba 
&H_V= \sum_\RR \sum_{i,a,b,\Delta\RR,j\sigma}V_{ai,bj\sigma}(\Delta \RR) \eta^\dag_{\RR,a,i} 
c_{\RR+\Delta\RR, b,j,\sigma} +\text{h.c.}\nonumber\\ 
&V_{ai,bj\sigma}(\Delta\RR) =\sum_{i'}U^{\eta}_{i,i'\sigma}V^c_{ab,i'j}(\Delta\RR) 
\label{eq:app:hyb_term_eta}
\ea 
where $V^c_{ab,i'j}$ has been introduced in Eq.~\ref{eq:app:hyb_term_c}.

For convenience, we introduce the molecular orbital 
\ba 
&w_{\RR,a,i} = \frac{1}{{v_{a,i}}} \sum_{\Delta\RR,bj\sigma} V_{ai,bj\sigma}(\Delta\RR) c_{\RR+\Delta\RR,b,j\sigma}\nonumber\\ 
&v_{a,i}^2 = \sum_{\Delta\RR,bj\sigma}|V_{ai,bj\sigma}(\Delta\RR)|^2
\label{eq:app:def_w}
\ea 
% where we note that $v_i$ does not depend on the sublattice index $a$ since two Ce atoms are equivalent. 
$\{w_{\RR,a,i}\}_{\RR,a,i}$ are not orthogonal and follow
\ba 
&\{w_{\RR,a,i}, w^\dag_{\RR',b,j}\} = A_{\RR ai, \RR' bj} \nonumber\\ 
&A_{\RR ai, \RR' bj} =\frac{1}{v_{a,i}v_{a,j}}\sum_{\Delta\RR, cs\sigma}V_{ai,cs\sigma}(\Delta\RR) V^*_{bj,cs\sigma}(\RR+\Delta\RR-\RR')
\ea 
We can observe that, in the $w_{\RR,a,i}$ electron basis, the hybridization term can be written in a more compact form 
\ba 
H_V = \sum_{\RR, ai} v_{a,i}\eta_{\RR,a,i}^\dag w_{\RR,a,i}
+\text{h.c.}
\label{eq:app:hyb_eta_w_basis}
\ea  

We now aim to perform SW transformation to our interacting Hamiltonian. We first separate our interacting Hamiltonian into the unperturbed part ($H_0$) and the perturbation ($H_1$) 
\ba 
&H = H_0+H_1\nonumber\\ 
&H_0 = H_{c}^{\text{AFM}} +\sum_{\RR,a}
H'_{loc}(\RR,a) ,\quad 
H'_{loc}(\RR,a) = \sum_i E_{a,i} |\RR,a,i\rangle \langle \RR,a,i| + \frac{U}{2}\bigg[ |\RR,a,0\rangle \langle \RR,a,0|
+\sum_{i>j} |\RR,a,i,j\rangle \langle \RR,a,i,j|\bigg]  \nonumber\\ 
&H_1 = \sum_{\RR,ai} v_{a,i} 
\bigg[ |\RR,a,i\rangle \langle \RR,a,0| 
+\sum_{j\ne i} |\RR,a,i,j\rangle \langle \RR,a,j|
\bigg] 
w_{\RR,a,i}  +\text{h.c.} 
\label{eq:app:h_anderson_decomp}
\ea 
$H_0$ includes the dispersion of $c$ electrons in the AFM phase ($H_{c}^{\text{AFM}}$), and the local Hamiltonian of $f$ electrons. $H_1$ corresponds to the $fc$ hybridization term $H_v$. 
States with three or more $f$ electrons are dropped due to their high energies. We thus write the Hamiltonian via $|\RR,a,0\rangle, |\RR,a,i,j\rangle$ to reflect the fact that the new Hamiltonian does not involve the states with more than $2$ electron. 
% We note that, in Eq.~\ref{eq:app:h_anderson_decomp}, we only consider the empty state, singly occupied states, and doubly occupied states of $f$ electrons. 

Moreover, in Eq.~\ref{eq:app:h_anderson_decomp}, we have rewritten $H_{loc}$ with operators $|\RR,a,i\rangle, |\RR,a,0\rangle , |\RR,a,i,j\rangle$. We can show that two definitions are equivalent when acting on the empty, singly occupied, and doubly occupied states. 
 In other words, 
\ba 
&H'_{loc}(\RR,a) \sim  H_{loc}[f_{\RR,a}^\dag,f_{\RR,a}] 
\ea 
where $H'_{loc}(\RR,a)$ has been introduced in Eq.~\ref{eq:app:h_anderson_decomp}. $\sim$ means two Hamiltonian are equivalent when acting on the empty, singly occupied, and doubly occupied states. 
We prove the equivalence by showing
\ba 
&H_{loc}[f_{\RR,a}^\dag, f_{\RR,a}] |\RR,a,i\rangle = E_i|\RR,a,i\rangle 
,\quad H_{loc}'(\RR,a) |\RR,a,i\rangle =
\bigg( \sum_{i'} E_{i'} |\RR,a,i'\rangle \langle \RR,a,i'|\bigg)  |\RR,a,i\rangle = E_i|\RR,a,i\rangle 
, \nonumber\\ 
\nonumber\\ 
&
H_{loc}[f_{\RR,a}^\dag, f_{\RR,a}] |\RR,a,0\rangle \approx \frac{U}{2}|\RR,a,0\rangle,\quad H_{loc}'(\RR,a) |\RR,a,0\rangle =
\bigg(\frac{U}{2} |\RR,a,0\rangle \langle \RR,a,0|\bigg) |\RR,a,0\rangle = \frac{U}{2}|\RR,a,0\rangle 
\nonumber\\ 
\nonumber\\ 
&
H_{loc}[f_{\RR,a}^\dag, f_{\RR,a}] |\RR,a,i,j\rangle \approx \frac{U}{2}|\RR,a,i,j\rangle,\quad H_{loc}'(\RR,a) |\RR,a,i,j\rangle =
\bigg(\frac{U}{2} \sum_{i'>j'}|\RR,a,i',j'\rangle \langle \RR,a,i',j'|\bigg) |\RR,a,i,j\rangle = \frac{U}{2}|\RR,a,i,j\rangle 
\ea 
In the definition of $H_1$ in Eq.~\ref{eq:app:h_anderson_decomp}, we have also replaced $\eta^\dag_{\RR,a,i}$ operator via 
\ba 
\tilde{\eta}^\dag_{\RR,a,i} =  |\RR,a,i\rangle \langle \RR,a,0| 
+\sum_{j\ne i} |\RR,a,i,j\rangle \langle \RR,a,j|
\ea 
We now show two definitions are equivalent when acting on the empty and singly occupied states. 
% Such a state has been ignored due to its high energy.
The equivalence can be seen by showing
\ba 
&\eta^\dag_{\RR,a,i} |\RR,a,0\rangle = |\RR,a,i\rangle,\quad \tilde{\eta}^\dag_{\RR,a,i}|\RR,a,0\rangle =
\bigg( 
|\RR,a,i\rangle \langle \RR,a,0| 
\bigg) |\RR,a,0\rangle = |\RR,a,i\rangle  \nonumber\\ 
&\eta^\dag_{\RR,a,j}|\RR,a,j\rangle = |\RR,a,i,j\rangle ,\quad \tilde{\eta}^\dag_{\RR,a,i}|\RR,a,i\rangle =
\bigg( 
|\RR,a,i,j\rangle \langle \RR,a,j| 
\bigg) |\RR,a,j\rangle = |\RR,a,i,j\rangle ,\quad i\ne j
\ea

% We also comment that, since $c$ electrons describe the low-energy gapless degrees of freedom, we always treat $c$ electron as our low-energy excitations. Therefore, we only rewrite $f$-electron operators with bra and ket notations (in order to more easily perform SW transformations), but keep using the operator formula of $c$ electrons. 

We can then utilize the Eq.~\ref{eq:app:sw_eff_h} and Eq.~\ref{eq:app:h_anderson_decomp} to perform Schrieffer-Wolff transformations. Since $c$ electrons describe the low-energy gapless degrees of freedom, we always treat $c$ electrons as our low-energy excitations. Therefore, we only rewrite $f$-electron operators with bra and ket notations (in Eq.~\ref{eq:app:h_anderson_decomp}), but keep using the operator formula of $c$ electrons since it always maps a low-energy $c$ states to another low-energy $c$ states. 

In practice, the effective Hamiltonian $H_{SW}$ (Eq.~\ref{eq:app:sw_eff_h}) from Schrieffer-Wolff transformations comes from the virtual process
\ba 
|L,J\rangle \rightarrow |H,M\rangle \rightarrow |L,I\rangle 
\ea 
where $|L,I\rangle$ labels the $I$-th low-energy state and $|H,M\rangle$ labels the $M$-th high-energy state. 
In the current model, only the following three types of virtual processes can be generated by $H_1$ ($fc$ hybridization terms)
\ba 
&|\RR,a,i\rangle \rightarrow |\RR,a,0\rangle \rightarrow |\RR,a,j \rangle \nonumber\\ 
&|\RR,a,i\rangle \rightarrow |\RR,a,j,i\rangle \rightarrow |\RR,a,i \rangle ,\quad j\ne i\nonumber\\
&|\RR,a,i\rangle \rightarrow |\RR,a,j,i\rangle \rightarrow |\RR,a,j \rangle ,\quad j\ne i\nonumber\\
\ea 
We now consider the contributions from each virtual process. 

For, 
\ba 
&|L,J\rangle = | \RR,a,j\rangle, \quad |L,I\rangle = |\RR,a,i\rangle \nonumber\\ 
&|H,M\rangle = |\RR,a,0\rangle 
\ea 
the contributions generated by Schrieffer-Wolff transformation (Eq.~\ref{eq:app:sw_eff_h}) is
\ba 
H_{SW}^1 =\sum_{\RR,a,i,j} \frac{1}{2} v^*_{a,j}  v_{a,i} \bigg( \frac{1}{-\frac{U}{2}} + \frac{1}{-\frac{U}{2}}\bigg) |\RR,a,i\rangle  w_{\RR,a,i} w_{\RR,a,j}^\dag \langle\RR,a,j| = \frac{-2}{U}\sum_{\RR,a,i,j} v_{a,j}^*v_{a,i} w_{\RR,a,i}w_{\RR,a,j}^\dag |\RR,a,i\rangle \langle \RR,a,j|
\ea 
When acting on the low-energy subspace (singly occupied states), we have the following equivalence
\ba 
|\RR,a,i\rangle \langle \RR,a,j| \sim \eta_{\RR,a,i}^\dag \eta_{\RR,a,j}
\ea 
Then we have 
\ba 
H_{SW}^1 =\frac{-2}{U}\sum_{\RR,a,i,j} v_{a,j}^*v_{a,i} w_{\RR,a,i}w_{\RR,a,j}^\dag \eta_{\RR,a,i}^\dag \eta_{\RR,a,j}
\label{eq:app:konod_sw_1}
\ea 

For 
\ba 
&|L,J\rangle = | \RR,a,i\rangle, \quad |L,I\rangle = |\RR,a,i\rangle \nonumber\\ 
&|H,M\rangle = |\RR,a,i,j\rangle 
\ea 
with $i\ne j$, 
the contributions generated by Schrieffer-Wolff transformation (Eq.~\ref{eq:app:sw_eff_h}) is
\ba 
H_{SW}^2 =\sum_{\RR,a,i,j\ne i} \frac{1}{2} v^*_{a,j}  v_{a,j} \bigg( \frac{1}{-\frac{U}{2}} + \frac{1}{-\frac{U}{2}}\bigg) w_{\RR,a,j}^\dag |\RR,a,i\rangle   \langle\RR,a,i| w_{\RR,a,j} = \frac{-2}{U}\sum_{\RR,a,i,j\ne i} v_{a,j}^*v_{a,j} w_{\RR,a,j}^\dag w_{\RR,a,j} |\RR,a,i\rangle \langle \RR,a,i|
\ea 
When acting on the low-energy subspace (singly occupied states), we have the following equivalence
\ba 
\sum_{i}(1-\delta_{i,j}) |\RR,a,i\rangle \langle \RR,a,i| \sim \eta_{\RR,a,j} \eta_{\RR,a,j}^\dag 
\ea 
This can be seen from 
\ba 
&\bigg( \sum_{i}(1-\delta_{i,j}) |\RR,a,i\rangle \langle \RR,a,i| \bigg) |\RR,a,j\rangle =0 ,\quad
\eta_{\RR,a,j} \eta_{\RR,a,j}^\dag |\RR,a,j\rangle = 0
\nonumber\\ 
&\bigg( \sum_{i}(1-\delta_{i,j}) |\RR,a,i\rangle \langle \RR,a,i| \bigg) |\RR,a,i'\rangle = |\RR,a,i'\rangle ,\quad 
\eta_{\RR,a,j} \eta_{\RR,a,j}^\dag|\RR,a,i'\rangle = |\RR,a,i'\rangle,\quad \text{for }
i'\ne j 
\ea 
Then we have 
\ba 
H_{SW}^2 =\frac{-2}{U}\sum_{\RR,a,j} v_{a,j}^*v_{a,j} w_{\RR,a,j}w_{\RR,a,j}^\dag \eta_{\RR,a,j} \eta_{\RR,a,j}^\dag 
\label{eq:app:konod_sw_2}
\ea

For 
\ba 
&|L,J\rangle = | \RR,a,j\rangle, \quad |L,I\rangle = |\RR,a,i\rangle \nonumber\\ 
&|H,M\rangle = |\RR,a,i,j\rangle 
\ea 
with $i\ne j$, 
the contributions generated by Schrieffer-Wolff transformation (Eq.~\ref{eq:app:sw_eff_h}) is
\ba 
H_{SW}^3 =&\sum_{\RR,a,i\ne j} \frac{1}{2} v^*_{a,j}  v_{a,i} \bigg( \frac{1}{-\frac{U}{2}} + \frac{1}{-\frac{U}{2}}\bigg)  (-1)w_{\RR,a,j}^\dag |\RR,a,i\rangle   \langle\RR,a,j| w_{\RR,a,i} \nonumber\\ 
=&\frac{2}{U} \sum_{\RR,a,i\ne j}  v^*_{a,j}  v_{a,i} w_{\RR,a,j}^\dag w_{\RR,a,i}|\RR,a,i\rangle   \langle\RR,a,j|
\ea 
where the additional minus sign can be seen from the definition of $H_1$ (Eq.~\ref{eq:app:h_anderson_decomp}), and the fact that $|\RR,a,i,j\rangle = -|\RR,a,j,i\rangle$. 
When acting on the low-energy subspace (singly occupied states), we have the following equivalence
\ba 
|\RR,a,i\rangle \langle \RR,a,j| \sim \eta_{\RR,a,i}^\dag \eta_{\RR,a,j}
\ea 
Then we have 
\ba 
H_{SW}^3 =\frac{2}{U}\sum_{\RR,a,i\ne j} v_{a,j}^*v_{a,i} w_{\RR,a,j}^\dag w_{\RR,a,i} \eta_{\RR,a,i}^\dag \eta_{\RR,a,j}
\label{eq:app:konod_sw_3}
\ea 

We can combining all three contributions (Eq.~\ref{eq:app:konod_sw_1}, Eq.~\ref{eq:app:konod_sw_2}, Eq.~\ref{eq:app:konod_sw_2}), and find 
\ba 
H_{SW} = H_{SW}^1 +H_{SW}^2 +H_{SW}^2 = -\frac{2}{U}\sum_{\RR,a,i,j} v_{a,j}^* v_{a,i}\bigg(  \eta_{\RR,a,i}^\dag w_{\RR,a,i} w_{\RR,a,j}^\dag \eta_{\RR,a,j} + w_{\RR,a,j}^\dag \eta_{\RR,a,j} \eta_{\RR,a,i}^\dag w_{\RR,a,i}
\bigg) 
\ea 
The low-energy effective Hamiltonian, which describes a Kondo lattice model, now becomes
\ba 
H_{KL}= &  H_c^{AFM}+H_{\text{Kondo}}  +H_{f,0} \nonumber\\  
H_{\text{Kondo}} =& H_{SW}=
\frac{-2}{U} \sum_{\RR}\sum_{a }\sum_{i,j } v_{a,i}^*v_{a,j}\bigg( 
\eta_{\RR,a,j}^\dag w_{\RR,a,j} w_{\RR,a,i}^\dag \eta_{\RR,a,i}
+ w_{\RR,a,i}^\dag
\eta_{\RR,a,i}\eta_{\RR,a,j}^\dag w_{\RR,a,j}
\bigg) \nonumber\\ 
H_{f,0}=& \sum_{\RR,a,i}E_{i}\eta_{\RR,a,i}^\dag \eta_{\RR,a,i}
\label{eq:ham_kondo_full}
\ea 
where the Kondo coupling term comes from the Schrieffer-Wolff transformations. $H_{f,0}$ is the atomic Hamiltonian of $f$ electrons. We can obtain it by noting the following equivalence (when acting on the low-energy singly occupied states)
\ba 
E_i|\RR,a,i\rangle \langle \RR,a,i| \sim E_i \eta_{\RR,a,i}^\dag \eta_{\RR,a,i}
\ea 

Since the Hamiltonian is defined in the low-energy subspace with singly occupied $f$ states. Therefore, we also introduce the following constraints (which enforce the filling of $f$ electrons of each Ce atom to be 1)
\ba 
\sum_{i} \eta_{\RR,a,i}^\dag \eta_{\RR,a,i} = 1 
\label{eq:constraint_eta_fill}
\ea

\subsection{Mean-field theory of Kondo phase}
\label{sec:app:kondo_mf}
To understand the Kondo effect, we perform mean-field calculations on the Hamiltonian defined in Eq.~\ref{eq:ham_kondo_full}. In this section, we briefly describe the mean-field approach.

We first rewrite the Kondo coupling term in the conventional ordering ($\eta^\dag \eta w^\dag w$)
\ba 
&H_{\text{Kondo}}  \nonumber\\
= & \frac{-2}{U} \sum_{\RR} 
\sum_{a, i,j} v_{a,i}^* v_{a,j}\bigg[  \eta_{\RR,a,j}^\dag \eta_{\RR,a,i} (A_{\RR aj, \RR ai} -w_{\RR,a,i}^\dag w_{\RR,a,j}) +  (\delta_{i,j}-\eta_{\RR,a,j}^\dag \eta_{\RR,a,i}) w_{\RR,a,i}^\dag w_{\RR,a,j} 
\bigg]
\nonumber\\ 
=& \frac{4}{U} \sum_{\RR} 
\sum_{a,i,j} v_{a,i}^* v_{a,j} \eta_{\RR,a,j}^\dag \eta_{\RR,a,i}  w_{\RR,a,i}^\dag w_{\RR,a,j}
-\frac{2}{U}\sum_{\RR}\sum_{a,i,j} v_{a,i}^* v_{a,j} A_{\RR a j, \RR a i} \eta_{\RR,a,j}^\dag \eta_{\RR,a,i} 
-\frac{2}{U}\sum_{\RR}\sum_{a,i} v_{a,i}^* v_{a,i} w_{\RR,a,i}^\dag w_{\RR,a,i}  
\ea 
where the first term describes a four-fermion Kondo interaction term, and the remaining two terms are one-body terms.

We can then perform a mean-field decoupling of $H_{\text{Kondo}}$
\ba 
H_{\text{Kondo}}^{MF} = &-\frac{4}{U}\sum_{\RR}
\sum_{i,j,a }v_{a,i}^* v_{a,j}
\bigg( 
 \chi_{\eta w,a,j} w_{\RR,a,i}^\dag \eta_{\RR,a,i} 
 +\chi_{w \eta,a,i}\eta_{\RR,a,j}^\dag w_{\RR,a,j} \nonumber\\ 
 &
 -\chi_{\eta \eta ,a,ji} w_{\RR,a,i}^\dag w_{\RR,a,j}
 -\chi_{ww,a, ij}\eta_{\RR,a,j}^\dag \eta_{\RR,a,i} - 
 \chi_{\eta w,a,j} \chi_{w \eta ,a,i } 
 +\chi_{\eta \eta ,a,ji}\chi_{ww,a, ij} 
\bigg)  \nonumber\\ 
&+\frac{-2}{U}\sum_{\RR}\sum_{i,j,a} v_{a,i}^* v_{a,j} A_{\RR aj, \RR ai} \eta_{\RR,a,j}^\dag \eta_{\RR,a,i} 
+\frac{-2}{U}\sum_{\RR}\sum_{i} v_{a,i}^* v_{a,i} w_{\RR,a,i}^\dag w_{\RR,a,i} 
\label{eq:mf_kondo_couple}
\ea 
where the mean-fields are 
\ba 
&\chi_{\eta w,a,i} = \frac{1}{N}\sum_{\RR} \langle \eta_{\RR,a,i}^\dag w_{\RR,a,i}\rangle \nonumber\\ 
&\chi_{w \eta, a,i} = \frac{1}{N}\sum_{\RR} \langle w_{\RR,a,i}^\dag \eta_{\RR,a,i}\rangle = \chi_{\eta w,a,i}^* \nonumber\\ 
&\chi_{w w, a,ij} = \frac{1}{N}\sum_{\RR} \langle w_{\RR,a,i}^\dag w_{\RR,a,j}\rangle \nonumber\\ 
&\chi_{\eta\eta, a,ij} = \frac{1}{N}\sum_{\RR} \langle \eta_{\RR,a,i}^\dag \eta_{\RR,a,j}\rangle 
\label{eq:mf_eq_chi}
\ea 
In addition, we also introduce the Lagrangian multiplier $\lambda_a$ to enforce the following filling constraints of $f$ electrons (see Eq.~\ref{eq:constraint_eta_fill})
\ba 
\frac{1}{N}\sum_{\RR,i} \langle \eta_{\RR,a,i}^\dag \eta_{\RR,a,i}\rangle = 1 
\label{eq:mf_eq_lam} 
\ea 
This leads to the following mean-field Hamiltonian
\ba 
&H^{MF} = H_c +H_{\text{Kondo}}^{MF} + H_f' \nonumber\\ 
&H_f' = \sum_{\RR,a,i} E_{i}\eta_{\RR,a,i}^\dag \eta_{\RR,a,i}  + \sum_{\RR,a} \lambda_{a}(\sum_{i} \eta_{\RR,a,i}^\dag \eta_{\RR,a,i}-1)
\label{eq:mf_ham_kondo}
\ea 
The expectation values $\langle \rangle $ in Eq.~\ref{eq:mf_eq_chi} and Eq.~\ref{eq:mf_eq_lam} are taken with respect to the ground state of the mean-field Hamiltonian $H^{MF}$. The mean-field equations (Eq.~\ref{eq:mf_eq_chi}) and the filling constraints (Eq.~\ref{eq:mf_eq_lam}) are then solved self-consistently~\cite{PhysRevLett.131.166501}. 

At the mean-field level, the development of the Kondo effect is characterized by 
\ba 
\chi_{\eta w, a, i } \ne 0 
\ea 
or equivalently $\chi_{ w\eta , a, i }\ne 0$. Whenever $\chi_{\eta w,a,i} \ne 0$, $f$ electron hybridizes with $c$ electron and becomes an effective itinerant degree of freedom.

% Moreover, whenever we have $\chi_{\eta w, a, i } \ne 0 $, the mean-field Hamiltonian breaks the $U(1)$ gauge symmetry (Eq.~\ref{eq:gauge_transf}). The following term, which appears in the $H_{\text{Kondo}}^{MF}$, is no longer invariant under $U(1)$ gauge transformation in the Kondo phase
% \ba 
% \chi_{\eta w,a,j} w_{\RR,a,i}^\dag \eta_{\RR,a,i}  \rightarrow \chi_{\eta w,a,j} w_{\RR,a,i}^\dag \eta_{\RR,a,i} e^{i\theta_{\RR,a}} \ne \chi_{\eta w,a,j} w_{\RR,a,i}^\dag \eta_{\RR,a,i} \quad, \quad \text{if}\quad \chi_{\eta w,a,j} \ne 0 
% \ea 
% Therefore, the mean-field Hamiltonian of the Kondo phase (with non-zero $\chi_{\eta w, a,i}$) does not have $U(1)$ gauge symmetry (Eq.~\ref{eq:gauge_transf}). Equivalently speaking, the Kondo phase corresponds to a Higgs phase (breaking the $U(1)$ gauge symmetry at the mean-field level). 

\section{Kondo effect in the AFM phase}
\label{sec:app:kondo_afm}
In this section, we demonstrate even with AFM ordering of $c$ electrons a Kondo effect can still develop.  
Due to the existence of $P\cdot\mathcal{T}$ symmetry in the AFM phase (where $P$ denotes inversion symmetry and $\mathcal{T}$ denotes time-reversal symmetry), $c$ electrons can still form two-fold degenerate Kramers' doublets. The doublets of $c$ electrons can then screen the $f$-local moments. 

\begin{figure}
    \centering
    \includegraphics[width=0.6\textwidth]{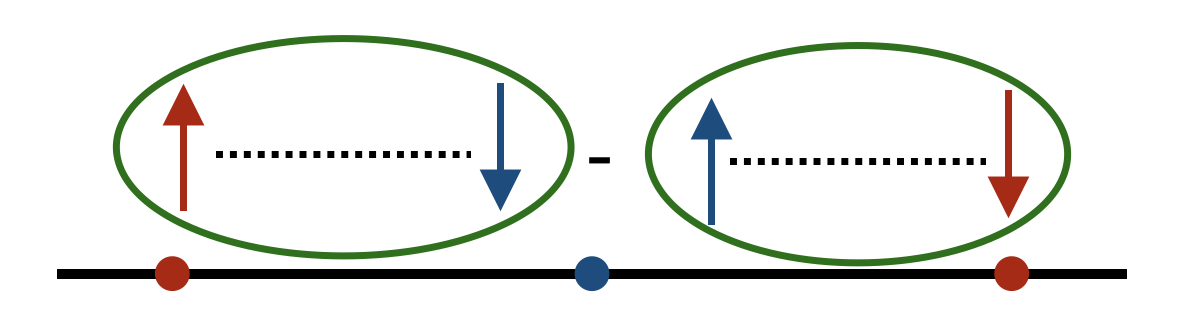}
    \caption{Illustration of the Kondo effect in the AFM phase. Red arrows denote the spins of $c$ electrons, and blue arrows denote the spins of $f$ electrons. The ground state is a Kondo-singlet state (Eq.~\ref{eq:toy_model_kondo_singlet}).   }
    \label{fig:1d_non_local_kondo}
\end{figure}

We first illustrate the formation of the Kondo singlet in the AFM phase via a three-site spinful toy model. The Hamiltonian of the toy model is
\ba 
H = \sum_{i=1,2,\sigma}(-\mu) c_{i,\sigma}^\dag c_{i,\sigma}+\sum_{\sigma_1,\sigma_2} m_z (c_{1,\sigma_1}^\dag \sigma^z_{\sigma_1\sigma_2} c_{1,\sigma_2} -c_{2,\sigma_1}^\dag \sigma^z_{\sigma_1\sigma_2} c_{2,\sigma_2}) + J\sum_{i,j,\mu}S^\mu c_{i,\sigma}\sigma^\mu_{\sigma,\sigma'}c_{j,\sigma'}.
\label{eq:app:toy_model_kondo}
\ea 
 $c_{i,\sigma}$ describes two $c$ electrons that develop an antiferromagnetic order as characterized by a non-zero $m_z$. Two $c$ electrons couple to the $f$ local moment $S^\mu$ via a Kondo coupling term $(J>0)$. We can observe that $c$ electrons develop an AFM order due to the $m_z$ term which splits the energy of spin $\up$ and spin $\dn$ states of $c$ electrons at the same site. However, due to the existence of the $P\cdot\mathcal{T}$ symmetry, $c_{1,\up}$ and $c_{2,\dn}$ electrons are still degenerate (similarly, $c_{1,\dn}$ and $c_{2,\up}$ electrons are also degenerate). This degeneracy allows the formation of the Kondo singlet. 
% Depending on which degenerate set of $c$ electrons are near the Fermi energy, we could form the non-local Kondo singlet.
For example, when $c_{\RR,1,\up}$ and $c_{\RR,2,\dn}$ electrons are close to the Fermi energy (this can be realized by setting $\mu \sim m_z$), the many-body ground state of Hamiltonian in Eq.~\ref{eq:app:toy_model_kondo} is a Kondo singlet and can be written as (in the limit of large magnetic ordering $\mu = m_z\rightarrow \infty$)
\ba 
|KS\rangle = \frac{ c_{\RR,1,\up}^\dag f_{\RR,\dn}^\dag - c_{\RR,2,\dn}^\dag f_{\RR,\up}^\dag}{\sqrt{2}} |0\rangle 
\label{eq:toy_model_kondo_singlet}
\ea 
as also illustrated in Fig.~\ref{fig:1d_non_local_kondo}. Clearly, such formation of the Kondo singlet is not affected by the AFM ordering and only relies on the existence of $P\cdot\mathcal{T}$ symmetry. 

For what follows, we will show that a similar mechanism can be applied to the CeCo$_2$P$_2$, where the $P\cdot \mathcal{T}$ symmetry also promotes the development of the Kondo effect.

\subsection{Filling-enforced metallic phase of the $c$ electrons}
\label{sec:app:filling_enforce_metal} 
Besides the Kramers double, the development of the Kondo effect also requires a finite density of states of $c$ electrons near the Fermi energy. 
In this section, we show that the $c$ electron band is always metal in the AFM phase. For the symmetry group (126.386) of the AFM phase, all the spinful representations at high symmetry points $A,M,R$ are four-dimensional. Therefore, if the total fillings of the systems $n$ follow $n \text{ mod } 4 \ne 0 $, then the system has to be metal. 

In the current model, each Ce has $11$ electrons (without including the Ce $f$ electrons which provide the local moments), each Co has $9$ electrons, and each P has $5$ electrons. Then, for each unit cell, we have in total $11\times 2 + 9 \times 4 + 5 \times 4 = 78$ where $78 \text{ mod } 4 =2 \ne 0 $. Then, $c$ electrons are metallic. The filling-enforced metallic nature of the $c$ electrons combined with the $P\cdot\mathcal{T}$ protected Kramers-doublet guarantee the development of the Kondo effect since we could always have $c$ electron density of states near the Fermi energy no matter how strong the magnetic order is.

\subsection{$P\cdot \mathcal{T}$ symmetry and the Kondo effect in the AFM phase}
\label{sec:app:inv_break}
% Before showing the development of the Kondo effect, we first show that the ordering of Co atoms will not induce an ordering of Ce atoms. From a symmetry point of view, the $P\cdot \mathcal{T}$ symmetry of the AFM phase already indicates that Ce $f$ electrons are paramagnetic. This is because Ce atoms remain invariant under $P$ transformation, but $\mathcal{T}$ will flip the spin of the Ce electrons. 
% To further understand the paramagnetic nature of Ce electrons, we pick one Ce atom and show that the coupling between Ce electrons and AFM-ordered Co electrons will not induce a Ce ordering. Each Ce atom has eight nearest-neighbor Co atoms, where 4 Co atoms are located at the plane above the Ce plane and 4 Co atoms are located at the plane below the Ce plane (Fig.~\ref{fig:lat_mag_structure}).
% The Co atoms within the same plane develop ferromagnetism. However, the upper 4 Co atoms and lower 4 Co atoms have opposite spin directions. Thus the contributions to the Ce electrons from upper and lower Co atoms cancel with each other. This then indicates Ce $f$ electrons remain paramagnetism. 
We now show the robustness of our Kondo phase against AFM ordering. In this section, we consider the conventional cell of the PM phase, which is equivalent to the primitive cell of the AFM phase. We start with the following simplified Kondo lattice model
\ba 
H_{KL} = H_{c}^{\text{AFM}} + H_{\text{Kondo}} +H_{f,0}
\label{eq:kl_model_simp}
\ea 
For the $c$-electron part, we take the simplified two-orbital model that we introduced in Sec.~\ref{app:sec:flat_band} with an additional mean-field term that describes the magnetic ordering
\ba 
H_{c}^{\text{AFM}} = H_c + H_{c,MF}
\ea 
where $H_c$ is the non-interacting part of the Co $(d_{x^2-y^2},d_{z^2})$ orbitals introduced in Eq.~\ref{eq:app:simplified_tbp} and $H_{c,MF}$ denotes the mean-field contributions that describes the AFM ordering
\ba 
H_{c,MF} = U_dm\sum_{\RR,a,i,\sigma} s{(\bm{R}+\rr_a)}\bigg(\sigma c_{\RR,a,i,\sigma }^\dag c_{\RR,a,i,\sigma}\bigg)
\label{eq:simplified_mf_afm_ham_c}
\ea 
$s(\bm{R}+\rr_a) = \pm 1 $ for the electrons of the Co$_{1,2}$ (Co$_{3,4}$) atoms. 
$m$ is the mean-field parameter of the AFM order, and $U_d$ is the strength of Coulomb repulsion. 
% To demonstrate the stability of the Kondo phase against the magnetic ordering, we will treat $U_dm$ as our tuning parameters. 
Since the Hamiltonian in Eq.~\ref{eq:simplified_mf_afm_ham_c} only depends on the product of $m$ and $U_d$, we will treat $U_dm$ as our tuning parameters here. 
As we increase $U_dm$, the magnetic ordering of $d$ orbitals becomes stronger. 
% As we show later, no matter how strong this magnetic ordering is, the Kondo state could still develop. 

For the $f$ electron, we will only keep the lowest two-fold degenerate $f$ electron states of the atomic problem to simplify the considerations (see Sec.~\ref{sec:f_atom}). Then we have 
\ba
H_{f,0} = \sum_{\RR}E_1(\eta_{\RR,1}^\dag \eta_{\RR,1} + \eta_{\RR,2}^\dag \eta_{\RR,2})
\ea 
The Kondo coupling is derived from the Schrieffer-Wolff transformation (Sec~\ref{sec:app:sw_transf}) with the nearest-neighbor coupling between $\eta_{\RR,1},\eta_{\RR,2}$ and $d_{x^2-y^2},d_{z^2}$ states.

\begin{figure}
    \centering
    \includegraphics[width=0.6\textwidth]{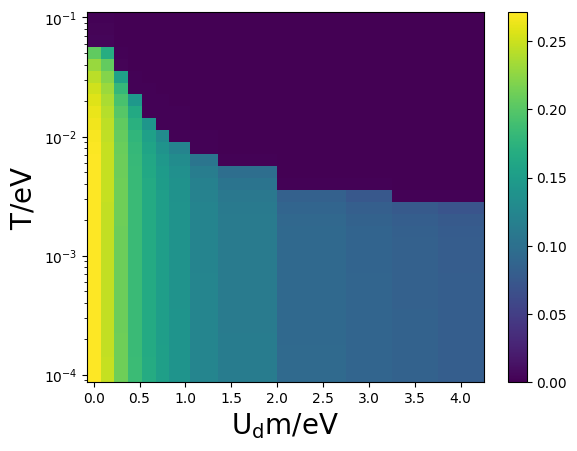}
    \caption{Maximum value of hybridization fields ($\text{max}|\chi_{\eta w}|$) at various temperatures and tuning parameters $U_dm$. }
    \label{fig:mag_kondo_simp_mod}
\end{figure}

We then use the mean-field approaches (Sec.~\ref{sec:app:kondo_mf}) to solve the Kondo Hamiltonian by fixing the filling of $f$ electrons to be $1$ for each Ce atom. For the $d$ electrons, each unit cell contains 4 Co atoms, each Co atom has 2 $d$ orbitals, and each $d$ orbital has 2 spin flavors. This results in a total of $4 \times 2 \times 2 = 16$ $d$ orbitals per unit cell (conventional cell of the PM phase). We then fill 14 $d$ electrons per unit cell. 
We treat $mU_d$ as the tuning parameters and solve the mean-field theory at Kondo coupling strength $J_K=t^2/U= 0.1$eV (with $t$ denoting the nearest-neighbor hybridization between $d_{x^2-y^2}$ orbitals and $\eta_{1/2}$ $f$ orbitals) and at different temperatures.
We show the evolution of the mean fields ($fc$ expectation value) $\text{max}|\chi_{\eta w}|$ in Fig.~\ref{fig:mag_kondo_simp_mod}. 
We can observe that, even with strong magnetic ordering (large $Um_d$), a non-zero Kondo temperature (the highest temperature where we have a non-zero $|\chi_{\eta w}|$) could always be found.

We emphasize that even though the strong magnetic ordering splits the spin $\sigma$ states and spin $-\sigma$ states of the $d$ electrons from the same Co layer. The spin $\sigma$ states of one Co layer and the spin $-\sigma$ states of the nearby Co layer are still degenerate due to $P\cdot \mathcal{T}$ symmetry. These degenerate Kramer-doublet states are unaffected by the magnetic ordering and promote the formation of the Kondo effect. 

Moreover, due to the presence of the $P \cdot \mathcal{T}$ symmetry, magnetic ordering of the $c$ electrons does not induce magnetic ordering of the $f$ electrons. 
This contrasts with the conventional expectation, where $c$-electron magnetism can induce $f$-electron magnetic order through a Hartree-Fock contribution.  
In our case, the site symmetry group of Ce atom has a $P \cdot \mathcal{T}$ symmetry with 
\begin{align}
    P \cdot \mathcal{T} \, \bm{J}^\mu_{\RR, a} \, (P \cdot \mathcal{T})^{-1} = -\bm{J}^\mu_{\RR, a},
\end{align}
where $\bm{J}^{\mu}_{\RR,a}$ denotes the total angular momentum of the $f$ electrons at site $\RR$ and sublattice $a \in \{\mathrm{Ce}_1, \mathrm{Ce}_2\}$. 
As a result, the expectation value $\langle \bm{J}^\mu_{\RR,a} \rangle$ must vanish as long as the $P \cdot \mathcal{T}$ symmetry is preserved. 
In other words, the magnetic ordering of the $c$ electrons, which respects the $P \cdot \mathcal{T}$ symmetry, cannot induce a magnetic order in the $f$ electrons. This is also important for the formation of the Kondo effect, since the magnetic ordering of $f$ electrons will also suppress the Kondo effect.

\begin{figure}
    \centering
    \includegraphics[width=0.5\linewidth]{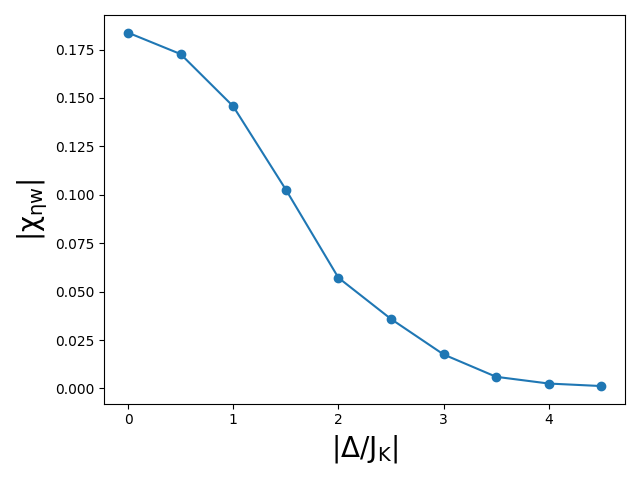}
    \caption{Maximum value of hybridization strength at fixed $\beta =10^3 eV^{-1}$ and $U_d m=1$eV, and various $P\cdot\mathcal{T}$ symmetry breaking strength $\Delta$ (noramlzied by the Kondo coupling strength $J_K$). }
    \label{fig:inv_break}
\end{figure}

For the purpose of discussion, we also consider the case in which the $P \cdot \mathcal{T}$ symmetry is broken. 
We consider the following term, which introduce an energy degeneracy of Co $d$ electrons between different layers and breaks the $P \cdot \mathcal{T}$ symmetry:
\begin{align}
    H_{c,\Delta} = \Delta \sum_{\RR,\sigma,i}
    \bigg( 
    \sum_{a\in \{Co_1,Co_2\} }c_{\RR,a,i,\sigma}^\dag c_{\RR,a,i,\sigma} 
    - \sum_{a\in \{Co_3,Co_4\} }c_{\RR,a,i,\sigma}^\dag c_{\RR,a,i,\sigma} 
    \bigg) 
\end{align}
We perform calculations with varying strengths of $P \cdot \mathcal{T}$ symmetry breaking, in the low-temperature limit $\beta = 10^3~\text{eV}^{-1}$, and with a fixed magnetic ordering strength of $m U_d = 1.0~\text{eV}$. The result is shown in \cref{fig:inv_break}, where we observe that a finite $\Delta$ gradually reduces the hybridization strength and finally destroys the Kondo effect at sufficiently large symmetry breaking term.

% We also mention the difference between the conventional Kondo effect and the Kondo effect here. In the conventional Kondo effect, the Karmers' doublets come from $\mathcal{T}$ that acts locally on the electrons. Therefore, the spin $\up$ electron and spin $\dn$ electron at the same site are degenerate. Then, once the magnetic ordering is developed on $c$ electrons, the degeneracy will be lifted. 
% Here, the Karmers' doublets come from the $P\cdot \mathcal{T}$ symmetry (where $\mathcal{T}$ symmetry is broken due to magnetic ordering) which involves spatial transformation $P$. Therefore, the spin $\up$ electrons of one layer of Co atoms degenerate with the spin $\dn$ electrons of another layer of Co atoms. As a consequence, such Kramers' doublets imposed by $P\cdot \mathcal{T}$ are not affected by the magnetic ordering, as long as $P\cdot \mathcal{T}$ symmetry is preserved. 

\subsection{Development of Kondo effect in the DFT model} 
\label{app:sec:kondo_band}
In the previous section, we demonstrated the stability of Kondo effect against magnetic order in a simplified model that only considered two $d$ orbitals of Co atoms and spin-$1/2$ states of the $f$ orbitals. In this section, we support our previous findings by including all the $c$-electron orbitals ($d$ orbitals of Ce atoms, $d$ orbitals of Co atoms, $p$ orbitals of P atoms) and all the $f$ orbitals.
% We now show that the Kondo-singlet formation is not affected by the AFM ordering of the $c$ electrons in the complete model (where all orbitals have been included) derived from DFT.
The model we study takes the same formula as Eq.~\ref{eq:kl_model_simp} which is 
\ba 
H_{KL} = H_{c}^{\text{AFM}} + H_{\text{Kondo}} +H_{f,0}
\ea 
 $H_{c}^{\text{AFM}}$ is derived from DFT calculations in the AFM phase with all $c$ electron orbitals being considered. $H_{\text{Kondo}}$ is obtained from the Scrieffer Wolff transformations (Sec.~\ref{sec:app:sw_transf}) using the the parameters obtained from DFT calculations. $H_{f,0}$ is the atomic Hamiltonian obtained DFT calculations where all $f$ orbitals have been included as we introduced in Eq.~\ref{eq:ham_kondo_full}.

% Since $P\cdot \mathcal{T}$ still exists in the AFM phase, the $c$ electrons can still form Karmers' doublets to screen the $f$ local moments. 

To investigate the Kondo effect, we first study the hybridization function. 
For the single-orbital Anderson model defined as
\ba 
H_{\text{single-orbital Anderson}} = \sum_{\kk,\sigma} \epsilon_\kk c_{\kk,\sigma}^\dag c_{\kk,\sigma}  + V\sum_{\RR,\sigma}(c_{\kk,\sigma}^\dag f_{\kk,\sigma} + \text{h.c.}) + \sum_\RR \frac{U}{2}\bigg( \sum_{\sigma} f_{\RR,\sigma}^\dag f_{\RR,\sigma} - 1 \bigg)^2 \, , 
\ea 
we can introduce the following hybridization function ~\cite{coleman2015introduction}
\ba 
\Delta(\epsilon) = \frac{1}{N}  \pi \sum_{\kk}|V|^2 \delta(\epsilon-\epsilon_\kk) = \pi V^2 \rho(\epsilon)
\label{eq:app:single_orb_hyb}
\ea 
$V$ is the $fc$ hybridization strength, $\epsilon_\kk$ is the dispersion of the conduction electrons (in the non-Kondo phase), $N$ is the total number of unit cells, and $\rho(\epsilon)$ is the density of states of $c$ electron bands. Physically, we can observe that the hybridization function is proportional to the density of states of $c$ electrons and the square of hybridization strength between $f$ and $c$ electrons. Thus it effectively characterizes the strength of the Kondo effect~\cite{coleman2015introduction}.

We can generalize the hybridization function to the multi-orbital case. To do so, we first rewrite the Kondo hybridization in the momentum space (Eq.~\ref{eq:app:hyb_term_eta}), which gives
\ba 
\label{eq:def_hyb_ham}
H_{V} = \sum_{\kk, ab,i,j,\sigma}V_{ai,bj\sigma}(\kk) \eta_{\kk,a,i}^\dag c_{\kk,b,j,\sigma}+\text{h.c.},\quad 
V_{ai,bj\sigma}(\kk) = \sum_{\Delta\RR}V_{ai,bj\sigma}(\Delta\RR)e^{-i\kk\cdot(\Delta\RR+\rr_{b} -\rr_a)}
\ea 
We also introduce the band basis of $c$ electrons 
\ba 
\label{eq:def_c_ham}
H_c^{AFM} = \sum_{\kk,n}\epsilon_{\kk,n} \gamma_{\kk,n}^\dag \gamma_{\kk,n},\quad \gamma_{\kk,n} = \sum_{bj\sigma}U^{c,*}_{bj\sigma,n}(\kk)c_{\kk,bj\sigma}.
\ea  
We then generalize the single-orbital formula of the hybridization function to the multi-orbital formula by considering the contributions from each orbital
\ba 
\Delta_{ai,bj\sigma}(\omega) = \frac{\pi}{N} \sum_{\kk,n}|V_{ai,bj\sigma}(\kk)|^2 \bigg( |U_{bj\sigma,n}^c(\kk) |^2 \delta(\epsilon-\epsilon_{\kk,n})
\bigg) 
\ea 
$|V_{ai,bj\sigma}(\kk)|^2$ is the square of the momentum-dependent hybridization strength between $\eta_{\kk,ai}$ and $c_{\kk,bj\sigma}$. $|U^c_{bj\sigma,n}(\kk)|^2$ denotes the orbital weight of $c_{\kk,bj\sigma}$ electrons. We comment that, the strength of Kondo effect is related to both the density of states of $c$ electrons and the hybridization strength between $f$ and $c$ electrons. We therefore can construct the hybridization function $\Delta_{ai,bj\sigma}$ by combining the density of states and hybridization strength. Therefore, $\Delta_{ai,bj\sigma}$ characterizes the strength of the Kondo effect. 
% hybridization strength between $\eta_{\kk,ai}$ fermions and $c_{\kk,bj\sigma}$ fermions. 

We now discuss the properties of $\Delta_{ai,bj\sigma}(\omega)$. We focus on $ \{\eta_{\RR,i,a}
\}_{i=1,2,3,4,5,6}$ fermions which corresponds to the low energy $f$ states with $J=5/2$. $f$ states with $J=7/2$ have much higher energy (0.3eV higher) due to the spin-orbit coupling terms and are less relevant to the Kondo effect. 
As for the $c$ electrons, we separate $c$ electrons into three sets based on their orbital and sublattice indices
\ba 
&S_I=\{(a,i) | a \in \{Co_1,Co_2,P_1,P_2\}\}\nonumber\\ 
&S_{II}=\{(a,i) | a \in \{Co_3,Co_4,P_3,P_4\}\}\nonumber\\ 
&S_{III}=\{(a,i) | a \in \{Ce_1,Ce_2\}\}.
\ea 
% For completeness, we consider all the $c$ electrons including $d$ orbitals of Ce atoms, $d$ orbitals of Co atoms, and $p$ orbitals of P atoms, since all these orbitals could hybridize with $f$ electrons. 
Electrons of set I correspond to one Co-P layer (layer I) and develop ferromagnetism within this layer. Electrons of set II correspond to another Co-P layer (layer II) and develop ferromagnetism within this layer, but with opposite spin directions compared to the electrons in layer I (Fig.~\ref{fig:lat_mag_structure}). Electrons of set $III$ correspond to the $d$ electrons of Ce, which do not develop magnetic order.  

We now discuss the hybridization strength between each set of $c$ electrons and the low-energy $\{\eta_{\RR,i,a}\}_{i=1,2,3,4,5,6}$ fermions. We let 
\ba 
&\Delta_{I\up} (\omega)=  \sum_{i\in\{1,...,6\},bj \in S_I}\Delta_{(Ce_a,i),bj\up} (\omega),\quad   \Delta_{I\dn} (\omega) =  \sum_{i\in\{1,...,6\},bj \in S_I}\Delta_{(Ce_a,i),bj\dn} (\omega)\nonumber\\ 
&\Delta_{II\up}  (\omega)=  \sum_{i\in\{1,...,6\},bj \in S_{II}}\Delta_{(Ce_a,i),bj\up} (\omega),\quad   \Delta_{II\dn} (\omega) =  \sum_{i\in\{1,...,6\},bj \in S_{II}}\Delta_{(Ce_a,i),bj\dn} (\omega)\nonumber\\ 
&\Delta_{III\up} (\omega) =  \sum_{i\in\{1,...,6\},bj \in S_{III}}\Delta_{(Ce_a,i),bj\up} (\omega),\quad   \Delta_{III\dn} (\omega) =  \sum_{i\in\{1,...,6\},bj \in S_{III}}\Delta_{(Ce_a,i),bj\dn} (\omega)
\label{eq:app:hyb_fun_c_channel}
\ea

\begin{figure}
    \centering
    \includegraphics[width=1.0\textwidth]{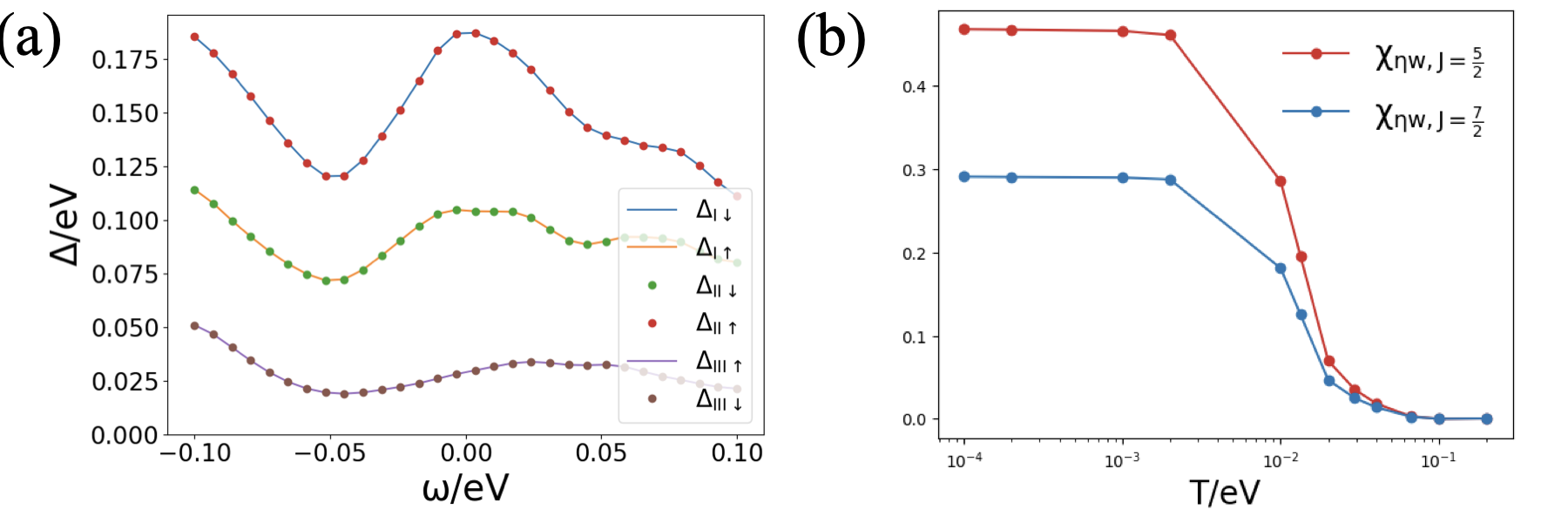}
    \caption{(Left) Hybridization functions from different $c$-electron channels (Eq.~\ref{eq:app:hyb_fun_c_channel}). 
    (Right) Evolution of the hybridization fields as a function of temperature.
    % Hybridization functions of different $f$ states (Eq.~\ref{eq:app:hyb_eta_w_basis}). 
    }
    \label{fig:hyb_orb_dep}
\end{figure}

As shown in Fig.~\ref{fig:hyb_orb_dep} (a), we observe that, due to the magnetic ordering, the contribution from spin $\up$ electrons and spin $\dn$ electrons from the same Co-P layers are different, where the contribution from one of the spin sectors is suppressed.
However, contributions from the set $I$ electrons of spin $\up$ and the set $II$ electrons of spin $\dn$ are degenerate due to the $P\cdot \mathcal{T}$ symmetry. These contributions are also most relevant. 
Moreover, the contribution from Ce $d$ electrons are less relevant. 

% We also provide a quantitative comparison. We consider the average value of the hybridization function within a small energy cutoff $\omega_0=10$meV near the Fermi energy
% \ba 
% \tilde{\Delta}_{\alpha\sigma} = \frac{1}{2\omega_0}\int_{-\omega_0}^{\omega_0}\Delta_{\alpha\sigma}(\omega)d\omega 
% \ea 
% We find 
% \ba 
% &\tilde{\Delta}_{I\up}=\tilde{\Delta}_{II,\dn}\approx 0.187\text{eV} \nonumber\\ 
% &\tilde{\Delta}_{I\dn}=\tilde{\Delta}_{II\up}\approx
% 0.104\text{eV} \nonumber\\ &\tilde{\Delta}_{III\up}=\tilde{\Delta}_{III\dn}\approx
% 0.029\text{eV} 
% \ea 
% We can observe that the spin $\dn$ ($\up$) contribution from layer I (II) Co-P atoms are suppressed due to the AFM, and is only 55\% of the spin $\up$ ($\dn$) contribution from layer I (II) Co-P atoms. The spin $\up$ and spin $\dn$ contributions from different Co-P layers are the same due to the $P\cdot\mathcal{T}$ symmetry
% \ba 
% &\Delta_{I\up}(\omega) = \Delta_{II\dn}(\omega),\quad \tilde{\Delta}_{I\up}=\tilde{\Delta}_{II\dn}\nonumber\\ 
% &\Delta_{II\up}(\omega) = \Delta_{I\dn}(\omega),\quad \tilde{\Delta}_{II\up}=\tilde{\Delta}_{I\dn}
% \ea 

In conclusion, we find that spin $\up$ $c$ electrons of layer I Co-P atoms and spin $\dn$ electrons of layer II Co-P atoms, which correspond to the majority carriers of the systems near the Fermi energy, are mostly relevant to the Kondo effect. These two types of electrons are degenerate due to $P\cdot \mathcal{T}$ symmetry which favors the formation of Kondo singlet, and their degeneracy is not affected by the magnetic ordering as long as $P\cdot \mathcal{T}$ symmetry is preserved.

To demonstrate the existence of the Kondo phase, we perform mean-field calculations at a relatively large Hubbard interaction of $f$ electrons ($U=12$eV). This value is much larger than the maximum hopping strength of the systems ($\sim$ 1 eV) and is comparable to the typical value of the Hubbard interactions for $f$ electrons ($\sim$ 10 eV). 
To show the development of the Kondo effect, we perform calculations at various temperatures. We use $\chi_{\eta w, a, i}$ (as defined in Eq.~\ref{eq:mf_eq_chi}) to characterize the strength of the Kondo effect. Whenever a non-zero $\chi_{\eta w,a, i}$ develops, the Kondo effect emerges. 
In Fig.~\ref{fig:hyb_orb_dep} (b), we show the evolution of $\chi_{\eta w, J=\frac{5}{2}} ,\chi_{\eta w,J=\frac{7}{2}}$ as a function of temperature. $\chi_{\eta w, J=\frac{5}{2}} ,\chi_{\eta w,J=\frac{7}{2}}$ denote the contributions from $J=\frac{5}{2}$ $f$ states (corresponding to $\eta_{\RR,a,i=1,...,6}$ fermions) and $J=\frac{7}{2}$ $f$ states (corresponding to $\eta_{\RR,a,i=7,...,14}$ fermions) respectively
\ba 
&\chi_{\eta w, J=\frac{5}{2}} = \sum_{j=1,2,3,4,5,6} |\chi_{\eta, w, Ce_a,i}|\nonumber \nonumber\\ 
&\chi_{\eta w, J=\frac{5}{2}} = \sum_{j=7,8,9,10,11,12,13,14} |\chi_{\eta, w, Ce_a,i}|
\ea 
with $\chi_{\eta w,a,i}$ defined in Eq.~\ref{eq:mf_eq_chi}. 

From Fig.~\ref{fig:hyb_orb_dep} (b), we can observe, that the system starts to develop the Kondo effect at around $T\sim 0.1$eV. However, since mean-field calculations usually underestimate quantum fluctuations, the Kondo temperature of the real system will be much smaller. We also observe that both $J=5/2,J=7/2$ states contribute to the formation of Kondo states. Since $J=5/2$ states have lower energy at the atomic limit (due to spin-orbit coupling, Sec.~\ref{sec:app:spin_obit_coupling}), they are more relevant to the Kondo effect and their contributions are stronger.

\begin{figure}
    \centering
    \includegraphics[width=0.5\textwidth]{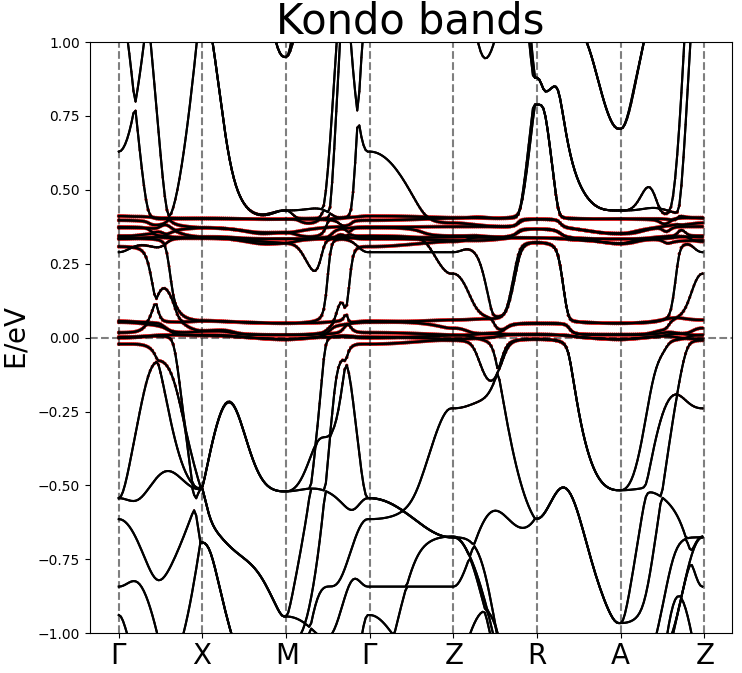}
    \caption{Band structure in the Kondo phase ($T=10^{-4}$eV), where red dots mark the orbital weights of $f$ electrons.  }
    \label{fig:kondo_band}
\end{figure}

At low temperatures, after the development of Kondo state, additional $f$-based excitations emerge. In Fig.~\ref{fig:kondo_band}, we show the band structure of the low-temperature Kondo phase by diagonalizing the mean-field Kondo Hamiltonian (Eq.~\ref{eq:mf_ham_kondo}). We can observe the presence of the Kondo excitation (marked by red dots). We observe two sets of Kondo bands (one near Fermi energy and the other one near $E\sim 0.3$eV). The first set of Kondo bands corresponds to the $J=\frac{5}{2}$ $f$-electron states whereas the second set of Kondo bands corresponds to the $J=\frac{7}{2}$ $f$-electron states. As we discussed in Sec.~\ref{sec:app:spin_obit_coupling}, the energy gap between $J=\frac{5}{2}$ and $J=\frac{7}{2}$ states in the atomic limit is created by spin-orbit coupling. The value of this SOC gap in the atomic limit is $\sim \frac{7\lambda}{2} \approx 0.3$eV which is also consistent with the band structures shown in Fig.~\ref{fig:kondo_band}. 

\subsection{Hybridization function}
\label{sec:app:hyb}

\begin{figure}
    \centering
    \includegraphics[width=0.8\linewidth]{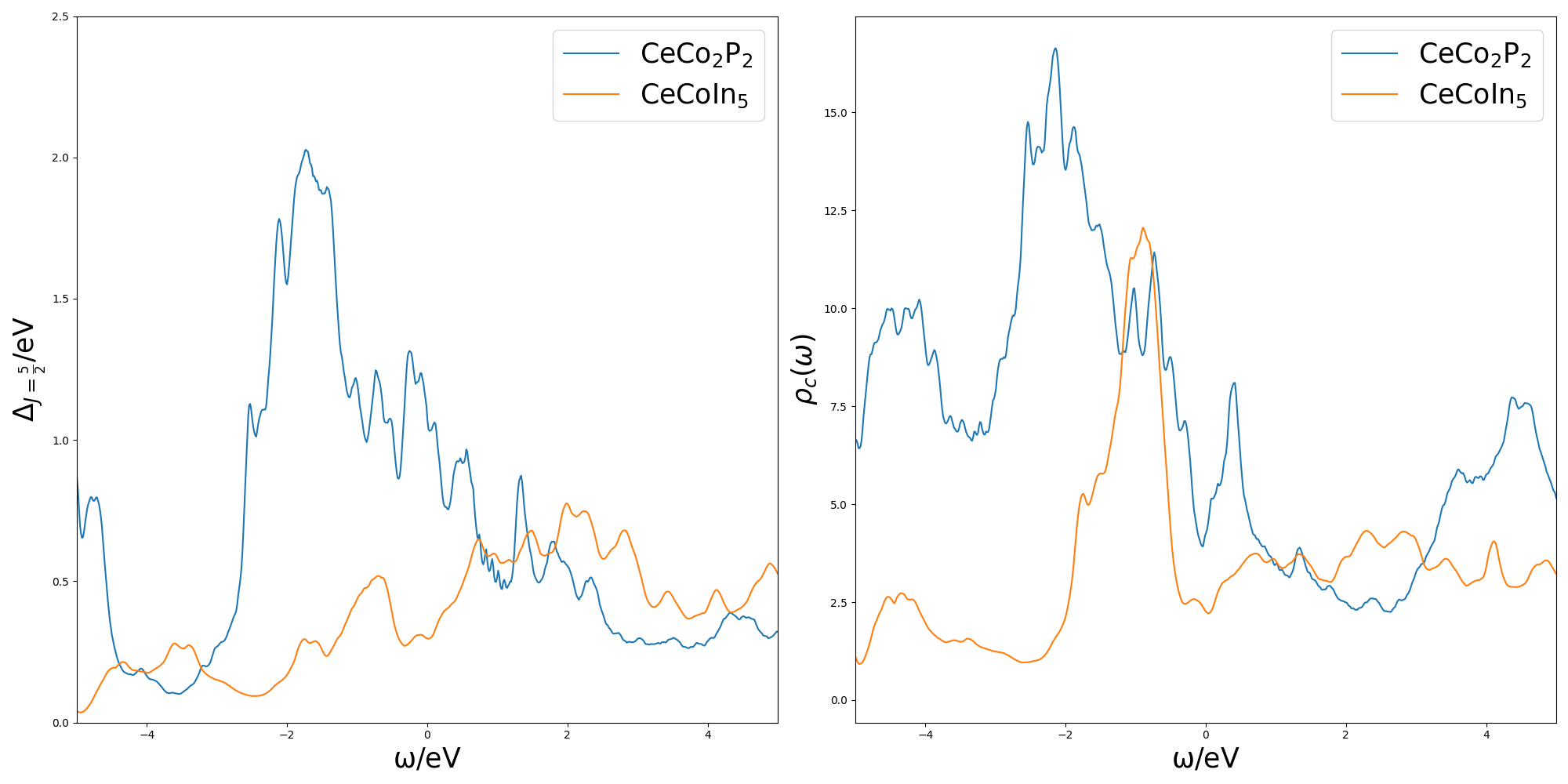}
    \caption{(Left) Hybridization functions of the CeCo$_2$P$_2$ in the AFM phase, and CeCoIn$_5$. 
    (Right) Conduction electron density of states of CeCo$_2$P$_2$ in the AFM phase, and CeCoIn$_5$. 
    }
    \label{fig:compare_hyb}
\end{figure}

In this section, we further investigate the hybridization function by comparing our system with the well-studied heavy-fermion compound CeCoIn$_5$. 
By analyzing the hybridization functions of CeCo$_2$P$_2$ and CeCoIn$_5$, we find that the relatively high Kondo temperatures observed in CeCo$_2$P$_2$ can be attributed to its stronger hybridization strength.

Our focus is on the hybridization function associated with the low-energy $f$-electron state characterized by total angular momentum $J = 5/2$. 
For CeCo$_2$P$_2$, the corresponding hybridization function can be obtained as follows:
\begin{align}
    \Delta_{J=5/2}(\omega)  =  \frac{\pi}{N} \sum_{a, i=1,...,6}\delta_{a,Ce_1}\sum_{\kk,n}\bigg| \sum_{bj\sigma} V_{ai,bj\sigma}(\kk) U^c_{bj\sigma,n}(\kk) \bigg|^2  \delta(\epsilon-\epsilon_{\kk,n})
\end{align}
Here, $V_{ai,bj\sigma}(\kk)$ denotes the $fc$ hybridization, and $U^c_{bj\sigma,n}(\kk)$ are the eigenvectors of the $c$-electron bands, as defined in \cref{eq:def_hyb_ham,eq:def_c_ham}. 
We consider only the contributions from the Ce$_1$ atom, since both Ce atoms are equivalent. 
Furthermore, we restrict the analysis to the lowest 6 $f$-electron states with total angular momentum $J = 5/2$. 
The equivalent hybridization function for the $J = 5/2$ $f$-orbitals in CeCoIn$_5$ can be defined in the same manner. 

In Fig.~\ref{fig:compare_hyb},  we compare the hybridization functions and the density of states of the $c$ electrons, $\rho_c(\omega)$, for the two compounds (CeCo$_2$P$_2$ and CeCoIn$_5$). 
The $c$-electron density of states is defined as
\begin{align}
    \rho_c(\omega) = \frac{1}{N}\sum_{\kk,n}\delta(\epsilon-\epsilon_{\kk,n})
\end{align}
We conclude that the hybridization function of CeCo$_2$P$_2$ at the Fermi energy is much larger than that of CeCoIn$_5$, with the latter being approximately 30\% of the former. 
This enhanced hybridization can be attributed to the larger $c$-electron density of states in CeCo$_2$P$_2$, which allows more $c$ electrons to participate in the hybridization. 
Such enhanced hybridization is expected to result in a higher Kondo temperature, which is qualitatively consistent with experimental observations: the Kondo temperature of CeCo$_2$P$_2$ is estimated to be around 100K, much higher than that of CeCoIn$_5$, which is approximately 40K
~\cite{petrovic2001heavy,RevModPhys.92.011002}.

Finally, we comment that the lattice constants may vary with temperature, as reported in Ref.~\cite{reehuis1998antiferromagnetic}, implying a potential temperature dependence of the hybridization functions. 
Such variations could, in turn, influence the effective Kondo temperatures. 
Nonetheless, our results suggest that the relatively high Kondo temperature of CeCo$_2$P$_2$ may be understood via its relatively strong hybridization function.

\subsection{eDMFT calculations} 
\label{sec:edmft}
\begin{figure}
    \centering
    \includegraphics[width=0.5\linewidth]{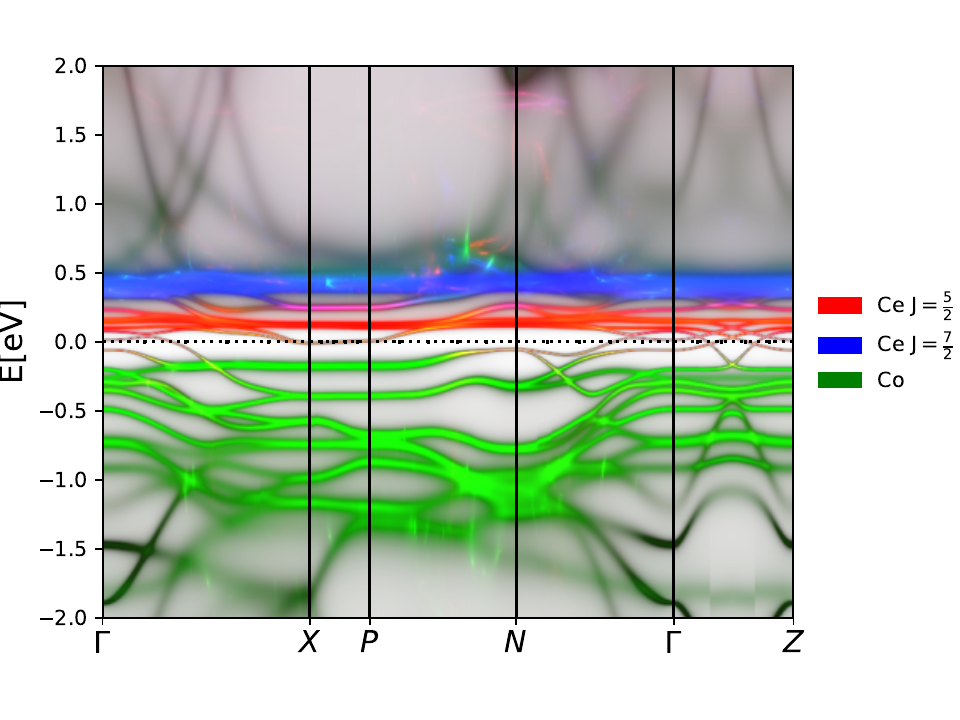}
    \caption{Momentum-resolved spectral function in the AFM phase. Colors indicate the spectral weight contribution from different electronic components.}
    \label{fig:edmft_k_spec}
\end{figure}

To further demonstrate the coexistence of the Kondo effect and antiferromagnetism (AFM), we perform embedded dynamical mean-field theory (eDMFT) calculations\cite{RevModPhys.78.865,RevModPhys.68.13,PhysRevLett.115.256402,PhysRevLett.115.196403}, combining DFT with DMFT. The calculations are carried out in the AFM phase at $T = 100$ K, using lattice structures taken from the experimental results of Ref.~\cite{REEHUIS1990961}. In Fig.\ref{fig:edmft_k_spec}, we present the $k$-resolved interacting spectral function, where the $J=5/2$ $f$-electron states are observed near the Fermi energy. Similar features are seen in the local density of states (DOS) shown in Fig.~\ref{fig:edmft_k_spec}. Both results indicate the development of the Kondo effect within the AFM phase. 

Our eDMFT results at $T=100$~K suggest that the Kondo temperature within the theoretical calculations is higher than 100~K. We note, however, that the lattice constants can vary with temperature in experiments, as reported in Ref.~\cite{reehuis1998antiferromagnetic}. 
Such variations, which are not captured in the present theoretical calculations, could also affect the Kondo temperature. Therefore, an accurate theoretical determination of the Kondo temperature would require accounting for the temperature dependence of the lattice structure, which would involve rather costly calculations. We thus leave this for future work.
Based on our current calculations, we can conclude that: (1) the Kondo effect coexists with antiferromagnetism; and (2) the Kondo temperature is relatively high. This relatively large Kondo temperature can be understood in terms of the strong hybridization function discussed near Fig.~\ref{fig:compare_hyb}.

\begin{figure}
    \centering
    \includegraphics[width=0.5\linewidth]{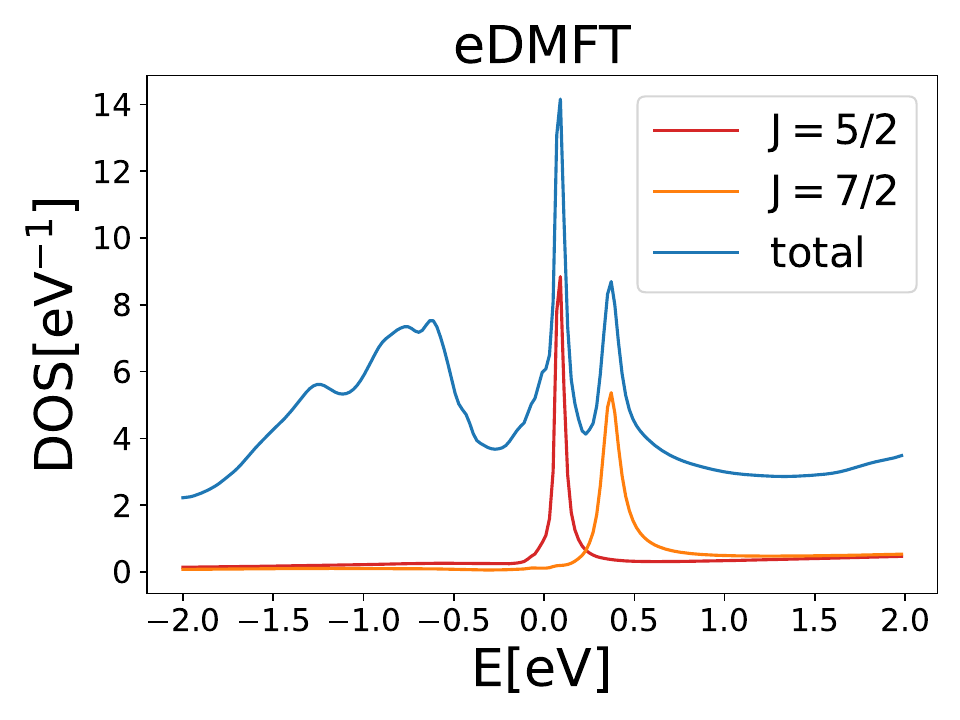}
    \caption{Density of states (DOS) of Ce $4f$ electrons with $J=5/2$ and $J=7/2$, along with the total electron density of states.}
    \label{fig:edmft_dos}
\end{figure}

% \subsection{Kondo nodal line}
\section{Band topology of the Kondo phase}
We discuss the band topology of the anti-ferromagnetic Kondo phase in this Appendix. In this section, we will use the conventional cell of the PM phase which is the primitive cell of the AFM phase. 

\subsection{Glide mirror-$z$ symmetry}
We discuss the properties of glide mirror-$z$ symmetry. In this section, we consider the conventional unit cell. In the non-magnetic phase, the system exhibits mirror-$z$ symmetry with the mirror plane located at the Ce plane in~\cref{fig:lat_mag_structure}, which is symmorphic. However, in the magnetic phase, the mirror-$z$ symmetry maps Co$_{1,2}$ atoms to Co$_{3,4}$ atoms. Since Co$_{1,2}$ and Co$_{3,4}$ atoms possess opposite magnetic moments, mirror-$z$ is no longer a symmetry operation. However, by combining mirror-$z$ with translational symmetry, we obtain a glide mirror-$z$ transformation defined as
 \ba 
 h = \{ M_z |1/2,1/2,1/2\} 
 \ea
 which characterizes a mirror $z$ transformation followed by a $(1/2,1/2,1/2)$ shifting. glide mirror-$z$ transformation is a symmetry operation of the magnetic phase.

It is worth noting that, in the non-magnetic phase, the $(1/2,1/2,1/2)$ translation is also a symmetry operation, meaning that there is no nonsymmorphic symmetry in this phase. The magnetic order breaks both the $(1/2,1/2,1/2)$ translational symmetry and the mirror-$z$ symmetry but preserves the glide mirror-$z$ symmetry ${M_z |1/2,1/2,1/2}$.

\subsection{Nodal line in the Kondo phase}
We demonstrate that, in the AFM Kondo phase, the glide mirror-$z$ symmetry protects the nodal line at $k_z=1/2$ plane. 
Acting on $c$ and $f$ electrons, we have
\ba 
&h c_{\RR,a,i,\sigma} h^{-1} = \sum_{b,j,\sigma'} U^h_{ai \sigma,bj\sigma'} c_{h(\RR+\rr_a) -\rr_b, b, j,\sigma'}\nonumber\\ 
&h f_{\RR,a,i,\sigma}h^{-1}=\sum_{b,j,\sigma'} U^{h,f}_{ai \sigma,bj\sigma'} f_{h(\RR+\rr_a) -\rr_b, b, j,\sigma'}.
\label{eq:glide_mz_property}
\ea 
$U^h_{ai\sigma,bj\sigma'}, U^{h,f}_{ai\sigma,bj\sigma'}$ are unitary matrices characterize the glide mirror $z$ transformation. 
In the momentum space, we have
\ba 
h c_{\kk,a,i,\sigma} h^{-1} =& \frac{1}{\sqrt{N}}\sum_{\RR,j,\sigma'} U^h_{ai\sigma,bj\sigma'}
c_{h(\RR+\rr_a)-\rr_b,b,j,\sigma'}e^{-i2\pi \kk\cdot(\RR+\rr_a)} \nonumber\\ 
=&\frac{1}{N}\sum_{\RR,\kk', b,j,\sigma'} U^h_{ai\sigma,bj\sigma'}
c_{\kk',b,j} e^{i2\pi \kk'\cdot( h(\RR+\rr_a) )}e^{-i2\pi \kk\cdot(\RR+\rr_a)} \nonumber\\ 
=&\sum_{b,j,\sigma'} U_{ai\sigma,bj\sigma'}^h c_{M_z\kk, b, j,\sigma'}e^{i2\pi M_z\kk\cdot\Delta \rr } \nonumber\\ 
h f_{\kk,a,i,\sigma} h^{-1} 
=&\sum_{b,j,\sigma'} U_{ai\sigma,bj\sigma'}^{h,f} f_{M_z\kk, b, j,\sigma'}e^{i2\pi M_z\kk\cdot\Delta \rr} 
\ea 
where $\Delta \rr = (1/2,1/2,1/2)$.

We now show the transformation matrix (defined in Eq.~\ref{eq:glide_mz_property}) for each orbital. 
For the $d$@Co orbitals, we have
\ba 
U^{h,Co}_{ai\sigma,bj\sigma'} = \begin{bmatrix}
0 & 1 \\ 
1 & 0 \\ 
& & 0 & 1 \\
& & 1 & 0 
\end{bmatrix}_{a,b} \otimes 
\begin{bmatrix}
1 \\ 
& -1 \\ 
& & -1 \\ 
& & & 1 \\ 
& & & & 1 
\end{bmatrix}_{i,j} \otimes 
\begin{bmatrix}
    i \\ 
    & -i 
\end{bmatrix}_{\sigma, \sigma'}
\ea 
where indices $a,b$ denote four Co atoms, $i,j$ denote five $d$-orbitals ($d_{z^2},d_{xz},d_{yz}, d_{x^2-y^2},d_{xy}$), and $\sigma,\sigma'$ denote the spin indices ($\up,\dn$).
The transformation matrix for the $p@$P orbitals are 
\ba 
U^{h,P}_{ai \sigma,bj\sigma'} = \begin{bmatrix}
    0 & 1 \\ 
    1 & 0 \\ 
    & & 0 & 1 \\ 
    & & 1 & 0 
\end{bmatrix}_{a,b}
\otimes
\begin{bmatrix}
    -1 \\ 
    & 1 \\ 
    & & 1 
\end{bmatrix}_{i,j}
\otimes 
\begin{bmatrix}
    i \\ 
    & -i 
\end{bmatrix}_{\sigma, \sigma'}
\ea 
where indices $a,b$ denote four P atoms, $i,j$ denote three $p$-orbitals ($p_z,p_x,p_y$), and $\sigma,\sigma'$ denote the spin indices ($\up,\dn$).
The transformation matrix for the $d$@Ce are 
\ba 
U_{ai \sigma,bj \sigma'}^{h,d@Ce} = 
\begin{bmatrix}
    & 1 \\ 
    1 & 
\end{bmatrix}_{a,b}\begin{bmatrix}
1 \\ 
& -1 \\ 
& & -1 \\ 
& & & 1 \\ 
& & & & 1 
\end{bmatrix}_{i,j} 
\otimes 
\begin{bmatrix}
    i \\ 
    & -i 
\end{bmatrix}_{\sigma, \sigma'}
\ea 
where indices $a,b$ denote two Ce atoms, $i,j$ denote five $d$-orbitals ($d_{z^2},d_{xz},d_{yz}, d_{x^2-y^2},d_{xy}$), and $\sigma,\sigma'$ denote the spin indices ($\up,\dn$). The transformation for the $d$@Ce are 
\ba 
U_{ai \sigma,bj \sigma'}^{h,f} = 
\begin{bmatrix}
    & 1 \\ 
    1 & 
\end{bmatrix}_{a,b}\begin{bmatrix}
-1 \\ 
& 1 \\ 
& & 1 \\ 
& & & -1 \\ 
& & & & -1 \\ 
&&&&& 1 \\ 
&&&&&& 1 
\end{bmatrix}_{i,j} 
\otimes 
\begin{bmatrix}
    i \\ 
    & -i 
\end{bmatrix}_{\sigma, \sigma'}
\ea 
where indices $a,b$ denote two Ce atoms, $i,j$ denote seven $f$-orbitals ($f_{z^3}, f_{xz^2},f_{yz^2}, f_{z(x^2-y^2)}, f_{xyz}, f_{x(x^2-3y^2)}, f_{y(3x^2-y^2)}$), and $\sigma,\sigma'$ denote the spin indices ($\up,\dn$).

There are two mirror-$z$ invariant planes labeled by $\{(k_x,k_y,1/2)\}_{k_x,k_y}$ and $\{(k_x,k_y,0)\}_{k_x,k_y}$. At the mirror-$z$-invariant planes, we have
\ba 
&h c_{(k_x,k_y,1/2),a,i,\sigma}h^{-1} = \sum_{b,j,\sigma'}U_{ai\sigma,bj\sigma'}^h 
c_{(k_x,k_y,1/2), b, j,\sigma'} e^{i2\pi(k_x,k_y,-1/2)\cdot(\Delta\rr ) +i2\pi(0,0,1)\cdot \rr_b }  \nonumber\\ 
&h c_{(k_x,k_y,0),a,i,\sigma}h^{-1} = \sum_{b,j,\sigma'}U_{ai\sigma,bj\sigma'}^h 
c_{(k_x,k_y,0), b, j,\sigma'} e^{i2\pi(k_x,k_y,0)\cdot(\Delta\rr )} \nonumber\\ 
&h f_{(k_x,k_y,1/2),a,i,\sigma}h^{-1} = \sum_{b,j,\sigma'}U_{ai\sigma,bj\sigma'}^{h,f} 
f_{(k_x,k_y,1/2), b, j,\sigma'} e^{i2\pi(k_x,k_y,-1/2)\cdot(\Delta\rr ) +i2\pi(0,0,1)\cdot \rr_b }  \nonumber\\ 
&h f_{(k_x,k_y,0),a,i,\sigma}h^{-1} = \sum_{b,j,\sigma'}U_{ai\sigma,bj\sigma'}^{h,f} 
f_{(k_x,k_y,0), b, j,\sigma'} e^{i2\pi(k_x,k_y,0)\cdot(\Delta\rr )} 
\label{eq:glide_mz_mirro_plane}
\ea 
Here, we focus on the $k_z=1/2$ plane, where we observe the formation of the nodal line. 
On the $k_z=1/2$ plane, we can introduce the electron operators which correspond to the eigenvectors of the glide mirror-$z$ symmetry (Eq.~\ref{eq:glide_mz_mirro_plane}).
The operators that form the eigenbasis of glide mirror-$z$ transformation ($h$) are given in Tab.~\ref{tab:app:glide_mz_eigen}.

\begin{center}
\begin{table}
\begin{tabular}{C|C}
\text{Operator} & \text{Eigenvalue}\\ \hline 
\frac{1}{\sqrt{2}}(c_{(k_x,k_y,1/2),Co_{1},i,\sigma} + c_{(k_x,k_y,1/2),Co_{2},i,\sigma} ) &is^{M_z}_{d,i}\sigma e^{i\pi(k_x+k_y)}\\ 
\frac{1}{\sqrt{2}}(c_{(k_x,k_y,1/2),Co_{1},i,\sigma} - c_{(k_x,k_y,1/2),Co_{2},i,\sigma} )& -is^{M_z}_{d,i}\sigma e^{i\pi(k_x+k_y)}
\\ 
\frac{1}{\sqrt{2}}(c_{(k_x,k_y,1/2),Co_{3},i,\sigma} + c_{(k_x,k_y,1/2),Co_{4},i,\sigma} ) &-is^{M_z}_{d,i}\sigma e^{i\pi(k_x+k_y)}\\ 
\frac{1}{\sqrt{2}}(c_{(k_x,k_y,1/2),Co_{3},i,\sigma} - c_{(k_x,k_y,1/2),Co_{4},i,\sigma} )& is^{M_z}_{d,i}\sigma e^{i\pi(k_x+k_y)}
\\ 
\frac{1}{\sqrt{2}}(-ie^{i2\pi z}c_{(k_x,k_y,1/2),P_1,i,\sigma} + c_{(k_x,k_y,1/2),P_2,i,\sigma} ) &  s_{p,i}^{M_z} e^{i\pi(k_x+k_y)}\sigma \\ 
\frac{1}{\sqrt{2}}(ie^{i2\pi z}c_{(k_x,k_y,1/2),P_1,i,\sigma} + c_{(k_x,k_y,1/2),P_2,i,\sigma} )&  -i s_{p,i}^{M_z} e^{i\pi(k_x+k_y)}\sigma \\ 
\frac{1}{\sqrt{2}}(ie^{i2\pi z}c_{(k_x,k_y,1/2),P_3,i,\sigma} +
c_{(k_x,k_y,1/2),P_4,i,\sigma})&  is_{p,i}^{M_z}\sigma e^{i\pi(k_x+k_y)}\\ 
\frac{1}{\sqrt{2}}(-ie^{i2\pi z}c_{(k_x,k_y,1/2),P_3,i,\sigma} +
c_{(k_x,k_y,1/2),P_4,i,\sigma})&  -is_{p,i}^{M_z}\sigma e^{i\pi(k_x+k_y)}
\\
\frac{1}{\sqrt{2}}(c_{(k_x,k_y,1/2),Ce_1,i,\sigma} + ic_{(k_x,k_y,1/2),Ce_2,i,\sigma})& 
is_{d,i}^{M_z}\sigma e^{i\pi(k_x+k_y)} \\ 
\frac{1}{\sqrt{2}}(c_{(k_x,k_y,1/2),Ce_1,i,\sigma} - ic_{(k_x,k_y,1/2),Ce_2,i,\sigma})&-
is_{d,i}^{M_z}\sigma e^{i\pi(k_x+k_y)}  
\\ 
\frac{1}{\sqrt{2}}(f_{(k_x,k_y,1/2),Ce_1,i,\sigma} + if_{(k_x,k_y,1/2),Ce_2,i,\sigma})&
is_{f,i}^{M_z}\sigma e^{i\pi(k_x+k_y)} \\ 
\frac{1}{\sqrt{2}}(f_{(k_x,k_y,1/2),Ce_1,i,\sigma} - if_{(k_x,k_y,1/2),Ce_2,i,\sigma})&-
is_{f,i}^{M_z}\sigma e^{i\pi(k_x+k_y)} \\
\hline 
\end{tabular}
\caption{Eigenbasis of glide mirror $z$ transformation ($h$). We have also introduced the mirror-$z$ eigenvalues for the spinless orbitals. Specifically, $s^{M_z}_{d,i} = 1, -1, -1, 1, 1$ for $i=1,2,3,4,5$ respectively, $s^{M_z}_{p,i} = -1,1,1$ for $i=1,2,3$ respectively, and $s^{M_z}_{f,i} = -1, 1,1, -1,-1,-1,1,1$ for $i=1,2,3,4,5,6,7$ respectively. }
\label{tab:app:glide_mz_eigen}
\end{table}
\end{center}

From Tab.~\ref{tab:app:glide_mz_eigen}, we conclude that bands at $k_z=1/2$ plane have either $+ie^{i\pi(k_x+k_y)}$ or $-ie^{i\pi(k_x+k_y)}$ eigenvalue under glide mirror-$z$ transformations ($h$). Moreover, due to the $P\cdot \mathcal{T}$ symmetry, the bands are also two-fold degenerate at each momentum. We now discuss the $h$-eigenvalues of two electrons that are connected by $P\cdot\mathcal{T}$ symmetry. We note
\ba 
&(P\cdot\mathcal{T}) c_{\RR,a,i,\sigma} (P\cdot\mathcal{T})^{-1} = \sum_{b,j,\sigma'} U^{P\cdot\mathcal{T}}_{ai \sigma,bj\sigma'} c_{P(\RR+\rr_a) -\rr_b, b, j,\sigma'}\nonumber\\ 
&(P\cdot\mathcal{T})f_{\RR,a,i,\sigma}(P\cdot\mathcal{T})^{-1}=\sum_{b,j,\sigma'} U^{P\cdot\mathcal{T},f}_{ai \sigma,bj\sigma'} f_{P(\RR+\rr_a) -\rr_b, b, j,\sigma'} \nonumber\\ 
&(P\cdot\mathcal{T})i(P\cdot\mathcal{T})^{-1} =- i 
\ea 
where $U^{P\cdot\mathcal{T}},U^{P\cdot\mathcal{T},f}$ are transformation matrices. For $d@$Co orbitals, we have
\ba 
U^{P\cdot\mathcal{T},Co}_{ai\sigma,bj\sigma'} =
\begin{bmatrix}0 & 0 & 0 & 1 \\ 
0 & 0 & 1 & 0 \\
0 & 1 & 0 & 0 \\ 
1 & 0 & 0 & 0 
\end{bmatrix}_{a,b} \oplus \mathbb{I}_{i,j}\oplus [i\sigma^y]_{\sigma,\sigma'} 
\ea 
For $p@$P orbitals, we have
\ba 
U^{P\cdot\mathcal{T},P}_{ai\sigma,bj\sigma'} =
\begin{bmatrix}0 & 0 & 0 & 1 \\ 
0 & 0 & 1 & 0 \\
0 & 1 & 0 & 0 \\ 
1 & 0 & 0 & 0 
\end{bmatrix}_{a,b} \oplus [-\mathbb{I}]_{i,j}\oplus [i\sigma^y]_{\sigma,\sigma'}
\ea 
For $d@$Ce orbitals, we have 
\ba 
U^{P\cdot\mathcal{T},d@Co}_{ai\sigma,bj\sigma'} =
\begin{bmatrix}1 & 0  \\ 
0&1
\end{bmatrix}_{a,b} \oplus \mathbb{I}_{i,j}\oplus [i\sigma^y]_{\sigma,\sigma'}
\ea 
For $f@$Ce orbitals, we have 
\ba 
U^{P\cdot\mathcal{T},f}_{ai\sigma,bj\sigma'} =
\begin{bmatrix}1 & 0  \\ 
0&1
\end{bmatrix}_{a,b} \oplus [-\mathbb{I}]_{i,j}\oplus [i\sigma^y]_{\sigma,\sigma'}
\ea 

We then investigate the $P\cdot\mathcal{T}$ transformations using the eigenbasis of glide mirror-$z$ transformation (Tab.~\ref{tab:app:glide_mz_eigen}). The results are shown in Tab.~\ref{tab:app:pt_transf}. We note that, at $k_z=1/2$ plane, two electron operators connected by $P\cdot\mathcal{T}$ always have the same glide mirror-$z$ eigenvalues. 

\begin{center}
\begin{table}
\begin{tabular}{C|C}
\text{Operator: }\gamma & \text{Operator after transformation: } (P\cdot \mathcal{T})\gamma (P\cdot \mathcal{T})^{-1}\\ \hline 
\frac{1}{\sqrt{2}}(c_{(k_x,k_y,1/2),Co_{1},i,\sigma} + c_{(k_x,k_y,1/2),Co_{2},i,\sigma} ) &\frac{\sigma}{\sqrt{2}}(c_{(k_x,k_y,1/2),Co_{4},i,-\sigma} + c_{(k_x,k_y,1/2),Co_{3},i,-\sigma} ) 
\\ 
\frac{1}{\sqrt{2}}(c_{(k_x,k_y,1/2),Co_{1},i,\sigma} - c_{(k_x,k_y,1/2),Co_{2},i,\sigma} )& \frac{\sigma}{\sqrt{2}}(c_{(k_x,k_y,1/2),Co_{4},i,-\sigma} - c_{(k_x,k_y,1/2),Co_{3},i,-\sigma} )
\\ 
\frac{1}{\sqrt{2}}(c_{(k_x,k_y,1/2),Co_{3},i,\sigma} + c_{(k_x,k_y,1/2),Co_{4},i,\sigma} ) &
\frac{\sigma}{\sqrt{2}}(c_{(k_x,k_y,1/2),Co_{2},i,-\sigma} + c_{(k_x,k_y,1/2),Co_{1},i,-\sigma} ) 
\\ 
\frac{1}{\sqrt{2}}(c_{(k_x,k_y,1/2),Co_{3},i,\sigma} - c_{(k_x,k_y,1/2),Co_{4},i,\sigma} )&
\frac{\sigma}{\sqrt{2}}(c_{(k_x,k_y,1/2),Co_{2},i,-\sigma} - c_{(k_x,k_y,1/2),Co_{1},i,-\sigma} )
\\ 
\frac{1}{\sqrt{2}}(-ie^{i2\pi z}c_{(k_x,k_y,1/2),P_1,i,\sigma} + c_{(k_x,k_y,1/2),P_2,i,\sigma} ) & 
\frac{\sigma}{\sqrt{2}}(ie^{-i2\pi z}c_{(k_x,k_y,1/2),P_4,i,-\sigma} + c_{(k_x,k_y,1/2),P_3,i,-\sigma} ) 
\\ 
\frac{1}{\sqrt{2}}(ie^{i2\pi z}c_{(k_x,k_y,1/2),P_1,i,\sigma} + c_{(k_x,k_y,1/2),P_2,i,\sigma} )&  
\frac{\sigma}{\sqrt{2}}(-ie^{-i2\pi z}c_{(k_x,k_y,1/2),P_4,i,-\sigma} + c_{(k_x,k_y,1/2),P_3,i,-\sigma} )
\\ 
\frac{1}{\sqrt{2}}(ie^{i2\pi z}c_{(k_x,k_y,1/2),P_3,i,\sigma} +
c_{(k_x,k_y,1/2),P_4,i,\sigma})&  
\frac{\sigma}{\sqrt{2}}(-ie^{-i2\pi z}c_{(k_x,k_y,1/2),P_2,i,-\sigma} +
c_{(k_x,k_y,1/2),P_1,i,-\sigma})
\\ 
\frac{1}{\sqrt{2}}(-ie^{i2\pi z}c_{(k_x,k_y,1/2),P_3,i,\sigma} +
c_{(k_x,k_y,1/2),P_4,i,\sigma})&  
\frac{\sigma}{\sqrt{2}}(ie^{-i2\pi z}c_{(k_x,k_y,1/2),P_2,i,-\sigma} +
c_{(k_x,k_y,1/2),P_1,i,-\sigma})
\\
\frac{1}{\sqrt{2}}(c_{(k_x,k_y,1/2),Ce_1,i,\sigma} + ic_{(k_x,k_y,1/2),Ce_2,i,\sigma})& 
\frac{\sigma}{\sqrt{2}}(c_{(k_x,k_y,1/2),Ce_1,i,-\sigma} - ic_{(k_x,k_y,1/2),Ce_2,i,-\sigma})
\\ 
\frac{1}{\sqrt{2}}(c_{(k_x,k_y,1/2),Ce_1,i,\sigma} - ic_{(k_x,k_y,1/2),Ce_2,i,\sigma})&
\frac{\sigma}{\sqrt{2}}(c_{(k_x,k_y,1/2),Ce_1,i,-\sigma} + ic_{(k_x,k_y,1/2),Ce_2,i,-\sigma})
\\ 
\frac{1}{\sqrt{2}}(f_{(k_x,k_y,1/2),Ce_1,i,\sigma} + if_{(k_x,k_y,1/2),Ce_2,i,\sigma})&
\frac{\sigma}{\sqrt{2}}(f_{(k_x,k_y,1/2),Ce_1,i,-\sigma} - if_{(k_x,k_y,1/2),Ce_2,i,-\sigma})
\\ 
\frac{1}{\sqrt{2}}(f_{(k_x,k_y,1/2),Ce_1,i,\sigma} - if_{(k_x,k_y,1/2),Ce_2,i,\sigma})&\frac{\sigma}{\sqrt{2}}(f_{(k_x,k_y,1/2),Ce_1,i,-\sigma} + if_{(k_x,k_y,1/2),Ce_2,i,-\sigma}) \\
\hline 
\end{tabular}
\caption{Transformation properties of the glide-mirror-$z$ eigenstates.  }
\label{tab:app:pt_transf}
\end{table}
\end{center}

\begin{figure}
    \centering
    \includegraphics[width=0.4\textwidth]{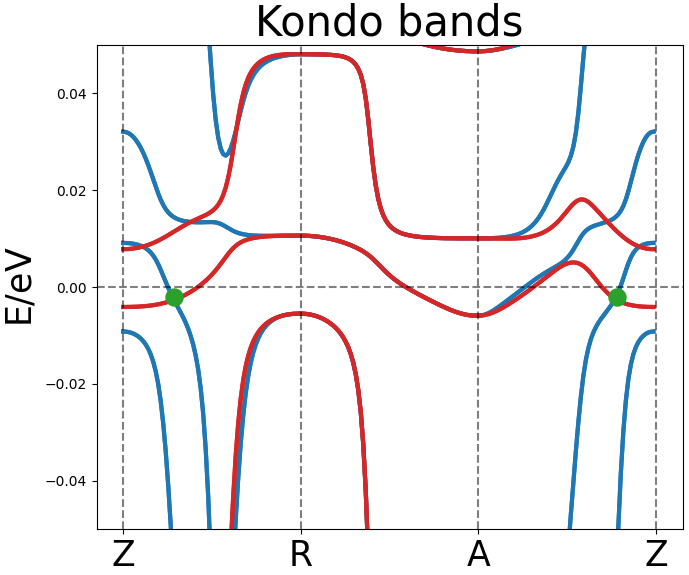}
    \caption{Band structure of the Kondo phase ($T=10^{-4}$eV) in a narrow energy window. The same band structure has also been shown in Fig.~\ref{fig:kondo_band} in a larger energy window.   
    Blue and red mark the bands with glide mirror-$z$ eigenvalues $\pm ie^{i\pi(k_x+k_y)}$ respectively. Green dots mark the positions of the nodal line that appears near Fermi energy.   }
    \label{fig:kondo_band_nodal_line}
\end{figure}

Therefore, when two bands with different glide mirror-$z$ eigenvalues form nodes at $k_z=1/2$ plane, the nodes will be symmetry protected. 
As we show in Fig.~\ref{fig:kondo_band_nodal_line}, there is a nodal line near Fermi energy formed by two bands with different glide mirror-$z$ eigenvalues. 

\subsection{Irreducible representation of the Kondo bands}
\begin{figure}
    \centering
    \includegraphics[width=0.4\textwidth]{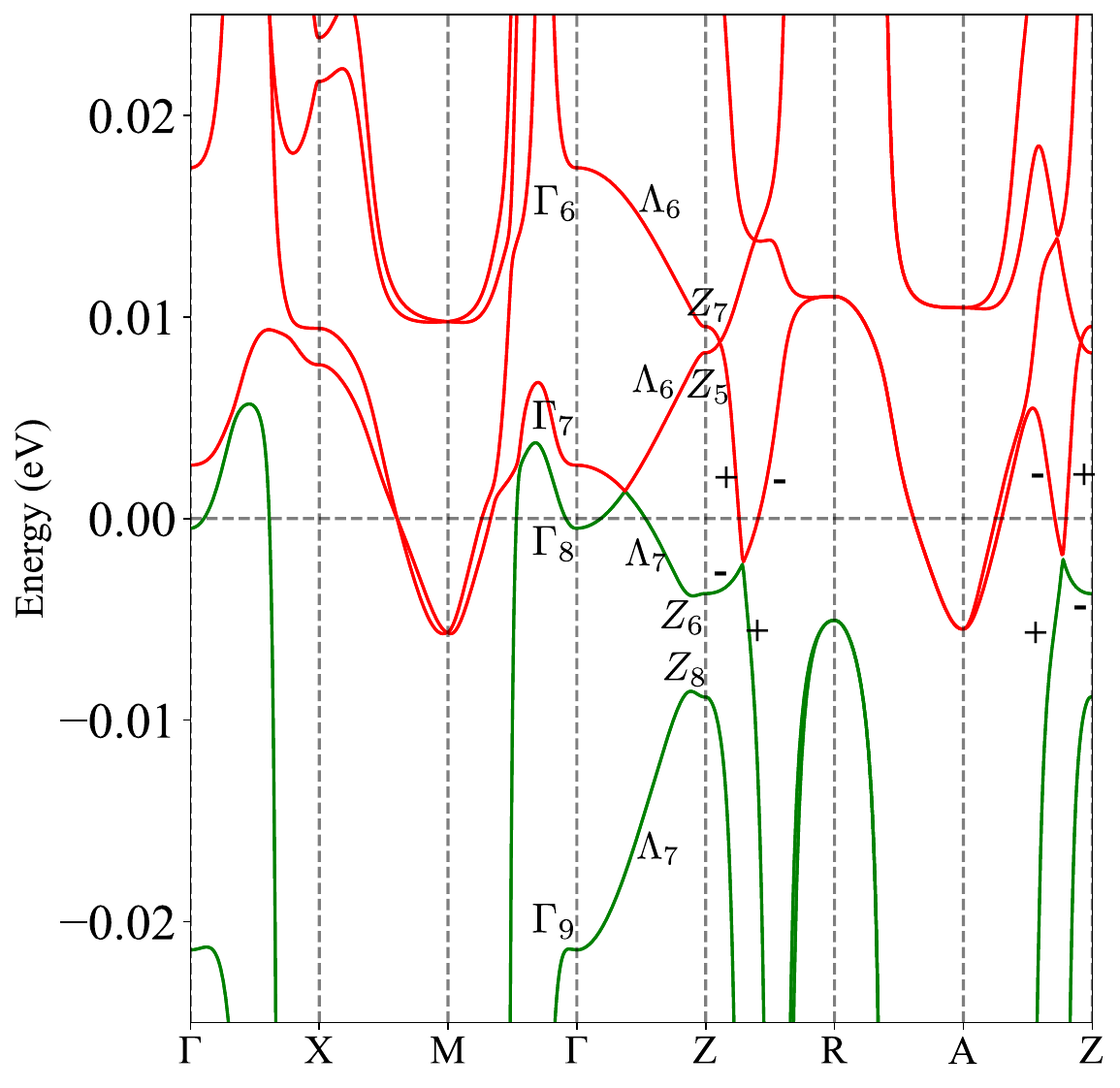}
    \caption{Band structure of the Kondo phase ($T=10^{-4}$eV). The same band structure with different energy windows has also been shown in Fig.~\ref{fig:kondo_band} and Fig.~\ref{fig:kondo_band_nodal_line}. We have also labeled the irreducible representation of bands at high-symmetry point $\Gamma,Z$, and along high-symmetry line $\Gamma$-$Z$. }
    \label{fig:kondo_band_top} 
\end{figure}
To further understand the topology of the Kondo phase, we investigate the irreducible representation of the bands~\citeBCS. The results are shown in Fig.~\ref{fig:kondo_band_top}. We observe that, besides the nodal line, there is also a $C_{4z}$-protected Dirac node near the Fermi energy (from two bands with irreducible representation $\Lambda_6$ and $\Lambda_7$) along $\Gamma$-Z line as shown in Fig.~\ref{fig:kondo_band_top}.

Finally, we discuss various possibilities of removing the node and nodal line by breaking symmetries or fine-tuning the bands. 
\begin{itemize}
    \item We can realize an insulator state by breaking the glide mirror-$z$ and $C_{4z}$ symmetries which will gap out both the Dirac node and the nodal line. The resulting state is a trivial insulator. We remark that $P\cdot\mathcal{T}$ symmetry alone cannot protect non-trivial topological states in magnetic space groups, while $P$ and $\mathcal{T}\cdot\bm{\tau}$ (where $\tau$ denotes translational transformation) can\cite{elcoro2021magnetic, peng2022topological}. 
    \item We could remove the Dirac node by switching the $\Gamma_7$ state and $\Gamma_8$ state at $\Gamma$ point. In this case, if we further break glide mirror-$z$ symmetry and gap out the nodal line, we realize an axion insulator~\cite{qi2008topological, qi2011topological, fang2012bulk} (protected by the inversion symmetry) with a non-trivial $Z_4$ index which equals to $2$. 
    \item We could remove the nodal line by switching $Z_7$ state with $Z_6$ state at $Z$ point. In this case, the Dirac node has also been removed (note that $Z_7$ state can connect to $\Gamma_8$ state along $\Gamma-Z$ line via $\Lambda_6$ state). We then realize an obstructed atomic insulator~\cite{xu2021filling, xu2021three} with obstructed positions at 8$f$ (at the middle of two nearest-neighbor Co atoms):
    \ba 
    (\pm \frac{1}{4},\pm \frac{1}{4},\pm \frac{1}{4})
    \ea 
\end{itemize}

\section{Surface state}
\label{app:sec:surface_state}
In this appendix, we discuss the surface states of the system. 

\subsection{Symmetry properties} 
For a given single-particle Hamiltonian with hopping matrix $h_{\kk,i\sigma,j\sigma'}$ where $i,j$ are orbital indices and $\sigma,\sigma'$ is the spin index. For fixed $k_x,k_y$, we can treat the system as an effective 1D system along $z$ direction with Hamiltonian 
\ba 
h_{k_z,i\sigma,j\sigma'}^{1D, (k_x,k_y)} = h_{(k_x,k_y,k_z),i\sigma,j\sigma'}
\ea 
Hence, we can examine the symmetry properties and the corresponding surface states of the 1D system with $k_x$ and $k_y$ lying within the nodal line (near the $\Gamma$ point), as well as outside the nodal line. We also comment that, here we are mainly focus on the surface state obtained from the effective 1D system. The 2D topology could potentially also produce surface states. However, as we will show later, the surface states can be merged into the gap, we do not expect the 2D topology to give an in-gap state. 

In general, the relevant symmetries of the effective 1D system at a generic $k_x,k_y$ point are $\{M_z|1/2,1/2,1/2\},P\cdot\mathcal{T}$. 
We also remark that, if we focus on the effective 1D system, $\{M_z|\frac{1}{2},\frac{1}{2},\frac{1}{2}\}$ is no longer a nonsymmorphic symmetry for this 1D system. 
For the effective 1D system, the only maximal Wyckoff positions ($2a$) are 
\ba 
R_z=1/4,\quad R_z=3/4 \, .
\ea 
The non-maximal Wyckoff positions $(4b)$ are 
\ba 
R_z= 0+z ,\quad R_z = 0-z, \quad R_z = 1/2 + z,\quad R_z = 1/2-z
\ea 
Co atoms are located at the maximal Wyckoff positions $(2a)$. Ce atoms are located at the non-maximal Wyckoff positions with $z=0$. 
In general, we have three situations as shown in~\cref{tab:ebr_1d} , where we have also provided the corresponding irreducible representations at $k_z=0,1/2$. 
\begin{table}[!h]
\begin{center}
\begin{tabular}{c|c|c|c|c  }
  EBR & Wyckoff position  &  Orbital & Irreps at $k_z=0$ & Irreps at $k_z=1/2$  \\
   \hline 
 $EBR_{+}$ & $2a$  & $s\up @\frac{1}{4}$,  $s \dn @\frac{3}{4}$ &$i,-i$ & $i\lambda_{k_x,k_y},i\lambda_{k_x,k_y}$\\ 
  \hline 
  $EBR_{-}$  & $2a$   & $s\dn @\frac{1}{4}$,  $s \up @\frac{3}{4}$ & $i,-i$ & $-i\lambda_{k_x,k_y},-i\lambda_{k_x,k_y}$\\ 
  \hline 
$EBR_{+} \oplus EBR_{-}  $ & $4b$   &  \shortstack{$s\up @z$, $s\dn @(-z)$\\ $s \up  @(1/2+z)$, $s \dn  @(1/2-z)$} &
  $i,-i,i,-i$ & $i\lambda_{k_x,k_y},-i\lambda_{k_x,k_y},i\lambda_{k_x,k_y},-i\lambda_{k_x,k_y}$
\end{tabular}    
\caption{Symmetry properties of the effective 1D system. $s\sigma$ suggest the corresponding orbital has the same symmetry property as $s$ orbital with spin $\sigma$, and $\lambda_{k_x,k_y}=e^{i\pi(k_x+k_y)}$. }
\label{tab:ebr_1d}
\end{center}
\end{table} 

There are two types of elementary band representations (EBRs), denoted by $EBR_+,EBR_-$, induced by two different types of orbitals at $2a$ positions as shown in Tab.~\ref{tab:ebr_1d}. 
Electrons at non-maximal Wyckoff positions can be decomposed into $EBR_+ \oplus EBR_-$. 
We can therefore define the following quantities to characterize the ground states
\ba 
C = N_+ - N_-
\label{eq:app:sym_ind_1d}
\ea 
where 
\ba 
&2N_+ = \text{ number of filled bands with eigenvalues $i\lambda_{k_x,k_y}$ at $k_z=1/2$}\nonumber\\ 
&2N_- = \text{ number of filled bands with eigenvalues $-i\lambda_{k_x,k_y}$ at $k_z=1/2$} \, . 
\ea 
$N_{\pm}$ corresponds to the number of $EBR_\pm $. The additional factor $2$ comes from the $P\cdot\mathcal{T}$-protected two-fold degeneracy of each band. 

\subsection{Effective model of nodal line}
\label{app:sec:fd_model}
We now discuss the nodal-line and the related surface states. To simplify the problem, we take an effective $fd$-orbital model with only $d_{x^2-y^2}$-orbital at the Co atom and $f_{xyz}$-orbital at the Ce atom, where both orbitals appear near the Fermi energy, to understand the surface states. As we will show later, this simplified model reproduces the correct nodal line as we observed in our mean-field solutions (~\cref{fig:kondo_band_nodal_line}).  The advantage of this simplified model is that we can remove other bands that do not contribute to the nodal line. The full Hamiltonian takes the form of 
\ba 
&H = H_c +H_f +H_{fc}
\ea 
where $H_c,H_f$ are the hopping terms of $c$ and $f$ electrons respectively, and $H_{fc}$ denote the $fc$ hybridization term. We now discuss each term of the Hamiltonian. 

\begin{figure}
    \centering
    \includegraphics[width=0.8\textwidth]{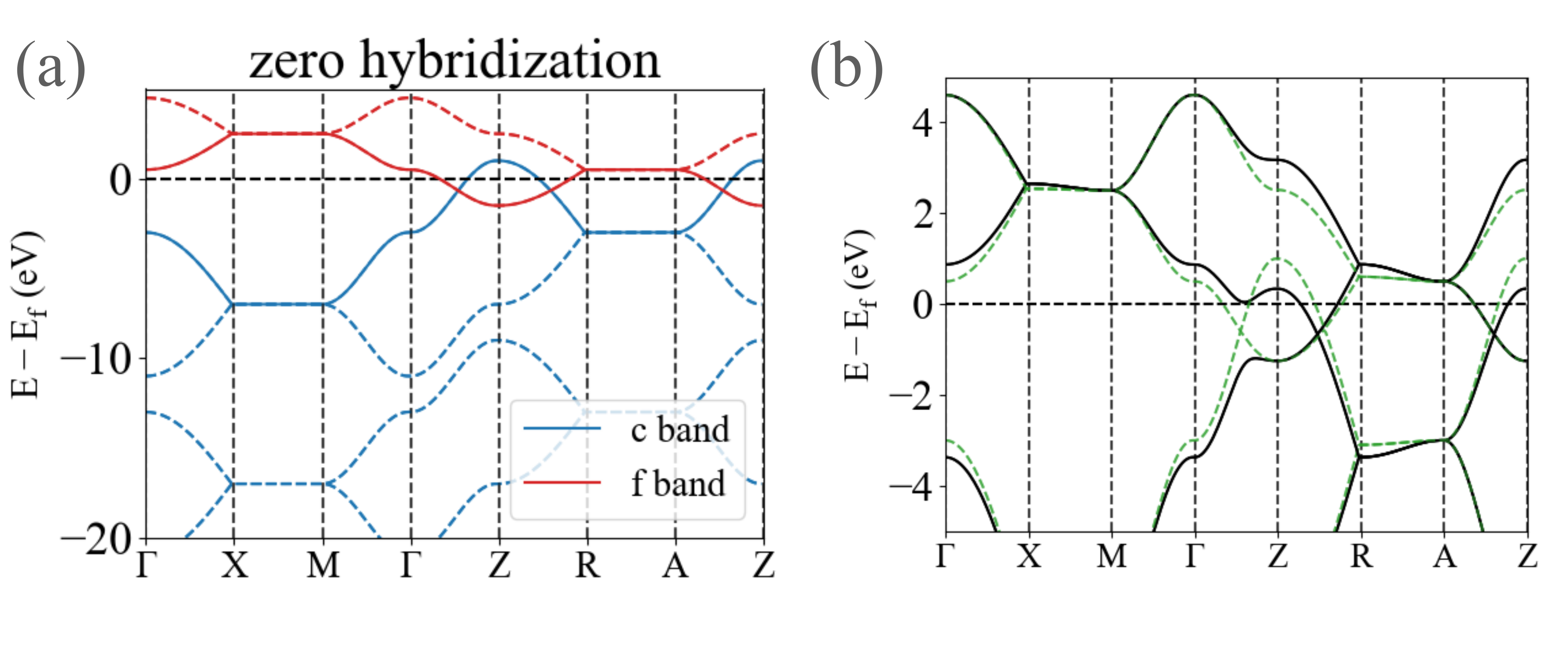}
    \caption{ (a) Band structures of the $fd$-orbital model in the limit of zero hybridization between $f$ and $d$ orbitals (\cref{eq:simp_model_c_block} and~\cref{eq:simp_model_f_block}). The parameters are $\epsilon_d =-10, M=5$, $\epsilon_f=1.5$ $t_{d,z}=t_d=-1, t_{f}=t_{f,z}=0.5$. (b) Zoom-in view of the band structures near the Fermi energy with finite hybridization (\cref{eq:simp_model_full}). Green dashed lines show the band structure with $C_{4z}$ symmetry where the system develops a Dirac node along $\Gamma$-$Z$ line and a nodal line at $k_z=1/2$ plane. Black solid lines show the band structure with a $C_{4z}$ symmetry-breaking term, where the Dirac node has been gapped out. The corresponding parameters are $-t_{fc,1}=t_{fc,C4}=0.3$. }
    \label{fig:simp_fd_model_zero_disp}
\end{figure}

The Hamiltonian of the $c$-block can be written as
\ba 
&H_{c} = \sum_{\kk ,ij,\sigma} \psi_{c,\kk,i,\sigma}^\dag \psi_{c,\kk,j,\sigma} h^c_{\kk, ij,\sigma}
,\quad 
\psi_{c,\kk,j,\sigma} = 
\begin{bmatrix}
    c_{\kk,Co_1,d_{x^2-y^2},\sigma} & 
     c_{\kk,Co_2,d_{x^2-y^2},\sigma}&
      c_{\kk,Co_3,d_{x^2-y^2},\sigma} & 
       c_{\kk,Co_4,d_{x^2-y^2},\sigma}
\end{bmatrix}_j^T
\nonumber\\ 
&h^c_{\kk,ij,\sigma} = \small{
\begin{bmatrix}
    \epsilon_d -M\sigma +2 t_{d,z}\cos(2\pi k_z) & 4t_d \cos(\frac{2\pi k_x}{2})\cos(\frac{2\pi k_y}{2})  \\ 
    4t_d \cos(\frac{2\pi k_x}{2})\cos(\frac{2\pi k_y}{2}) & \epsilon_d-M\sigma + 2t_{d,z}\cos(2\pi k_z) \\ 
  &&   \epsilon_d+M\sigma +2 t_{d,z}\cos(2\pi k_z) & 4t_d \cos(\frac{2\pi k_x}{2})\cos(\frac{2\pi k_y}{2})  \\ 
  &&  4t_d \cos(\frac{2\pi k_x }{2})\cos(\frac{2\pi k_y}{2}) & \epsilon_d+M\sigma + 2t_{d,z}\cos(2\pi k_z)  
\end{bmatrix}_{ij}}
\label{eq:simp_model_c_block}
\ea 
where $\epsilon_d$ denotes the on-site energy term, $t_d$ denotes the in-plane hopping, $t_{d,z}$ denotes the $z$-direction hopping and $M$ characterizes the splitting induced by the magnetic order. We can diagonalize the $h_{\kk,ij,\sigma}^c$ which gives the following eigenvalues and eigenstates
\ba 
&E_{c,\kk,1/2,\sigma} = \epsilon_d - M\sigma +2t_{d,z}\cos(2\pi k_z) \pm  4t_d\cos(2\pi k_x /2)\cos(2\pi k_y/2) \nonumber\\ 
&\gamma_{c,\kk,1/2,\sigma} = \frac{1}{\sqrt{2}}( c_{\kk,Co_1,d_{x^2-y^2},\sigma}  \pm   c_{\kk,Co_2,d_{x^2-y^2},\sigma} )\nonumber\\ 
&E_{\kk,3/4,\sigma} = \epsilon_d + M\sigma + 2t_{d,z}\cos(2\pi k_z) \pm 4t_d\cos(2\pi k_x /2)\cos(2\pi k_y/2)  \nonumber\\ 
&\gamma_{c,\kk,3/4,\sigma} = \frac{1}{\sqrt{2}}( c_{\kk,Co_3,d_{x^2-y^2},\sigma}  \pm   c_{\kk,Co_4,d_{x^2-y^2},\sigma}  )
\ea 
Without loss of generality, we take $M>0,t_d<0,t_{d,z}<0$. This leads to  $P\cdot\mathcal{T}$-protected two-fold degenerate states with energy $E_{c,\kk,4,\uparrow},E_{c,\kk,2,\downarrow}$ near Fermi energy. These states at $k_z=1/2$ have glide mirror-$z$ eigenvalues $e^{i\pi(k_x+k_y)}$. We show the band structures in \cref{fig:simp_fd_model_zero_disp} (a), where the two-fold degenerate bands described by $\gamma_{c,\kk,4,\up},\gamma_{c,\kk,2,\dn}$ are characterized by solid blue line. To simplify the problem, we could also drop $ c_{\kk,Co_1,d_{x^2-y^2},\up} ,c_{\kk,Co_2,d_{x^2-y^2},\up} ,c_{\kk,Co_3,d_{x^2-y^2},\dn} ,c_{\kk,Co_4,d_{x^2-y^2},\dn}$ electron. These electrons form high-energy bands near $E\sim -15$eV in \cref{fig:simp_fd_model_zero_disp} (a). We can also adiabatically shift the energy of these bands to negative infinity by gradually increasing the strength of magnetic ordering $M$. We therefore obtain the following new Hamiltonian for $c$ electrons
\ba 
&H'_{c} = \sum_{\kk ,ij,\sigma} \psi_{c,\kk,i,\sigma}^{'\dag} \psi'_{c,\kk,j,\sigma} h^{c'}_{\kk, ij,\sigma}\nonumber\\
&
\psi'_{c,\kk,j,\up} = 
\begin{bmatrix}
    c_{\kk,Co_1,d_{x^2-y^2},\dn} & 
     c_{\kk,Co_2,d_{x^2-y^2},\dn}&
      % c_{\kk,Co_3,d_{x^2-y^2},\sigma} & 
      %  c_{\kk,Co_4,d_{x^2-y^2},\sigma}
\end{bmatrix}_j^T,\quad 
\psi'_{c,\kk,j,\dn} = 
\begin{bmatrix}
    c_{\kk,Co_3,d_{x^2-y^2},\up} & 
     c_{\kk,Co_4,d_{x^2-y^2},\up}&
\end{bmatrix}_j^T
\nonumber\\ 
&h^{'c}_{\kk,ij,\sigma} = 
\begin{bmatrix}
    \epsilon_d +M +2 t_{d,z}\cos(2\pi k_z) & 4t_d \cos(2\pi k_x /2)\cos(2\pi k_y/2)  \\ 
    4t_d \cos(2\pi k_x /2)\cos(2\pi k_y/2) & \epsilon_d+M + 2t_{d,z}\cos(2\pi k_z) 
\end{bmatrix}_{ij}
\label{eq:simp_model_c_block_2}
\ea 

The Hamiltonian of $f$ block reads
\ba 
&H_f = \sum_{\kk,ij,\sigma} \psi_{f,\kk,i,\sigma}^\dag \psi_{f,\kk,j,\sigma} h_{\kk,ij,\sigma}^f ,\quad \psi_{f,\kk,j,\sigma} = \begin{bmatrix}
    f_{\kk,Ce_1, f_{xyz},\sigma} & f_{\kk,Ce_2, f_{xyz},\sigma} 
\end{bmatrix}_j^T \nonumber\\ 
&h_{\kk,ij,\sigma}^f = 
\begin{bmatrix}
    \epsilon_f + 2t_{f,z}\cos(2\pi k_z) & -4e^{i\sigma 2\pi k_z/2}t_f \cos(2\pi k_x /2)\cos(2\pi k_y/2) \\
    -4e^{-i\sigma 2\pi k_z/2}t_f \cos(2\pi k_x /2)\cos(2\pi k_y/2) & \epsilon_f +2t_{f,z}\cos(2\pi k_z) 
\end{bmatrix}_{ij}
\label{eq:simp_model_f_block}
\ea 
where $\epsilon_f$ is the on-site energy of $f$ electrons, $t_{f,z}$ and $t_f$ are symmetry-allowed hopping of $f$ electrons. We can diagonalize $H_{\kk,ij,\sigma}^f$. The corresponding eigenvalues and eigenvectors are 
\ba 
&E_{f,\kk,1/2,\sigma} = \epsilon_f + 2t_{f,z}\cos(2\pi k_z) \pm 4t_f \cos(2\pi k_x /2)\cos(2\pi k_y/2) \nonumber\\ 
&\gamma_{f,\kk,1/2,\sigma} = \frac{1}{\sqrt{2}}\left( \mp e^{-i2\pi k_z\sigma/2}
f_{\kk,Ce_1,f_{xyz},\sigma}  + f_{\kk,Ce_2,f_{xyz},\sigma}
\right)
\ea 
We take $t_f>0$ such that we have a two-fold degenerate state ($E_{f,\kk,2,\sigma}$) near the Fermi energy. These states also have glide mirror-$z$ eigenvalues $-e^{i\pi (k_x+k_y)}$ at $k_z=-1/2$. 
We show the band structures in~\cref{fig:simp_fd_model_zero_disp} (a), where the two-fold degenerate bands described by $\gamma_{f,\kk,2,\up},\gamma_{f,\kk,2,\dn}$ are characterized by the solid red line. Here, we comment that the parameters are fine-tuned aiming to reproduce the band structures of the the mean-field calculations (~\cref{app:sec:kondo_band}). In addition, we have also taken relatively large hopping parameters of $f$-electrons (large $t_{f,c},t_{f}$). The large $t_{f,z}$ will cause the $f$-bands at the $k_z=0$ plane to be shifted away from the Fermi energy. The large $t_f$ will cause the upper $f$ bands ($E_{f,\kk,1,\sigma}$, the red dashed line of~\cref{fig:simp_fd_model_zero_disp} (a)) at $k_z=1/2$ plane to be shifted away from the Fermi energy. 
This adjustment will not change the topological properties, such as nodal lines, of the band structure. However, it will simplify the analysis by shifting irrelevant bands away from the Fermi energy.

 We now summarize the band structure of the systems in the zero $fc$ hybridization limit (~\cref{fig:simp_fd_model_zero_disp} (a)). First, for the $c$ blocks, we have in total $8$ bands. Due to the magnetic ordering $M$, half of the bands will be shifted away from the Fermi energy and can be ignored. In addition, the interlayer hoping $t_d$ will split the remaining four bands into two two-fold-degenerate groups. One group has $\{M_z|\frac{1}{2},\frac{1}{2},\frac{1}{2}\}$ eigenvalue $e^{i\pi(k_x+k_y)}$ at $k_z=1/2$ plane and appear near the Fermi energy (blue solid line in~\cref{fig:simp_fd_model_zero_disp} (a)). The other group has $\{M_z|\frac{1}{2},\frac{1}{2},\frac{1}{2}\}$ eigenvalue $-e^{i\pi(k_x+k_y)}$ at $k_z=1/2$ plane and has been shifted away from the Fermi energy. 
 As for the $f$ block, the hopping term $t_f$ will split 4 $f$ bands into two two-fold-degenerate groups. One group has $\{M_z|\frac{1}{2},\frac{1}{2},\frac{1}{2}\}$ eigenvalue $e^{i\pi(k_x+k_y)}$ at $k_z=\pi$ plane and appear near the Fermi energy (red solid line in~\cref{fig:simp_fd_model_zero_disp} (a)). The other group has $\{M_z|\frac{1}{2},\frac{1}{2},\frac{1}{2}\}$ eigenvalue  $-e^{i\pi(k_x+k_y)}$ at $k_z=\pi$ plane and has been shifted away from the Fermi energy by the $t_f$ term. 
 Therefore, in the zero-hybridization limit, $c$ electrons create a two-fold degenerate band near Fermi energy with $\{M_z|\frac{1}{2},\frac{1}{2},\frac{1}{2}\}$ eigenvalue $e^{i\pi(k_x+k_y)}$ at $k_z=1/2$ plane, and $f$ electrons create a two-fold degenerate band near Fermi energy with $\{M_z|\frac{1}{2},\frac{1}{2},\frac{1}{2}\}$ eigenvalue $-e^{i\pi(k_x+k_y)}$ at $k_z=1/2$ plane. 
These two bands possess different $\{M_z|\frac{1}{2},\frac{1}{2},\frac{1}{2}\}$ eigenvalues, resulting in a nodal line at the $k_z=1/2$ plane. This nodal line persists even after the introduction of $f$$c$ hybridization due to the glide-mirror-$z$ symmetry. Finally, we comment that the nodal-line here differs from the hourglass fermion~\cite{Wang2016}.

We now introduce the $fc$ hybridization term. 
We consider the following symmetry-allowed $fc$ hybridization term
% \ba 
% &H_{fc} = \sum_{\kk,ij,\sigma} \psi_{c,\kk,i,\sigma}^\dag \psi_{f,\kk,j,\sigma}  h^{cf}_{\kk,ij,\sigma} +\text{h.c.}\nonumber\\ 
% &h^{cf}_{\kk,ij,\up} = 
% \begin{bmatrix}
%     -2t_{fc,2}e^{-\frac{i}{4}k_z} \cos(k_y/2) &2t_{fc,2}e^{\frac{i}{4}k_z}\cos(k_x/2) \\ 
%     -2t_{fc,2}e^{-i\frac{k_z}{4}}\cos(k_x/2) 
%     & 2t_{fc,2}e^{i\frac{k_z}{4}}\cos(k_y/2) \\ 
%     -2e^{i\frac{k_z}{4}}(t_{fc,1}+t_{fc,C4})\cos(k_x/2) &
%     -2e^{-i\frac{k_z}{4}}(-t_{fc,1}+t_{fc,C4})\cos(k_y/2) \\
%     2(t_{fc,C4}-t_{fc,1})e^{i\frac{k_z}{4}}\cos(k_y/2) & 2(t_{fc,C4}+t_{fc,1})e^{-i\frac{k_z}{4}}\cos(k_x/2) 
% \end{bmatrix}_{ij} \nonumber\\ 
% &h^{fc}_{\kk,ij,\dn} = 
% \begin{bmatrix}
%     -2(t_{fc,C4}-t_{fc,1})e^{-\frac{i}{4}k_z} \cos(k_y/2) &-2(t_{fc,C4}+t_{fc,1})e^{\frac{i}{4}k_z}\cos(k_x/2) \\ 
%     2t(t_{fc,C4}+t_{fc,1})e^{-i\frac{k_z}{4}}\cos(k_x/2) 
%     & 2(t_{fc,C4}-t_{fc,1})e^{i\frac{k_z}{4}}\cos(k_y/2) \\ 
%     2e^{i\frac{k_z}{4}}t_{fc,2}\cos(k_x/2) &
%     -2e^{-i\frac{k_z}{4}}t_{fc,2}\cos(k_y/2) \\
%     2t_{fc,2}e^{i\frac{k_z}{4}}\cos(k_y/2) & -2t_{fc,2}e^{-i\frac{k_z}{4}}\cos(k_x/2) 
% \end{bmatrix}_{ij} 
% \ea 
\ba 
&H_{fc} = \sum_{\kk,ij,\sigma} \psi_{c,\kk,i,\sigma}^{'\dag} \psi_{f,\kk,j,\sigma}  h^{cf}_{\kk,ij,\sigma} +\text{h.c.}\nonumber\\ 
&h^{cf}_{\kk,ij,\up} = 
\begin{bmatrix}
    -2e^{i\frac{k_z}{4}}(t_{fc,1}+t_{fc,C4})\cos(k_x/2) &
    -2e^{-i\frac{k_z}{4}}(-t_{fc,1}+t_{fc,C4})\cos(k_y/2) \\
    2(t_{fc,C4}-t_{fc,1})e^{i\frac{k_z}{4}}\cos(k_y/2) & 2(t_{fc,C4}+t_{fc,1})e^{-i\frac{k_z}{4}}\cos(k_x/2) 
\end{bmatrix}_{ij} \nonumber\\ 
&h^{fc}_{\kk,ij,\dn} = 
\begin{bmatrix}
    -2(t_{fc,C4}-t_{fc,1})e^{-\frac{i}{4}k_z} \cos(k_y/2) &-2(t_{fc,C4}+t_{fc,1})e^{\frac{i}{4}k_z}\cos(k_x/2) \\ 
    2t(t_{fc,C4}+t_{fc,1})e^{-i\frac{k_z}{4}}\cos(k_x/2) 
    & 2(t_{fc,C4}-t_{fc,1})e^{i\frac{k_z}{4}}\cos(k_y/2) 
\end{bmatrix}_{ij} 
\ea 
where we have introduced an additional term ($t_{fc,C4}$) that breaks $C_{4z}$ symmetry. As we show later, this additional term could gap out the Dirac crossing along $\Gamma$-$Z$ line, but will not affect the formation of the nodal line. This allows us to avoid the effect of the Dirac node when we study the surface states of the nodal line.

Therefore, the full Hamiltonian can be defined as 
\ba 
&H = H_c' +H_f +H_{fc} = \sum_{\kk,ij,\sigma}h_{\kk,ij,\sigma} \psi_{\kk,i,\sigma}^\dag \psi_{\kk,j,\sigma} \nonumber\\ 
&\psi_{\kk,i,\up} = \begin{bmatrix}
      c_{\kk,Co_3,d_{x^2-y^2},\up} & 
       c_{\kk,Co_4,d_{x^2-y^2},\up} & 
         f_{\kk,Ce_1, f_{xyz},\up} & f_{\kk,Ce_2, f_{xyz},\up} 
\end{bmatrix}_i^T \nonumber\\
&
\psi_{\kk,i,\dn} = \begin{bmatrix}
      c_{\kk,Co_1,d_{x^2-y^2},\dn} & 
       c_{\kk,Co_2,d_{x^2-y^2},\dn} & 
         f_{\kk,Ce_1, f_{xyz},\dn} & f_{\kk,Ce_2, f_{xyz},\dn} 
\end{bmatrix}_i^T 
\nonumber\\ 
&h_{\kk,ij,\sigma} =
\begin{bmatrix}
    h_{\kk,\sigma}^{'c} & h^{cf}_{\kk,\sigma} \\
    h^{fc,\dag}_{\kk,\sigma} & h^{f}_{\kk,\sigma}
\end{bmatrix}
\label{eq:simp_model_full}
\ea 
The band structures are shown in Fig.~\ref{fig:simp_fd_model_zero_disp} (b), where we can observe the formation of nodal lines at $k_z=1/2$ plane. Without $C_{4z}$ symmetry breaking, the system also develops a Dirac node along $\Gamma-Z$ line. To simplify the analysis of the surface states, we consider a non-zero $C_{4z}$ symmetry-breaking term to gap out this Dirac node.

\subsection{Surface states}
We now discuss the surface states of the $fd$ model we introduced in~\cref{eq:simp_model_full}. For given $k_x,k_y$ points, we can treat the system as an effective 1D system with the hopping matrix 
\ba 
h_{k_z,i\sigma,j\sigma'}^{1D, (k_x,k_y)} = h_{\kk,i\sigma,j\sigma'}
\ea 

We now calculate the quantities we introduced in~\cref{eq:app:sym_ind_1d}. 
For the $k_x,k_y$ outside the nodal line (near $M$ point), we have $C=0$ ($N_+=1,N_-=1$). 
The ground state of the effective 1D system is adiabatically connected to the atomic insulators where all the $c$ orbitals are filled. Therefore, we do not expect any non-trivial surface state for the momentum points outside the nodal line.

Inside the nodal line, we have $C=-2$ with $N_+ = 0$, and $N_-=2$. 
% Even though we have atoms at the maximal Wyckoff positions, we could still develop the ground state may still correspond to an obstructed atomic insulator (OAI). 
Since $N_-=2$, the filled bands can be decomposed into 
\ba 
\text{filled bands } \sim  EBR_- \oplus EBR_- 
\ea 
However, the EBR induced by $\psi_{c,\kk}'$ electrons are
\ba 
\psi_{c,\kk}' \sim EBR_- \oplus EBR_+
\ea 
The EBR induced by $\psi_{f,\kk}$ electrons are (see Tab.~\ref{tab:ebr_1d})
\ba 
\psi_{f,\kk} \sim EBR_- \oplus EBR_+
\ea 
Therefore, one of the $EBR_-$ formed by the filled bands must come from the $f$ orbitals. $f$-orbital is located at non-maximal Wyckoff positions $4b$ and $EBR_-$ corresponds to the Wannier states at maximal Wayckoff positions $2a$. Therefore, we could conclude that $f$ orbitals develop molecular orbitals located at positions distinct from the Ce atoms.

\begin{figure}
    \centering
    \includegraphics[width=0.7\textwidth]{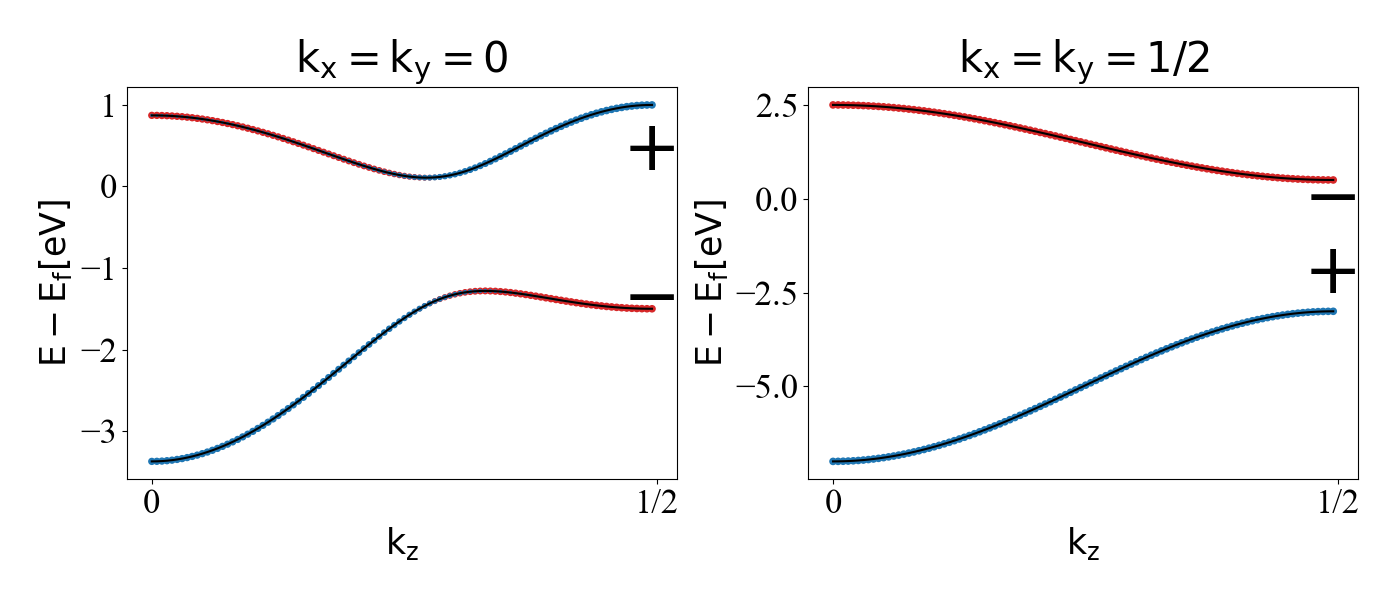}
    \caption{Dispersions of the effective 1D model in~\cref{eq:simp_model_full} at $k_x=k_y=0$ (inside the nodal line) and $k_x=k_y=1/2$ (outside the nodal line). Each band is two-fold degenerate due to $P\cdot\mathcal{T}$ symmetry. Red and blue mark the orbital weights of $f$ and $c$ electrons respectively. $+,-$ label the $\{M_z|\frac{1}{2},\frac{1}{2},\frac{1}{2}\}$ eigenvalues of the bands at $k_z=1/2$.  }
    \label{fig:simp_model_1d_disp}
\end{figure}

The formation of the molecular orbitals for the $k_x,k_y$ inside the nodal line could then introduce surface states. 
To gain a better understanding of the surface state, we investigate the low-energy effective Hamiltonian of the 1D system. The relevant low-energy degrees of freedom (electrons that appear near Fermi energy) are $\gamma_{c,\kk,4,\up},\gamma_{c,\kk,2,\dn}$ (solid blue lines in Fig.~\ref{fig:simp_fd_model_zero_disp} (a)), and $\gamma_{f,\kk,2,\sigma}$ (solid red lines in Fig.~\ref{fig:simp_fd_model_zero_disp} (a)). 
Transforming to the real space, we find 
\ba 
&\gamma_{c,(k_x,k_y,R_z),4,\up} 
= \frac{1}{\sqrt{2}}\left(
c_{(k_x,k_y,R_z),Co_3,d_{x^2-y^2},\up} 
-c_{(k_x,k_y,R_z),Co_4,d_{x^2-y^2},\up}  \right) \nonumber\\
&\gamma_{f,(k_x,k_y,R_z),2,\up} 
= \frac{1}{\sqrt{2}}\left(  f_{(k_x,k_y,R_z),Ce_1, f_{xyz},\up}+  f_{(k_x,k_y,R_z),Ce_2, f_{xyz},\up}\right) \nonumber\\ 
&
\gamma_{c,(k_x,k_y,R_z),2,\dn} 
\frac{1}{\sqrt{2}}\left(c_{(k_x,k_y,R_z),Co_1,d_{x^2-y^2},\dn} -c_{(k_x,k_y,R_z),Co_2,d_{x^2-y^2},\dn}  \right) \nonumber\\
&\gamma_{f,(k_x,k_y,R_z),2,\dn} 
\frac{1}{\sqrt{2}}\left(  f_{(k_x,k_y,R_z),Ce_1, f_{xyz},\dn}+f_{(k_x,k_y,R_z-1),Ce_2, f_{xyz},\dn} \right)
\label{eq:real_space_def_gam_fc}
\ea 
For the effective 1D system, $\gamma^\dag_{c,(k_x,k_y,R_z),4,\up}$ creates an electron at $R_z=3/4$, since it is formed by the $d$ orbitals of Co$_3$,Co$_4$. $\gamma_{f,(k_x,k_y,R_z),2,\up}$ creates an electron at $1/4$ since it corresponds to the bonding state of $f_{(k_x,k_y,R_z),Ce_1, f_{xyz},\up}$ 
and $f_{(k_x,k_y,R_z),Ce_2, f_{xyz},\up}$.  
$\gamma^\dag_{c,(k_x,k_y,R_z),2,\up}$ creates an electron at $R_z=1/4$, since it is formed by the $d$ orbitals of Co$_1$,Co$_2$. $\gamma_{f,(k_x,k_y,R_z)2,\dn}$ creates an electron at $3/4$ since it corresponds to the bonding state of $f_{(k_x,k_y,R_z),Ce_1, f_{xyz},\dn}$ 
and $f_{(k_x,k_y,R_z-1),Ce_2, f_{xyz},\dn}$. 
The Wannier center of the electron operator can also be obtained by calculating the corresponding Berry phase. We can already observe that $\gamma_{f,(k_x,k_y,R_z),2,\sigma}$ denotes the molecular orbital formed by $f$ electrons. 
Furthermore, we show the band structures and also the orbital weights for the effective 1D system in~\cref{fig:simp_model_1d_disp}. 
We find that, outside the nodal line, $f$ orbitals are fully empty and thus are irrelevant. This indicates the absence of the surface states.

\begin{figure}
    \centering
    \includegraphics[width=0.7\textwidth]{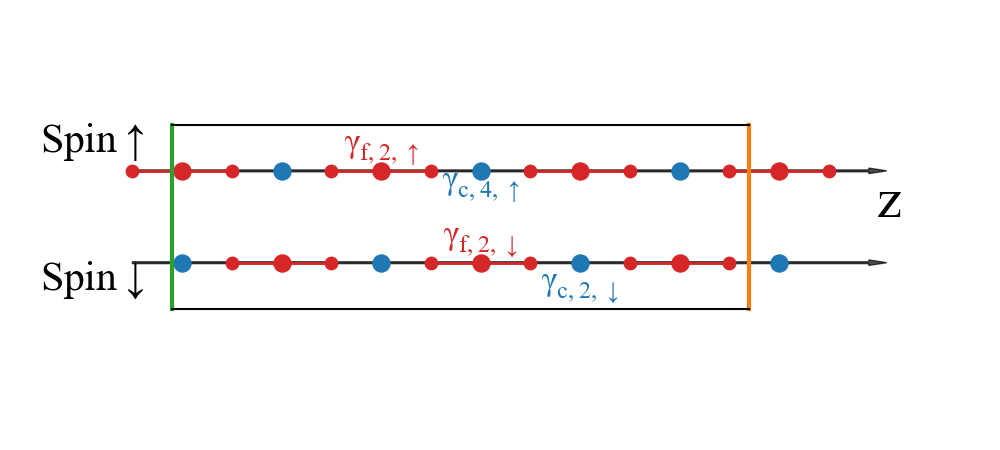}
    \caption{Position of the electron orbitals for the effective 1D system for spin $\up$ and spin $\dn$ sector.  $\gamma_{f,(k_x,k_y,R_z),2,\up/dn}$ is located at $n \pm 1/4$ with $n\in\mathbb{Z}$ (big red dots). $\gamma_{c,(k_x,k_y,R_z),4,\up}$ ($\gamma_{c,(k_x,k_y,R_z),2,\dn}$) is located at $n+3/4$ ($n+1/4$), marked by blue dots in the upper (lower) axis. However, from~\cref{eq:real_space_def_gam_fc}, $\gamma_{f,(k_x,k_y,R_z),2,\sigma}$ is a molecular orbital. $\gamma_{f,(k_x,k_y,R_z),2,\up}$ is formed by the superposition of $f_{(k_x,k_y,R_z),Ce_1, f_{xyz},\up}$ at $n+0$ and $f_{(k_x,k_y,R_z),Ce_2, f_{xyz},\up}$ at $n+1/2$ (small red dots in the upper axis).  $\gamma_{f,(k_x,k_y,R_z),2,\dn}$ is formed by the superposition of $f_{(k_x,k_y,R_z-1),Ce_2, f_{xyz},\dn}$ at $n-1/2$ and $f_{(k_x,k_y,R_z),Ce_1, f_{xyz},\dn}$ at $n$ (small red dots in the lower axis). The solid line represents the slab geometry along $z$ direction, where green denotes the first type of boundary (Boundary I, left) and orange denotes the second type of boundary (Boundary II, right). 
    }
    \label{fig:1d_boundary_up}
\end{figure}

We now discuss the surface states. Analyzing the 1D band structures shown in ~\cref{fig:simp_model_1d_disp} (left panel), we observe a band inversion between the $f$-molecular orbital and the $c$ orbital. This inversion suggests that the occupied band exhibits orbital contributions from both $f$ and $c$ orbitals. Consequently, a surface state could potentially emerge if the boundary cut through either the molecular $f$ orbital or the bond formed between the $f$ and $c$ orbitals.

We next consider the open boundaries as illustrated in~\cref{fig:1d_boundary_up}, where the real-space positions of the relevant orbitals are also marked. Generally, there are two types of open boundaries, each distinguished by different colors in~\cref{fig:1d_boundary_up}. Both types of boundaries cut through the molecular orbitals ($\gamma_{f,(k_x,k_y,R_z),2,\up}$) of $f$ electrons in the spin $\up$ sector. For the spin $\dn$ sector, both boundary types slice through the bonds linking the $f$ and $c$ orbitals. 

Furthermore, we could consider a boundary that cuts through the molecular $f$ orbitals in the spin $\dn$ sector and the $f$-$c$ bonds in the spin $\up$ sector. However, these boundary configurations can be derived by applying a $P\cdot\mathcal{T}$ transformation to our current boundaries, rendering them equivalent.

\begin{figure}
    \centering
    \includegraphics[width=0.7\textwidth]{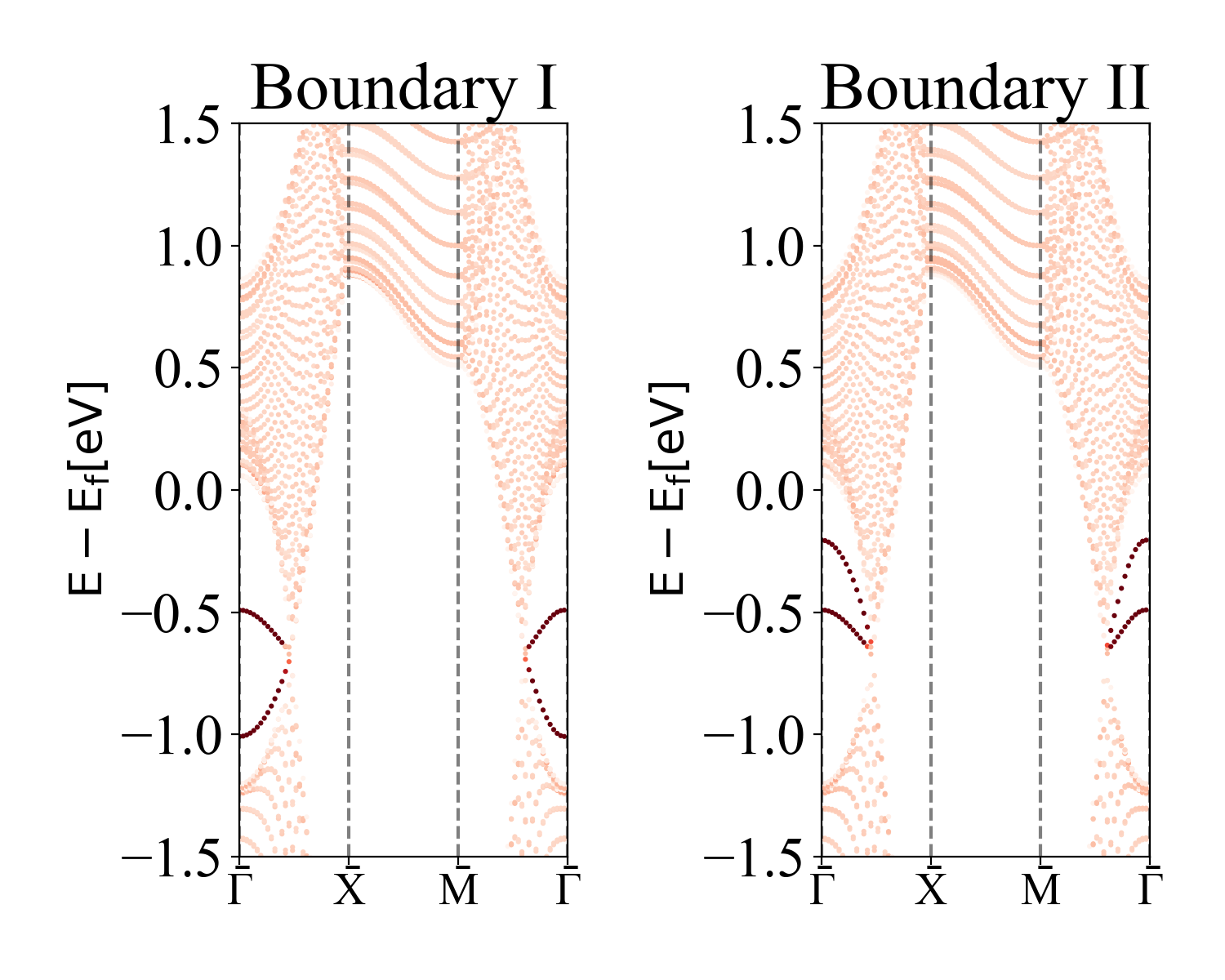}
    \caption{Surface bands calculated via Hamiltonian given by~\cref{eq:simp_model_full} with open boundary along $z$ directions. Two types of boundaries are illustrated in Fig.~\ref{fig:1d_boundary_up} by solid and dashed lines. The color denotes the weights of the bands on the surfaces. We can observe two surface states inside the nodal lines. }
    \label{fig:surface_state}
\end{figure}

\begin{figure}
    \centering
    \includegraphics[width=0.7\textwidth]{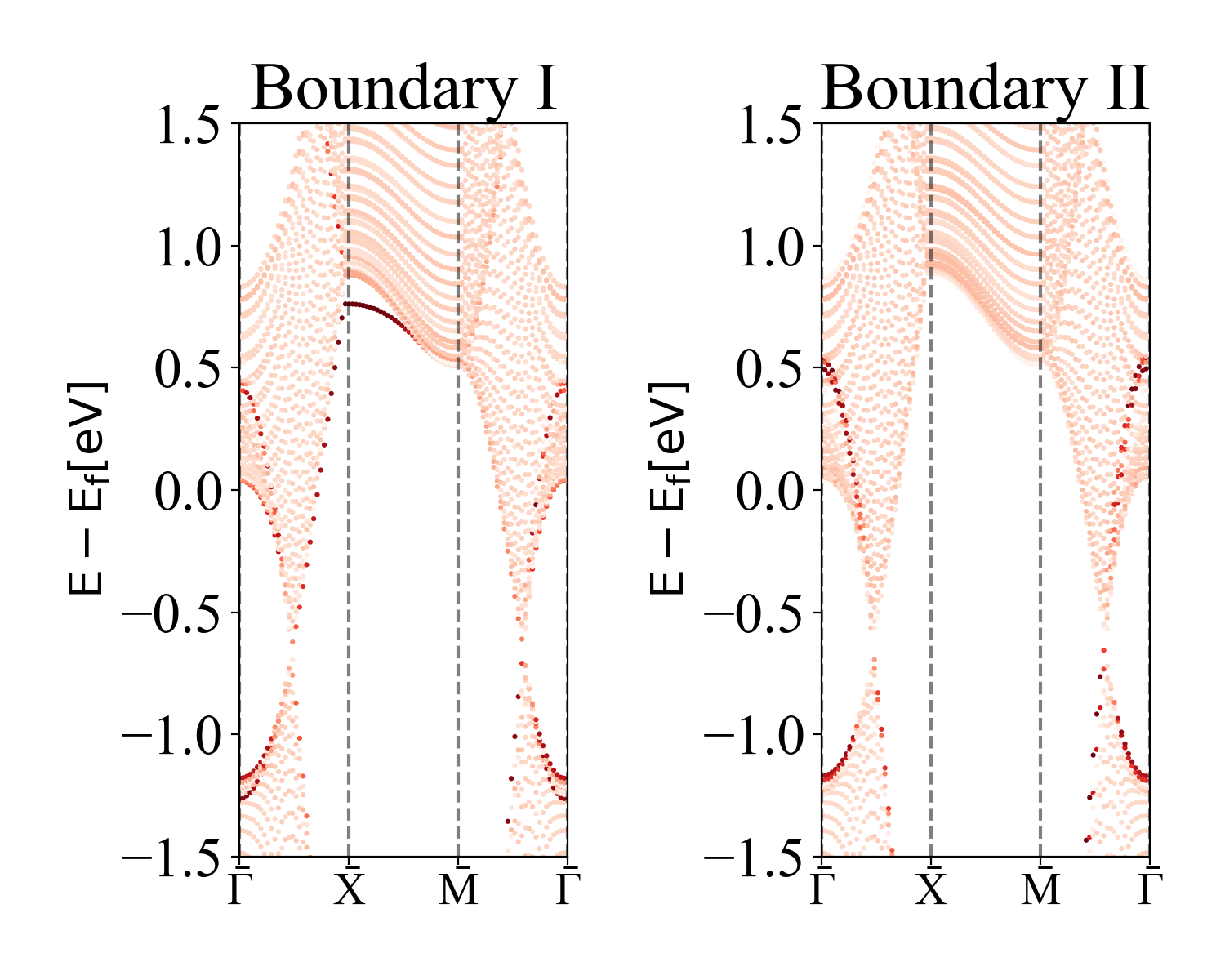}
    \caption{Surface bands with open boundary along $z$ directions. Additional boundary terms defined in ~\cref{eq:boundary_term_I,eq:boundary_term_II,eq:boundary_term_III,eq:boundary_term_IV} have been introduced with $m_1=1.5,m_2=1.3,m_3=-1.0,m_4=0.4$. We can observe the surface states merge into the bulk spectrum.}
    \label{fig:surface_state_boundary_term}
\end{figure}

In conclusion, we expect two surface states to emerge (each spin sector gives one surface state). This expectation has been confirmed by numerical calculations (\cref{fig:surface_state}) using the Hamiltonian given by~\cref{eq:simp_model_full}.  
To confirm our previous analysis, we now show that we can merge the surface states to the bulk spectrum by adding additional boundary terms. 

We first consider the spin $\up$ sector. Boundary I (green boundary in~\cref{fig:1d_boundary_up}) cuts through the $f$ molecular orbitals, and then we have an $f$ operator located at the boundary 
$
f_{(k_x,k_y,0),Ce_2, f_{xyz},\up}
$. 
This $f$ operator could still hybridize with $\gamma_{c,(k_x,k_y,R_z=0),4,\up}$ operator nearby which leads to a surface state. Therefore, we can consider the following boundary term
\ba 
m_1 (
f_{(k_x,k_y,0),Ce_2, f_{xyz},\up}^
\dag 
f_{(k_x,k_y,0),Ce_2, f_{xyz},\up}
  + 
  c_{(k_x,k_y,0),Co_4, d{x^2-y^2},\up}^\dag 
   c_{(k_x,k_y,0),Co_4, d{x^2-y^2},\up})
\label{eq:boundary_term_I}
\ea 
to shift the energy of the corresponding surface states. 

Boundary II (orange boundary~\cref{fig:1d_boundary_up}) cuts through the $f$ molecular molecular in the spin $\up$ sector, and then we have an $f$ operator located at the boundary $
f_{(k_x,k_y,R_n),Ce_1, f_{xyz},\up }
$ 
(where $R_n$ denotes the corresponding unit cell at the boundary). It will hybridize with the nearby $c$ electron $c_{(k_x,k_y,R_n-1),Co_4, d{x^2-y^2},\up}$ and creates a surface state. 
We can consider the following boundary term 
\ba 
m_2 (c_{(k_x,k_y,R_n-1),Co_4, d{x^2-y^2},\up}^\dag c_{(k_x,k_y,R_n-1),Co_4, d{x^2-y^2},\up} + f_{(k_x,k_y,R_n),Ce_1, f_{xyz},\up }^\dag f_{(k_x,k_y,R_n),Ce_1, f_{xyz},\up })
\label{eq:boundary_term_II}
\ea 
to shift the energy of the corresponding surface states.

We next consider the spin $\dn$ sector. Boundary I cuts through the bonds formed by $f_{(k_x,k_y,-1),Ce_2, f_{xyz},\dn }$ and $c_{(k_x,k_y,0),Co_2, d{x^2-y^2},\dn}$. Then we have a $c_{(k_x,k_y,0),Co_2, d{x^2-y^2},\dn}$ operator located at the boundary which gives a surface state. 
We can consider the following boundary term 
\ba 
m_3 (c_{(k_x,k_y,0),Co_2, d{x^2-y^2},\dn}^\dag 
c_{(k_x,k_y,0),Co_2, d{x^2-y^2},\dn})
\label{eq:boundary_term_III}
\ea 
to shift the energy of the corresponding surface states. 

Boundary II cuts through the bonds formed by $f_{(k_x,k_y,R_n),Ce_2, f_{xyz},\dn }$ and $c_{(k_x,k_y,R_n+1),Co_2, d{x^2-y^2},\dn}$ in the spin $\dn$ sector. Then we have $f_{(k_x,k_y,R_n),Ce_2, f_{xyz},\dn }$ operator located at the boundary which gives a surface state. 
We can consider the following boundary term 
\ba 
m_4 (f_{(k_x,k_y,R_n),Ce_2, f_{xyz},\dn }^\dag f_{(k_x,k_y,R_n),Ce_2, f_{xyz},\dn })
\label{eq:boundary_term_IV}
\ea 
to shift the energy of the corresponding surface states. 

In~\cref{fig:surface_state_boundary_term}, we show the surface states with boundary terms defined in~\cref{eq:boundary_term_I,eq:boundary_term_II,eq:boundary_term_III,eq:boundary_term_IV}. We can observe the additional boundary terms have shifted the energy of the surface states and the surface states have merged into the bulk spectrum. 
% These also indicate that surface states
% are not topologically protected in-gap states. 

\begin{figure}
    \centering
    \includegraphics[width=0.8\textwidth]{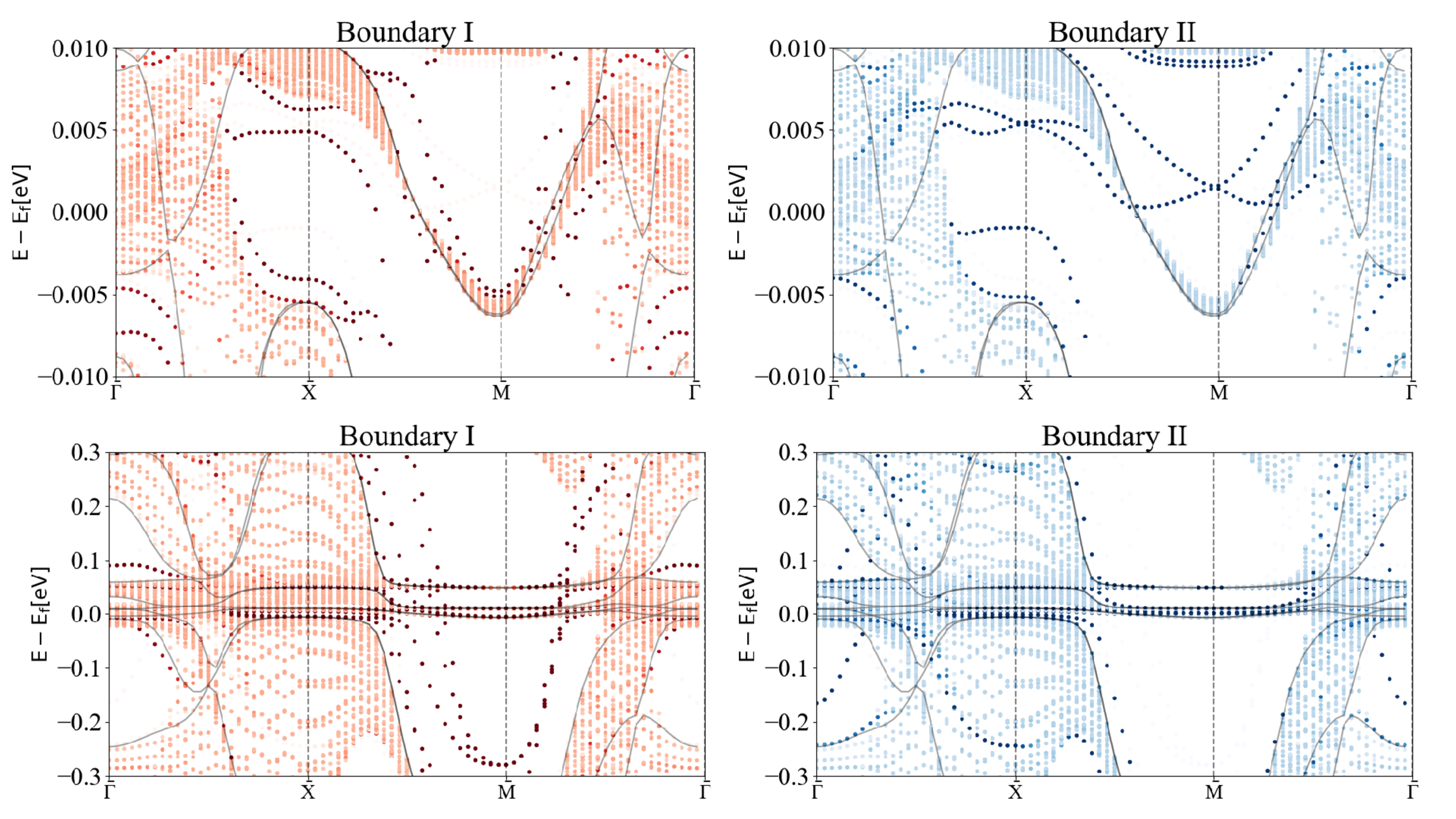}
    \caption{Surface states calculation with open boundary along $z$ directions. Bands marked by dark red and dark blue denote the bands with large weights on the surface. The upper and lower panels represent the same surface spectrum, each displayed within distinct energy windows. In the bottom figures, we can observe several narrow surface bands that are related to the narrow bandwidth of the $f$ electrons.  }
    \label{fig:realistic_model_surface}
\end{figure}

Finally, we discuss the surface states of the realistic model obtained from the mean-field calculations (\cref{app:sec:kondo_band}).  
For the realistic model, there are more than one $d$ orbitals per Co, more than one $f$ orbitals per Ce and we also have $p$ orbitals from P atoms. Therefore, we cannot use $N_+,N_-$ to conclude whether $f$ forms obstructed orbitals or not. We take an open boundary along $z$ directions and plot the surface bands in ~\cref{fig:realistic_model_surface}. We observe multiple surface states emerge. Here, we provide several possible origins of the surface states. 

\begin{figure}
    \centering
    \includegraphics[width=0.8\textwidth]{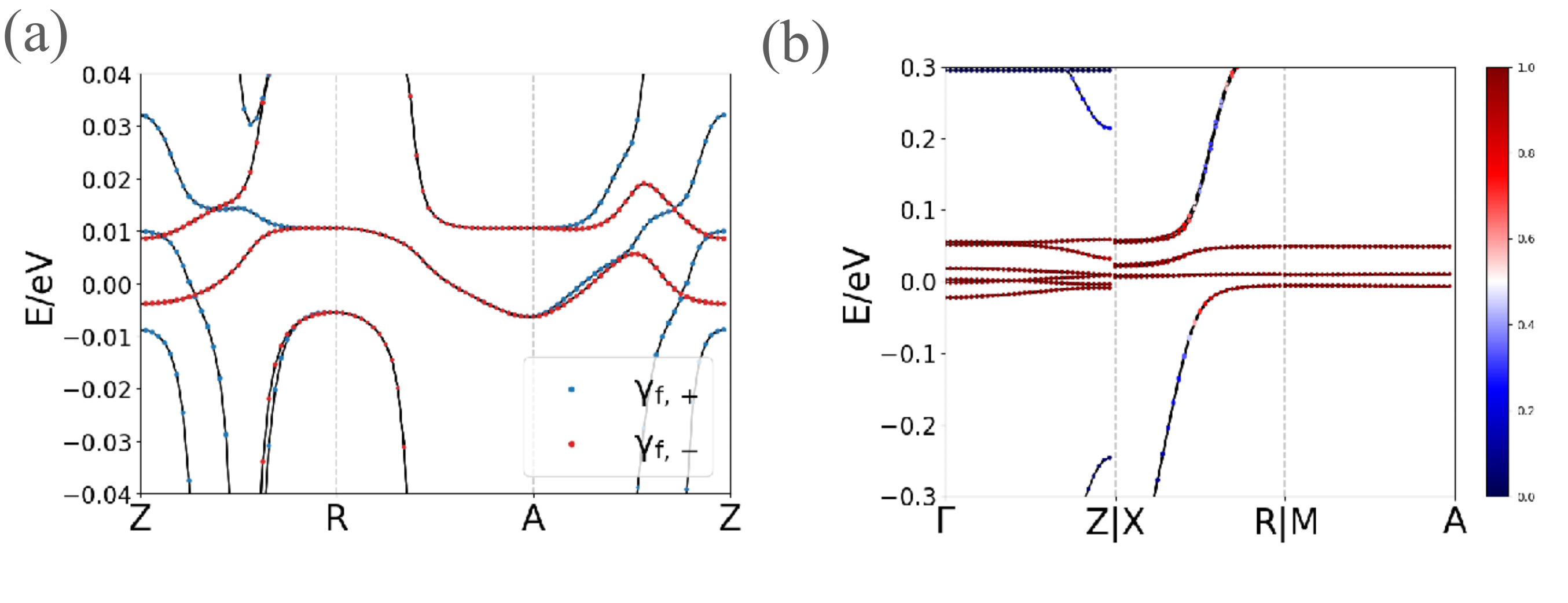}
    \caption{Band structures of the Kondo phase obtained from mean-field calculations (~\cref{app:sec:kondo_band}). (a) Red and blue denote the orbital weights of two molecular orbitals of $f$ electrons(~\cref{eq:mol_orb_f}).  (b) Red and blue denote the orbital weights of $f$ and $c$ electrons respectively, where we can observe a band inversion happens along $X$-$R$.}
    \label{fig:real_model_disp}
\end{figure}

From the previous simple model, we observe that there are two mechanisms that can generate surface states
\begin{itemize}
    \item $f$ orbitals form molecular orbitals. 
    \item Band inversion between $f$ electrons and $c$ electrons. 
\end{itemize} 
These two mechanisms may also produce surface states in the realistic model. However, the details of the surface states are different since the simple model does not include all the orbitals.

We first show the obstructed nature of $f$ bands in the realistic model. We introduce the following molecular orbitals (for the effective 1D system) in a similar manner in~\cref{eq:real_space_def_gam_fc}
\ba 
&\gamma_{f,\pm,\alpha, (k_x,k_y,R_z),\up} =\frac{1}{\sqrt{2}}\left( \mp s_\alpha f_{(k_x,k_y,R_z),Ce_1,\alpha,\up}+f_{(k_x,k_y,R_z),Ce_2,\alpha,\up}\right)\nonumber\\
&\gamma_{f,\pm,\alpha, (k_x,k_y,R_z),\dn} =\frac{1}{\sqrt{2}}\left( \pm s_\alpha f_{(k_x,k_y,R_z),Ce_1,\alpha,\dn}+f_{(k_x,k_y,R_z-1),Ce_2,\alpha,\dn}\right)
\label{eq:mol_orb_f}
\ea 
where 
\ba 
& \alpha = 
f_{z^3}, f_{xz^2},f_{yz^2}, f_{z(x^2-y^2)}, f_{xyz}, f_{x(x^2-3y^2)}, f_{y(3x^2-y^2)}\nonumber\\ 
&s_{f_{z^3}, f_{xz^2},f_{yz^2}, f_{z(x^2-y^2)}, f_{xyz}, f_{x(x^2-3y^2)}, f_{y(3x^2-y^2)}} = -1,1,1,-1,-1,1,1 
\ea 
For the effective 1D system $\gamma_{f, \pm, \alpha,(k_x,k_y,R_z),\sigma} $ is equivalent to an $s$-orbital with spin $\mp \sigma$ at $R_z=1/4$, and $\gamma_{f, \pm, \alpha,(k_x,k_y,R_z),\sigma} $ is equivalent to an $s$-orbital with spin $\pm\sigma$ at $R_z=3/4$.  
% \$gamma_{f, \pm, \alpha,(k_x,k_y,R_z),\up} $ and $\gamma_{f, \pm, \alpha,(k_x,k_y,R_z),\up} $ are connected by $P\cdot \mathcal{T}$ symmetry. 
If there is a large enough energy splitting between $\gamma_{f,+,\alpha,(k_x,k_y,R_z),\sigma}$ and $\gamma_{f,-,\alpha,(k_x,k_y,R_z),\sigma}$ states, we could conclude there is a formation of molecular orbitals. 

In~\cref{fig:real_model_disp} (a), we plot the orbital weights of $\gamma_{f,+, \alpha,(k_x,k_y,R_z),\sigma}$ and $\gamma_{f,-,\alpha,(k_x,k_y,R_z),\sigma}$.
We can observe the splitting between $\gamma_{f,+, \alpha,(k_x,k_y,R_z),\sigma}$ bands and $\gamma_{f,-,\alpha,(k_x,k_y,R_z),\sigma}$ bands near $k_x=k_y=0$, which indicates the $f$-electrons tend to form molecular orbitals in the ground state. However, the corresponding surface states may merge into the bulk spectrum. 

Secondly, as we show in~\cref{fig:real_model_disp} (b), a band inversion between $f$ and $c$ electrons also appears in the realistic model. However, in the realistic model, the band inversion appears at $k_x=0,k_y=1/2$ ($X$-$R$), but is absent at $k_x=k_y=0$. This is because our simple model overestimates the $z$-direction hopping of $d_{x^2-y^2}$ orbitals which creates the band inversion at $k_x=k_y=0$ in the simple model, and has ignored the $d_{xz},d_{yz}$ orbitals which creates the band inversion at $k_x=0,k_y=1/2$ in the realistic model. 

Finally, there are also strong bonds formed by $p_z$ orbitals of P atoms (as we have also discussed in~\cref{sec:app:LCP}, see also~\cref{fig:pz_orb_ce_la}). Cutting $p_z$ bonds could also create surface states. 

In conclusion, our study demonstrates that various mechanisms can create surface states in this system. However, it is important to note that the existence of these surface states is not protected by symmetries, allowing them to potentially merge into the bulk spectrum. 
Moreover, it is worth mentioning that, due to the narrow bandwidth and correlation effects of $f$-electron systems, DFT calculations may not accurately capture all the details of the band structures. 
Despite the complexity of real materials, our research outlines several potential mechanisms for the generation of surface states observed in experiments, and shows there is not a topological reason for their appearances.

\section{Schrieffer-Wolff transformation}
\label{sec:app:sw_transf}
In this section, we discuss the Schrieffer-Wolff (SW) transformation~\cite{PhysRev.149.491, coleman2015introduction} of generic many-body systems. 
In Sec.~\ref{sec:app:klm}, we use the Schrieffer-Wolff transformation to obtain the Kondo lattice model from the periodic Anderson model. In Sec.~\ref{sec:app:flat_band_magnetism}, we also use the Schrieffer-Wolff transformation to demonstrate the flat bands could induce a type-A antiferromagnetic order.

We consider an unperturbed Hamiltonian
\ba 
H_0 = \sum_{L,i} E_{L,i} |L,i\rangle \langle L,i| +\sum_{H,i}E_{H,i}|H,i\rangle \langle H,i| 
\ea 
where $|L,i\rangle$ and $|H,i\rangle$ denote the eigenstate in the low-energy and high-energy subspaces. These eigenvalues and eigenvectors are obtained by directly diagonalizing $H_0$. We then introduce the following projection operators
\ba 
P_L = \sum_i |L,i\rangle \langle L,i| ,\quad P_H =\sum_i |H,i\rangle \langle H,i| 
\ea 
We then introduce the perturbation term
\ba 
H_1 = \sum_{ij}V_{ij}|L,i\rangle \langle H,j| +\text{h.c.}
\label{eq:SW_transform_H1_def}
\ea 
The perturbation term maps a low-energy state to a high-energy state or a high-energy state to a low-energy state. We then find
\ba 
&P_L H_1 P_H = \sum_{ij}V_{ij}|L,i\rangle \langle H,j| \nonumber\\ 
&P_HH_1P_L =  \sum_{ij}V^*_{ij}|H,j\rangle \langle L,i|
\label{eq:SW_transform_H1_two_part}
\ea 
We next aim to obtain an effective Hamiltonian defined in the low-energy subspace that takes into account the effects of the high-energy subspace in perturbation theory.

We consider the following operators
\ba 
S = \sum_{i,j}V_{ij}\frac{|L,i\rangle \langle H,j|}{E_{L,i}-E_{H,j}} - \sum_{i,j}V^*_{ij}\frac{|H,j\rangle \langle L,i|}{E_{L,i}-E_{H,j}} 
\ea 
We perform a unitary transformation correspondingly 
\ba 
H' = e^{S}(H_0+H_1)e^{-S} = H_0+H_1 + [S,H_0]+[S,H_1] + \frac{1}{2} [S,[S,H_0]] +....
\label{eq:SW_eff_H}
\ea 
We note that 
\ba 
&[S,H_0] \nonumber\\ 
=& \sum_{i,j} V_{ij}E_{H,j}\frac{|L,i\rangle \langle H,j| H,j\rangle \langle H,j|}{E_{L,i}-E_{H,j}} - \sum_{i,j} V_{ij}E_{L,i}\frac{|L,i\rangle \langle L,i| L,i\rangle \langle H,j|}{E_{L,i}-E_{H,j}}  \nonumber\\
& -\sum_{i,j} V^*_{ij}E_{L,i}\frac{|H,j\rangle \langle L,i| L,i\rangle \langle L,i|}{E_{L,i}-E_{H,j}}
+ \sum_{i,j} V^*_{ij}E_{H,j}\frac{|H,j\rangle \langle H,j|H,j\rangle \langle L,i|}{E_{L,i}-E_{H,j}}\nonumber\\ 
=&\sum_{i,j}-V_{ij}|L,i\rangle \langle H,j| - \sum_{i,j}-V_{ij}^*|H,j\rangle \langle L,i| \nonumber\\ 
=&-H_1
\ea 
Therefore, the new effective Hamiltonian now becomes
\ba 
H' \approx H_0 +H_1 -H_1 + [S,H_1] - \frac{1}{2}[S,H_1] =H_0 +\frac{1}{2}[S,H_1] 
\ea 
We note that
\ba 
&[S,H_1] \nonumber\\ 
=& \sum_{i,i', j} V_{ij} \frac{|L,i\rangle \langle H,j| }{E_{L,i}-E_{H,j}} V_{i'j}^* |H,j\rangle \langle L,i'| 
- \sum_{i,j,j'} V_{ij}^* \frac{|H,j\rangle \langle L,i| }{E_{L,i}-E_{H,j}} V_{ij'}|L,i\rangle \langle H,j'| \nonumber\\ 
&- \sum_{i,i',j}V_{ij}|L,i\rangle \langle H,j| (-V^*_{i'j})\frac{|H,j\rangle \langle L,i'|}{E_{L,i'}-E_{H,j}} 
- \sum_{i,j,j'}V_{ij}^*|H,j\rangle \langle L,i| V_{ij'}\frac{|L,i\rangle \langle H,j'|}{E_{L,i}-E_{H,j'}}\nonumber\\ 
=&\sum_{i,i',j} V_{ij}V_{i'j}^*\bigg( \frac{1}{E_{L,i}-E_{H,j}} + \frac{1}{E_{L,i'}-E_{H,j}}\bigg) 
|L,i\rangle \langle L,i'| 
+\sum_{i,j,j'} V_{ij}^*V_{ij'}\bigg( \frac{-1}{E_{L,i}-E_{H,j}} + \frac{-1}{E_{L,i}-E_{H,j'}}\bigg) 
|H,j\rangle \langle H,j'| 
\ea 
Using Eq.~\ref{eq:SW_eff_H}, we obtain
\ba 
&H'\nonumber\\ 
\approx  &\sum_{L,i} E_{L,i} |L,i\rangle \langle L,i| +\sum_{H,i}E_{H,i}|H,i\rangle \langle H,i| \nonumber\\ 
& +\frac{1}{2}\sum_{i,i',j} V_{ij}V_{i'j}^*\bigg( \frac{1}{E_{L,i}-E_{H,j}} + \frac{1}{E_{L,i'}-E_{H,j}}\bigg) 
|L,i\rangle \langle L,i'| 
+\frac{1}{2}\sum_{i,j,j'} V_{ij}^*V_{ij'}\bigg( \frac{-1}{E_{L,i}-E_{H,j}} + \frac{-1}{E_{L,i}-E_{H,j'}}\bigg) 
|H,j\rangle \langle H,j'| 
\ea 

Finally, we only keep the term in the low-energy subspace, which gives
\ba 
H_{eff} &= P_L H'P_L = H_0' +H_{SW} \nonumber\\ 
H_0'=&P_L H_0 P_L =  \sum_{i} E_{L,i} |L,i\rangle \langle L,i| \nonumber\\ 
H_{SW} = & \frac{1}{2} \sum_{i,j,m }V_{im}V_{jm}^*\left( \frac{1}{E_{L,i}-E_{H,m}} + \frac{1}{E_{L,j} -E_{H,m}}
\right) |L,i\rangle \langle L,j|
\label{eq:app:sw_eff_h}
\ea 
$H_0'$ denotes the contribution from the original unperturbed Hamiltonian and $H_{SW}$ denotes the additional contribution generated from the Schrieffer-Wolff transformation. In practice, the additional term $H_{SW}$ describes the following virtual process
\ba 
|L,j\rangle \xrightarrow{V_{jm}^* |H,m\rangle \langle L,j| }|H,m\rangle\xrightarrow{V_{im} |L,i\rangle \langle H,m| } |L,i\rangle 
\ea 
where 
$H_1$ maps a low-energy state $|L,j\rangle$ to a high-energy state $|H,m\rangle$ and then map the high-energy state $|H,m\rangle$ back to a low-energy state $|L,i\rangle$

\section{Flat-band ferromagnetism/anti-ferromagnetism}
\label{sec:app:flat_band_magnetism}
In this section, we demonstrate that a relatively flat band that is atomic along $z$ direction and non-atomic along $x,y$ directions leads to a type-A antiferromagnetism (in-plane magnetism and out-of-plane antiferromagnetism). 

Several comments are in order:
\begin{itemize} 
    \item In this section, we will pick a toy model to demonstrate the ferromagnetism/antiferromagnetism tendency induced by flat bands. 
    \item We will consider the case of a half-filled single flat band to simplify the considerations. 
    \item We will ignore the contributions of the other dispersive bands. The flat band, when it appears near Fermi energy, induces a large density of states. We expect the leading-order instability to be generated by flat bands instead of dispersive bands. 
    % as long as the interactions are not strong enough to also change the band structure of dispersive bands drastically. 
    \item We comment that the real materials are more complicated, in which the flat bands are not perfectly flat and the dispersive bands could also make contributions. However, based on the toy model we considered in this appendix and the more realistic DFT model we discussed in Appendix~\ref{sec:app:afm_c}, we could conclude that flat bands (atomic along $z$ direction and non-atomic along $xy$ direction) indeed favor the type-A antiferromagnetism. 
    % which is consistent with the experimental observations. 
\end{itemize}

In Sec.~\ref{sec:app:2d_ferro}, we will show that a 2D flat band which is non-atomic along $xy$ directions leads to an in-plane ferromagnetism. In Sec.~\ref{sec:app:3d_afm}, we show that, a weak hopping along $z$ directions leads to an antiferromagnetic coupling along $z$ direction and stabilizes a type-A antiferromagnetism.

\subsection{Ferromagnetism from 2D non-atomic flat band}
\label{sec:app:2d_ferro}
In this section, we provide more understanding of the ferromagnetism induced by flat bands, based on a single-flat-band toy model. We use $d_{\kk,\alpha,\sigma}$ to denote the electron operators with momentum $\kk$, orbital $\alpha$ and spin $\sigma$. 

We assume a flat band (with two-fold degeneracy due to $SU(2)$ spin symmetry) can be described by the following electron operators in the band basis
\ba 
\gamma_{\kk,\sigma} = \sum_{\alpha}V^*_{ \alpha }(\kk) d_{\kk,\alpha,\sigma } 
\ea 
 $V_{\alpha } (\kk) $ describes the wavefunction of the flat band. We assume $V_{\alpha}(\kk)$ is $k_z$-independent since the flat band is atomic along $z$ direction. 
 
We first consider perfect flat bands with the kinetic term in the band basis taken to be
 \ba 
H_0 = \sum_{\kk,\sigma} E_0 \gamma_{\kk,\sigma}^\dag \gamma_{\kk,\sigma} 
 \ea 
 To simplify the calculations, we consider the case where the flat band is half-filled at the non-interacting limit with $E_0=0$. 
 We next consider the interaction term. We take the following density-density interactions characterized by interaction matrix $U_{\alpha\gamma}$ with $U_{\alpha\gamma}=U_{\gamma\alpha}>0$ (we also drop the Hund's coupling term here, since it is usually small compared to the density-density interactions)
 \ba 
 H_U = \sum_{\alpha, \RR, \sigma,\sigma'  }\frac{ U_{\alpha\gamma} }{2} \bigg(P:d_{\RR,\alpha,\sigma} ^\dag   d_{\RR,\alpha,\sigma}:P \bigg) \bigg(P:d_{\RR,\gamma,\sigma'} ^\dag d_{\RR,\gamma,\sigma'}:P\bigg) 
 \ea 
 where $P$ is the projection operator that projects the electron operator to the band basis
 \ba 
 Pd_{\kk,\alpha,\sigma}P = V_\alpha(\kk)\gamma_{\kk,\sigma}
 % P\gamma_{\kk,\sigma}^\dag P = \gamma_{\kk,\sigma}
 % ,\quad Pb_{\kk,i,\sigma} P = Pa_{\kk,i,\sigma} P=0
 \ea 
 and $::$ denotes normal ordering. 
%  For a given operator $O$, the normal ordering is defined as
%  % (with half-filled flat bands)
%  \ba 
% : {O}: = {O} -\langle O\rangle_0 
%  \ea 
%  where $\langle \rangle_0$ denotes the expectation value taken with respect to the non-interacting Hamiltonian $H_0$. 
 We consider the case of $E_0=0$. Then we have $\langle \gamma_{\kk,\alpha,\sigma}^\dag \gamma_{\kk+\qq,\gamma,\sigma}\rangle_0=\delta_{\qq,0}\delta_{\alpha,\gamma}$ which corresponds to the half-filling of flat bands. 
 The projected density operators can then be written as 
 \ba 
&:Pd_{\RR,\alpha,\sigma} ^\dag   d_{\RR,\alpha,\sigma}P:\nonumber\\ 
= &\sum_{\kk,\qq}\frac{e^{+i\qq\cdot(\RR+\rr_\alpha)}}{N}V^*_{\alpha}(\kk) V_{\alpha}(\kk+\qq)\gamma_{\kk,\alpha,\sigma}^\dag 
\gamma_{\kk+\qq,\alpha,\sigma}- \langle \sum_{\kk,\qq}\frac{e^{+i\qq\cdot(\RR+\rr_\alpha)}}{N}V^*_{\alpha}(\kk) V_{\alpha}(\kk+\qq)\gamma_{\kk,\alpha,\sigma}^\dag 
\gamma_{\kk+\qq,\alpha,\sigma} \rangle_0 \nonumber\\ 
=& \sum_{\kk,\qq}\frac{e^{+i\qq\cdot(\RR+\rr_\alpha)}}{N}V^*_{\alpha}(\kk) V_{\alpha}(\kk+\qq)\bigg( \gamma_{\kk,\alpha,\sigma}^\dag 
\gamma_{\kk+\qq,\alpha,\sigma}-\frac{1}{2}\delta_{\qq,0}\bigg)
 \ea 
 
The interaction term can be written as
\ba 
H_U = \sum_{\alpha,\gamma,\kk,\kk',\qq, \sigma,\sigma'}\frac{U_{\alpha\gamma} }{2N} V^*_{\alpha }(\kk)V_{\alpha }(\kk+\qq)V_{\gamma}^*(\kk'+\qq)V_{\gamma}(\kk') (\gamma_{\kk,\sigma}^\dag \gamma_{\kk+\qq,\sigma}-\frac{\delta_{\qq,0}}{2})( \gamma_{\kk'+\qq,\sigma' }^\dag\gamma_{\kk',\sigma'}-\frac{\delta_{\qq,0}}{2})
\ea 

The total Hamiltonian is
\ba 
H=&H_0 +H_U \nonumber\\ 
=&\sum_{\alpha,\gamma,\kk,\kk',\qq, \sigma,\sigma'}\frac{U_{\alpha\gamma} }{2N} V^*_{\alpha }(\kk)V_{\alpha }(\kk+\qq)V_{\gamma}^*(\kk'+\qq)V_{\gamma}(\kk') (\gamma_{\kk,\sigma}^\dag \gamma_{\kk+\qq,\sigma}-\frac{\delta_{\qq,0}}{2})( \gamma_{\kk'+\qq,\sigma' }^\dag\gamma_{\kk',\sigma'}-\frac{\delta_{\qq,0}}{2})
\label{eq:app:ham_flat_band}
% \nonumber\\
\ea 
where we take $E_0=0$. 

 We now aim to rewrite the Hamiltonian in a more compact formula. We can perform an eigen-decomposition of the interaction matrix $U_{\alpha\gamma}$
\ba 
U_{\alpha\gamma}= \sum_{n=1,...,n_{orb}} u_{n} v_{\alpha n} v^*_{\gamma n }
\label{eq:app:eigen_coluomb}
\ea 
with $u_{n}$ and $v_{\alpha n}$ are eigenvalues and eigenvectors of $U_{\alpha\gamma}$ matrix. $n_{orb}$ denotes the number of orbitals. The Hamiltonian now becomes (using Eq.~\ref{eq:app:ham_flat_band}, and Eq.~\ref{eq:app:eigen_coluomb})
\ba 
H= &\sum_{n,\qq} \frac{u_n}{2N} 
\bigg[ 
\sum_{\kk,\sigma,\alpha}v_{\alpha n}V^*_\alpha(\kk)V_\alpha(\kk+\qq)(\gamma_{\kk,\sigma}^\dag \gamma_{\kk+\qq,\sigma} -\frac{\delta_{\qq,0}}{2})
\bigg]
\bigg[ 
\sum_{\kk',\sigma',\gamma}v^*_{\gamma n}V_\gamma(\kk')V^*_\gamma(\kk'+\qq)
(\gamma_{\kk'+\qq,\sigma'}^\dag \gamma_{\kk',\sigma'} -\frac{\delta_{\qq,0}}{2})
\bigg]
\ea  
Therefore, we find 
\ba 
H = \sum_{n,q} \frac{u_n}{2}O_{n,\qq}O_{n,\qq}^\dag 
\ea 
with 
\ba 
&O_{n,\qq} =\frac{1}{\sqrt{N}}
\sum_{\kk,\sigma,\alpha}v_{\alpha n}V^*_\alpha(\kk)V_\alpha(\kk+\qq)(\gamma_{\kk,\sigma}^\dag \gamma_{\kk+\qq,\sigma} 
-\frac{\delta_{\qq,0}}{2})
\ea 

We now discuss the values of $v_n$ which are the eigenvalues of $U_{\alpha\gamma}$. In general, we expect $v_n$ to be positive. The negative $v_n$ indicates an attractive interaction in certain channels. This could then lead to additional instability such as orbital ordering. To observe this, we take a two-orbital case as an example. The interaction matrix of a two-orbital model can be written as
\ba 
U_{\alpha\gamma} = 
\begin{bmatrix}
    U_1 & U_2 \\
    U_2 & U_1 
\end{bmatrix}_{\alpha\gamma}
\ea 
where $U_1,U_2$ denotes intra and inter-orbital interactions. Obviously, the eigenvalues of the interaction matrix are $U_1+U_2,U_1-U_2$. If $U_2 > U_1$, one of the eigenvalues becomes negative. To observe its consequence, we could write the interaction term via density operators as
\ba 
&\sum_\RR\bigg[ \frac{U_1}{2}(\nu_{\RR,1}^2+\nu_{\RR,2}^2) + U_2 \nu_{\RR,1}\nu_{\RR,2}
\bigg] 
= 
\sum_\RR\bigg[ \frac{U_1+U_2}{4}(\nu_{\RR,1}+\nu_{\RR,2})^2+\frac{U_1-U_2}{4}(\nu_{\RR,1}-\nu_{\RR,2})^2
\bigg] 
\ea 
with the density operator defined as 
\ba 
&\nu_{\RR,\alpha} = \sum_{\sigma} :d_{\RR,\alpha\sigma}^\dag d_{\RR,\alpha\sigma}:
\ea 
We note that, if $U_1-U_2 < 0$, the system minimizes the energy by only filling one of the two orbitals (maximizing $|\nu_{\RR,1}-\nu_{\RR,2}|$), which could induce an orbital ordering. 

However, for the generic system, the diagonal components of $U_{\alpha\gamma}$ (which corresponds to the intra-orbital repulsion) are usually stronger than the off-diagonal components (which correspond to the inter-orbital repulsion). The strong enough diagonal components of $U_{\alpha\gamma}$ matrix could make all the eigenvalues positive $u_n>0$. In the current case, since we are mostly interested in the formation of magnetism (instead of orbital ordering), we will assume $u_n \ge 0$ for all $n$. 

When $u_n>0$, 
\ba 
H =  \sum_{n,q} \frac{u_n}{2}O_{n,\qq}O_{n,\qq}^\dag 
\ea 
becomes a positive semi-definite Hamiltonian. The state $|\phi\rangle$ that satisfy
\ba 
O_{n,\qq}^\dag |\phi\rangle = 0 ,\quad \forall n ,\qq 
\ea 
would be the ground state~\cite{PhysRevB.103.205415}.

Since we assume the system is an effective 2D system where $V_{\alpha}(\kk)$ is $k_z$ independent,
we treat $xy$ directions and $z$ direction separately.
We use $\kk^{xy},\qq^{xy}$ to denote the in-plane components of the vector with $\kk^{xy}=(k_x,k_y), \qq^{xy}=(q_x,q_y)$. Therefore, we can rewrite $O_{\alpha,\qq}^\dag$ as 
\ba 
O^\dag_{n,\qq} = \frac{1}{\sqrt{N}}\sum_{\kk^{xy},\alpha, \sigma}  
v_{\alpha n}^*  V_{\alpha }^*(\kk^{xy}+\qq^{xy}) V_{\alpha}(\kk^{xy}) \sum_{k_z} (\gamma_{(\kk^{xy}+\qq^{xy},k_z+q^z),\sigma}^\dag \gamma_{(\kk^{xy},k^z),\sigma} - \frac{\delta_{\qq^{xy},0}\delta_{q^z,0} }{2})
\ea 
Since the system is atomic along $z$ direction, we can perform Fourier transformation along $z$ direction with
\ba 
\gamma_{(\kk^{xy},R_z), \sigma} =\frac{1}{\sqrt{L_z}} \sum_{\RR} \gamma_{(\kk^{xy},k_z),\sigma}e^{ik_z R_z}
\label{eq:ft_z}
\ea 
where $L_\alpha$ is the system size along $\alpha$ direction with the total number of unit cells $N=L_xL_yL_z$. Then we find
\ba 
O^\dag_{n, (\qq^{xy},R^z)} =& \frac{1}{\sqrt{L_z}} \sum_{q_z}O^\dag_{n,(\qq^{xy},q^z)}e^{-iq_zR_z}\nonumber\\ 
=&  \frac{1}{\sqrt{N_{xy}}}\sum_{\kk^{xy},\alpha, \sigma}  
v_{\alpha n}^*  V_{\alpha }^*(\kk^{xy}+\qq^{xy}) V_{\alpha}(\kk) (\gamma_{(\kk^{xy}+\qq^{xy},R_z),\sigma}^\dag \gamma_{(\kk^{xy},R_z),\sigma} -\frac{\delta_{\qq^{xy},0} }{2})
\ea 
where $N_{xy}=L_xL_y$. 
The Hamiltonian now can be written as 
\ba 
H_0 +H_U = \sum_{n,\qq^{xy},R_z} \frac{u_n}{2}O_{n,(\qq^{xy},R_z)}O_{n,(\qq^{xy},R_z)}^\dag 
\ea 
where we can observe that the Hamiltonian of each layer is decoupled. 
We now solve the positive semidefinite Hamiltonian of each $R_z$ layer 
\ba
H_{R_z} =  \sum_{n,\qq^{xy}} \frac{u_n}{2}O_{\alpha,(\qq^{xy},R^z)} O^\dag_{\alpha, (\qq^{zy},R_z)}
\ea 

We now prove that the following states are the ground states of $H_{R_z}$
\ba 
|R_z \rangle = \prod_{\kk^{xy}}\gamma_{(\kk^{xy},R_z), \up}^\dag |0\rangle
\label{eq:gnd_rz} 
\ea 
where $|0\rangle$ is the vacuum state. We prove it by showing
\ba 
 O^\dag_{\alpha, (\qq^{zy},R_z)} |R_z\rangle =0 
\ea 
Written explicitly, we have 
\ba 
&O_{n,(\qq^{xy},R_z)}^\dag |R_z\rangle \nonumber\\ 
=&\frac{1}{\sqrt{N_{xy}}}
 \sum_{\kk^{xy},\alpha, \sigma}  
v_{\alpha n}^*  V_{\alpha }^*(\kk^{xy}+\qq^{xy}) V_{\alpha}(\kk)( \gamma_{(\kk^{xy}+\qq^{xy},R_z),\sigma}^\dag \gamma_{(\kk^{xy},R_z),\sigma} -\frac{1}{2}\delta_{\qq^{xy},0})\prod_{\kk^{xy}}\gamma_{(\kk^{xy},R_z), \up}^\dag |0\rangle 
\nonumber\\ 
=&\frac{1}{\sqrt{N_{xy}}}
 \sum_{\kk^{xy},\alpha}  
v_{\alpha n}^*  V_{\alpha }^*(\kk^{xy}+\qq^{xy}) V_{\alpha}(\kk) (\frac{\delta_{\qq^{xy},0}}{2} -\gamma_{(\kk^{xy},R_z),\up} \gamma_{(\kk^{xy}+\qq^{xy},R_z),\up}^\dag )\prod_{\kk^{xy}}\gamma_{(\kk^{xy},R_z), \up}^\dag |0\rangle \nonumber\\ 
&+\frac{1}{\sqrt{N_{xy}}}
 \sum_{\kk^{xy},\alpha,\sigma = \dn}  
v_{\alpha n}^*  V_{\alpha }^*(\kk^{xy}+\qq^{xy}) V_{\alpha}(\kk) (-\frac{\delta_{\qq^{xy},0}}{2} )\prod_{\kk^{xy}}\gamma_{(\kk^{xy},R_z), \up}^\dag |0\rangle
\nonumber\\ 
=&0
\ea

Therefore, $|R_z\rangle$ is the ground states of $H_{R_z}$.
The ground state in Eq.~\ref{eq:gnd_rz} defines a state where the spin $\up$ flavor of the flat band is fully filled. Due to the $SU(2)$ rotation symmetry, all the other states obtained by acting $SU(2)$ symmetry on the $|R_z\rangle $ are also ground states. The states obtained by $SU(2)$ transformation can be written as
\ba 
| R_z, \theta(R_z), \phi (R_z) \rangle = \hat{R}_{R_z}(\theta(R_z),\phi(R_z) ) | R_z\rangle
\ea 
where $\hat{R}_{R_z}(\theta,\phi)$ describes the following $SU(2)$ symmetry transformation
\ba 
\hat{R}_{R_z}(\theta,\phi)\gamma_{\kk^{xy},R_z,\up}^\dag \hat{R}_{R_z}^\dag(\theta,\phi) = \cos(\theta) e^{i\phi} \gamma_{\kk^{xy},R_z,\up}^\dag  + \sin(\theta) \gamma_{\kk^{xy},R_z,\dn}^\dag
\label{eq:def_rot_op}
\ea 

Therefore a generic ground state of the whole system can be obtained by taking the tensor product of the ground state of each layer
\ba 
|\theta,\phi\rangle = \bigotimes_{R_z} | R_z ,\theta(R_z), \phi(R_z)\rangle 
\ea 
where $\theta,\phi$ are functions of $R_z$ that characterize the $SU(2)$ rotation of each layer. Since the system is decoupled along $z$ direction (in the case of zero $z$-direction hopping), we observe the relative spin directions of different layers can be arbitrary. In other words, there is no $z$ direction magnetic order.

\subsubsection{Orthonormal basis}
 The ground states of the Hamiltonian characterized by
\ba 
|\theta,\phi\rangle  =  \bigotimes_{R_z}|R_z,\theta(R_z),\phi(R_z)\rangle 
\ea 
have a large degeneracy. Moreover, the set of ground states,
$\{|\theta,\phi\rangle | \theta\in [0,\pi),\phi\in[0,2\pi)\}$, does not form an orthonormal basis of 
the ground-state manifold. 

For future convenience, we now aim to find the orthonormal basis of the ground-state manifold. For layer $R_z$, the ground state is characterized by 
\ba 
|R_z, \theta(R_z),\phi(R_z)\rangle 
\ea 
The Hamiltonian has a $SU(2)$ symmetry for each layer. The generators of $SU(2)$ group (or the spin operators) are 
\ba 
S^{x,y,z}_{R_z} =\sum_{\kk^{xy},\sigma,\sigma'} \gamma_{(\kk^{xy},R_z),\sigma}^\dag \frac{ \sigma^{x,y,z}_{\sigma\sigma'}}{2}\gamma_{(\kk^{xy},R_z),\sigma'}
\label{eq:lared_spin_operator}
\ea 
where $\sigma^{x,y,z}$ are Pauli matrices. 
We find the ground states of layer $R_z$ which are characterized by $\{|R_z,\theta(R_z),\phi(R_z)\rangle\}_{\theta(R_z),\phi(R_z)}$ form a spin $S=(N_{xy}/2)$ representation of the $SU(2)$ group, where $N_{xy}$ denotes number of unit cells for each layer. 
The spin $S=(N_{xy}/2)$ representation corresponds to the ferromagnetic state where all the filled flat-band electrons have the same spin direction. 
To show the ground states indeed form $S=(N_{xy}/2)$ representation, we calculate 
\ba 
&\sum_{\mu}(S^{\mu}_{R_z})^2 |R_z, \theta(R_z),\phi(R_z)\rangle  = \sum_{\mu}(S^{\mu}_{R_z})^2 \hat{R}_{R_z}(\theta(R_z),\phi(R_z))|R_z, 0,0\rangle  =  \hat{R}_{R_z}(\theta(R_z),\phi(R_z))\sum_{\mu}(S^{\mu}_{R_z})^2 |R_z, 0,0\rangle  \nonumber\\
=&S(S+1) \hat{R}_{R_z}(\theta(R_z),\phi(R_z)) |R_z, 0,0\rangle  = S(S+1)|R_z,\theta(R_z),\phi(R_z)\rangle 
\label{eq:su2_rep_lar_state}
\ea 
where we use the fact that the Casimir operator $\sum_{\mu}(S^{\mu}_{R_z})^2$ commutes with the $SU(2)$ rotation operator $\hat{R}_{R_z}(\theta(R_z),\phi(R_z))$. From Eq.~\ref{eq:su2_rep_lar_state}, we conclude the ground states form $S=(N_{xy}/2)$ representation. 

Since the degenerate ground states form an $S=(N_{xy}/2)$ representation, we can then use the eigenstates of $S^z_{R_z}$ operator and the ladder operators to build the orthonormal basis of the ground-state manifolds. We introduce the following orthonormal states to characterize the ground-state manifolds
\ba 
&\{|R_z,M\rangle \}_{M=-S}^{M=S}\nonumber\\ 
&S_{R_z}^z|R_z,M\rangle = M|R_z,M\rangle
,\quad \sum_\mu 
(S_{R_z}^\mu)^2|R_z,M\rangle = S(S+1)|R_z,M\rangle
\ea 
The above orthonormal basis can be obtained by acting ladder operators on the $|R_z,\theta(R_z)=0,\phi(R_z)=0\rangle$. In other words, we can let 
\ba 
&|R_z,M=S \rangle = |R_z,\theta(R_z)=0,\phi(R_z)=0\rangle  \nonumber\\ 
&|R_z,M-1\rangle = \frac{S_{R_z}^-}{\sqrt{S(S+1)-M(M-1)}}|R_z,M\rangle 
\label{eq:normalization_condition_for_mz_state}
\ea 
where $S_{R_z}^- = S_{R_z}^+ -iS_{R_z}^-$ is the spin ladder operator.

We now prove we can indeed construct $|R_z,M\rangle$ via Eq.~\ref{eq:normalization_condition_for_mz_state}. 
We first prove that $|R_z,\theta(R_z)=0,\phi(R_z)=0\rangle$ is the eigenstate of $S_{R_z}^z$ operator with eigenvalue $S$:
\ba 
S_{R_z}^z |R_z,\theta(R_z)=0,\phi(R_z)=0\rangle =&\frac{1}{2}\sum_{\kk^{xy},\sigma}\sigma \gamma_{(\kk^{xy},R_z),\sigma}^\dag \gamma_{(\kk^{xy},R_z),\sigma} \prod_{\qq^{xy}}\gamma_{(\qq^{xy},R_z),\up}^\dag|0\rangle = \frac{N_{xy}}{2} 
\prod_{\qq^{xy}}\gamma_{(\qq^{xy},R_z),\up}^\dag|0\rangle\nonumber\\ 
=&S |R_z,\theta(R_z)=0,\phi(R_z)=0\rangle 
\ea  
Therefore, $|R_z,\theta(R_z)=0,\phi(R_z)=0\rangle =|R_z,M=S\rangle$. 

We next show that, for a given $|R_z,M\rangle$, we can generate $|R_z,M-1\rangle$ by acting spin ladder operator ($S_{R_z}^- = S_{R_z}^+ -iS_{R_z}^-$) on $|R_z,M\rangle$. Via spin commutation relation, we find
\ba 
&S^z_{R_z} \bigg( \frac{S_{R_z}^-}{\sqrt{S(S+1)-M(M-1)}}|R_z,M\rangle  \bigg) = 
\bigg( S_{R_z}^{-} S_{R_z}^z - S_{R_z}^{-}\bigg) 
\frac{1}{\sqrt{S(S+1)-M(M-1)}}|R_z,M\rangle\nonumber\\ 
=&\bigg( \frac{S_{R_z}^-}{\sqrt{S(S+1)-M(M-1)}} (M-1)|R_z,M\rangle  \bigg)  \nonumber\\ 
=&(M-1) \bigg( \frac{S_{R_z}^-}{\sqrt{S(S+1)-M(M-1)}}|R_z,M\rangle  \bigg) 
\ea 
where the additional $\frac{1}{\sqrt{S(S+1)-M(M-1)}}$ factor ensures the normalization condition $\langle R_z,M| R_z,M\rangle =1 $. Therefore, we prove that, we can construct $|R_z,M-1\rangle$ state from $|R_z,M\rangle$ state via Eq.~\ref{eq:normalization_condition_for_mz_state} (second line). 
% Suppose $|R_z,M\rangle, |R_z,M+1\rangle$ is normalized, we now prove $|R_z,M-1\rangle$ is also normalized
% \ba 
% \langle R_z,M-1|R_z,M-1\rangle =\frac{1}{S(S+1)-M(M-1)} 
% \langle R_z,M|S_{R_z}^+S_{R_z}^-  |R_z,M\rangle = 
% \ea 

We can also rewrite Eq.~\ref{eq:normalization_condition_for_mz_state} as following
\ba 
&|R_z, M\rangle =\alpha_{S,M}  (S_{R_z}^-)^{S-M} |R_z,S\rangle \nonumber\\ 
&\alpha_{S,S}=1\nonumber\\ 
&\alpha_{S,M} = \bigg(\prod_{i=M}^{S-1}\frac{1}{\sqrt{S(S+1)-i(i+1)}}\bigg),\quad \text{ for } M<S
\label{eq:1flatband:gnd_state_spin_ladder}
\ea 
where $\alpha_{S,M}$ is the normalization factor to ensure $\langle R_z,M|R_z,M\rangle=1$.

% Since $|R_z,M\rangle$ are eigenstates of $S_{R_z}^z$ with different eigenvalues, they are orthogonal. However, $|R_z,M\rangle$ is not a normalized state. We now compute its norm. We first note that from the definition
% \ba 
% &\langle R_z,S |R_z,S\rangle = 1 \nonumber\\ 
% &\langle R_z,S-1| R_z,S-1\rangle = \langle R_z,S | S_{R_z}^+ S_{R_z}^- |R_z,S\rangle = \langle R_z,S| S_{R_z}^- S_{R_z}^+
% \ea 

Taking the tensor product of the states from different layers, we can introduce the following orthogonal basis for the ground-state manifold
\ba 
\{ |m\rangle   \},\quad \quad |m\rangle = \bigotimes_{R_z} |R_z, m(R_z)\rangle 
\label{eq:1flatband:gnd_basis}
\ea 
where we have defined $m$ as a function of $R_z$ with $m(R_z)$ characterizes the states of layer $R_z$. 

\subsubsection{Charge $\pm 1$ excitation state}
After finding the ground states, we can also calculate the charge $ \pm 1$ excitation. The charge $\pm 1$ excitation will be useful when we study the $z$-direction correlation induced by weak $z$-direction hopping via perturbation theory. 

Since different layers are decoupled, we could treat each layer separately. 
The charge $-1$ excitation state 
with in-plane momentum $\kk^{xy}$, layer index $R_z$ and spin $\sigma$ can be described by 
\ba 
\gamma_{(\kk^{xy},R_z),\sigma} | R_z ,m(R_z) \rangle 
\label{eq:hole_exct_gamma_def}
\ea 

We now show that the state in Eq.~\ref{eq:hole_exct_gamma_def} is an eigenstate of the interaction Hamiltonian by proving
\ba 
[(H_0 +H_U), \gamma_{(\kk^{xy},R_z),\sigma}] | R_z ,m(R_z) \rangle  = E^{hole}_{\kk^{xy},R_z} \gamma_{(\kk^{xy}),\sigma} |R_z ,m(R_z) \rangle 
\label{eq:1flatband:derive_hole_exct_1}
\ea 
with $E^{hole}_{\kk^{xy}} $ the corresponding energy of the excitation state. 

We first calculate
 \ba 
 &[H_U+H_0,\gamma_{\kk^{xy},R_z,\sigma}]  
 \ea 
We note that 
\ba 
&[O^\dag_{n,(\qq^{xy},R_z)}, \gamma_{\pp^{xy},R_z',\sigma''}] \nonumber\\ 
=
&
\frac{1}{\sqrt{N_{xy}}}\sum_{\kk^{xy},\alpha, \sigma}  
v_{\alpha n}^*  V_{\alpha }^*(\kk^{xy}+\qq^{xy}) V_{\alpha}(\kk) [\gamma_{(\kk^{xy}+\qq^{xy},R_z),\sigma}^\dag \gamma_{(\kk^{xy},R_z),\sigma}, \gamma_{(\pp^{xy},R_z'),\sigma''}]\nonumber\\
=&\delta_{R_z,R_z'}\frac{1}{\sqrt{N_{xy}}}\sum_{\alpha}v_{\alpha n}^*
V_{\alpha }^*(\pp^{xy})V_{\alpha }(\pp^{xy} - \qq^{xy})(-\gamma_{(\pp^{xy} -\qq^{xy},R_z),\sigma''})
\ea 
and then
\ba 
&[O_{n,(\qq^{xy},R_z)}O^\dag_{n,(\qq^{xy},R_z)}, \gamma_{\pp^{xy},R_z',\sigma''}]\nonumber\\ 
=&\delta_{R_z,R_z'}\frac{1}{N_{xy}}\sum_{\alpha,\gamma}v_{\alpha n}^* v_{\gamma n} \sum_{\kk}
V_{\alpha }^*(\pp^{xy})V_{\alpha }(\pp^{xy} - \qq^{xy})V_{\gamma}(\kk^{xy}+\qq^{xy})V_{\gamma}^*(\kk_{xy}) \nonumber \\ 
&[\gamma_{(\kk^{xy},R_z),\sigma}^\dag \gamma_{(\kk^{xy}+\qq^{xy},R_z),\sigma},
(-\gamma_{(\pp^{xy} -\qq^{xy},R_z),\sigma''})] \nonumber\\ 
=& \delta_{R_z,R_z'}\frac{1}{N_{xy}}\sum_{\alpha,\gamma}v_{\alpha n}^* v_{\gamma n} 
V_{\alpha }^*(\pp^{xy})V_{\alpha }(\pp^{xy} - \qq^{xy})V_{\gamma}(\pp^{xy})V_{\gamma}^*(\pp^{xy}-\qq^{xy}) \gamma_{(\pp^{xy},R_z),\sigma''} 
\ea 

Then we have 
\ba 
&[H_U+H_0,\gamma_{\pp^{xy},R_z,\sigma}]  \nonumber\\ 
=& \sum_{\qq^{xy},n}\frac{u_{n}}{2N_{xy}} \sum_{\alpha,\gamma}v_{\alpha n}^* v_{\gamma n} 
V_{\alpha }^*(\pp^{xy})V_{\alpha }(\pp^{xy} - \qq^{xy})V_{\gamma}(\pp^{xy})V_{\gamma}^*(\pp^{xy}-\qq^{xy}) \gamma_{(\pp^{xy},R_z),\sigma}  \nonumber\\ 
=& E_{\pp^{xy},R_z}^{hole} \gamma_{(\pp^{xy},R_z),\sigma} 
\ea 
where 
\ba 
& E_{\pp^{xy}}^{hole}  = \sum_{\qq^{xy},n}\frac{u_{n}}{2N_{xy}} \sum_{\alpha,\gamma}v_{\alpha n}^* v_{\gamma n} 
V_{\alpha }^*(\pp^{xy})V_{\alpha }(\pp^{xy} - \qq^{xy})V_{\gamma}(\pp^{xy})V_{\gamma}^*(\pp^{xy}-\qq^{xy})
\label{eq:1fltaband:hole_exct_en}
\ea 

Therefore, we have proved that
\ba 
[(H_0 +H_U),\gamma_{(\kk^{xy},R_z),\sigma}] | R_z ,m(R_z)\rangle  = E^{hole}_{\kk^{xy}} \gamma_{(\kk^{xy},R_z),\sigma} |R_z ,m(R_z)  \rangle
\ea 
This indicates, as long as, 
\ba 
\gamma_{(\kk^{xy},R_z), \sigma'} |R_z ,m(R_z)\rangle   \ne 0 
\label{eq:1flatband:valid_hole_exct_2}
\ea 
$\gamma_{(\kk^{xy},R_z), \sigma'} |R_z ,m(R_z)\rangle$ describes a charge $-1$ excitation state with energy $E^{hole}_{\kk^{xy}}$. 

We now aim to find the situation where Eq.~\ref{eq:1flatband:valid_hole_exct_2} holds. We evaluate the following equations explicitly
\ba 
\gamma_{(\kk^{xy},R_z), \sigma'} |R_z ,m(R_z)\rangle 
\ea 
We first note the following commutation relations 
\ba 
% &[S^-_{R_z},H] =0\nonumber\\ 
& [\gamma_{(\kk^{xy},R_z),\up}^\dag,S^-_{R_z}] =-\gamma_{(\kk^{xy},R_z),\dn}^\dag ,\quad [\gamma_{(\kk^{xy},R_z),\up},S^-_{R_z}] =0 \nonumber\\ 
& [\gamma_{(\kk^{xy},R_z),\dn}^\dag,S^-_{R_z}] =0 ,\quad [\gamma_{(\kk^{xy},R_z),\dn},S^-_{R_z}] =\gamma_{(\kk^{xy},R_z),\up} 
\ea 
and 
\ba 
&[(S^-_{R_z})^{n},H] =0\nonumber\\ 
& [\gamma_{(\kk^{xy},R_z),\up}^\dag,(S^-_{R_z})^{n}] =-n(S^-_{R_z})^{n-1}\gamma_{(\kk^{xy},R_z),\dn}^\dag ,\quad [\gamma_{(\kk^{xy},R_z),\up},(S^-_{R_z})^{n}] =0 \nonumber\\ 
& [\gamma_{(\kk^{xy},R_z),\dn}^\dag,(S^-_{R_z})^{n}] =0 ,\quad [\gamma_{(\kk^{xy},R_z),\dn},(S^-_{R_z})^{n}] =n(S^-_{R_z})^{n-1}\gamma_{(\kk^{xy},R_z),\up} 
\label{eq:commutation_gamma_spin}
\ea

Combining Eq.~\ref{eq:1flatband:gnd_state_spin_ladder} and Eq.~\ref{eq:commutation_gamma_spin}, we find
\ba 
&\gamma_{(\kk^{xy},R_z), \up} |R_z,m(R_z)\rangle = \alpha_{S,m(R_z)}\bigg(1-\delta_{m(R_z),-S}\bigg)
\gamma_{(\kk^{xy},R_z), \up} (S_{R_z}^{-})^{S-m(R_z)} |R_z,S\rangle \nonumber\\ 
=&  \alpha_{S,m(R_z)} \bigg(1-\delta_{m(R_z),-S}\bigg)(S_{R_z}^{-})^{S-m(R_z)} \gamma_{(\kk^{xy},R_z),\up} |R_z,S\rangle  \nonumber\\ 
\nonumber\\ 
&\gamma_{(\kk^{xy},R_z),\dn} |R_z,m(R_z)\rangle =  \alpha_{S,m(R_z)}\bigg(1-\delta_{m(R_z),S}\bigg)
\gamma_{(\kk^{xy},R_z), \dn}(S_{R_z}^{-})^{S-m(R_z)} |R_z,S\rangle \nonumber\\ 
=& \alpha_{S,m(R_z)}\bigg(1-\delta_{m(R_z),S}\bigg)(S_{R_z}^{-})^{S-m(R_z)} \gamma_{(\kk^{xy},R_z),\dn}  |R_z,S\rangle  \nonumber\\ 
&+ \alpha_{S,m(R_z)}(S-m(R_z))\bigg(1-\delta_{m(R_z),S}\bigg)(S_{R_z}^{-})^{S-m(R_z)-1} \gamma_{(\kk^{xy},R_z),\up}  |R_z,S\rangle \nonumber\\ 
=& \alpha_{S,m(R_z)}(S-m(R_z))(S_{R_z}^{-})^{S-m(R_z)-1} \gamma_{(\kk^{xy},R_z),\up} |R_z,S\rangle 
\label{eq:1flatband:hole_exct_state}
\ea 
Since 
\ba 
|R_z,S\rangle = \prod_{\kk} \gamma_{(\kk^{xy},R_z),\up}^\dag|0\rangle 
\ea 
then
\ba 
 \gamma_{(\kk^{xy},R_z),\up} |R_z,S\rangle  \ne 0 
\ea 
We thus conclude the following states give the charge $-1$ excitation state
\ba 
&\gamma_{\kk^{xy},R_z,\up} |R_z,m(R_z)\rangle \ne 0,\quad \text{with } m(R_z)\ne -S\nonumber\\ 
&\gamma_{\kk^{xy},R_z,\dn} |R_z,m(R_z)\rangle \ne 0,\quad \text{with } m(R_z) \ne S 
\ea

We next discuss the charge $+1$ excitation. Similarly, we show that the following state describes the charge $+1$ excitation 
\ba 
 \gamma^\dag_{(\kk^{xy},R_z),\sigma} | R_z ,m(R_z) \rangle 
\ea 
by proving that 
\ba 
[(H_0 +H_U), \gamma^\dag_{(\kk^{xy},R_z),\sigma}] |R_z,m(R_z) \rangle = E^{ele}_{\kk^{xy}}  \gamma^\dag_{(\kk^{xy},R_z),\sigma} |R_z,m(R_z) \rangle 
\label{eq:1flatband:derive_hole_exct_1}
\ea 
with $E^{ele}_{\kk^{xy}} $ the energy of the excitation state.

We now calculate
 \ba 
 &[H_U+H_0,  \gamma^\dag_{\kk^{xy},R_z,\sigma}]   \ea 
We first show 
\ba 
&[O^\dag_{n,(\qq^{xy},R_z)}, \gamma^\dag_{\pp^{xy},R_z',\sigma''}] \nonumber\\ 
=
&
\frac{1}{\sqrt{N_{xy}}}\sum_{\kk^{xy},\alpha, \sigma}  
v_{\alpha n}^*  V_{\alpha }^*(\kk^{xy}+\qq^{xy}) V_{\alpha}(\kk) [\gamma_{(\kk^{xy}+\qq^{xy},R_z),\sigma}^\dag \gamma_{(\kk^{xy},R_z),\sigma}, \gamma^\dag_{(\pp^{xy},R_z'),\sigma''}]\nonumber\\
=&\delta_{R_z,R_z'}\frac{1}{\sqrt{N_{xy}}}\sum_{\alpha}v_{\alpha n}^*
V_{\alpha }^*(\pp^{xy}+\qq^{xy})V_{\alpha }(\pp^{xy})\gamma_{(\pp^{xy} +\qq^{xy},R_z),\sigma''}
\ea 
and then
\ba 
&[O_{n,(\qq^{xy},R_z)}O^\dag_{n,(\qq^{xy},R_z)}, \gamma^\dag_{\pp^{xy},R_z',\sigma''}]\nonumber\\ 
=&\delta_{R_z,R_z'}\frac{1}{N_{xy}}\sum_{\alpha,\gamma}v_{\alpha n}^* v_{\gamma n} \sum_{\kk}
V_{\alpha }^*(\pp^{xy}+\qq^{xy})V_{\alpha }(\pp^{xy} )V_{\gamma}(\kk^{xy}+\qq^{xy})V_{\gamma}^*(\kk_{xy}) \nonumber \\ 
&[\gamma_{(\kk^{xy},R_z),\sigma}^\dag \gamma_{(\kk^{xy}+\qq^{xy},R_z),\sigma},
\gamma^\dag_{(\pp^{xy}+\qq^{xy},R_z),\sigma''}] \nonumber\\ 
=& \delta_{R_z,R_z'}\frac{1}{N_{xy}}\sum_{\alpha,\gamma}v_{\alpha n}^* v_{\gamma n} V_{\alpha }^*(\pp^{xy}+\qq^{xy})V_{\alpha }(\pp^{xy} )V_{\gamma}(\pp^{xy}+\qq^{xy})V_{\gamma}^*(\pp_{xy})
\gamma_{(\pp^{xy},R_z),\sigma''}^\dag 
\ea

Then we have 
\ba 
[H_U+H_0,  \gamma^\dag_{\pp^{xy},R_z',\sigma''}]  | R_z ,m(R_z) \rangle 
= E_{\pp^{xy}}^{ele} \gamma^\dag_{\pp^{xy},R_z',\sigma''} | R_z ,m(R_z) \rangle 
\ea 
where the energy function is
\ba 
E_{\pp^{xy}}^{ele} =\sum_{\qq^{xy},n}\frac{u_n}{2N_{xy}}\sum_{\alpha,\gamma,n}v_{\alpha n}^* v_{\gamma n} V_{\alpha }^*(\pp^{xy}+\qq^{xy})V_{\alpha }(\pp^{xy} )V_{\gamma}(\pp^{xy}+\qq^{xy})V_{\gamma}^*(\pp_{xy})
\label{eq:1fltaband:ele_exct_en}.
\ea 
Therefore, we prove $ \gamma^\dag_{\pp^{xy},R_z',\sigma''} | R_z ,m(R_z) \rangle $ describes a charge $+1$ excitation state.
% We also find that 
% where we also find (from Eq.~\ref{eq:1fltaband:hole_exct_en} and Eq.~\ref{eq:1fltaband:ele_exct_en})
% \ba 
% E_{\pp^{xy},R_z}^{ele} = E_{\pp^{xy},R_z}^{hole} 
% \ea 

We note that a valid electron excitation needs to satisfy
\ba 
\gamma^\dag_{\pp^{xy}, R_z', \sigma} |R_z,m(R_z)\rangle   \ne 0 
\label{eq:1flatband:valid_hole_exct}
\ea 
We now find the condition in which Eq.~\ref{eq:1flatband:valid_hole_exct} holds by evaluating
\ba 
\gamma^\dag_{\pp^{xy}, R_z', \sigma} |R_z,m(R_z)\rangle
\ea 
Using the commutation relation in Eq.~\ref{eq:commutation_gamma_spin} and the definition of $|R_z,m(R_z)\rangle$ in Eq.~\ref{eq:1flatband:gnd_state_spin_ladder} , we find
\ba 
&\gamma_{\kk^{xy},R_z,\up}^\dag |R_z,m(R_z)\rangle 
= - \alpha_{S,m(R_z)} (S-m(R_z))(S_{R_z}^-)^{S-m(R_z)-1} 
\gamma_{\kk^{xy},R_z,\dn}^\dag |R_z,S\rangle \nonumber\\ 
&\gamma_{\kk^{xy},R_z,\dn}^\dag |R_z,m(R_z)\rangle 
=   \alpha_{S,m(R_z)}\bigg(1-\delta_{m(R_z),-S}\bigg)(S_{R_z}^-)^{S-m(R_z)} 
\gamma_{\kk^{xy},R_z,\dn}^\dag |R_z,S\rangle
\label{eq:1flatband:charge_exct_state}
\ea 
Since 
\ba 
|R_z,S\rangle = \prod_{\kk} \gamma_{(\kk^{xy},R_z),\up}^\dag|0\rangle 
\ea 
then
\ba 
 \gamma_{(\kk^{xy},R_z),\dn}^\dag |R_z,S\rangle  \ne 0 
\ea 
We thus conclude the following states give the charge $+1$ excitation state
\ba 
&\gamma_{\kk^{xy},R_z,\up}^\dag |R_z,m(R_z)\rangle \ne 0,\quad \text{with } m(R_z)\ne S\nonumber\\ 
&\gamma_{\kk^{xy},R_z,\dn}^\dag |R_z,m(R_z)\rangle \ne 0,\quad \text{with } m(R_z) \ne -S 
\ea

From Eq.~\ref{eq:1flatband:hole_exct_state} and Eq.~\ref{eq:1flatband:charge_exct_state}, we can introduce the following charge $ \pm 1 $ excitation state
\ba 
&|m(R_z), \kk^{xy},R_z, h\rangle =  (S^-_{R_z})^{S-m(R_z)} \gamma_{\kk^{xy},R_z,\up} |R_z,S\rangle ,\quad m(R_z) \in\{-S,-S+1,...,S-1\} 
% \bigotimes_{R_z' \ne R_z} |R_z',M(R_z')\rangle
\nonumber\\ 
&|m(R_z) ,\kk^{xy},R_z,e\rangle =  (S^-_{R_z})^{S-m(R_z)} \gamma_{\kk^{xy},R_z,\dn}^\dag |R_z,S\rangle ,\quad m(R_z) \in\{-S+1,...,S-1,S\}
% \bigotimes_{R_z' \ne R_z} |R_z',M(R_z')\rangle
\label{eq:1flatband:charge_pm1_exct_state}
\ea 
with 
\ba 
&\gamma_{(\kk^{xy},R_z), \up} |R_z,m(R_z)\rangle = 
\alpha_{S,m(R_z)} \bigg(1-\delta_{m(R_z),-S}\bigg) 
|m(R_z),\kk^{xy},R_z,h\rangle  
\nonumber\\ 
&\gamma_{(\kk^{xy},R_z),\dn} |R_z,m(R_z)\rangle   = \alpha_{S,m(R_z)}(S-m(R_z))
|m(R_z)+1,\kk^{xy},R_z,h\rangle \nonumber\\ 
&\gamma_{\kk^{xy},R_z,\up}^\dag |R_z,m(R_z)\rangle 
= - \alpha_{S,m(R_z)} (S-m(R_z)) |m(R_z)+1 ,\kk^{xy},R_z,e\rangle
\nonumber\\ 
&\gamma_{\kk^{xy},R_z,\dn}^\dag |R_z,m(R_z)\rangle 
=   \alpha_{S,m(R_z)}\bigg(1-\delta_{m(R_z),-S}\bigg)|m(R_z) ,\kk^{xy},R_z,e\rangle
\label{eq:1flatband:act_gamma_on_gnd}
\ea 
where $\alpha_{S,m}$ has been introduced in Eq.~\ref{eq:1flatband:gnd_state_spin_ladder}. 

We also comment that  $|m(R_z),\kk^{xy},R_z,h/e\rangle$ is not normalized state. 
We now calculate the norm of $|m(R_z),\kk^{xy},R_z,h\rangle$
\ba 
&\langle m(R_z),\kk^{xy},R_z,h|m(R_z),\kk^{xy},R_z,h\rangle = \langle R_z,S| \gamma_{\kk^{xy},R_z,\up}^\dag (S_{R_z}^+)^{S-m(R_z)}(S_{R_z}^-)^{S-m(R_z)}
\gamma_{\kk^{xy},R_z,\up}|R_z,S\rangle \nonumber\\ 
=& \langle R_z,S| \gamma_{\kk^{xy},R_z,\up}^\dag (S_{R_z}^+)^{S-m(R_z)-1}(2S^z_{R_z})(S_{R_z}^-)^{S-m(R_z)-1}
\gamma_{\kk^{xy},R_z,\up}|R_z,S\rangle\nonumber\\ 
&
+  \langle R_z,S| \gamma_{\kk^{xy},R_z,\up}^\dag (S_{R_z}^+)^{S-m(R_z)-1}S_{R_z}^-S_{R_z}^+(S_{R_z}^-)^{S-m(R_z)-1}
\gamma_{\kk^{xy},R_z,\up}|R_z,S\rangle
\nonumber\\ 
=& \langle R_z,S| \gamma_{\kk^{xy},R_z,\up}^\dag (S_{R_z}^+)^{S-m(R_z)-1}(2S^z_{R_z})(S_{R_z}^-)^{S-m(R_z)-1}
\gamma_{\kk^{xy},R_z,\up}|R_z,S\rangle\nonumber\\ 
&
+  \langle R_z,S| \gamma_{\kk^{xy},R_z,\up}^\dag (S_{R_z}^+)^{S-m(R_z)-2}(2S^z_{R_z}) S_{R_z}^+(S_{R_z}^-)^{S-m(R_z)-1}
\gamma_{\kk^{xy},R_z,\up}|R_z,S\rangle \nonumber\\ 
&
+  \langle R_z,S| \gamma_{\kk^{xy},R_z,\up}^\dag (S_{R_z}^+)^{S-m(R_z)-2} (S_{R_z}^-)(S_{R_z}^+)^2(S_{R_z}^-)^{S-m(R_z)-1}
\gamma_{\kk^{xy},R_z,\up}|R_z,S\rangle \nonumber\\ 
=& \sum_{m=1}^{S-m(R_z)}
\langle R_z,S |\gamma_{\kk^{xy},R_z,\up}^\dag (S_{R_z}^+)^{S-m(R_z)-m}2(S_{R_z}^z)(S_{R_z}^{+})^{m-1}(S_{R_z}^-)^{S-m(R_z)-1}\gamma_{\kk^{xy},R_z,\up}  |R_z,S\rangle 
\nonumber\\ 
&+\langle R_z,S |\gamma_{\kk^{xy},R_z,\up}^\dag S_{R_z}^- (S_{R_z}^+)^{S-m(R_z)} (S_{R_z}^{-})^{S-m(R_z)-1}\gamma_{\kk^{xy},R_z,\up}|R_z,S\rangle \nonumber\\ 
=&\sum_{m=1}^{S-m(R_z)}(2m(R_z)+2m-1)\langle m(R_z)+1,\kk^{xy},R_z,h
|m(R_z)+1,\kk^{xy},R_z,h\rangle \nonumber\\ 
=& (S^2-m(R_z)^2)\langle m(R_z)+1,\kk^{xy},R_z,h
|m(R_z)+1,\kk^{xy},R_z,h\rangle 
\ea 
We note that
\ba 
\langle S, \kk^{xy},R_z,h|S,\kk^{xy},R_z,h\rangle = \langle R_z,S| \gamma_{\kk^{xy},R_z,\up}^\dag 
\gamma_{\kk^{xy},R_z,\up}|R_z,S\rangle =1
\ea 
Thus, we define the norm as 
\ba 
&\beta_{m(R_z)} = \langle m(R_z),\kk^{xy},R_z,h|m(R_z),\kk^{xy},R_z,h\rangle = \prod_{m=m(R_z)}^{S-1}(S^2-m^2),\quad m(R_z)<S \nonumber\\
&\beta_S = 1 
\label{eq:norm_hole_exct}
\ea 

We then calculate the norm of $|M,\kk^{xy},R_z,e\rangle$
\ba 
&\langle m(R_z),\kk^{xy},R_z,e|m(R_z),\kk^{xy},R_z,e\rangle = \langle R_z,S| \gamma_{\kk^{xy},R_z,\dn} (S_{R_z}^+)^{S-m(R_z)}(S_{R_z}^-)^{S-m(R_z)}
\gamma_{\kk^{xy},R_z,\dn}^\dag|R_z,S\rangle \nonumber\\ 
=& \sum_{m=1}^{S-m(R_z)}
\langle R_z,S |\gamma_{\kk^{xy},R_z,\dn}(S_{R_z}^+)^{S-m(R_z)-m}2(S_{R_z}^z)(S_{R_z}^{+})^{m-1}(S_{R_z}^-)^{S-m(R_z)-1}\gamma_{\kk^{xy},R_z,\dn}^\dag  |R_z,S\rangle 
\nonumber\\ 
&+\langle R_z,S | \gamma_{\kk^{xy},R_z,\dn} S_{R_z}^- (S_{R_z}^+)^{S-m(R_z)} (S_{R_z}^{-})^{S-m(R_z)-1}\gamma_{\kk^{xy},R_z,\dn}^\dag |R_z,S\rangle \nonumber\\ 
=&\sum_{m=1}^{S-m(R_z)}(2m(R_z)+2m-1)\langle m(R_z)+1,\kk^{xy},R_z,e
|m(R_z)+1,\kk^{xy},R_z,e\rangle \nonumber\\ 
=& (S^2-m(R_z)^2)\langle m(R_z)+1,\kk^{xy},R_z,e
|m(R_z)+1,\kk^{xy},R_z,e\rangle 
\ea 
Since
\ba 
\langle S,\kk^{xy},R_z,e|S,\kk^{xy},R_z,e\rangle=\langle R_z,S| \gamma_{\kk^{xy},R_z,\dn} \gamma_{\kk^{xy},R_z,\dn}^\dag |R_z,S\rangle = 1
\ea 
we have 
\ba 
\langle m(R_z),\kk^{xy},R_z,e|m(R_z),\kk^{xy},R_z,e\rangle = \beta_{m(R_z)}
\label{eq:norm_ele_exct}
\ea 
For future reference, we also note that by acting $S_{R_z}^+$, we could generate $|m(R_z)+1,\kk^{xy},R_z,e/h\rangle$ state from $|m(R_z),\kk^{xy},R_z,e/h\rangle$ state. To observe this, we show
\ba 
&S_{R_z}^+|m(R_z),\kk^{xy},R_z,h\rangle = S^+_{R_z} (S^-_{R_z})^{S-m(R_z)} \gamma_{\kk^{xy},R_z,\up} |R_z,S\rangle \nonumber\\
 =& \bigg( 2S_{R_z}^z (S^{-}_{R_z})^{S-m(R_z)-1} + (S^{-}_{R_z})^{1}S^+_{R_z}(S^{-}_{R_z})^{S-m(R_z)-1}\bigg)\gamma_{\kk^{xy},R_z,\up} |R_z,S\rangle \nonumber\\
 =&\sum_{m=1}^{S-m(R_z)} 2(m(R_z)+m-\frac{1}{2})(S^{-}_{R_z})^{S-m(R_z)-1}\gamma_{\kk^{xy},R_z,\up} |R_z,S\rangle 
 + (S_{R_z}^{-})^{S-m(R_z)}S_{R_z}^+ \gamma_{\kk^{xy},R_z,\up} |R_z,S\rangle 
 \nonumber\\ 
 =&(S^2-m(R_z)^2)
|m(R_z)+1,\kk^{xy},R_z,h\rangle 
% \nonumber\\ 
% \nonumber\\ 
% &|m(R_z)+1,\kk^{xy},R_z,h\rangle = \frac{2}{S^2-M^2}S_{R_z}^+|m(R_z),\kk^{xy},R_z,h\rangle 
\ea 
Similarly,
\ba 
&S_{R_z}^+|m(R_z),\kk^{xy},R_z,e\rangle  =S^+_{R_z} (S^-_{R_z})^{S-m(R_z)} \gamma_{\kk^{xy},R_z,\dn}^\dag |R_z,S\rangle \nonumber\\ 
=&\sum_{m=1}^{S-m(R_z)} 2(m(R_z)+m-\frac{1}{2})(S^{-}_{R_z})^{S-m(R_z)-1}\gamma_{\kk^{xy},R_z,\dn}^\dag |R_z,S\rangle 
 + (S_{R_z}^{-})^{S-m(R_z)} S_{R_z}^+ \gamma_{\kk^{xy},R_z,\dn}^\dag |R_z,S\rangle 
 \nonumber\\ 
 =&(S^2-m(R_z)^2)
|m(R_z)+1,\kk^{xy},R_z,e\rangle 
\ea 
In summary, we have 
\ba 
&|m(R_z)+1,\kk^{xy},R_z,h\rangle = \frac{1}{S^2-m(R_z)^2}S_{R_z}^+|m(R_z),\kk^{xy},R_z,h\rangle \nonumber\\ 
& |m(R_z)+1,\kk^{xy},R_z,e\rangle = \frac{1}{S^2-m(R_z)^2}S_{R_z}^+|m(R_z),\kk^{xy},R_z,e\rangle
\label{eq:1flatband:m_to_m+1_state}
\ea

\subsection{$z$ direction hopping and out-of-plane antiferromagnetism}
\label{sec:app:3d_afm}
In the previous section, we discuss the 2D flat bands and the resulting in-plane ferromagnetic state. In the real system, even though the flat band is atomic along $z$ direction, there could still be a weak hopping along $z$ direction which will then generate out-of-plane antiferromagnetism.

We now introduce a weak $z$-direction hopping to the flat bands, such that different layers of the system are no longer decoupled. The additional $z$ direction hopping we introduced (written in the original basis $d$) is
\ba
H_z = \sum_{\langle \RR,\RR'\rangle_z,\sigma  } t_{z,\alpha\gamma} d_{\RR,\alpha,\sigma}^\dag d_{\RR',
\gamma,\sigma} = \sum_{\kk,\sigma  } t_{z,\alpha\gamma} \cos(k_z)d_{\kk,\alpha,\sigma}^\dag d_{\kk,\gamma,\sigma} 
\ea 
where $\langle \RR,\RR'\rangle_z $ denotes nearest neighbors along $z$ directions and $t_{z,\alpha\gamma}$ is the hopping strength. 

Transforming to band basis and projecting to the flat bands, we have 
\ba 
H_{z,\gamma}= &PH_zP  \nonumber\\
 = &\sum_{\kk}\bigg[\sum_{\alpha\gamma}t_{z,\alpha\gamma}V_{\alpha}^*(\kk^{xy})V_{\gamma}(\kk^{xy})\bigg] \cos(k_z)\gamma_{\kk,\sigma}^\dag \gamma_{\kk,\sigma} \nonumber\\ 
=&\sum_{\kk}f_{z,\kk^{xy}}\cos(k_z)\gamma_{\kk,\sigma}^\dag \gamma_{\kk,\sigma} 
\label{eq:z_direction_hopping}
\ea 
where we utilize the fact that the wavefunction of flat bands $V_{\alpha}^*$ only depends on $\kk^{xy}$ and introduce 
\ba 
f_{z,\kk^{xy}} = \sum_{\alpha\gamma}t_{z,\alpha\gamma}V_{\alpha}^*(\kk^{xy})V_{\gamma}(\kk^{xy})
\ea

We next treat the model with Schrieffer-Wolff transformation. We take 
\ba 
H =&h_0 +h_1 \nonumber\\ 
h_0 =&\sum_{R_z}H_{R_z}= \sum_{R_z} \sum_{n,\qq^{xy}} \frac{u_n}{2}O_{n,(\qq^{xy},R_z)}O_{n,(\qq^{xy},R_z)}^\dag \nonumber\\ 
h_1 = &  h_{HL} +h_{LH} \nonumber\\ 
h_{HL} = &P_H \bigg[\sum_{\kk^{xy},\langle R_z,R_z'\rangle }  f_{z,\kk^{xy}} \gamma_{\kk^{xy},R_z,\sigma}^\dag \gamma_{\kk^{xy},R_z',\sigma} \bigg] P_L\nonumber\\ 
h_{LH} =& P_L \bigg[\sum_{\kk^{xy},\langle R_z,R_z'\rangle }  f_{z,\kk^{xy}} \gamma_{\kk^{xy},R_z,\sigma}^\dag \gamma_{\kk^{xy},R_z',\sigma} \bigg] P_H
\ea 
with $h_0$ is the unperturbed Hamiltonian and $h_1$ is the perturbation (Eq.~\ref{eq:z_direction_hopping}). We have also separated the perturbation term into two parts: $h_{HL}$ and $h_{LH}$ (similarly as Eq.~\ref{eq:SW_transform_H1_two_part}). The projection operators are defined as 
\ba 
&P_L = \sum_{\{m\} } |m\rangle \langle m| \nonumber\\ 
& P_H = \mathbb{I} -P_L
\label{eq:z_hop_sw_transf_proj}
\ea 
where  $P_L$ is the projection operator of the low-energy space formed by the ground states $|m\rangle$ of $h_0$ (Eq.~\ref{eq:1flatband:gnd_basis}) and $P_H$ is the projection operator of the high-energy states (excitation states of $h_0$, Eq.~\ref{eq:1flatband:act_gamma_on_gnd}). Via Schrieffer-Wolff transformation, we will show that, the additional hopping along $z$ direction ($h_1$) lifts the degeneracy of the ground states (Eq.~\ref{eq:1flatband:gnd_basis}) of the original Hamiltonian $h_0$.

In practice, we note that the $z$-direction hopping $h_1$ could map a low-energy state to a high-energy state by creating an electron excitation at layer $R_z$ and a hole excitation at layer $R_z'$. $h_1$ could also map a high-energy state to a low-energy state by reversing this procedure. To observe the effect of $h_1$, we consider each term of $h_1$ separately 
\ba 
&h_{\kk^{xy},R_z,R_z',\sigma}=f_{z,\kk^{xy}}\gamma_{\kk^{xy},R_z,\sigma}^\dag \gamma_{\kk^{xy},R_z',\sigma} \nonumber\\ 
&
h_1 = \sum_{\kk^{xy},\langle R_z,R_z'\rangle }h_{\kk^{xy},R_z,R_z',\sigma}
\label{eq:app:h1_def_decompose_each_lar}
\ea 
% Then acting $h_{\kk^{xy},R_z,R_z'}$ on the ground state will creates a charge $-1$ states at layer $R_z'$ and a charge $+1$ state at layer $R_z$. 

We mention that, for each layer, $h_1$ effectively created a charge $\pm 1$ excitation. More specifically, 
acting $h_1$ on the ground state will create a charge $-1$ state at one layer and a charge $+1$ state at another layer. 
We comment that $h_1$ will not create a Goldstone excitation. A Goldstone excitation for the in-plane ferromagnetic state requires the particle-hole excitation at the same layer which takes the form of $\gamma_{\kk^{xy},R_z,\sigma}^\dag \gamma_{\kk^{xy},R_z,\sigma'}$. Such types of excitation will not be created by simply acting $h_1$ on the ground states.

Since we focus on the case of nearest-neighbor hopping along $z$ direction, we have $R_z =R_z' \pm 1$. We first consider the case with $R_z =R_z'+1$. By acting $h_{\kk^{xy}, R_z,R_z+1,\sigma}$ on the ground state and using Eq.~\ref{eq:1flatband:act_gamma_on_gnd}, we could obtain
\ba 
&h_{\kk^{xy},R_z,R_z+1,\up}|m\rangle  \nonumber\\ 
=&\bigotimes_{R_z''=1}^{R_z-1}|R_z'', m(R_z'')\rangle
\bigg[ f_{z,\kk^{xy}}\gamma_{\kk^{xy},R_z,\up}^\dag \gamma_{\kk^{xy},R_z+1,\up} |R_z,m(R_z)\rangle |R_z+1,m(R_z+1)\rangle \bigg] 
\bigotimes_{R_z''=R_z+2}^{L_z}|R_z'', m(R_z'')\rangle
\nonumber\\ 
= &\bigotimes_{R_z''=1}^{R_z-1}|R_z'', M(R_z'')\rangle\bigg[ -\alpha_{S,m(R_z)} \alpha_{S,m(R_z+1)}
(-1)^{N_{xy}}f_{z,\kk^{xy}}(1-\delta_{ m(R_z+1),-S})(S-m(R_z))  \bigg]
\nonumber\\ 
&\bigg( 
|m(R_z)+1,R_z,\kk^{xy},e \rangle 
|m(R_z+1),R_z+1,\kk^{xy},h\rangle \bigg) \bigotimes_{R_z''=R_z+2}^{L_z}|R_z'', M(R_z'')\rangle|\Phi_a\rangle 
\nonumber \\ 
\nonumber\\ 
&h_{\kk^{xy},R_z,R_z+1,\dn}|m\rangle  \nonumber\\ 
=&\bigotimes_{R_z''=1}^{R_z-1}|R_z'', m(R_z'')\rangle
\bigg[ f_{z,\kk^{xy}}\gamma_{\kk^{xy},R_z,\dn}^\dag \gamma_{\kk^{xy},R_z+1,\dn} |R_z,m(R_z)\rangle |R_z+1,m(R_z+1)\rangle \bigg] 
\bigotimes_{R_z''=R_z+2}^{L_z}|R_z'', m(R_z'')\rangle\nonumber\\ 
= &\bigotimes_{R_z''=1}^{R_z-1}|R_z'', m(R_z'')\rangle\bigg[ \alpha_{S,m(R_z)} \alpha_{S,m(R_z+1)}
(-1)^{N_{xy}}f_{z,\kk^{xy}}(S-m(R_z+1))(1-\delta_{m(R_z),-S)})  \bigg] 
\nonumber\\ 
&\bigg( 
|m(R_z),R_z,\kk^{xy},e \rangle 
|m(R_z+1)+1,R_z+1,\kk^{xy},h\rangle \bigg) \bigotimes_{R_z''=R_z+2}^{L_z}|R_z'', m(R_z'')\rangle|\Phi_a\rangle 
\label{eq:1flatband:act_h1_gg_on_low_1}
\ea 
where the additional $(-1)^{N_{xy}}$ factor comes from the fermion anticommutation relation. Similarly, for the case of $R_z =R_z' -1 $
\ba 
&h_{\kk^{xy},R_z,R_z-1,\up}|m\rangle  \nonumber\\ 
= &\bigotimes_{R_z''=1}^{R_z-1}|R_z'', m(R_z'')\rangle\bigg[ 
- \alpha_{S,m(R_z)} \alpha_{S,m(R_z-1)}(-1)^{N_{xy}+1}f_{z,\kk^{xy}}(1-\delta_{ m(R_z'),-S})(S-m(R_z))
\bigg] 
\nonumber\\ 
&\bigg( 
|m(R_z-1),R_z-1,\kk^{xy},h\rangle
|m(R_z)+1,R_z,\kk^{xy},e \rangle 
 \bigg) \bigotimes_{R_z''=R_z+1}^{L_z}|R_z'', M(R_z'')\rangle|\Phi_a\rangle 
\nonumber \\ 
\nonumber\\ 
&h_{\kk^{xy},R_z,R_z-1,\dn}|m\rangle  \nonumber\\ 
= &\bigotimes_{R_z''=1}^{R_z-2}|R_z'', M(R_z'')\rangle
\bigg[ 
\alpha_{S,m(R_z)} \alpha_{S,m(R_z-1)}
(-1)^{N_{xy}+1}f_{z,\kk^{xy}}(S-m(R_z-1))(1-\delta_{m(R_z),-S)}) \bigg] \nonumber\\
& \bigg( 
|m(R_z-1)+1,R_z-1,\kk^{xy},h \rangle 
|m(R_z),R_z,\kk^{xy},e\rangle \bigg) 
\nonumber\\ 
&\bigotimes_{R_z''=R_z+1}^{L_z}|R_z'', m(R_z'')\rangle|\Phi_a\rangle 
\label{eq:1flatband:act_h1_gg_on_low_2}
\ea 
where the additional $(-1)^{N_{xy}+1}$ factor comes from the fermion anticommutation relation. 
% In a more compact format, we define 
We can rewrite Eq.~\ref{eq:1flatband:act_h1_gg_on_low_1} and Eq.~\ref{eq:1flatband:act_h1_gg_on_low_1} in a simpler formula by introducing 
\ba 
&|m,\kk^{xy}, R_z,R_z+1\rangle  = \bigotimes_{R_z''=1}^{R_z-1}|R_z'', m(R_z'')\rangle
\bigg( 
|m(R_z),R_z,\kk^{xy},e \rangle 
|m(R_z+1),R_z+1,\kk^{xy},h\rangle \bigg) \bigotimes_{R_z''=R_z+2}^{L_z}|R_z'', m(R_z'')\rangle
\nonumber\\ 
&|m,\kk^{xy}, R_z,R_z-1\rangle =\bigotimes_{R_z''=1}^{R_z-2}|R_z'', m(R_z'')\rangle \bigg( 
|m(R_z-1),R_z-1,\kk^{xy},h \rangle 
|m(R_z),R_z,\kk^{xy},e\rangle \bigg) \bigotimes_{R_z''=R_z+1}^{L_z}|R_z'', m(R_z'')\rangle
\ea 
where $|m,\kk^{xy},R_z,R_z'\rangle$ denotes the states obtained by creating a charge $+1$ excitation with momentum $\kk^{xy}$ at layer $R_z$ and a charge $-1$ excitation with momentum $\kk^{xy}$  at layer $R_z'$ on top of the ground state $|m\rangle$. 
% whose norm is 
% \ba 
% \ea 

Then we have 
\ba 
h_{\kk^{xy},R_z,R_z+1,\up} |m\rangle   =&-\alpha_{S,m(R_z)} \alpha_{S,m(R_z+1)}(-1)^{N_{xy}}f_{z,\kk^{xy}}(1-\delta_{ m(R_z+1),-S})(S-m(R_z)) \nonumber \\
&
\frac{S^+_{R_z}}{S^2-m(R_z)^2}|m, \kk^{xy},R_z,R_z+1\rangle   \nonumber\\ 
h_{\kk^{xy},R_z,R_z+1,\dn} |m\rangle 
=&\alpha_{S,m(R_z)} \alpha_{S,m(R_z+1)}(-1)^{N_{xy}}f_{z,\kk^{xy}}(S-m(R_z+1))(1-\delta_{m(R_z),-S)}) \nonumber\\ 
&\frac{S^+_{R_z+1}}{S^2-m(R_z+1)^2} |m,\kk^{xy},R_z,R_z+1\rangle  \nonumber\\ 
h_{\kk^{xy},R_z,R_z-1,\up} |m\rangle  
=&-\alpha_{S,m(R_z)} \alpha_{S,m(R_z-1)}(-1)^{N_{xy}+1}f_{z,\kk^{xy}}(1-\delta_{ m(R_z-1),-S})(S-m(R_z)) \nonumber\\ 
&\frac{S^+_{R_z}}{S^2-m(R_z)^2}|m,\kk^{xy},R_z,R_z-1\rangle  \nonumber\\ 
h_{\kk^{xy},R_z,R_z-1,\dn} |m\rangle
=&\alpha_{S,m(R_z)} \alpha_{S,m(R_z-1)}(-1)^{N_{xy}+1}f_{z,\kk^{xy}}(S-m(R_z-1))(1-\delta_{m(R_z),-S}) \nonumber\\ 
&\frac{S^+_{R_z-1}}{S^2-m(R_z-1)^2}|m,\kk^{xy},R_z,R_z-1\rangle
\label{eq:1flatband:act_h1_on_low_2} 
\ea 
where we have used the relation given in Eq.~\ref{eq:1flatband:m_to_m+1_state}. 

We then use the projection operator defined in Eq.~\ref{eq:z_hop_sw_transf_proj} and find (from Eq.~\ref{eq:app:h1_def_decompose_each_lar} and Eq.~\ref{eq:1flatband:act_h1_on_low_2})
\ba 
P_H h_1P_L =& \sum_{\kk^{xy}, M,R_z} 
\bigg\{ 
-
\alpha_{S,m(R_z)} \alpha_{S,m(R_z+1)}
(-1)^{N_{xy}}f_{z,\kk^{xy}}(1-\delta_{ m(R_z+1),-S})(S-M(R_z))  
\frac{S^+_{R_z}}{S^2-m(R_z)^2}|m, \kk^{xy},R_z,R_z+1\rangle  \langle m| \nonumber\\ 
&
+\alpha_{S,m(R_z)} \alpha_{S,m(R_z+1)}
(-1)^{N_{xy}}f_{z,\kk^{xy}}(S-m(R_z+1)) (1-\delta_{M(R_z),-S)})  \frac{S^+_{R_z+1}}{S^2-m(R_z+1)^2} |m,\kk^{xy},R_z,R_z+1\rangle  \langle m| \nonumber\\ 
&
-
\alpha_{S,m(R_z)} \alpha_{S,m(R_z-1)}
(-1)^{N_{xy}+1}f_{z,\kk^{xy}}(1-\delta_{ m(R_z-1),-S})(S-m(R_z)) \frac{S^+_{R_z}}{S^2-m(R_z)^2}|m,\kk^{xy},R_z,R_z-1\rangle   \langle m| \nonumber\\
 &+
 \alpha_{S,m(R_z)} \alpha_{S,m(R_z-1)}
 (-1)^{N_{xy}+1}f_{z,\kk^{xy}}(S-m(R_z-1))  (1-\delta_{m(R_z),-S}) \frac{S^+_{R_z-1}}{S^2-m(R_z-1)^2}|m,\kk^{xy},R_z,R_z-1\rangle  \langle m| \bigg\}
 \label{eq:app:ph_h1_pl}
\ea 
We can also find $P_H h_1P_L$ via Hermitian conjugation
\ba 
P_L h_1 P_H = (P_Hh_1P_L)^\dag 
\ea

Therefore, we have rewritten the perturbation term in a similar form as Eq.~\ref{eq:SW_transform_H1_def} 
\ba 
P_Hh_1P_L +P_Lh_1P_H 
=  \sum_{ij}V_{ij}|L,i\rangle \langle H,j| +\text{h.c.}
\label{eq:php_exp}
\ea 
where $|L,i\rangle$ denotes the ground states of $h_0$, and $|H,j\rangle$ denotes the excitation state 
\ba 
&|L,i\rangle\quad  \sim \quad |m\rangle \nonumber\\ 
&|H,j\rangle \sim  \quad S_{R_z}^+|m,\kk^{xy},R_z,R_z\pm 1\rangle ,\quad 
S_{R_z\pm 1 }^+|m,\kk^{xy},R_z,R_z\pm 1\rangle 
\ea 
% with $\beta_{m(R_z)}$ is the normalization factor to ensure $\langle H,j|H,j\rangle = 1$ as introduced in Eq.~\ref{eq:norm_hole_exct} and Eq.~\ref{eq:norm_ele_exct}. 
The coefficient $V_{ij}$ in Eq.~\ref{eq:php_exp} can be obtained from Eq.~\ref{eq:app:ph_h1_pl}.  

To perform the Schrieffer-Wolff transformation, we also need to obtain the energy of excitation state $|H,j\rangle$. We find
\ba 
h_0S_{R_z}^+
|m,\kk^{xy}, R_z,R_z+1\rangle  
= &\bigotimes_{R_z''=1}^{R_z-1}|R_z'', m(R_z'')\rangle
\bigg( H_{R_z}S_{R_z}^+
|m(R_z),R_z,\kk^{xy},e \rangle 
|m(R_z+1),R_z+1,\kk^{xy},h\rangle\nonumber\\ 
&
+ 
S_{R_z}^+|m(R_z),R_z,\kk^{xy},e \rangle H_{R_z+1}
|m(R_z+1),R_z+1,\kk^{xy},h\rangle
\bigg) 
\bigotimes_{R_z''=R_z+2}^{L_z}|R_z'', m(R_z'')\rangle\nonumber\\
=&(E_{\kk^{xy}}^{hole} + E_{\kk^{xy}}^{ele})S_{R_z}^+|m,\kk^{xy}, R_z,R_z+1\rangle
\nonumber\\ 
h_0S_{R_z}^+|m,\kk^{xy}, R_z,R_z-1\rangle
=&\bigotimes_{R_z''=1}^{R_z-2}|R_z'', m(R_z'')\rangle \bigg(H_{R_z-1}  
|m(R_z-1),R_z-1,\kk^{xy},h \rangle 
S_{R_z}^+|m(R_z),R_z,\kk^{xy},e\rangle 
\nonumber\\ 
&+ 
|m(R_z-1),R_z-1,\kk^{xy},h \rangle 
H_{R_z} S_{R_z}^+
|m(R_z),R_z,\kk^{xy},e\rangle 
\bigg) \bigotimes_{R_z''=R_z+1}^{L_z}|R_z'', m(R_z'')\rangle\sum_{R_z} H_{R_z} \nonumber\\ 
=&(E_{\kk^{xy}}^{hole} + E_{\kk^{xy}}^{ele})S_{R_z}^+|m,\kk^{xy}, R_z,R_z-1\rangle 
\nonumber\\ 
h_0S_{R_z+1}^+
|m,\kk^{xy}, R_z,R_z+1\rangle  
= &\bigotimes_{R_z''=1}^{R_z-1}|R_z'', m(R_z'')\rangle
\bigg( H_{R_z}
|m(R_z),R_z,\kk^{xy},e \rangle S_{R_z+1}^+
|m(R_z+1),R_z+1,\kk^{xy},h\rangle\nonumber\\ 
&
+ 
|m(R_z),R_z,\kk^{xy},e \rangle H_{R_z+1}S_{R_z+1}^+
|m(R_z+1),R_z+1,\kk^{xy},h\rangle
\bigg) 
\bigotimes_{R_z''=R_z+2}^{L_z}|R_z'', m(R_z'')\rangle\nonumber\\
=&(E_{\kk^{xy}}^{hole} + E_{\kk^{xy}}^{ele})S_{R_z+1}^+|m,\kk^{xy}, R_z,R_z+1\rangle
\nonumber\\ 
h_0S_{R_z-1}^+|m,\kk^{xy}, R_z,R_z-1\rangle
=&\bigotimes_{R_z''=1}^{R_z-2}|R_z'', m(R_z'')\rangle \bigg(H_{R_z-1}  S_{R_z-1}^+
|m(R_z-1),R_z-1,\kk^{xy},h \rangle 
|m(R_z),R_z,\kk^{xy},e\rangle 
\nonumber\\ 
&+ S_{R_z-1}^+
|m(R_z-1),R_z-1,\kk^{xy},h \rangle 
H_{R_z} 
|m(R_z),R_z,\kk^{xy},e\rangle 
\bigg) \bigotimes_{R_z''=R_z+1}^{L_z}|R_z'', m(R_z'')\rangle\sum_{R_z} H_{R_z} \nonumber\\ 
=&(E_{\kk^{xy}}^{hole} + E_{\kk^{xy}}^{ele})S_{R_z-1}^+|m,\kk^{xy}, R_z,R_z-1\rangle \nonumber\\ 
\ea 
where the charge $\pm 1$ exctiation energy is introduced in Eq.~\ref{eq:1fltaband:hole_exct_en} and Eq.~\ref{eq:1fltaband:ele_exct_en}.
% , and we have utilize Eq.~\ref{eq:1flatband:m_to_m+1_state}.

In summary, in the current problems, the high-energy states created by acting $h_1$ on the ground states of $h_0$ are 
\ba 
&S_{R_z}^+|m,\kk^{xy},R_z,R_z+1\rangle  ,\quad 
S_{R_z+1}^+|m,\kk^{xy},R_z,R_z+1\rangle   \nonumber \\
&S_{R_z}^+|m,\kk^{xy},R_z,R_z-1\rangle  ,\quad 
S_{R_z-1}^+|m,\kk^{xy},R_z,R_z-1\rangle  
\label{eq:high_energy_space_sw}
\ea 
with the excitation energies being
\ba 
&E_{\kk^{xy}}^{exct}
% ,\quad E_{\kk^{xy}}^{exct}, \nonumber\\ 
% &E_{\kk^{xy}}^{exct},\quad E_{\kk^{xy}}^{exct}
\ea 
for all four types of high-energy states, where we have also defined
\ba 
E_{\kk^{xy}}^{exct} = E_{\kk^{xy}}^{hole} + E_{\kk^{xy}}^{ele}
\ea 
We also mention that, in Eq.~\ref{eq:high_energy_space_sw}, we utilize the definition of Eq.~\ref{eq:1flatband:charge_pm1_exct_state}, such that all the excitation state can be obtained by acting $S^+_{R_z}, S_{R_z\pm 1}^+$ on the $|m,\kk^{xy},R_z,R_z\pm 1\rangle$. In Eq.~\ref{eq:1flatband:charge_pm1_exct_state}, we use $\gamma_{\kk^{xy},R_z,\up}$ operators and $\gamma_{\kk^{xy},R_z,\dn}^\dag$ operators to define the charge $\pm 1$ excitation state. 
We could also adapt other definitions (for example, we could use $\gamma_{\kk^{xy},R_z,\dn}$ and $\gamma_{\kk^{xy},R_z,\up}^\dag$ operators), such that all the excitation states can be written in different formats. 

However, we comment that the high-energy states (Eq.~\ref{eq:high_energy_space_sw}) are not normalized. We can evaluate its norm by using (Eq.~\ref{eq:1flatband:m_to_m+1_state}, Eq.~\ref{eq:norm_hole_exct} and Eq.~\ref{eq:norm_ele_exct})
\ba 
&\langle m,\kk^{xy},R_z,R_z+1|(S_{R_z}^+)^\dag S_{R_z}^+ |m,\kk^{xy},R_z,R_z+1\rangle \nonumber\\ 
=& \langle m(R_z),R_z,\kk^{xy},e |S_{R_z}^-S_{R_z}^+|m(R_z),R_z,\kk^{xy},e\rangle 
\langle m(R_z+1),R_z+1,\kk^{xy},h|m(R_z+1),R_z+1,\kk^{xy},h\rangle \nonumber\\ 
=&[S^2-m(R_z)^2]^2 \langle m(R_z)+1,R_z,\kk^{xy},e |m(R_z)+1,R_z,\kk^{xy},e\rangle 
\beta_{m(R_z+1)} \nonumber\\ 
=&[S^2-m(R_z)^2]^2\beta_{m(R_z)+1} \beta_{m(R_z+1)}  
= \eta_{m(R_z),m(R_z+1)}
\nonumber\\
\nonumber\\ 
&\langle m,\kk^{xy},R_z,R_z+1|(S_{R_z+1}^+)^\dag S_{R_z+1}^+ |m,\kk^{xy},R_z,R_z+1\rangle = [S^2-m(R_z+1)^2]\beta_{m(R_z+1)+1}\beta_{m(R_z)}
= \eta_{m(R_z+1),m(R_z)}
\nonumber\\ 
&\langle m,\kk^{xy},R_z,R_z-1|(S_{R_z}^+)^\dag S_{R_z}^+ |m,\kk^{xy},R_z,R_z-1\rangle = [S^2-m(R_z)^2]\beta_{m(R_z)+1}\beta_{m(R_z-1)}
= \eta_{m(R_z),m(R_z-1)}
\nonumber\\ 
&\langle m,\kk^{xy},R_z,R_z-1|(S_{R_z-1}^+)^\dag S_{R_z-1}^+ |m,\kk^{xy},R_z,R_z-1\rangle = [S^2-m(R_z-1)^2]\beta_{m(R_z-1)+1}\beta_{m(R_z)}
= \eta_{m(R_z-1),m(R_z)}
\ea 
where we have introduced the normalization factor
 \ba 
 \eta_{M,M'} = [S^2-M^2]^2 \beta_{M+1} \beta_{M'}
 \ea

Now we can perform Schrieffer-Wolff transformation (Eq.~\ref{eq:app:sw_eff_h})
\ba
H_{SW} = 
\frac{1}{2} \sum_{i,j,m }V_{im}V_{jm}^*\left( \frac{1}{E_{L,i}-E_{H,m}} + \frac{1}{E_{L,j} -E_{H,m}}
\right)  |L,i\rangle \langle L,j| 
\ea 

We first comment that the Schrieffer-Wolff transformation describes the virtual procedure from low-energy state $|L,i\rangle$ to high-energy state $|H,m\rangle $ and then back to low-energy state $|L,j\rangle$. Therefore, the following virtual procedures could be generated and contribute to the $H_{SW}$
\ba 
&|L,j\rangle \rightarrow |H,m\rangle \rightarrow |L,i\rangle \nonumber\\ 
\nonumber\\
I:&|m\rangle \rightarrow \frac{1}{\sqrt{\eta_{m(R_z),m(R_z')}}}S_{R_z}^+ |m,\kk^{xy},R_z,R_z'\rangle  \rightarrow |m\rangle \nonumber\\ 
II:&|m\rangle \rightarrow \frac{1}{\sqrt{\eta_{m(R_z'),m(R_z)}}}S_{R_z'}^+ |m,\kk^{xy},R_z,R_z'\rangle  \rightarrow |m\rangle \nonumber\\ 
III:&|m\rangle \rightarrow \frac{1}{\sqrt{\eta_{m(R_z),m(R_z')}}}S_{R_z}^+ |m,\kk^{xy},R_z,R_z'\rangle \rightarrow S_{R_z'}^{-}S_{R_z}^+ |m\rangle \nonumber\\ 
IV:&|m\rangle \rightarrow \frac{1}{\sqrt{\eta_{m(R_z'),m(R_z)}}}S_{R_z'}^+ |m,\kk^{xy},R_z,R_z'\rangle  \rightarrow S_{R_z}^{-}S_{R_z'}^+ |m\rangle \nonumber\\ 
&\text{with  }R_z' = R_z \pm 1 
\ea 

We now evaluate $H_{SW}$ with contributions from each virtual procedure. 

\subsubsection{Contribution from I}
We take 
\ba 
&|L,j\rangle = |m\rangle ,\quad |L,i\rangle =  |m\rangle \nonumber\\ 
&|H,n\rangle = \frac{1}{\sqrt{\eta_{m(R_z),m(R_z')}}}S_{R_z}^+ |m,\kk^{xy},R_z,R_z'\rangle  
\ea 
The corresponding $V_{in}$ (from Eq.~\ref{eq:app:ph_h1_pl} and Eq.~\ref{eq:php_exp}) is 
\ba 
V_{in} =  (V_{ni}^*) =
\begin{cases}
    -\alpha_{S,m(R_z)}\alpha_{S,m(R_z')}(-1)^{N_{xy}} f^*_{z,\kk^{xy}} (1- \delta_{m(R_z'),-S})(S-m(R_z))\frac{\sqrt{\eta_{m(R_z),m(R_z')}}}{S^2-m(R_z)^2} & R_z'=R_z+1 \\ 
    \alpha_{S,m(R_z)}\alpha_{S,m(R_z')}(-1)^{N_{xy}} f^*_{z,\kk^{xy}} (1- \delta_{m(R_z'),-S})(S-m(R_z))\frac{\sqrt{\eta_{m(R_z),m(R_z')}}}{S^2-m(R_z)^2} & R_z'=R_z-1 
\end{cases}
\ea 
The contribution to the $H_{SW}$ then reads
\ba 
&H_{SW}^I \nonumber\\ 
=& \sum_{m,\kk^{xy},R_z,R_z'=R_z \pm 1} \frac{1}{2}
\bigg|\alpha_{S,m(R_z)}\alpha_{S,m(R_z')}(-1)^{N_{xy}} f_{z,\kk^{xy}} (1- \delta_{m(R_z'),-S})(S-m(R_z))\frac{\sqrt{\eta_{m(R_z),m(R_z')}}}{S^2-m(R_z)^2} \bigg|^2  \nonumber\\ 
&\bigg(\frac{1}{-E_{\kk^{xy}}^{exct}}
+\frac{1}{-E_{\kk^{xy}}^{exct}}\bigg) 
|m\rangle \langle m|  \nonumber\\ 
=&- \sum_{m,\kk^{xy},R_z,R_z'=R_z\pm 1} \frac{|f_{z,\kk^{xy}}|^2}{E_{\kk^{xy}}^{exct}}  
\prod_{i=m(R_z)+1}^{S-1} \frac{S^2-i^2}{S(S+1)-i(i+1)}
\prod_{j=m(R_z')}^{S-1} \frac{S^2-j^2}{S(S+1)-j(j+1)} \nonumber\\ 
&\frac{1}{S(S+1)-m(R_z)(m(R_z)+1)}
(1-\delta_{m(R_z'),-S})(S-m(R_z))^2 |m\rangle \langle m|\nonumber\\ 
=&-\sum_{m,\kk^{xy},R_z,R_z'=R_z\pm 1}
\frac{|f_{z,\kk^{xy}}|^2}{E_{\kk^{xy}}^{exct}}\frac{
(S+m(R_z'))(S-m(R_z))}{4S^2} |m\rangle \langle m|
\label{eq:hsw_i}
\ea 

\subsubsection{Contribution from II}

We take 
\ba 
&|L,j\rangle = |m\rangle, \quad |L,i\rangle = |m\rangle \nonumber\\ 
&|H,n\rangle = \frac{1}{\sqrt{\eta_{m(R_z'),m(R_z)}}}S_{R_z'}^+ |m,\kk^{xy},R_z,R_z'\rangle 
\ea 
The corresponding $V_{in}$ (from Eq.~\ref{eq:app:ph_h1_pl} and Eq.~\ref{eq:php_exp}) is 
\ba 
V_{in} =  (V_{ni}^*) =
\begin{cases}
\alpha_{S,m(R_z)}\alpha_{S,m(R_z')}(-1)^{N_{xy}} f^*_{z,\kk^{xy}} (1- \delta_{m(R_z),-S})(S-m(R_z'))\frac{\sqrt{\eta_{m(R_z'),m(R_z)}}}{S^2-m(R_z')^2} & R_z'=R_z+1 \\ 
    -\alpha_{S,m(R_z)}\alpha_{S,m(R_z')}(-1)^{N_{xy}} f^*_{z,\kk^{xy}} (1- \delta_{m(R_z),-S})(S-m(R_z'))\frac{\sqrt{\eta_{m(R_z'),m(R_z)}}}{S^2-m(R_z')^2}& R_z'=R_z-1 
\end{cases}
\ea 
The contribution to the $H_{SW}$ then reads
\ba 
&H_{SW}^{II} \nonumber\\ 
=& \sum_{m,\kk^{xy},R_z,R_z'=R_z \pm 1} \frac{1}{2}
\bigg|\alpha_{S,m(R_z)}\alpha_{S,m(R_z')}(-1)^{N_{xy}} f_{z,\kk^{xy}} (1- \delta_{m(R_z),-S})(S-m(R_z'))\frac{\sqrt{\eta_{m(R_z'),m(R_z)}}}{S^2-m(R_z')^2} \bigg|^2  \nonumber\\ 
&\bigg(\frac{1}{-E_{\kk^{xy}}^{exct}}
+\frac{1}{-E_{\kk^{xy}}^{exct}}\bigg) 
|m\rangle \langle m|  \nonumber\\ 
=&- \sum_{m,\kk^{xy},R_z,R_z'=R_z\pm 1} \frac{|f_{z,\kk^{xy}}|^2}{E_{\kk^{xy}}^{exct}}  
\prod_{i=m(R_z')+1}^{S-1} \frac{S^2-i^2}{S(S+1)-i(i+1)}
\prod_{j=m(R_z)}^{S-1} \frac{S^2-j^2}{S(S+1)-j(j+1)} \nonumber\\ 
&\frac{1}{S(S+1)-m(R_z')(m(R_z')+1)}
(1-\delta_{m(R_z),-S})(S-m(R_z'))^2 |m\rangle \langle m|\nonumber\\ 
=&-\sum_{m,\kk^{xy},R_z,R_z'=R_z\pm 1}
\frac{|f_{z,\kk^{xy}}|^2}{E_{\kk^{xy}}^{exct}}
\frac{ (S+m(R_z))(S-m(R_z')) }{4S^2} |m\rangle \langle m| 
\label{eq:hsw_ii}
\ea

\subsubsection{Contribution from III}
We take 
\ba 
&|L,j\rangle = |m\rangle,\quad |L,i\rangle = |\tilde{m}\rangle\nonumber\\ 
&|H,n\rangle = \frac{1}{\sqrt{\eta_{m(R_z),m(R_z')}}}S_{R_z}^+ |m,\kk^{xy},R_z,R_z'\rangle ,\quad R_z'= R_z \pm 1 
\ea 
where $|\tilde{m}\rangle$ denotes the ground state obtained by acting  $S_{R_z'}^-S_{R_z}^+ $ on the $|m\rangle $
\ba 
&|\tilde{m} \rangle  =\frac{ S_{R_z'}^-S_{R_z}^+ }{\sqrt{[S(S+1)-m(R_z')(m(R_z')-1)][S(S+1)-m(R_z)(m(R_z)+1)]}} |m\rangle 
\label{eq:m_tilde_def}
\ea 
where the additional coefficient in the definition of $|\tilde{m}\rangle$ has been introduced to ensure $\langle \tilde{m}| \tilde{m}\rangle = 1 $ (see also Eq.~\ref{eq:normalization_condition_for_mz_state}). We also comment that, the contribution from $S_{R_z'}^{+}S_{R_z}^-|m\rangle $ can be obtained by switching the position indices $R_{z}'$ and $R_z$ in Eq.~\ref{eq:m_tilde_def}. 
In practice, use the notation introduced in Eq.~\ref{eq:1flatband:gnd_basis}, we find 
\ba 
&\tilde{m}(R) = 
\begin{cases}
    m(R) &  R\ne R_z, R \ne R_z'\\ 
    m(R)+1 & R=R_z   \\
    m(R)-1 & R= R_z'
\end{cases} 
\label{eq:def_tilde_m}
\ea 
The corresponding $V_{in},V_{jn}^*$ (from Eq.~\ref{eq:app:ph_h1_pl}) are then
\ba 
&V_{in} = (V_{ni}^*)  =
\begin{cases}
    -\alpha_{S,m(R_z)}\alpha_{S,m(R_z')}(-1)^{N_{xy}} f^*_{z,\kk^{xy}} (1- \delta_{m(R_z'),-S})(S-m(R_z))\frac{\sqrt{\eta_{m(R_z),m(R_z')}}}{S^2-m(R_z)^2} & R_z'=R_z+1 \\ 
    \alpha_{S,m(R_z)}\alpha_{S,m(R_z')}(-1)^{N_{xy}} f^*_{z,\kk^{xy}} (1- \delta_{m(R_z'),-S})(S-m(R_z))\frac{\sqrt{\eta_{m(R_z),m(R_z')}}}{S^2-m(R_z)^2} & R_z'=R_z-1 
\end{cases} \nonumber\\ 
&V_{nj} = 
\begin{cases}
\alpha_{S,\tilde{m}(R_z)}\alpha_{S,\tilde{m}(R_z')}(-1)^{N_{xy}} f_{z,\kk^{xy}} (1- \delta_{\tilde{m}(R_z),-S})(S-\tilde{m}(R_z'))\frac{\sqrt{\eta_{\tilde{m}(R_z'),\tilde{m}(R_z)}}}{S^2-\tilde{m}(R_z')^2} & R_z'=R_z+1 \\ 
    -\alpha_{S,\tilde{m}(R_z)}\alpha_{S,\tilde{m}(R_z')}(-1)^{N_{xy}} f_{z,\kk^{xy}} (1- \delta_{\tilde{m}(R_z),-S})(S-\tilde{m}(R_z'))\frac{\sqrt{\eta_{\tilde{m}(R_z'),\tilde{m}(R_z)}}}{S^2-\tilde{m}(R_z')^2}& R_z'=R_z-1 
\end{cases}
\ea 
The contribution to the $H_{SW}$ then reads
\ba 
&H_{SW}^{III} \nonumber\\ 
=&\sum_{m,\kk^{xy},R_z,R_z'=R_z\pm 1} \frac{1}{2}\bigg(\frac{1}{-E_{\kk^{xy}}^{exct}}
+\frac{1}{-E_{\kk^{xy}}^{exct}}\bigg) 
|m\rangle \langle \tilde{m}| \nonumber\\ 
&\alpha_{S,m(R_z)}\alpha_{S,m(R_z')}\alpha_{S,\tilde{m}(R_z)}\alpha_{S,\tilde{m}(R_z')}
(-1)|f_{z,\kk^{xy}}|^2  (1-\delta_{m(R_z'),-S})(1-\delta_{\tilde{m}(R_z),-S})(S-m(R_z))(S-\tilde{m}(R_z')) \nonumber\\ 
&\frac{
\sqrt{ \eta_{m(R_z),m(R_z')} \eta_{\tilde{m}(R_z'),\tilde{m}(R_z)}}
}{(S^2-m(R_z)^2)(S^2-\tilde{m}(R_z')^2)}\nonumber\\
=&\sum_{m,\kk^{xy},R_z,R_z'=R_z\pm 1} \frac{|f_{z,\kk^{xy}}|^2} {E_{\kk^{xy}}^{exct}}|m\rangle\langle \tilde{m}| 
(S-m({R_z}))(S-m({R_z'})-1) \nonumber\\ 
&\prod_{i=m(R_z)+1}^{S-1}\frac{S^2-i^2}{S(S+1)-i(i+1)}\prod_{j=m(R_z')}^{S-1}\frac{S^2-i^2}{S(S+1)-i(i+1)}  \nonumber\\ 
&\frac{1}{\sqrt{S(S+1)-m(R_z)(m(R_z)+1) }\sqrt{S(S+1)-m(R_z')(m(R_z')-1)} } \nonumber\\ 
=&\sum_{m,\kk^{xy},R_z,R_z'=R_z\pm 1} \frac{|f_{z,\kk^{xy}}|^2} {E_{\kk^{xy}}^{exct}}
\frac{\sqrt{S(S+1)-m(R_z)(m(R_z)+1)}\sqrt{S(S+1)+m(R_z')(m(R_z')-1)}}{4S^2}
|m\rangle\langle \tilde{m}| 
\label{eq:hsw_iii}
\ea

\subsubsection{Contribution from IV}
We take 
\ba 
&|L,j\rangle = |m\rangle,\quad |L,i\rangle = |\tilde{\tilde{m}}\rangle\nonumber\\ 
&|H,n\rangle = \frac{1}{\sqrt{\eta_{m(R_z'),m(R_z)}}}S_{R_z'}^+ |m,\kk^{xy},R_z,R_z'\rangle ,\quad R_z'= R_z \pm 1 
\ea 
where $|\tilde{\tilde{m}}\rangle$ denotes the ground state obtained by acting  $S_{R_z}^-S_{R_z'}^+ $ on the $|m\rangle $
\ba 
&|\tilde{\tilde{m}} \rangle  =\frac{ S_{R_z}^-S_{R_z'}^+ }{\sqrt{[S(S+1)-m(R_z)(m(R_z)-1)][S(S+1)-m(R_z')(m(R_z')+1)]}} |m\rangle 
\ea 
where the additional coefficient in the definition of $|\tilde{\tilde{m}}\rangle$ has been introduced to ensure $\langle \tilde{\tilde{m}}|\tilde{\tilde{m}}\rangle = 1 $ (see also Eq.~\ref{eq:normalization_condition_for_mz_state}). In practice, using the notation introduced in Eq.~\ref{eq:1flatband:gnd_basis}, we find 
\ba 
&\tilde{\tilde{m}}(R) = 
\begin{cases}
    m(R) &  R\ne R_z, R \ne R_z'\\ 
    m(R)-1 & R=R_z   \\
    m(R)+1 & R= R_z'
\end{cases} 
\label{eq:def_tildetilde_m}
\ea 
The corresponding $V_{in},V_{jn}^*$ (from Eq.~\ref{eq:app:ph_h1_pl}) are then
\ba 
&V_{in} = (V_{ni}^*)  =
\begin{cases}
    -\alpha_{S,m(R_z')}\alpha_{S,m(R_z)}(-1)^{N_{xy}} f^*_{z,\kk^{xy}} (1- \delta_{m(R_z),-S})(S-m(R_z'))\frac{\sqrt{\eta_{m(R_z'),m(R_z)}}}{S^2-m(R_z')^2} & R_z'=R_z-1 \\ 
    \alpha_{S,m(R_z')}\alpha_{S,m(R_z)}(-1)^{N_{xy}} f^*_{z,\kk^{xy}} (1- \delta_{m(R_z),-S})(S-m(R_z'))\frac{\sqrt{\eta_{m(R_z'),m(R_z)}}}{S^2-m(R_z')^2} & R_z'=R_z+1 
\end{cases} \nonumber\\ 
&V_{nj} = 
\begin{cases}
\alpha_{S,\tilde{\tilde{m}}(R_z')}\alpha_{S,\tilde{\tilde{m}}(R_z)}(-1)^{N_{xy}} f_{z,\kk^{xy}} (1- \delta_{\tilde{\tilde{m}}(R_z'),-S})(S-\tilde{\tilde{m}}(R_z))\frac{\sqrt{\eta_{\tilde{\tilde{m}}(R_z),\tilde{\tilde{m}}(R_z')}}}{S^2-\tilde{\tilde{m}}(R_z)^2} & R_z'=R_z-1 \\ 
    -\alpha_{S,\tilde{\tilde{m}}(R_z')}\alpha_{S,\tilde{\tilde{m}}(R_z)}(-1)^{N_{xy}} f_{z,\kk^{xy}} (1- \delta_{\tilde{\tilde{m}}(R_z'),-S})(S-\tilde{\tilde{m}}(R_z))\frac{\sqrt{\eta_{\tilde{\tilde{m}}(R_z),\tilde{\tilde{m}}(R_z')}}}{S^2-\tilde{\tilde{m}}(R_z)^2}& R_z'=R_z+1 
\end{cases}
\ea 
The contribution to the $H_{SW}$ then reads
\ba 
&H_{SW}^{IV} \nonumber\\ 
=&\sum_{m,\kk^{xy},R_z,R_z'=R_z\pm 1} \frac{1}{2}\bigg(\frac{1}{-E_{\kk^{xy}}^{exct}}
+\frac{1}{-E_{\kk^{xy}}^{exct}}\bigg) 
|m\rangle \langle \tilde{\tilde{m}}| \nonumber\\ 
&\alpha_{S,m(R_z')}\alpha_{S,m(R_z)}\alpha_{S,\tilde{\tilde{m}}(R_z')}\alpha_{S,\tilde{\tilde{m}}(R_z)}
(-1)|f_{z,\kk^{xy}}|^2  (1-\delta_{m(R_z),-S})(1-\delta_{\tilde{\tilde{m}}(R_z'),-S})(S-m(R_z'))(S-\tilde{\tilde{m}}(R_z)) \nonumber\\ 
&\frac{
\sqrt{ \eta_{m(R_z'),m(R_z)} \eta_{\tilde{\tilde{m}}(R_z),\tilde{\tilde{m}}(R_z')}}
}{(S^2-m(R_z')^2)(S^2-\tilde{\tilde{m}}(R_z)^2)}\nonumber\\
=&\sum_{m,\kk^{xy},R_z,R_z'=R_z\pm 1} \frac{|f_{z,\kk^{xy}}|^2} {E_{\kk^{xy}}^{exct}}|m\rangle\langle \tilde{\tilde{m}}| 
(S-m({R_z'}))(S-m({R_z})-1) \nonumber\\ 
&\prod_{i=m(R_z')+1}^{S-1}\frac{S^2-i^2}{S(S+1)-i(i+1)}\prod_{j=m(R_z)}^{S-1}\frac{S^2-i^2}{S(S+1)-i(i+1)} \nonumber\\ 
&\frac{1}{\sqrt{S(S+1)-m(R_z')(m(R_z')+1) }\sqrt{S(S+1)-m(R_z)(m(R_z)-1)} } \nonumber\\ 
=&\sum_{m,\kk^{xy},R_z,R_z'=R_z\pm 1} \frac{|f_{z,\kk^{xy}}|^2} {E_{\kk^{xy}}^{exct}}
\frac{\sqrt{S(S+1)-m(R_z')(m(R_z')+1)}\sqrt{S(S+1)+m(R_z)(m(R_z)-1)}}{4S^2}
|m\rangle\langle \tilde{\tilde{m}}| 
\label{eq:hsw_iv}
\ea

\subsubsection{Effective spin-spin interaction terms}
We combine Eq.~\ref{eq:hsw_i}, Eq.~\ref{eq:hsw_ii},  Eq.~\ref{eq:hsw_iii},  Eq.~\ref{eq:hsw_iv}, and find the additional term generated by Schrieffer-Wolff transformation is
\ba 
H_{SW} = &H_{SW}^{I}+H_{SW}^{II}+H_{SW}^{III}+H_{SW}^{IV}\nonumber\\ 
=&\sum_{m,\kk^{xy},R_z,R_z'=R_z\pm 1}
\frac{|f_{z,\kk^{xy}}|^2}{E_{\kk^{xy}}^{exct}}
\frac{ -S^2+m(R_z)m(R_z') }{2S^2} |m\rangle \langle m| \nonumber\\ 
&\sum_{m,\kk^{xy},R_z,R_z'=R_z\pm 1} \frac{|f_{z,\kk^{xy}}|^2} {E_{\kk^{xy}}^{exct}}
\bigg[ 
\frac{\sqrt{S(S+1)-m(R_z)(m(R_z)+1)}\sqrt{S(S+1)+m(R_z')(m(R_z')-1)}}{4S^2}
|m\rangle\langle \tilde{m}| \nonumber\\ 
&+
\frac{\sqrt{S(S+1)-m(R_z')(m(R_z')+1)}\sqrt{S(S+1)+m(R_z)(m(R_z)-1)}}{4S^2}
|m\rangle\langle \tilde{\tilde{m}}| 
\bigg] 
\ea 
where $|\tilde{m}\rangle ,|\tilde{\tilde{m}}\rangle$ states are defined via Eq.~\ref{eq:def_tilde_m} and Eq.~\ref{eq:def_tildetilde_m} respectively.

In a more compact formula, we find 
\ba 
&H_{SW}  \nonumber\\ 
=&\sum_{m,\kk^{xy},R_z,R_z'=R_z\pm 1}
\frac{|f_{z,\kk^{xy}}|^2}{E_{\kk^{xy}}^{exct}}
\frac{ -S^2+m(R_z)m(R_z') }{2S^2} |m\rangle \langle m| \nonumber\\ 
&+\sum_{m,\kk^{xy},R_z,R_z'=R_z\pm 1} \frac{|f_{z,\kk^{xy}}|^2} {E_{\kk^{xy}}^{exct}}
\bigg[ 
\frac{\sqrt{S(S+1)-m(R_z)(m(R_z)+1)}\sqrt{S(S+1)+m(R_z')(m(R_z')-1)}}{4S^2}
\bigg[ |m\rangle\langle \tilde{m}| 
+|\tilde{m}\rangle\langle {m}| \bigg]
\label{eq:org_sw_spinspin}
\ea 

We can rewrite Eq.~\ref{eq:org_sw_spinspin} in a more compact form via the following spin operators (Eq.~\ref{eq:lared_spin_operator})
\ba 
&S^{x,y,z}_{R_z} =\sum_{\kk^{xy},\sigma,\sigma'} \gamma_{\kk^{xy},R_z,\sigma}^\dag \frac{ \sigma^{x,y,z}_{\sigma\sigma'}}{2}\gamma_{\kk^{xy},R_z,\sigma'} \nonumber\\ 
&S^+_{R_z}= S_{R_z}^x +i S_{R_z}^y \nonumber\\
& S^-_{R_z}= S_{R_z}^x -i S_{R_z}^y 
\ea 

We find
\ba 
S_{R_z}^zS_{R_z'}^z|m\rangle =m(R_z)m(R_z') |m\rangle  
\ea 
Thus, acting on the ground state manifolds, we have
\ba 
\sum_m m(R_z)m(R_z') |m\rangle \langle m | = S_{R_z}^zS_{R_z'}^z
\label{eq:szsz_sw_spinspin}
\ea 

Also, we find
\ba 
&S_{R_z}^+S_{R_z'}^- | m\rangle  \nonumber\\
=&\bigg[
\bigotimes_{R_z''\ne R_z,R_z''\ne R_z'}|R_z'',m(R_z'')\rangle \bigg] S_{R_z}^+ |R_z,m(R_z)\rangle S_{R_z'}^-|R_z',m(R_z')\rangle \nonumber\\ 
=&\bigotimes_{R_z''\ne R_z,R_z''\ne R_z'}|R_z'',m(R_z'')\rangle \bigg] 
\sqrt{S(S+1)-m(R_z)(m(R_z)+1)} |m(R_z)+1\rangle 
\sqrt{S(S+1)-m(R_z')(m(R_z')-1)} |m(R_z')-1\rangle \nonumber\\ 
=&\sqrt{S(S+1)-m(R_z)(m(R_z)+1)}\sqrt{S(S+1)+m(R_z')(m(R_z')-1)}
|\tilde{m}\rangle 
\ea 
Therefore, we have the following identity
\ba 
\sum_m |\tilde{m} \rangle \langle m |  = \frac{1 }{\sqrt{S(S+1)-m(R_z)(m(R_z)+1)}\sqrt{S(S+1)+m(R_z')(m(R_z')-1)}} S_{R_z}^+ S_{R_z'}^-
\label{eq:spsm_sw_spinspin}
\ea 

In addition, 
\ba 
&S_{R_z}^-S_{R_z'}^+ | \tilde{m}\rangle  \nonumber\\
=&\bigg[
\bigotimes_{R_z''\ne R_z,R_z''\ne R_z'}|R_z'',m(R_z'')\rangle \bigg] S_{R_z}^- |R_z,m(R_z)+1\rangle S_{R_z'}^+|R_z',m(R_z')-1\rangle \nonumber\\ 
=&\bigotimes_{R_z''\ne R_z,R_z''\ne R_z'}|R_z'',m(R_z'')\rangle \bigg] 
\sqrt{S(S+1)-m(R_z)(m(R_z)+1)} |m(R_z)\rangle 
\sqrt{S(S+1)-m(R_z')(m(R_z')-1)} |m(R_z')\rangle \nonumber\\ 
=&\sqrt{S(S+1)-m(R_z)(m(R_z)+1)}\sqrt{S(S+1)+m(R_z')(m(R_z')-1)}
|{m}\rangle 
\ea 
Therefore, we have the following identity
\ba 
\sum_m |{m} \rangle \langle \tilde{m} |  = \frac{1 }{\sqrt{S(S+1)-m(R_z)(m(R_z)+1)}\sqrt{S(S+1)+m(R_z')(m(R_z')-1)}} S_{R_z}^- S_{R_z'}^+
\label{eq:smsp_sw_spinspin}
\ea 

Combining Eq.~\ref{eq:org_sw_spinspin}, Eq.~\ref{eq:szsz_sw_spinspin}, Eq.~\ref{eq:spsm_sw_spinspin} and Eq.~\ref{eq:smsp_sw_spinspin}, 
we could rewrite the $H_{SW}$ with the spin operators as
\ba 
H_{SW} =&\sum_{R_z,R_z'=\pm 1}J
\bigg[ -S^2+
S_{R_z}^z S_{R_z'}^z + \frac{1}{2}S_{R_z}^+S_{R_z'}^-+\frac{1}{2}S_{R_z}^-S_{R_z'}^+
\bigg] = \sum_{R_z,R_z'=\pm 1}J
\bigg[ -S^2+\bm{S}_{R_z}\cdot \bm{S}_{R_z'}
\bigg] \nonumber\\ 
J  =& \sum_{\kk}\frac{ |f_{z,\kk^{xy}}|^2 }{2E_{\kk}^{exct} S^2}  > 0 
\ea  
Therefore, we could observe that the $z$-direction hopping generates inter-layer Heisenberg-types of antiferromagnetic interactions ($J>0$). The antiferromagnetic interactions then stabilize the following type-A antiferromagnetic state
\ba 
\bigotimes_{R_z} |R_z, (-1)^{R_z}S\rangle 
\ea 
We also note that, by combining all terms, $H^{I}_{SW},H^{II}_{SW},H^{III}_{SW},H^{IV}_{SW}$, we recover the SU(2) symmetry of the system.

Finally, we comment that, we could also introduce an in-plane dispersion to the flat bands. However, as we demonstrated in Ref.~\cite{flat_band_mag}, the in-plane ferromagnetic coupling generated by quantum geometry is proportional to $U$, and the antiferromagnetic coupling generated by the dispersion is proportional to $D^2/U$ with $D$ the bandwidth of the system. Therefore, the in-plane dispersion will not destroy the in-plane ferromagnetism as long as the dispersion is weak.

\end{document}